\documentclass[12pt,          % font size: 11pt or 12pt
               phd,           % degree:    ms or phd
               doublespacing % spacing: onehalfspacing or doublespacing
               ]{ncsuthesis}

\usepackage{booktabs}  % professionally typeset tables
\usepackage{amsmath}%,amssymb,amsfonts}
\usepackage{textcomp}  % better copyright sign, among other things
\usepackage{lipsum}    % filler text
\usepackage{subfig}    % composite figures
\usepackage{adjustbox}

\usepackage{skt}

\RequirePackage[
			style=alphabetic,%numeric-comp,%authoryear-comp,%
			sorting=nyt,%ynt					
			hyperref=true, %	
			firstinits=true,%
			backend=bibtex,
			natbib=true,
			url=false,
			isbn=false,
			maxnames=2, %for et al to be used
			maxalphanames=1, %to avoid printing a + for every et al in the abbreviation
			doi=false]{biblatex}

    \renewcommand*{\intitlepunct}{}
    \DefineBibliographyStrings{german}{in={}}
    \DefineBibliographyStrings{english}{in={}}
    \DeclareNameAlias{sortname}{last-first}
    \DeclareNameAlias{default}{last-first}
	
\DeclareFieldFormat[article,periodical]{volume}{\mkbibbold{#1}}
\makeatletter

\newrobustcmd*{\parentexttrack}[1]{%
  \begingroup
  \blx@blxinit
  \blx@setsfcodes
  \blx@bibopenparen#1\blx@bibcloseparen
  \endgroup}

\AtEveryCite{%
  }

\makeatother
\renewcommand{\cite}[1]{\parencite{#1}}

\renewbibmacro{in:}{%
  \ifentrytype{article}{}{%
  \printtext{\bibstring{in}\intitlepunct}}}
  
\AtEveryBibitem{\clearfield{month}}

\AtEveryBibitem{\clearfield{language}}
 \defbibheading{myheading}[BIBLIOGRAPHY]{
 \chapter*{#1}
 \markboth{#1}{#1}}

\usepackage{dcolumn}% Align table columns on decimal point
\usepackage{bm}% bold math
\usepackage{cancel}
\usepackage{verbatim}% multiline commenting
\usepackage{ifthen}
\usepackage{url}
\usepackage{sectsty}
\usepackage{balance} 
\usepackage{graphicx} %eps figures can be used instead
\usepackage{lastpage}
\usepackage[format=plain,justification=RaggedRight,singlelinecheck=false,font=small,labelfont=bf,labelsep=space]{caption} 
\usepackage{fancyhdr}
\usepackage{abbrevs}
\usepackage{etoolbox}
\robustify{\DateMark} % after having loaded abbrevs

\usepackage{units} %Needed to solve bug from citation Hydrodynamics in 21/2 dimensions
\usepackage[sharp]{easylist} %used for brainstorming purposes 
\usepackage[table]{xcolor}
\usepackage{array,booktabs}
\usepackage{colortbl}
\newcolumntype{L}{@{}>{\kern\tabcolsep}l<{\kern\tabcolsep}}

\usepackage{comment}

\usepackage{listings}
\usepackage{color} %red, green, blue, yellow, cyan, magenta, black, white
\definecolor{mygreen}{RGB}{28,172,0} % color values Red, Green, Blue
\definecolor{mylilas}{RGB}{170,55,241}

\usepackage{titlesec}

\titleformat{\paragraph}
{\normalfont\normalsize\bfseries}{\theparagraph}{1em}{}
\titlespacing*{\paragraph}
{0pt}{3.25ex plus 1ex minus .2ex}{1.5ex plus .2ex}

\usepackage{placeins}
\dispositionformat{\bfseries}            % bold and serif

\headingformat{\large\MakeUppercase}   % All letters uppercase
\usepackage[Rejne]{fncychap}

\usepackage[
  pdfauthor={Carlos Pompeyo Ortiz},
  pdftitle={Rigidity of Microsphere Heaps},
  pdfcreator={pdftex},
  pdfsubject={NC State ETD Thesis},
  pdfkeywords={microfluidics, hard sphere, jamming, suspension, rigidity, friction, microscopy},
  colorlinks=true,
  linkcolor=black,
  citecolor=black,
  filecolor=black,
  urlcolor=black,
]{hyperref}

\usepackage{microtype}

\usepackage[T1]{fontenc}
\usepackage[adobe-utopia]{mathdesign}

\committeesize{4}

\chair{Brian A. Floyd}
\memberI{Paul Franzon}
\memberII{John Muth}
\memberIII{Chih-Hao Chang}   % unnecessary if committeesize=3
\student{Vikas}{Chauhan} % a full middle name
\program{Electrical Engineering}

\thesistitle{PhD Thesis on Code Modulated Interferometric Imaging System %\break 
using Phased Arrays}

\newcommand{\fref}[1]{Fig.~\ref{#1}}

\usepackage{color}
\graphicspath{{./Chapter-1/figs/}{./Chapter-2/figs/}{./Chapter-3/figs/}}%{./Chapter-4/figs/}{./Chapter-5/figs/}{./Chapter-6/figs/}}

\usepackage{calc}
\newlength{\chaptercapitalheight}
\newlength{\chapterfootskip}
\renewcommand{\bibname}{BIBLIOGRAPHY}

\newlength\graphht

\usepackage{pdflscape}
\usepackage{tikz}
\fancypagestyle{lscapedplain}{%
  \fancyhf{}
  \fancyfoot{%
    \tikz[remember picture,overlay]
      \node[outer sep=1cm,above,rotate=90] at (current page.east) {\thepage};}

}

\begin{document}
\pagestyle{plain}
%%---------------------------------------------------------------------------%%
\frontmatter

%% ------------------------------ Abstract ---------------------------------- %%
\begin{abstract}

%Millimeter-wave (mm-wave) imaging provides compelling capabilities for security screening, navigation, and bio-medical applications. Traditional scanned or focal-plane mm-wave imagers are bulky and costly. In contrast, phased-array hardware developed for mass-market wireless communications and automotive radar promise to be extremely low cost. 
This work presents techniques which can allow low-cost phased-array receivers to be reconfigured or re-purposed as interferometric imagers, removing the need for custom hardware and thereby reducing cost. Since traditional phased arrays power combine incoming signals prior to digitization, orthogonal code-modulation is applied to each incoming signal using phase shifters within each front-end and two-bit codes. These code-modulated signals can then be combined and processed coherently through a shared hardware path. Once digitized, visibility functions can be recovered through squaring and code-demultiplexing operations. Provided that codes are selected such that the product of two orthogonal codes is a third unique and orthogonal code, it is possible to demultiplex complex visibility functions directly. As such, the proposed system modulates incoming signals but demodulates desired correlations.

Firstly, we present the operation of the system, a validation of its operation using behavioral models of a traditional phased array and a benchmarking of the code-modulated interferometer against traditional interferometer using simulation results and sensitivity analysis. Secondly, for the proof of concept with a prototype, we present a simple CMI system operating in the license-free 60-GHz band using a four-element phased-array receiver developed for IEEE 802.11ad (WiGig) and packaged with compact antenna structures. The four-element 60-GHz phased array chip is wire-bonded onto a Rogers-5880 substrate board with on-board slot antennas, and a single 60-GHz output is measured using a power detector. This scalar measurement is then demodulated to obtain the interferometric visibilities. The four-element phased array is thinned to obtain a 13-pixel image and the system is demonstrated through a point-source detected at different locations.

Finally, the operation and capabilities of code-modulated interferometry (CMI) are demonstrated at 10-GHz using commercially-available phased arrays. A 33-pixel, eight-element prototype is created using two commercially-available ADAR1000 phased-array receivers from Analog Devices Inc. The chips are connected at board level to a patch antenna array. The serial interface is used to apply codes whereas the on-chip power detector and data converter are used for direct read out of the composite code-multiplexed imaging data. These are then processed off-line in MATLAB to reconstruct the image. The 33-pixel camera is demonstrated in hardware for point-source detection. Further to demonstrate the scalability of the concept, a 16-element, 169-pixels CMI imaging system is presented at 10-GHz using the four of the same commercially-available phased arrays from ADI. Two active point sources are imaged simultaneously to present the resolution of the system.

\end{abstract}

%% ---------------------------- Copyright page ------------------------------ %%
%% Comment the next line if you don't want the copyright page included.
\makecopyrightpage

%% -------------------------------- Title page ------------------------------ %%
\maketitlepage

%% -------------------------------- Dedication ------------------------------ %%
\begin{dedication}
 \centering To the Queen of Vrndavana:

\begin{figure}[h]
\begin{center}
   \includegraphics[width=.5\textwidth]{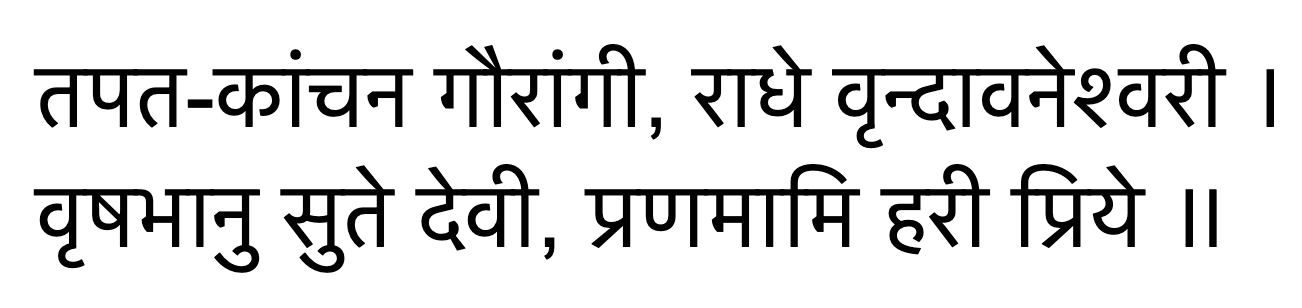}    
\end{center}
\vspace{-1.5em}
\end{figure} 

\textit{ 
tapta-kancana-gaurangi radhe vrndavanesvari \\
vrsabhanu-sute devi pranamami hari-priye  }

\

"I offer my respects to Radharani whose bodily complexion is like molten gold and who is the Queen of Vrndavana. You are the daughter of King Vrsabhanu, and You are very dear to Lord Krsna."
\end{dedication}

%% -------------------------------- Biography ------------------------------- %%
\begin{biography}
Vikas Chauhan was born in Sisana (District Baghpat), U.P., India. He grew up in Gurgaon (now Gurugram), Haryana, India.

Vikas received the B.E.(Hons.) degree in electrical and electronics engineering from Birla Institute of Technology and Science (BITS - Pilani), India, in Spring 2012. In Fall 2012, he enrolled in the Master of Science (M.S.) program in electrical engineering at North Carolina State University, Raleigh, NC, USA. In 2013, he joined Professor Brian Floyd's iNtegrated Circuits and Systems lab at NC State (iNCS$^2$) and transferred into the Ph.D. program in 2014. He received his en-route M.S. degree in 2014.  

Vikas completed two internships during his time at NC State University. He was an intern at Qualcomm Corp. R\&D, Qualcomm Technologies, Inc., San Diego, CA, USA in summer 2017 where he worked on milli-meter wave integrated circuit (mm-wave IC) design for 5G-communication. In spring 2019, he interned at imec USA Nanoelectronics Design Center Inc., FL, USA  where he worked on mm-wave IC design for imaging applications. He will be joining full-time as a researcher at imec USA in fall 2019. His research interests include mm-wave IC design and imaging system design. 
\end{biography}

%% ----------------------------- Acknowledgements --------------------------- %%
\begin{acknowledgements}
First, I would like to thank my advisor Dr. Brian Floyd for this untiring help, mentorship and support during my time at NC State University. I would also like to thank Dr. Paul Franzon, Dr. John Muth and Dr. Chih-Hao Chang for serving on my committee, and their valuable time and comments. 

I would like to thank everyone involved in  the CoMET and IRIS projects: Zhangjie Hong, Simon Sch\"onherr, Dr. Kevin Greene, Dr. Dong Gun Kam, Haekyo Seo and Dr. Zhengxin Tong who contributed in this research and working with whom has been a pleasure. I am grateful to Dr. Yi-Shin Yeh, Dr. Charley Wilson, Dr. Weihu Wang, Dr. Anirban Sarkar, Jeff Bonner-Stewart, Tiantong Ren, Yuan Chang, and all my other colleagues at iNCS$^2$ for technical discussions and their readiness to always help and support me. 

My friends and fellow Ph.D. students, Sandeep Hari, Dr. Deeksha Lal, Kirti Bhanushali, Munirah Boufarsan and Dr. Viswanath Ramesh for making my time inside and outside MRC enjoyable. My dearest friends Shalki Shrivastava, Nirmaan Aggarwal, Varun Joshi, Marian Mersmann, Crystal Hilton, and my other friends who made my life in NC memorable.  

I would like to thank my family for their love, support and patience. 

Lastly, I would like to thank everyone who has helped in one way or other in this research and my journey as a graduate student. 

The material in this thesis is based upon work partially supported by the DARPA Young Faculty Award (N66001-11-1-4144), the 2015 Chancellor's Innovation Fund at NCSU, the Army Small Business Innovation Research (SBIR) Program Office under contract no. W31P4Q-17-C-0020, the AFRL and DARPA under agreement FA8650-16-1-7629, GlobalFoundries (chip fabrication), and Analog Devices Inc. 

\end{acknowledgements}

\thesistableofcontents

\thesislistoftables

\thesislistoffigures

%%---------------------------------------------------------------------------%%
\mainmatter

\pagestyle{plain}
%\newgeometry{margin=1in,lmargin=1.25in,footskip=\chapterfootskip, includehead, includefoot}
\chapter{Introduction}
\label{chap-one}
Millimeter-wave (mm-wave) energy can penetrate clothing, bandages, packages, fog, and dust/snow storms due to their small wavelength. As such, mm-wave cameras operating in the range of 30-300 GHz can be used for numerous applications, including the following: 
\begin{itemize}
\item \emph{Concealed Object Detection:} Millimeter-waves go through clothing and are used in security portals at airports,courthouses, etc., to detect objects hidden on the body (weapons, drugs, smuggled goods) \cite{Hugu97}. The key need today is a lower cost solution which still provides necessary performance (\textit{i.e.}, resolution, sensitivity). Lower-cost cameras (<\$1k) could enlarge the market to include schools, shopping centers, train stations, prisons, etc.
\item \emph{Aircraft Navigation:} Millimeter-waves penetrate clouds and dust storms and can be used to improve visibility in poor conditions (e.g., helicopter detecting power lines through a dust storm). Key requirements for a camera mounted on an aircraft are to reduce size and weight (no lens). Interferometers are ideal for this application \cite{Pergande07}.
\item \emph{Imaging from UAVs:} UAVs or drones represents an emerging market, where mm-wave imaging could provide unique sensing capabilities, particularly if the imager can double as a radar or radio system.
\item \emph{Biomedical:} Millimeter-waves may also be used for biomedical imaging. For example, it may be possible to image through bandages to monitor the healing of dry wounds such as burns without removing the bandage. Millimeter-wave cameras could also measure skin temperature through clothing and allow for diagnosis of conditions related to poor circulation such as compartment syndrome or over- or under-resuscitation. High sensitivity and small cameras are required. These could be deployed to hospitals and doctors' offices.
\item \emph{Quality-Control:} Millimeter-waves can penetrate certain types of packaging material and a camera could be used to provide quality control within a manufacturing line to assess the condition or presence of a product.
\item \emph{Wireless Clothes Fitting:} Millimeter-wave cameras provide an ability to see through clothes and remotely measure a person's size for use in clothing stores. Key needs would include privacy and low system cost.
\end{itemize}
%people screening to detect concealed objects, through-bandage imaging of burns and wounds, through-clothing measurement of skin temperature in emergency medicine, aircraft navigation and surveillance systems, quality-control monitoring in manufacturing settings, and wireless clothes fitting. 

The imaging systems that work with ambient mm-wave energy are called passive imagers \cite{Intech}; other systems require external mm-wave illumination and are called active imagers \cite{Kuki06}. Every object above the absolute zero temperature emits radiation, called as black body radiation or thermal radiation that carry intrinsic information about the body and can be used for imaging. Since the passive imaging is non-intrusive and non-invasive, it is considered a safe way of inspection. Such an inspection is preferred at mm-wave  frequencies (94 GHz \cite{JasonPhD} or higher) for high resolution. Millimeter-wave imagers are widely used at airports for security screening \cite{TSA1,TSA2}. Due to all the above listed applications and advantages, there is an increasing interest in mm-wave imaging systems.  

A variety of mm-wave imaging systems have been developed to date \cite{Yujiri06,Yeom11,Lettington03,Dow96,Sheen2001,Brendan2013, Ahmed2013, Kolinko05, Millivision, Sinclair01, Sato09, Intech}. Existing passive mm-wave imagers are able to achieve a sensitivity as low as 0.28 K \cite{Sinclair01}, as well as a frame rate as high as 10-17 frames per second \cite{Yujiri06,Vizard06,Mizuno05}. Many of these use mechanical scanning \cite{Kolinko05, Millivision, Sinclair01, Sato09}, mirrors for scanning \cite{Vizard06, Yujiri06}, electronic beam forming and scanning or focal-plane arrays (FPA). These cameras can be bulky due to the use of an external lens \cite{Mizuno05, Sato09} and expensive due to the reliance on custom compound-semiconductor hardware for scanned solutions \cite{Rebeiz13, May2010} or the need for a large number of detecting elements in FPA solutions. Also, any kind of mechanical or electronic scanning may slow down the system making it unsuitable for video frame rates.  

An alternative approach which has the potential to reduce both camera volume and camera cost is to shift to interferometry.  Interferometry or synthetic aperture radiometer (SAR) \cite{Thompson08, Harvey00,Narayan96} eliminates the need for a lens since angle-of-arrival information is obtained through complex correlations or visibilities.  Furthermore, the number of required elements is reduced, since N elements can be used to obtain $^NC_2$ correlations.  A drawback of interferometry is the more complicated image processing arising from the need to determine all pairwise complex correlations and compute the Fourier transform.  Because of this, the ideal technology for realizing an interferometer would be one having both high-performance mm-wave transistors and dense low-power digital computation \cite{Pergande07}. These requirements can now be met in today's advanced silicon technologies, \textit{e.g.}, 130-nm SiGe BiCMOS, 65-nm CMOS and beyond, which feature transistors having $f_{T}$ and $f_{MAX}$ in excess of 200 GHz.

While it is possible to develop a highly-integrated mm-wave interferometer in silicon, such a customized circuit would require non-negligible development costs.  Also, if the total mm-wave camera market volume remains low, the unit cost would remain high. The low cost of imaging system is crucial for mass deployment in certain applications. In contrast to this, high-volume mm-wave applications are emerging \cite{Roe14,Pierre14,Shaheen19, Collaert_2020a,Collaert_2020b}, including 60-GHz phased arrays for short-range multi-Gb/s wireless links \cite{60STD,Floyd11,Harish10,Chuang14} and 77-GHz phased arrays for automotive radar systems \cite{radar13,Jlee10,Mitomo10,Hasch12}. Moreover, due to the increasing interest in 5G applications, more of such hardware will be available commercially for a fraction of cost today. These commercial-of-the-shelf (COTS) phased arrays have the potential to be used in other multi-antenna applications, allowing for substantial cost reduction.  In this work, we explore the possibility of turning a COTS phased array receiver into an interferometric imaging receiver using code-modulation.  

%In our approach of code-modulated interferometery (CMI), existing analog phased arrays can be used for the interferometer by orthogonal encoding of each incoming signal using the phase shifter. This code-multiplexing allows all incoming signals to be merged and processed through a shared hardware path. Downstream within the receiver, the complex visbility functions can then be demultiplexed using an appropriate code in the digital domain. 

Code-modulated interferometry (CMI) \cite{Spie16, IRISTMTT2021} is an alternative that employs thinned apertures to reduce the hardware requirement and repurposes communication phased arrays to reduce development costs. In CMI, the complex visibility samples are obtained by orthogonal code-modulation of individual signals using phase shifters, processing of combined signals using the shared receiver path, creation and detection of complex code-multiplexed visibilities using a scalar power detector, and  demodulating all individual visibilities using appropriate code products.

%is used to obtain an image with fewer detectors and without a lens. The image is obtained through a Fourier transform of measured complex cross-correlations between pairs of antennas (\textit{i.e.}, visibility functions). These visibilities contain angle-of-arrival information; hence, a lens is not required and the imager can be ``flat".  Additionally, the use of sparsely-filled antenna arrays in interferometry reduces the total number of antennas and receiver channels required for same number of pixels. For example, $^NC_2$ correlations can be obtained using only $N$ elements which correspond to up to 2x$^NC_2$ pixels with complex non-redundant visibilities (\textit{e.g.}, 2773 pixels can be obtained using 64 elements in a ``Y" antenna configuration [reference]). %This further reduces the cost and form factor of the imager significantly. 
%Further, 

The use of code-modulation within receivers has been studied in prior work in the context of either imaging or communications.  First, for imaging, code modulation has been used within correlating interferometers to eliminate unwanted performance artifacts such as LO feedthrough or spurious signals \cite{Thompson08}.  Second, for communications, code modulations have been applied within a receiver to allow multiple receiver element to share a common hardware path \cite{Heydari09}; however, the concepts in \cite{Heydari09} were not implemented within an N-element phased array.  Our demodulation approach which recovers correlations rather than signals further distinguishes our work over this prior art. 

Note that a similar coded imaging approach, referred to as compressive synthetic aperture interferometric radiometer, is presented in \cite{French18}. $M$ antenna channels are coded into $N$ measured signals ($N<<M$ receivers) using a passive microwave device such as a resonant cavity. In contrast to \cite{French18}, the CMI approach does not require bulky microwave cavities and enables application of software-programmable codes. Also, CMI enables the codes to be modified in software, providing flexibility to change the codes based on the application. Finally, CMI can employ a simple scalar (power) detector as a direct measurement, eliminating the need for complex correlators and hence compact, has a single receiver channel ($N=1$), requires only one scalar measurement from power detector, and has active coding with ability to upgrade codes as they evolve. The possible disadvantage in CMI compared to \cite{French18} is its lower sensitivity since all signals are combined into one shared signal for processing.

\section{Research Objectives}
The objectives of this work are to investigate code-modulated interferometry as a method to re-purpose phased arrays into imaging system and build the hardware to demonstrate the theoretical and behavioral model results, formulate sensitivity and evaluate image resolution. 

This document is organized into five chapters. Chapter \ref{chap-one} presents the introduction to the topic and motivation behind this work. Chapter \ref{chap-two} presents the theoretical investigation and modeling of code-modulated interferometry. Chapter \ref{chap-four} presents a millimeter-wave implementation using in-house 60 GHz phased array. The  millimeter-wave imaging prototype demonstrates the code-modulated interferometry for a simple four-element system. Chapter \ref{chap-three} presents an eight-element 10 GHz prototype of code-modulated interferometric imaging using commercially available phased arrays. Further, Chapter \ref{chap-three} discusses the scalability and expands the 10 GHz eight-element imager in to a true two-dimensional 16-elements imager and demonstrates the resolution of two point sources. %Chapter-\ref{chap-six} presents a compact ultra-wide-band low-noise amplifier which provides low noise and high gain across 24 to 44 GHz and compares it to a narrow-band 28 GHz LNA. Due to the dependency of system sensitivity and frame-rate on noise-figure and band-width of radio frequency front end, such wide-band receivers for 5G applications are best suited for imaging. 
Finally, Chapter \ref{chap-seven} presents the conclusions and summary of the completed work, and the scope for future work.

%%%%%%%%%%%%%% MWCL %%%%%%%%%%%%%%%

%Millimeter-wave signals can penetrate clothing, packaging, and atmospheric conditions such as fog, snow, and dust storms; thus, millimeter-wave (mmWave) imagers find applications in security screening, navigation, and obstacle detection in self-driving automobiles. Even though several mmWave imaging solutions exist, they rely on either a moving part, lenses, or a large number of receiver elements to obtain useful resolution \cite{Meng2018,Yeom11,Sheen2001,Ahmed2013}. Low-cost alternatives are still desirable for mass deployment.
 
%In prior work, the operation and capabilities of CMI have been demonstrated at 10-GHz \cite{IRIS2019} using commercially-available phased arrays; however, operation in the 10-GHz amateur-radio band is less desirable due to the larger wavelength. In this work, we present a simple 13-pixel CMI system operating in the license-free 60-GHz band using a four-element phased-array receiver developed for IEEE 802.11ad (WiGig) and packaged with compact antenna structures. %is different from the published work in two aspects. One, the implementation is at much higher 60 GHz frequency, and two, the in-house phased array chip has been packaged on board and described in this paper. Although the theoretical concept is independent of frequency of use, the implementation and packaging at milli-meter wave frequencies is challenging.  

\chapter{Code-modulated Interferometry}
\label{chap-two}

\section{Interferometry Fundamentals} \label{t1}

As is well known, interferometry is a technique used by radio astronomers to realize higher resolution telescopes using a sparse array of coherent detectors to sample an aperture \cite{Thompson08}.  Measurements are made by cross correlating the signals from spatially separated pairs of antennas (a baseline) with overlapping field-of-view (FOV) known as visibility samples. Measurements for different baselines, collectively known as the visibility function, $V(u,v)$, are related to the brightness distribution, $T_{\Omega}$ through a Fourier transform, as follows \cite{LevineBook,Levine88}

\begin{equation}
\label{eqn:vis}
V(u,v) = \int_{0}^{2\pi} \int_{0}^{\pi} T_{\Omega}(\theta,\phi) \ e^{j 2 \pi (u \cdot l +v \cdot m ) }\ sin\theta \  d\theta \ d\phi  \ ,
%V(u,v) = \int_{0}^{2\pi} \int_{0}^{\pi} T_{\Omega}(\theta,\phi). \ e^{j 2 \pi (u sin\theta cos\phi +v sin\theta sin\phi) }\ sin\theta \  d\theta \ d\phi  \ .
\end{equation}
where $l=sin\theta cos\phi$ and $m= sin\theta sin\phi$, and $u$ and $v$ baselines relate to two-dimensional antenna spacings per unit wavelength. $T_{\Omega}$ is obtained using a discrete inverse Fourier transform of the measured visibility samples. The unit samples in the brightness domain are inversely related to the unit visibility samples, as shown below, where $N$ and $M$ are total number of complex visibility samples in $u$ and $v$ domains respectively \cite{Thompson08}, 
    \begin{equation}
        N\Delta l=(\Delta u)^{-1} \;\;\ \textrm{and}  \;\;\ M\Delta m=(\Delta v)^{-1}  .
        \label{eqn:sampling} 
    \end{equation}
These determine the field-of-view (FOV) and resolution. For a linear array, resolution is $\Delta sin(\theta _{res})=(N\Delta u)^{-1} \approx \lambda /2D_{max}$ and FOV is $N\Delta sin(\theta)=(\Delta u)^{-1} \approx \lambda /D_{min}$ , where $D_{max}$ and $D_{min}$ are maximum and minimum antenna spacing respectively  \cite{LimPhD}. The maximum antenna spacing (and hence minimum angular resolution) is limited by the coherence requirement \cite{Lettington03, LevineBook} as follows
\begin{equation}
D_{max} sin\ \theta_{max} < \frac{c}{B} \ ,
\end{equation}
where $c$ is the velocity of light and $B$ is the bandwidth. More details about interferometry, Nyquist criterion for \textit{u-v} sampling, correlators and other requirements can be found in \cite{Thompson08, LimPhD, Levine88}.

    \begin{figure}[]
    \centering
    \includegraphics[width=0.9\textwidth]{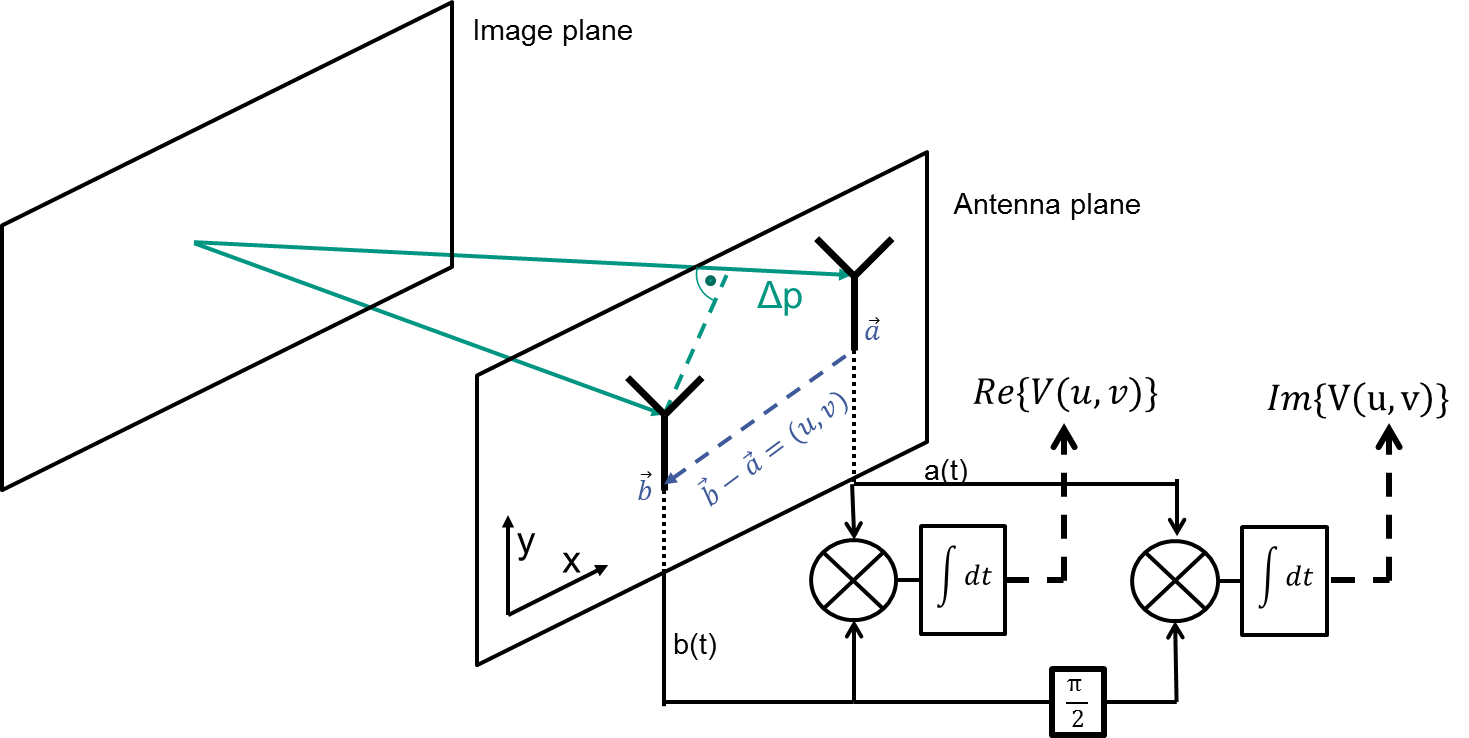}
    \caption{Schematic of the measurement of a visibility sample \cite{Simon18}.}
    \label{fig:correlator}
    \end{figure}

Interferometers must therefore evaluate (complex) correlations for each baseline and then compute the inverse Fourier transform to obtain the image.  Interferometry does not require a focusing lens, allowing for flat or conformal imagers.  Note that image resolution is related to the number of independent baselines, and measurements from redundant baselines can be averaged to reduce noise. 
 
Since an interferometer must measure each antenna signal and since the correlations are typically evaluated in the digital domain, an ideal interferometer array would include N receive chains in parallel, as shown in Fig.~\ref{fig:comparison}(a).  Each chain would generally include a front-end amplifier, a quadrature downconversion mixer, and baseband analog-to-digital converters (ADCs). A bank of digital complex correlators would then be used to evaluate all possible pairwise correlations.

    \begin{figure} [ht]
   \begin{center}
   \begin{tabular}{c c } 
   \includegraphics[height=5.5cm]{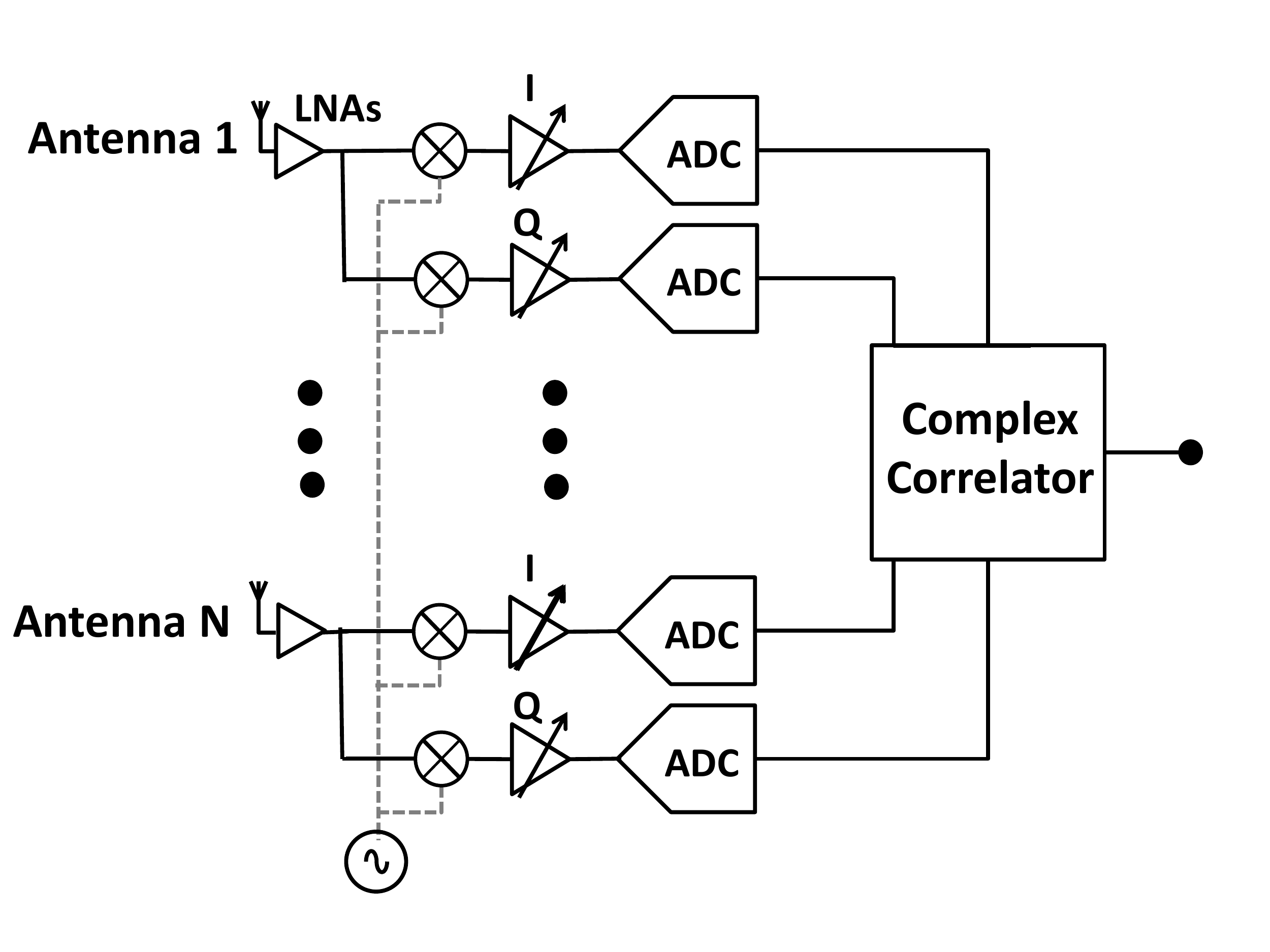} & \includegraphics[height=5.5cm]{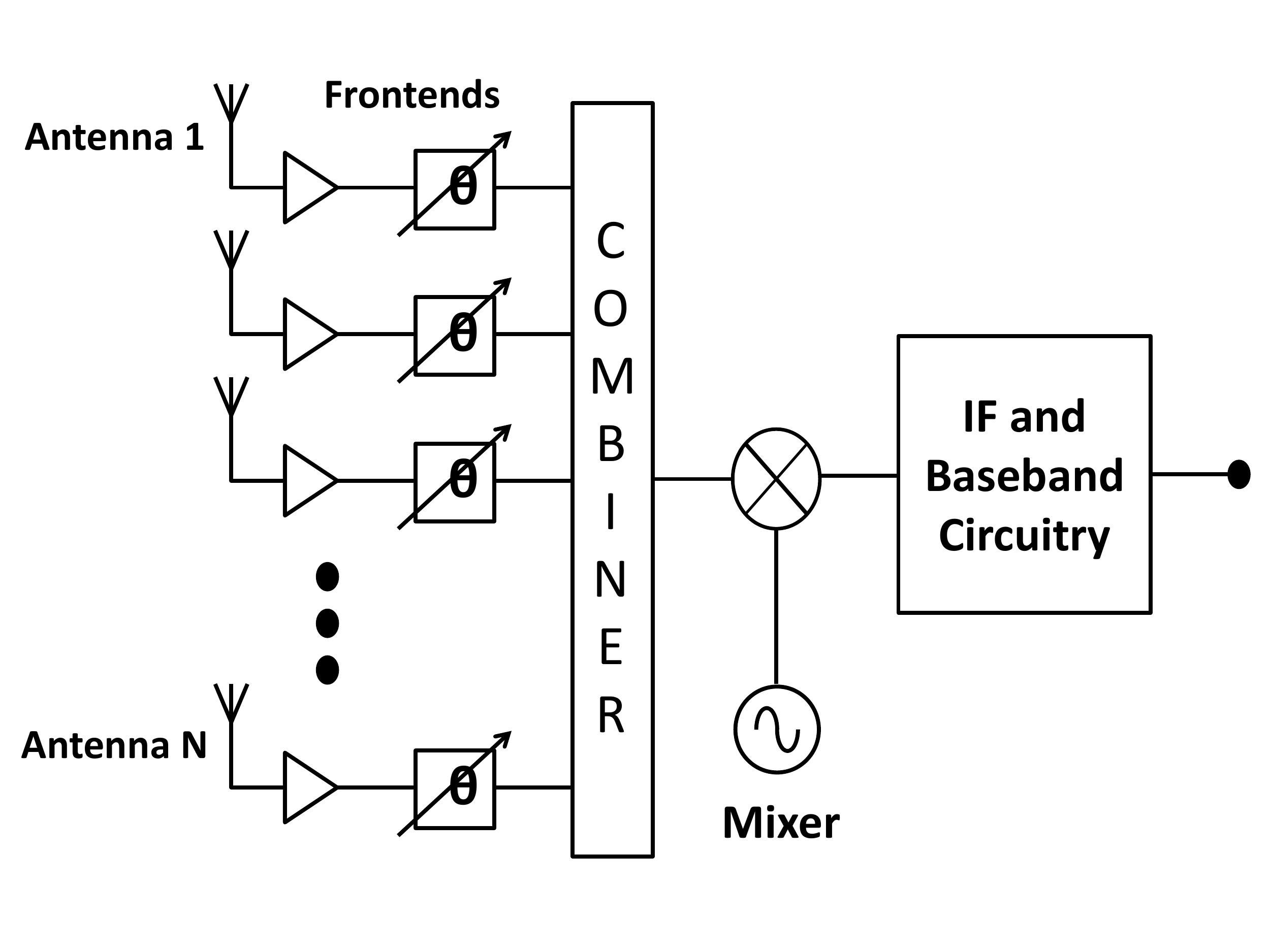}   \\ 
   \rule[-1ex]{0pt}{3.5ex}  (a) & (b)  \\	
   \end{tabular}
   \end{center}
   \label{fig:comparison}
   \caption{(a) Block diagram of a conventional interferometric array and (b) Block diagram of a conventional phased array employing RF phase shifting and combining.}
   \end{figure}
 
In contrast to this digital-combined array, Fig.~\ref{fig:comparison}(b) shows a simplified block diagram of a conventional phased array.  Each front-end element contains an RF amplifier and phase-shifter.  Individual signals are then added together, downconverted, and processed through to the digital domain. In comparing the phased array to the ideal interferometer array, we observe that the interferometer has individual hardware paths per element whereas the phased array has a shared hardware path beginning at the signal combiner. As a result, any attempt to use a phased array as an interferometer array must allow for the shared hardware path, while retaining necessary individual signal information and/or signal correlations. The proposed solution to overcome this limitation is by using code-modulation technique.

\section{Code-Modulated Interferometry with Phased Arrays}
\subsection{Modulation of Antenna Responses}
A phased array can be reconfigured into an interferometer array by applying orthogonal coding functions to each incoming signal using the phase shifters present within each front end. Code modulation allows multiple, individual radiometer data streams to be multiplexed onto a single hardware path.  As will be shown, we can then recover correlations between signals downstream through correlating the aggregate response with code products.

The speed of the codes relates to the rate of the signal or scene change.  In conventional code-modulated communications (\textit{e.g.}, CDMA), the data signals are modulated with a code running at a significantly higher frequency than the modulation or symbol rate. For imaging applications, the signals are being modulated proportional to the scene change; hence, the code rate can be kept low.

Mathematically, the phase modulation within each front-end can be represented as
\begin{equation}
\label{eq:modulated}
s_{n}(t)=A_{n}cos(\omega _{o}t+\theta _{n}-\phi _{n})\ ,
\end{equation}
where $s_{n}(t)$ represents the signal at the output of the $n^{th}$ phase shifter, $A_{n}$ and $\theta_{n}$ represent the amplitude and phase of the signal, and $\phi_{n}$ represents the phase modulation imparted. We can subdivide this expression into in-phase ($s_{i,n}$) and quadrature-phase ($s_{q,n}$) signal components which are then multiplied by in-phase ($i_{n}$) and quadrature-phase ($q_{n}$) codes, as follows:
\begin{equation}
\label{eq:modulated2}
s_{n}(t)=\sqrt{2}cos(\phi _{n})[A_{n} \tfrac{\sqrt{2}}{2} cos(\omega _{o}t+\theta _{n})]+\sqrt{2}sin(\phi _{n})[A_{n} \tfrac{\sqrt{2}}{2} sin(\omega _{o}t+\theta _{n})]=i_{n}s_{i,n}+q_{n}s_{q,n} \ .
\end{equation}
Provided that the phase shifter is at least two bits (true for nearly all phased arrays), then the phase shifter can be used to realize phase shifts of $\pm 45^{\circ}$ and $\pm 135^{\circ}$.  This turns $i_{n}$ and $q_{n}$ into simple $\pm 1$ codes (\textit{i.e.}, $\sqrt{2}cos(\pm 45) = 1$ and $\sqrt{2}cos(\pm 135) = -1$ ).  Note that the application of these ``two-level" codes allows us to separately modulate both in-phase and quadrature-phase portions of the signal, which will be used later to obtain both real and imaginary visibility functions.

All $N$ signals are then power combined in the phased array, turning the phased-array signal-combining operation into a code-multiplexing operation. The aggregate signal, $s_{sum}$, is represented as
\begin{equation}
\label{eq:summ}
s_{sum}=k_{c}\sum_{n} (i_{n}s_{i,n}+q_{n}s_{q,n}) \ .
\end{equation}
Note that this is a time-varying signal, where the time notation has been dropped (\textit{i.e.}, $s_{sum}(t)\rightarrow s_{sum}$.  In our summation, we assume that each path has identical gain and amplitude response, represented with $k_{c}$, generally obtained after calibration of the array.  Additionally, ``normalized summations" are used, where $k_{c}$ is allowed to be one, where $k_{c}$ becomes a scalar applied to all visibilities.

\subsection{Demodulation of Complex Visibility Functions} \label{subsec:comp_vis}
\begin{figure}[]
   \begin{center}
   \includegraphics[width=.8\textwidth]{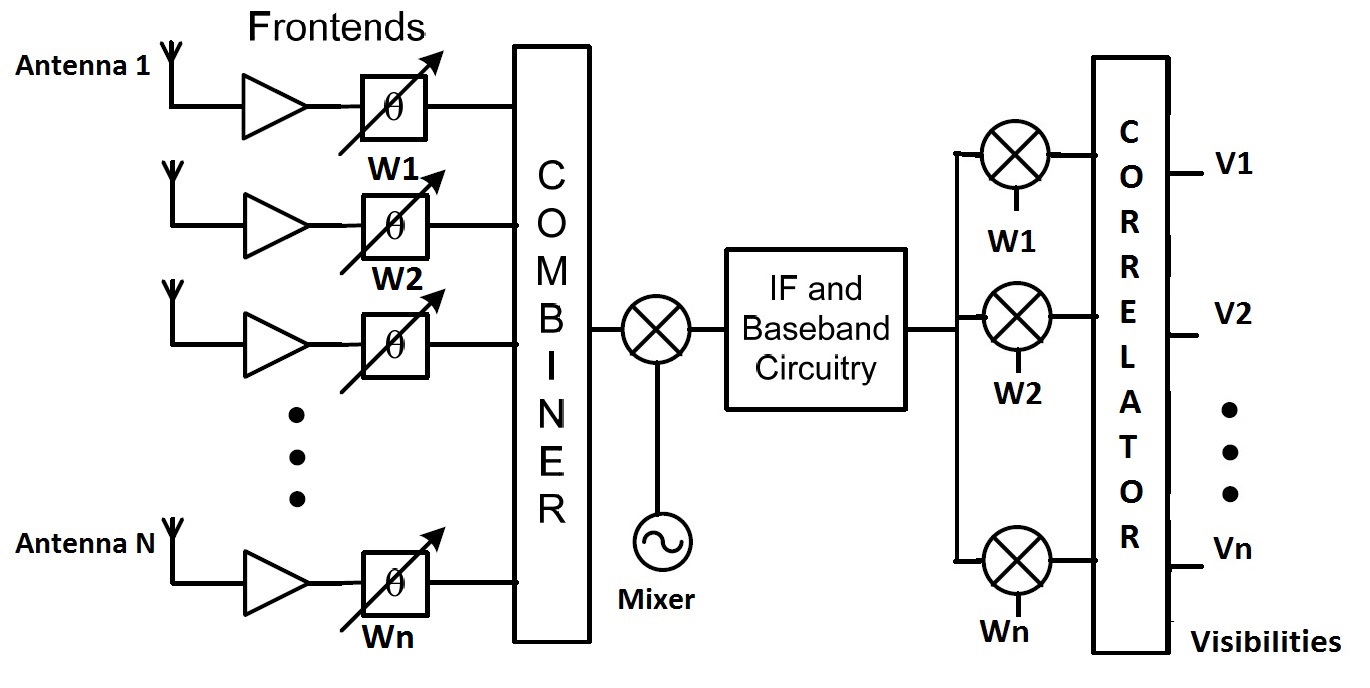}    
   \end{center}
   \caption{Block diagram of a code-modulated interferometric imaging system employing code-modulation within each phase shifter of a phased array and baseband visibility demodulation using a bank of correlators.}
   \label{fig:cmist1} 
   \end{figure}

Now that all individual antenna signals have been code multiplexed, they can then be coherently processed through a shared hardware path, including digitization in the ADC.  In a traditional code-multiplexing system, to obtain the original signal, we would multiply the code-multiplexed signal with the code corresponding to the signal component of interest and then integrate over the code period. In the case of passive imaging, the incoming signals are noise-like with zero mean and therefore would give zero output after demodulation. As a result, in our system, we defer integration until the final cross-correlation step, essentially merging the demultiplexing and correlation processes. Fig. \ref{fig:cmist1} shows the possible solution where the codes $W_1 - W_n$ are multiplied to each signal and demodulated at baseband before feeding into correlator. To simplify the demodulation of visibilities, a more compact solution is discussed in further sections.

  \begin{figure}[]
   \begin{center}
   \includegraphics[width=.95\textwidth]{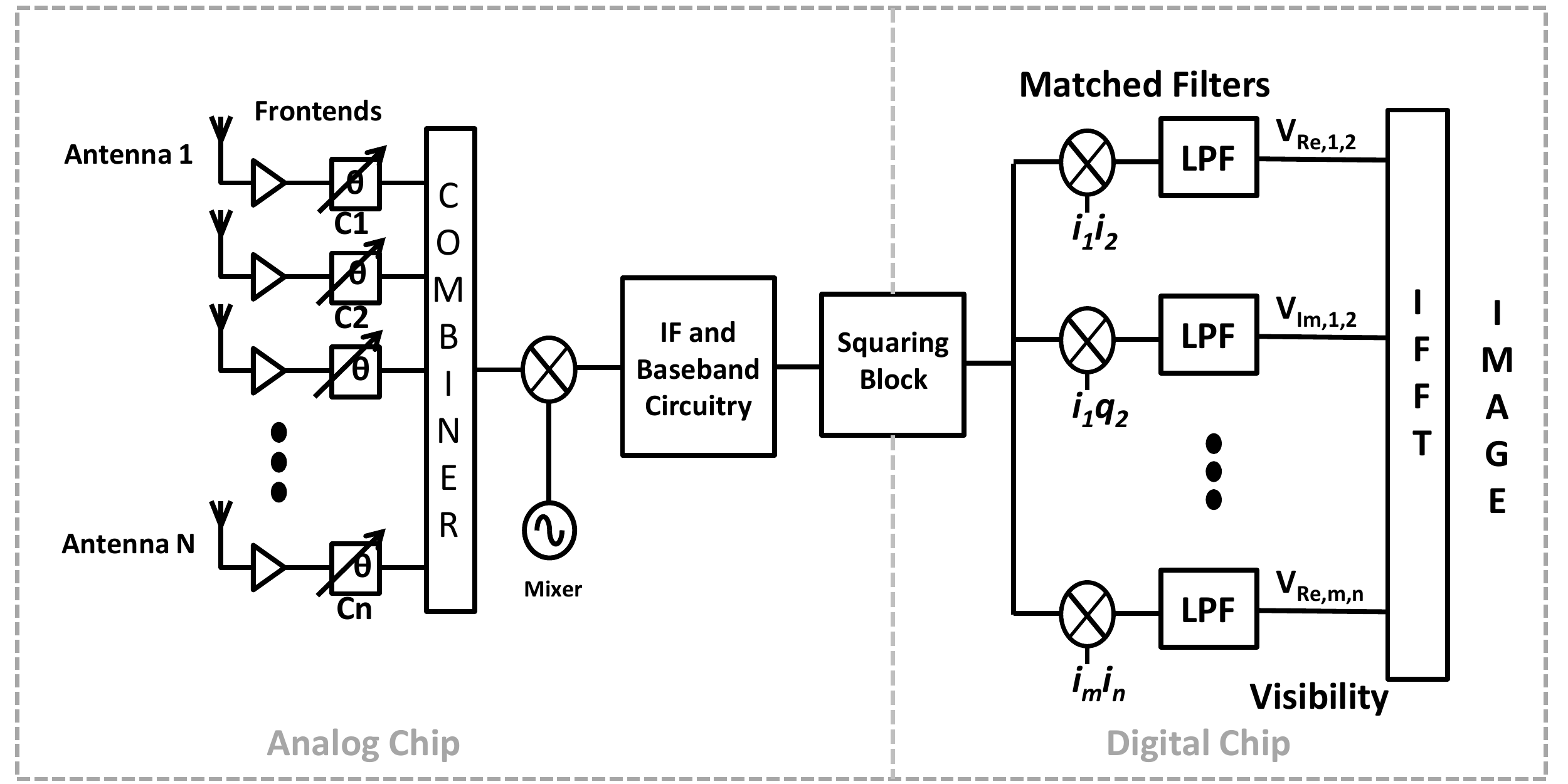}    
   \end{center}
   \caption{Block diagram of a code-modulated interferometric imaging system employing code-modulation within each phase shifter of a phased array and baseband visibility demodulation using code products.}    \label{fig:cmist}
   \end{figure} 
   
A block diagram of our proposed system which includes both the code modulation within each front-end and then the visibility demodulation within the digital domain is shown in Fig.~\ref{fig:cmist}.  For demultiplexing and demodulation, a signal ``squaring" operation is applied to the aggregate $s_{sum}$ signal. Such an operation can be an actual squaring circuit, such as an analog power detector, an RF mixer, a digital squaring operation, or another operation which creates an ``interference" of the code-multiplexed signal with itself.  The rationale for such a squaring operation is as follows.  In a conventional interferometer, the two signals of interest are correlated or interfered to obtain the visibility products.  In Fig.~\ref{fig:cmist}, the aggregate code-multiplexed signal is interfered with itself to obtain all possible visibility products, with the added result that these visibility products are now code modulated \cite{Provisional,ProvisionalUS}.  

The squared ``power" signal $p$ is represented as
\begin{equation}
\label{eq:square}
p= [s_{sum}]^2=\sum_{n}|s_n|^2 + 2\sum_ni_nq_ns_{i,n}s_{q,n} +2\sum_{n\neq m}(i_ni_ms_{i,n}s_{i,m} + q_nq_ms_{q,n}s_{q,m} + i_nq_ms_{i,n}s_{q,m})\ .
\end{equation}
The squared summation or power signal $p$ includes a summation of all of the ``self-powers" of the individual signals, a summation of the in-phase to quadrature-phase cross-products of individual signals which are in general orthogonal to one another and would average to zero, and a summation of all of the code-modulated cross-products between signal pairs.  

To demodulate the visibility samples, the power signal is correlated with code products $i_ni_m$ and/or $q_nq_m$ to obtain the real visibility samples.  Likewise, the power signal is correlated with code products $i_nq_m$ to obtain the imaginary visibility samples. Since we are correlating the square of the summation with a code product, all code products must be balanced and orthogonal, a property we refer to as ``Balanced Orthogonal Code Products" (BOCP).  In general, it is possible to have identical code products occur within a set of balanced orthogonal codes.  This would result in multiple visibility functions obtained at once or conflicting one with the other.  To avoid this, each code product must be balanced and orthogonal. 

The resulting visibilities obtained are as follows:
\begin{equation}
\label{eq:visibility_re}
v_{Re,n,m} = E(i_ni_m \ . \ p) = 2 \overline{s_{i,n}s_{i,m}\raisebox{2.5mm}{}} 
\end{equation}
\begin{equation}
\label{eq:visibility_im }
v_{Im,n,m} = E(i_nq_m \ . \ p) = 2 \overline{s_{i,n}s_{q,m}\raisebox{2.5mm}{}}
\end{equation}
where the $E(.)$ notation is used to denote the expectation or integration function.  In the derivations above, noise has not been included and there will be a component to these visibilities which relates to the average noise values within the system.  Additionally, codes have been assumed to be perfectly orthogonal; however, code skew can result in partial correlation between codes leading to residues remaining within the demodulated visibilities. 
%A behavioral model for the discussed approach will be designed in MATLAB\textsuperscript{\textregistered} SIMULINK\textsuperscript{\textregistered}.

Code modulation has traditionally found use in interferometry in two applications. Firstly, in phase-switching interferometry \cite{Sullivan91,Phaseswitch} wherein phase switching is applied to remove unwanted components (such as, constant total power terms) to obtain cross-correlations, and secondly in correlator-based interferometers to eliminate small offsets in correlator outputs that can result from imperfections in circuit operation or from spurious signals \cite{Thompson08}. The technique of multiplying each channel with orthogonal functions and demodulating with the products of those functions to remove unwanted components is similar to code-modulated interferometry proposed in this work. However, in this work the code-modulation and successive demodulation after squaring of combined signal is to reconfigure a phased array into an interferometer. Code modulation in this application is primarily to be able to obtain cross-correlations of all pairs of antennas simultaneously. Removing of unwanted and spurious signals is, however, an added advantage. Overall, CMI can be seen not just as an extension to, but a different application of a phase-switched interferometry.

Code-modulated interferometry has been implemented for an example four-element array and presented in detail in Sec. \ref{sec:four_element_CMI} for better understanding of extraction of visibilities. 

\section{Balanced-Orthogonal-Code-Products (BOCP)} \label{t1-preliminary}
As discussed in previous section, the proposed approach to code-modulated interferometry encodes incoming signals using orthogonal codes and decodes desired cross-correlations or visibilities using balanced-orthogonal \textit{code-products}. In particular, the codes must be chosen such that the product of any two codes from the set is a third unique code -- the code product. Here we discuss the requirements for a code-set to qualify as BOCP with examples of suitable code sets. 

\subsection{Rademacher Codes}
Rademacher codes \cite{Henderson64} or divide-by-two codes are inherently balanced, orthogonal, and BOCP code set. For example, for a code of length eight, the four Rademacher codes are
%\begin{center}
\begin{equation}
\label{eq:matrix_rad }
%$
\begin{bmatrix}
 R_0 \\
 R_1 \\
 R_2 \\
 R_3 \\
\end{bmatrix} %$
= 
%$
\begin{bmatrix}
1 &\ 1 &\ 1  &\ 1 &\ 1 &\ 1 &\ 1 &\ \ 1 \\
1&\ 1&\ 1&\ 1& -1& -1& -1& -1 \\
1&\ 1& -1& -1&\ 1&\ 1 &-1 &-1 \\
1& -1&\ 1& -1&\ 1& -1&\ 1& -1 \\
\end{bmatrix}
%$
\end{equation}
%\end{center}
For a code of length $L$, there are $( log_2{L}+1)$ Rademacher codes. The advantage of using Rademacher codes is the ease of generation, where each code can be easily generated using a frequency divider circuit. The drawback, however, is that the length of the codes increases exponentially with the total number of code-sets required, or the total number of elements in the receiver. For smaller number of elements, like in the case of four-element prototype discussed in Chapter \ref{chap-four}, Rademacher codes have been used for ease of implementation.

\subsection{Hadamard-Walsh Codes}
Walsh codes \cite{Walsh1923} are orthogonal code sets and have been used for the behavioral results in this section. For a fixed code length, the product of any two codes from a Walsh set gives a third code from the same set. For example, a complete set of Walsh code for a code of length eight is:

%\begin{center}
\begin{equation}
\label{eq:matrix_walsh }
%$
\begin{bmatrix}
 W_0 \\
 W_1 \\
 W_2 \\
 W_3 \\
 W_4 \\
 W_5 \\
 W_6 \\
 W_7 \\
\end{bmatrix} %$
= 
%$
\begin{bmatrix}
1 &\ 1 &\ 1  &\ 1 &\ 1 &\ 1 &\ 1 &\ \ 1 \\
1&\ 1&\ 1&\ 1& -1& -1& -1& -1 \\
1&\ 1& -1& -1& -1& -1 &\ 1 &\ 1 \\
1& \ 1& -1& -1&\ 1& \ 1& -1& -1 \\
1& -1& -1&\ 1& \ 1& -1& -1& \ 1 \\
1& -1& -1& \ 1& -1&\ 1 &\ 1 &-1 \\
1& -1&\ 1& -1& -1& \ 1& -1& \ 1 \\
1& -1&\ 1& -1&\ 1& -1&\ 1& -1 \\
\end{bmatrix}
%$
\end{equation}
%\end{center}

Walsh codes can be generated using Rademacher codes \cite{Henderson64}, and thus Rademacher codes can be seen as a subset of Walsh codes. For example, a complete set of Walsh code for a code of length eight generated using the Rademacher codes is $W_1=R_0$, $W_2=R_1$, $W_3=R_1R_2$, $W_4=R_2$, $W_5=R_2R_3$, $W_6=R_1R_2R_3$, $W_7= R_1R_3$ and $W_8=R_3$.

It is evident that different code pairs can result in same third code (\textit{e.g.}, $W_2W_3 = W_6W_7 = W_4$). The BOCP codes used to modulate should be selected such that unique codes are generated by product of any two codes. If $R1, R2$ and $R3$ are used, then only one out of remaining four ($R_1R_2, R_2R_3, R_1R_3$ and $R_1R_2R_3$) can be used.  For a code length of 1024, 30 such Walsh codes are available. A MATLAB\textsuperscript{\textregistered} code was written to identify BOCP code set for any given code length, given in Appendix \ref{Appendix_matlab_BOCP}.  For a system with $N$ antennas, we need $2N$ codes to be able to obtain complex visibility function from all baselines. With increasing number of antennas, the code length increases significantly and increases the integration time, reducing the frame rate of the imager. Similar challenge is faced in phase switched interferometry; more information can be found in \cite{Phaseswitch, Thompson08}.

\subsection{Gold Codes} \label{subsec:Gold_codes}
Uncorrelated codes are needed for code-modulated interferometry, where codes are used to control or modulate each phase shifter in the phased array. Hadamard-Walsh codes or Rademacher codes are possible candidates as explained in previous section, however, these codes must be finalized before designing the circuit, limiting system flexibility. Therefore a more generic code generator is desirable which is software programmable. For this purpose, Gold codes have been explored in this section because of their bounded correlation property and easy generation using maximum-length sequences (m-sequences). Gold codes have only three cross-correlation peaks which reduce in magnitude as the length of the code increases. Gold codes are also simple to realize through the modulo-2 addition (XOR) of two m-sequences \cite{Simon94}. In the following paragraphs, we first describe properties and simple circuits for m-sequence generators and then show how these can be used to generate Gold codes.

Maximum-length sequences are, by definition, the largest codes that can be generated using a number of shift registers placed in linear feedback, i.e., linear feedback shift registers (LFSR).  The m-sequence generated by a length of shift registers is decided by the placement of taps (i.e. XOR gates) in the feedback. For a length $n$ of shift registers, the largest code is of length $T= 2^n-1$. There can be multiple m-sequences for a given length of shift registers, depending upon where and how many XORs (or taps) are placed in the feedback. 

\begin{figure}
	\centering
	\includegraphics[clip,trim=0 0 0 0,width=.8\textwidth]{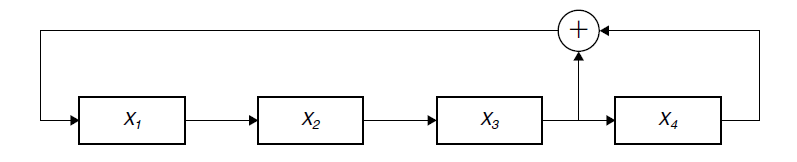}
	\caption{A four stage shift register with one tap generating an m-sequence \cite{Vijay10}.}
	\label{fig:lfsr}
\end{figure}

Fig. \ref{fig:lfsr} shows a simple block diagram of a four-stage m-sequence generator with the feedback applied by XORing of the outputs of registers three and four. This is denoted herein as a tap feedback of [3,4].  Note that for every set of feedback taps, if the tap positions are ``mirrored", an identical sequence reversed in time is generated.  For the circuit in Fig. \ref{fig:lfsr}, this mirrored sequence is obtained with [4,1] feedback.  This concept is easily extendable to larger sequences, with a summary of the number of registers, feedback locations, and number of possible m-sequences shown in Table \ref{tab:Mseq}.  

\begin{table}[]
    \caption{Feedback connections for m-sequences \cite{Vijay10}.}
    \centering
    \begin{tabular}{c}
\includegraphics[clip,trim=0 0 0 0,width=.95\textwidth]{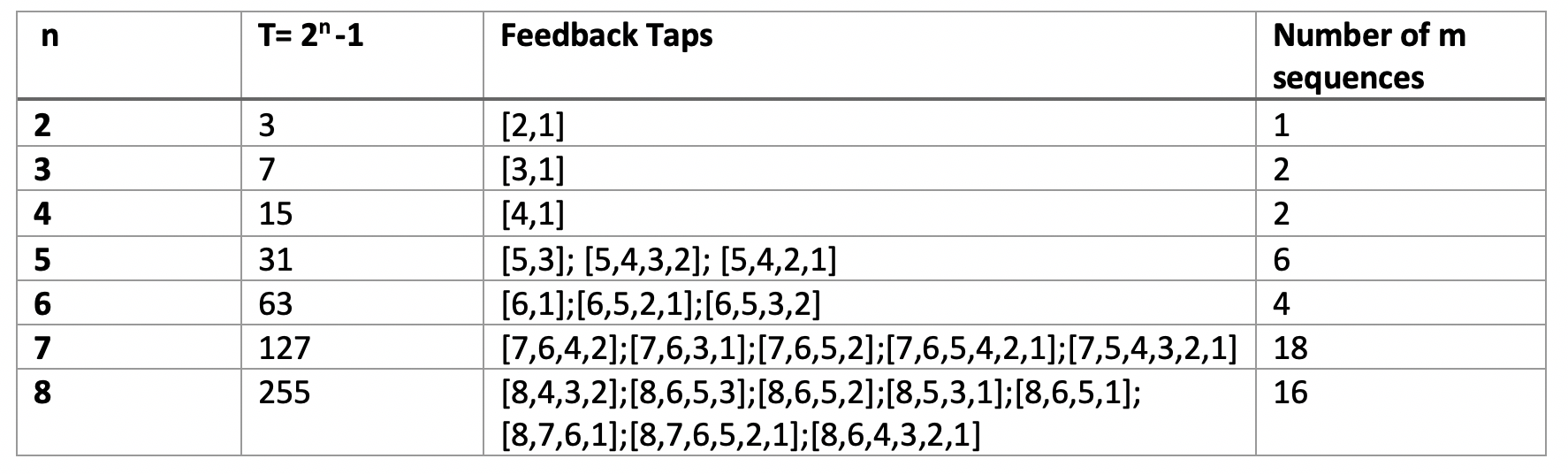}
    \end{tabular}
    \label{tab:Mseq}
\end{table}

\begin{figure}
	\centering
	\includegraphics[clip,trim=0 0 0 0,width=.8\textwidth]{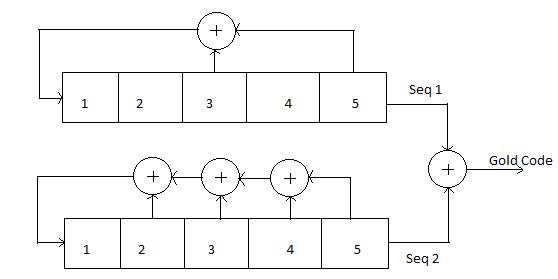}
	\caption{Two LFSR producing m-sequences Seq 1 and Seq 2 used to generate a Gold code \cite{Dinan96}.}
	\label{fig:gold}
\end{figure}

Two m-sequences are used to generate one set of Gold codes as shown in Fig. \ref{fig:gold}. One m-sequence is delayed and XORed with another to obtain a Gold code. Different Gold codes of the same set can be obtained by varying the phase delay. Since the two m-sequences are of same length $N$, all the generated Gold codes of the set are of length $N$. Each pair of m-sequence selected has specific properties and are called \emph{preferred pairs}. 

Fig. \ref{fig:gold} illustrates a Gold code generation using five-stage m-sequence generators. Sequence one (Seq1) is programmed to have feedback at [5,3] and Sequence two (Seq2) is programmed to have feedback at [5,4,3,2]. The length of the m-sequences and Gold codes = $2^5 -1 = 31$; thus, there are 31 Gold codes that can be generated by varying phase shifts in m-sequences. Example sequences are shown below in Table \ref{tab:gold_shifts}. Sequence 1 is the m-sequence for feedback of [5,3] whereas Sequence 2 is the m-sequence for feedback of [5,4,3,2]. The 0-Shift combination is the Gold code output when Seq1 and Seq2 are XORed without any phase delay between the two. The 1-Shift combination is the Gold code output when Seq1 and Seq2 are XORed with a single unit phase delay between the two.  Finally, the 30-Shift combination is the Gold code output when Seq1 and Seq2 are XORed with a 30-unit phase delay between the two. Circuit design for Gold codes is discussed in Appendix \ref{Appen:Gold_circuit}.

\begin{table}[]
    \caption{Gold codes using different sifts \cite{Dinan96} .}
    \centering
    \begin{tabular}{c c}
    \toprule
Sequence 1 & 11111 00011 01110 10100 00100 10110 0 \\
Sequence 2 & 11111 00100 11000 01011 01010 00111 0 \\
    \midrule
0 Shift combination & 00000 00111 10110 11111 01110 10001 0 \\
1 Shift combination &  	00001 01010 11110 00010 10000 11000 1 \\
30 Shift combination & 	10000 10001 00010 10001 10001 10101 1 \\
    \bottomrule
    \end{tabular}

    \label{tab:gold_shifts}
\end{table}

\section{Behavioral Models}
\subsection{One-dimensional CMI}

\begin{figure}[h]
	\centering
	\includegraphics[clip,trim=0 0 0 0,width=1\textwidth]{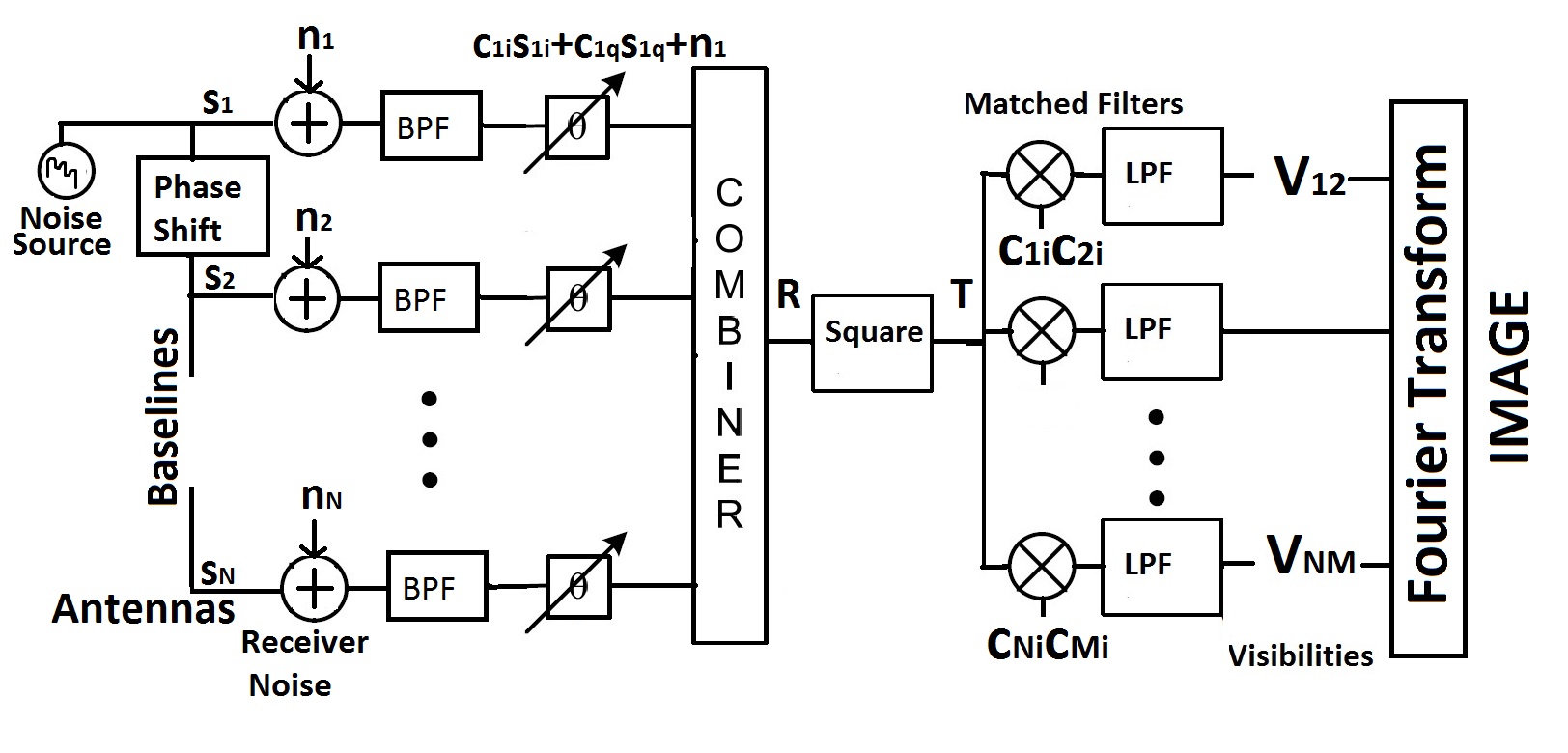}
	\caption{Behavioral model for one-dimensional code-modulated Interferometry. }
	\label{fig:Behav_model}
\end{figure}

\begin{figure}[h]
	\centering
	\includegraphics[clip,trim=0 0 0 0,width=1\textwidth]{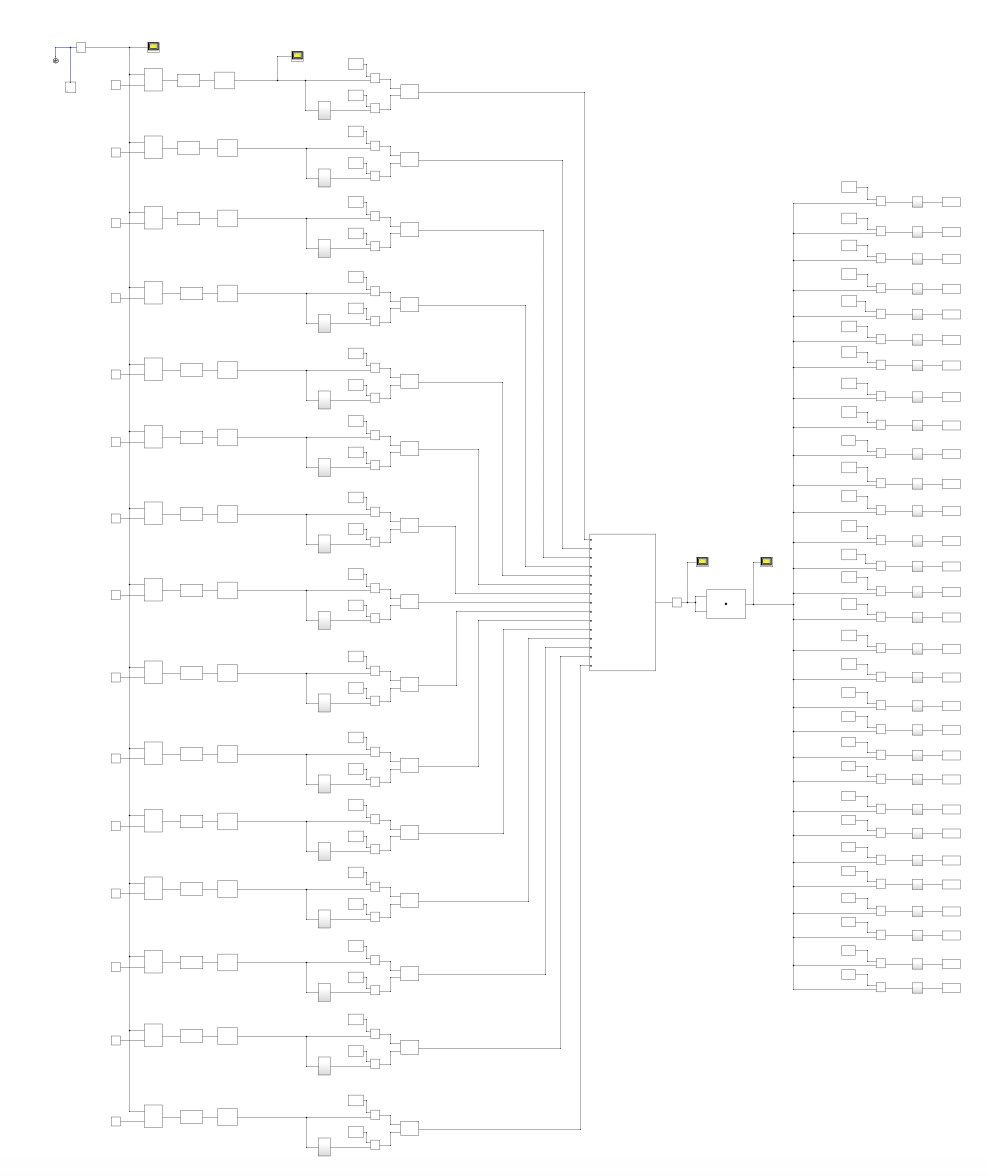}
	\caption{SIMULINK behavioral model for fifteen elements one-dimensional code-modulated interferometry. }
	\label{fig:simu1D}
\end{figure}

Several behavioral models are created to test the concept using MATLAB\textsuperscript{\textregistered} and SIMULINK\textsuperscript{\textregistered}. First, conventional interferometry is compared to code-modulated interferometry. Fig. \ref{fig:Behav_model} shows the block diagram of the behavioral model created to simulate a 1-D CMI. Fig.\ref{fig:simu1D} shows the behavioural model for an ideal 1-D array of fifteen antennas with wavelength spacing. To simulate a point source, a wide-band RF noise block is used. To model the path difference to different antenna elements, a phase shift is applied to the incoming noise corresponding to the relative geometry of the antennas with respect to the point source. For example, assuming a parallel wave-front, a wavefront from a point source at 20$^\circ$ azimuthal angle would incur a phase difference of $2\pi.D.sin(20^\circ)$ radian at the two receiving antennas placed at a distance $D.\lambda$ apart. As such, fifteen antennas with wavelength spacing receive the noise signal with a phase shift in multiples of $2\pi.sin(20^\circ)$ radian, with $1.2\pi.sin(20^\circ)$ rad. phase shift for second receiver to $14.2\pi.sin(20^\circ)$ rad. phase shift for fifteenth receiver.   

\begin{figure}[h]
	\centering
	\includegraphics[clip,trim=0 0 0 0,width=1\textwidth]{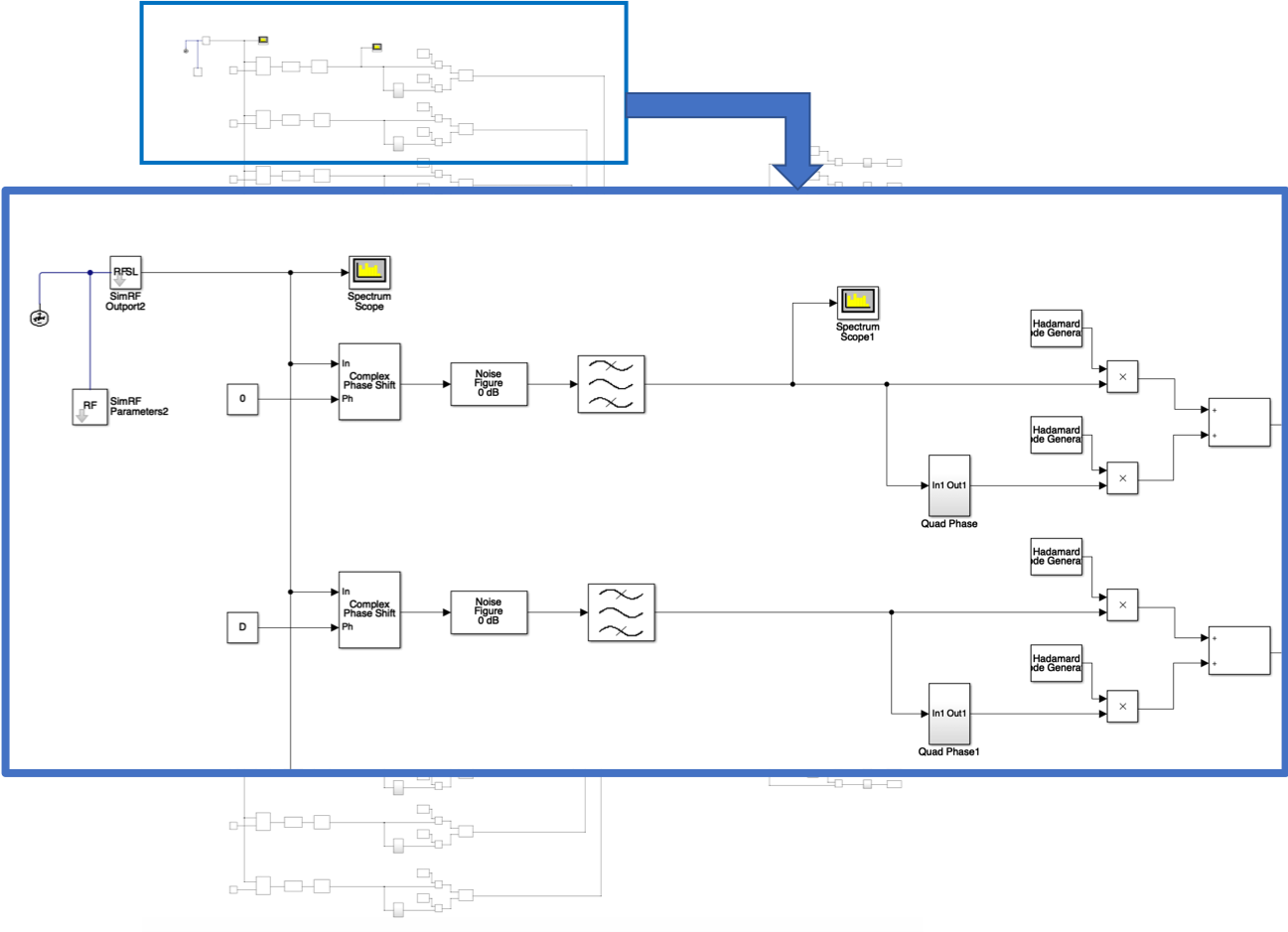}
	\caption{SIMULINK behavioral model for fifteen elements 1D CMI showing noise point source, front end model and code-modulation. }
	\label{fig:simu1DA}
\end{figure}

\begin{figure}[h]
	\centering
	\includegraphics[clip,trim=0 0 0 0,width=1\textwidth]{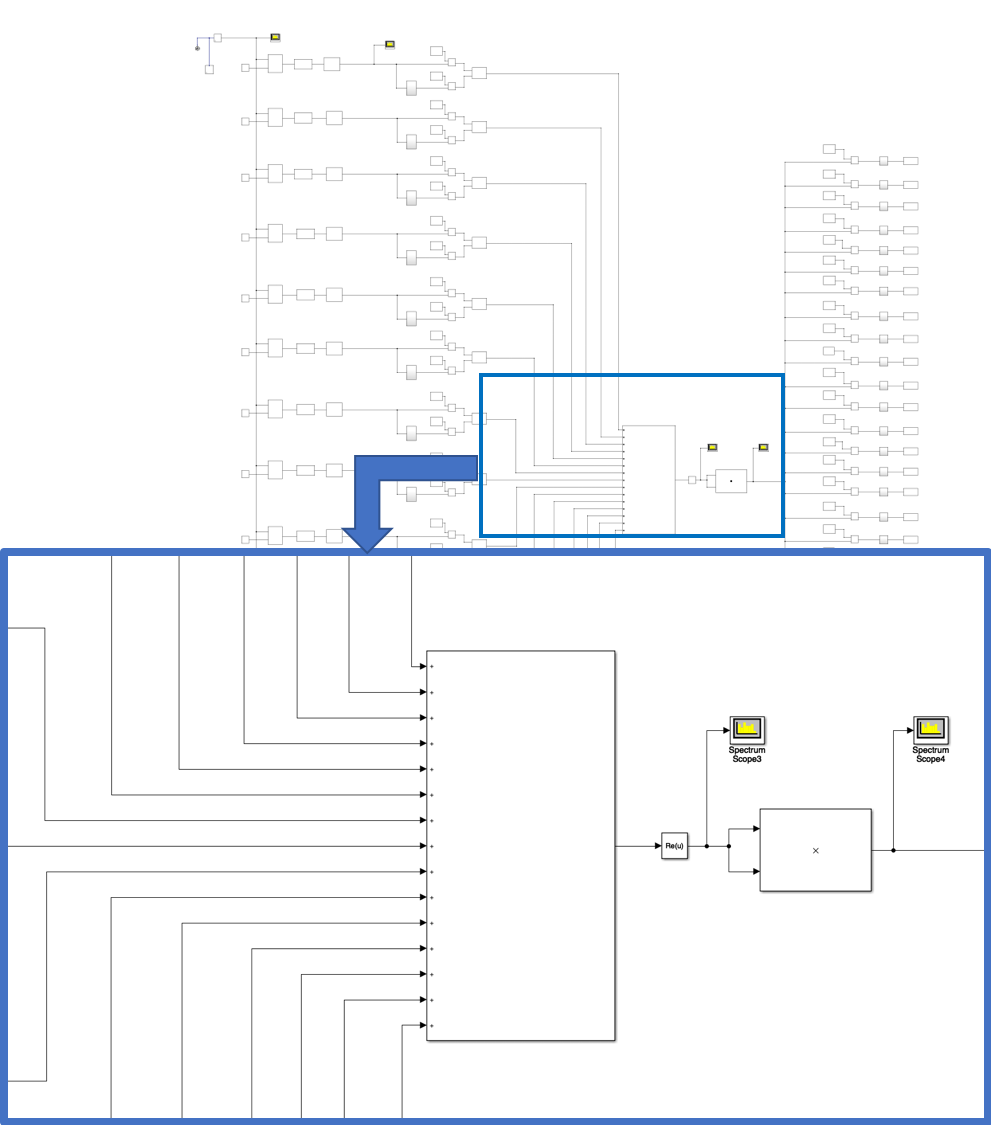}
	\caption{SIMULINK behavioral model for fifteen elements 1D CMI showing power combining and squaring operations.}
	\label{fig:simu1DB}
\end{figure}

\begin{figure}[h]
	\centering
	\includegraphics[clip,trim=0 0 0 0,width=1\textwidth]{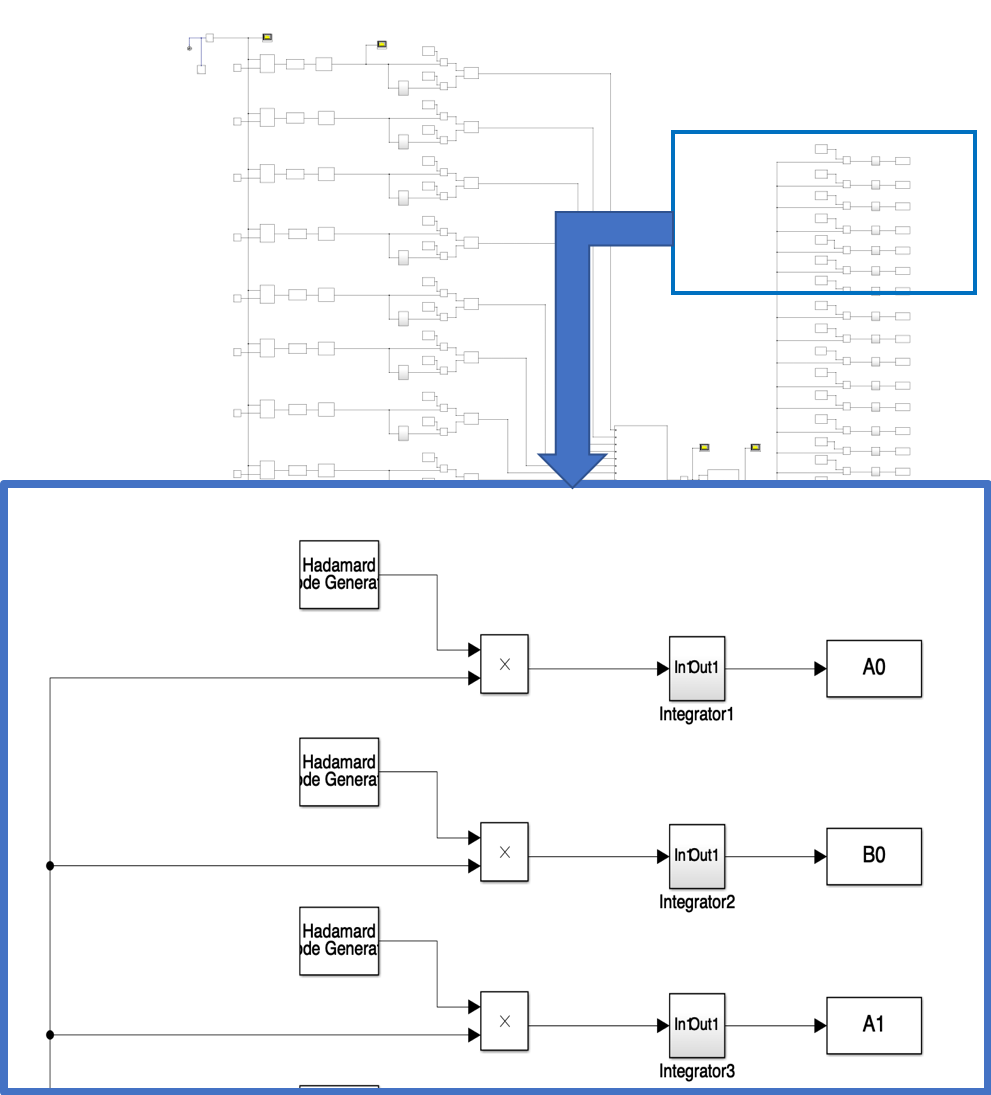}
	\caption{SIMULINK behavioral model for fifteen elements 1D CMI showing demodulation using matched filters (code-products) and integration.}
	\label{fig:simu1DC}
\end{figure}

The signals are then simulated with receiver noise for each element and a band pass filter to limit the bandwidth, as shown in Fig. \ref{fig:simu1DA}. Each signal is then code-modulated using $I$ and $Q$ codes to simultaneously extract complex visibilities, as discussed previously in Sec. \ref{subsec:comp_vis}. Hadamard-Walsh codes of length 1024 are used in the simulation which provide 30 BOCP codes. Note that 15 elements require $30$ BOCP codes for complex code-modulation, $15$ for $I$ and $15$ for $Q$ code-modulation, respectively. A MATLAB code is written to select a set of $30$ BOCP codes from a set of $1024$ Walsh codes of length 1024, given in Appendix \ref{Appendix_matlab_BOCP}. The code-modulation in a phased array can be achieved using a minimum of 2-bit phase shifter. All the signals from fifteen receivers are then power combined and squared, as shown in Fig. \ref{fig:simu1DB}. The squaring operation provides all the cross-correlations (real and imaginary visibilities) between all possible pairs of antennas. Fig. \ref{fig:simu1DC} shows the demodulation of the visibilities from squared signal using matched filters consisting of multiplication with respective code-products followed by an integration. The visibility samples are then saved in MATLAB workspace and the data is processed using MATLAB, including the Fourier transform to produce the brightness patterns.

\clearpage

\begin{figure}[h]
	\centering
	\includegraphics[clip,trim=0 0 0 0,width=.9\textwidth]{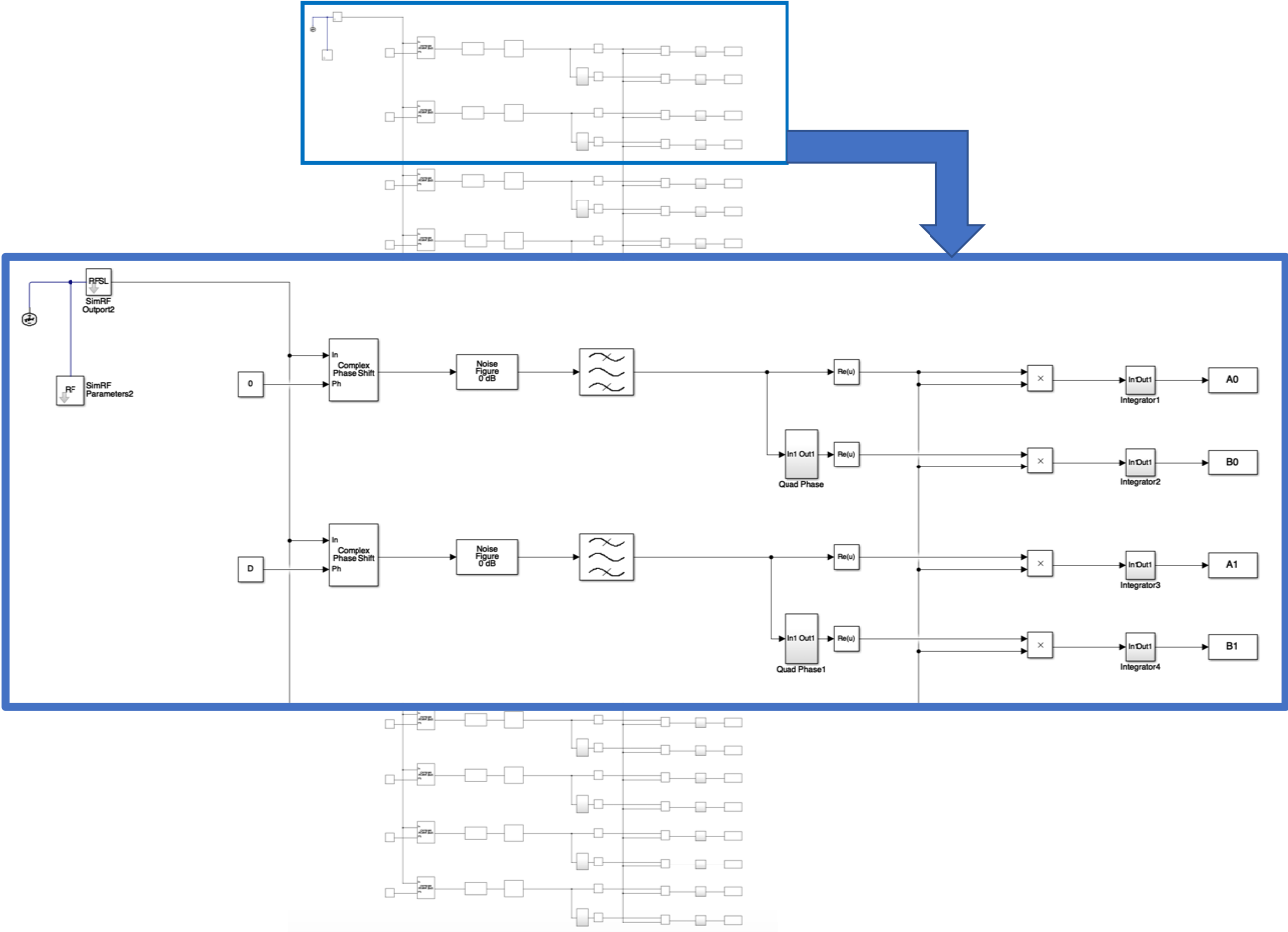}
	\caption{SIMULINK behavioral model for fifteen elements 1D basic interferometry showing a bank of complex correlators to obtain complex visibility samples.}
	\label{fig:simu1D_basic}
\end{figure}

Another SIMULINK model is created to simulate the conventional interferometry for comparison with CMI. As shown in Fig. \ref{fig:simu1D_basic}, the conventional interferometry is modeled using a similar approach; the model consists of a noise source, phase shifts to capture path length, a noise figure block and bandpass filter to model circuit front end, and a bank of complex correlators (pairwise multiplication and integration) to obtain complex visibilities. The visibility samples are once again saved in MATLAB workspace for data processing including Fourier transform to obtain brightness pattern.

\begin{figure}[h]
    \begin{center}
    \begin{tabular}{c c}  
    \includegraphics[height=4.5cm]{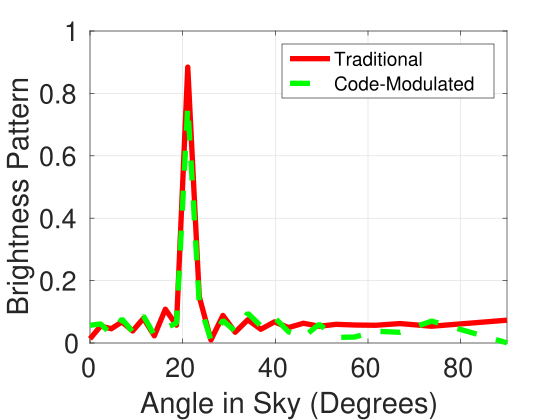} &      \includegraphics[height=4.5cm]{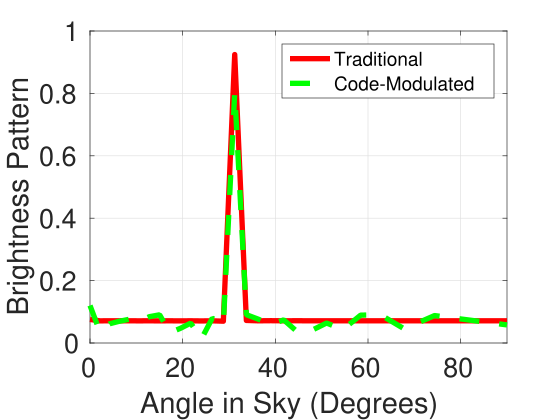} \\    
    \rule[-1ex]{0pt}{4.5ex}  (a) & (b) \\	
   \includegraphics[height=4.5cm]{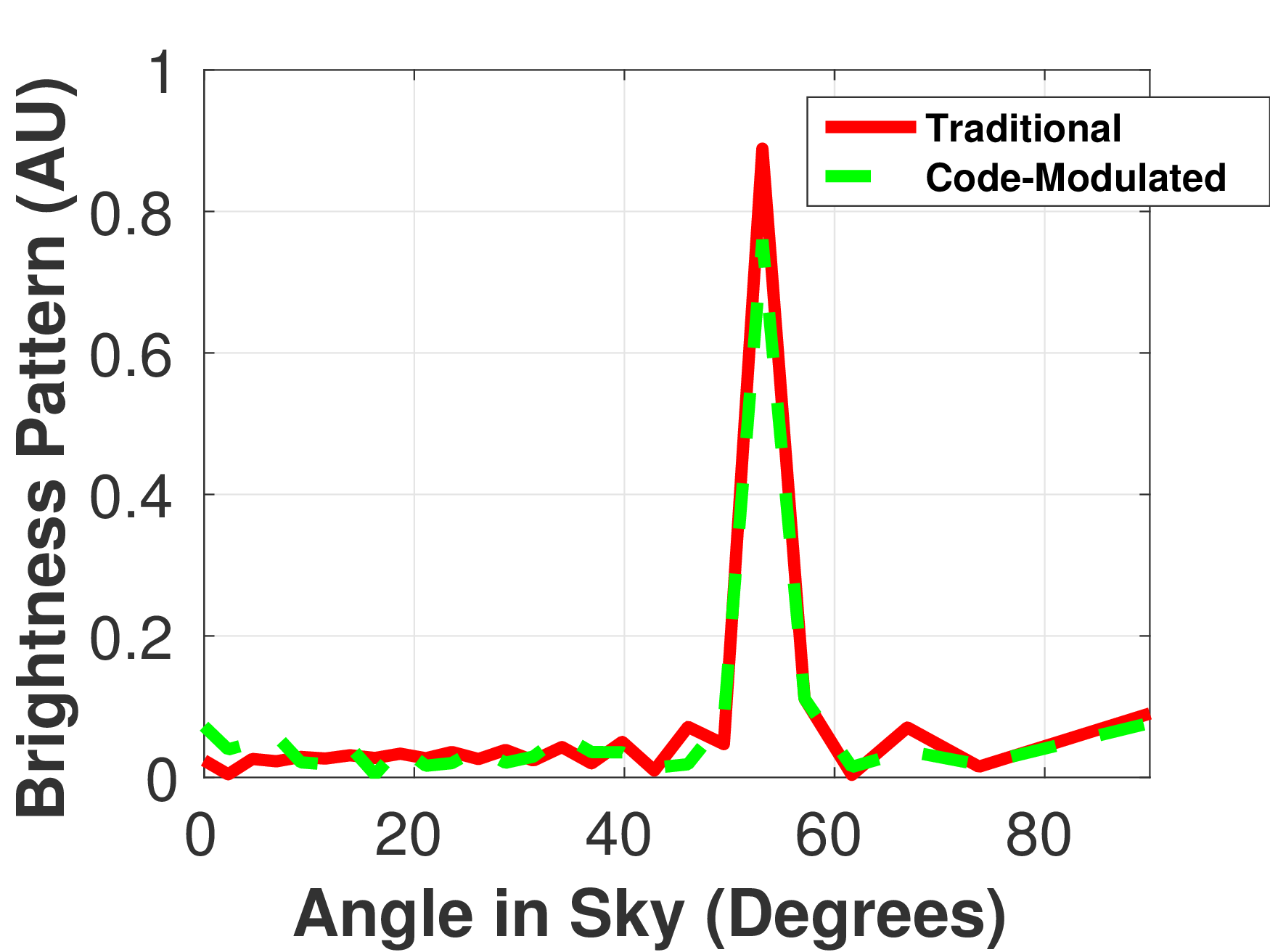} &
  \includegraphics[height=4.5cm]{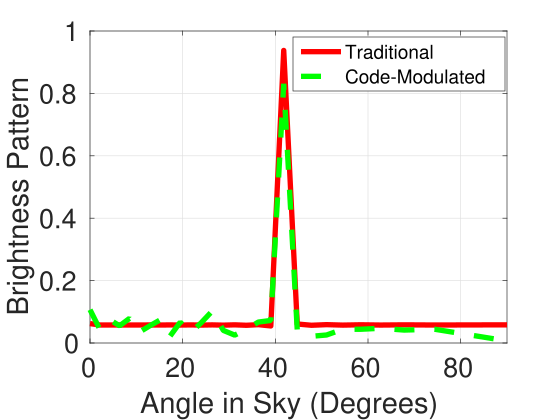} \\
   \rule[-1ex]{0pt}{4.5ex}  (c) & (d) \\
   \includegraphics[height=4.5cm]{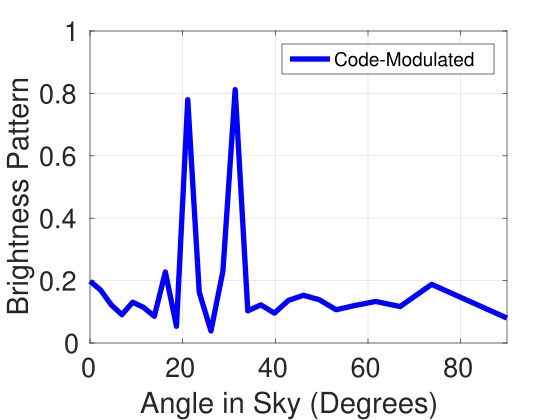} &
   \includegraphics[height=4.5cm]{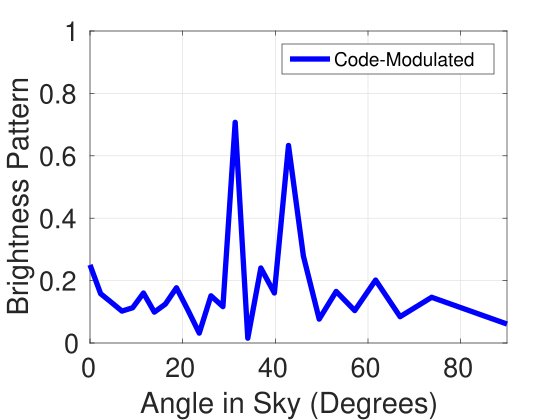} \\ 
    \rule[-1ex]{0pt}{4.5ex}  (e) & (f)  \\
    \end{tabular}
    \end{center}
    \caption{Behavioral modeling results comparing traditional to code-modulated interferometry: (a)-(d) shows the results of a point source at $20^{\circ}$, $30^{\circ}$, $40^{\circ}$, and $50^{\circ}$ elevation angles; (e) and (f) show code-modulated interferometry resolving two point sources at elevation angles $20^{\circ}$ and $30^{\circ}$ in (e), and $30^{\circ}$ and $40^{\circ}$ in (f).}
    \label{fig:point_source} 
    \end{figure}

The bright point source is then swept across elevation angles and the two approaches are compared. The point source consists of a 1-GHz signal with 2-MHz bandwidth to allow for faster computation.  The code modulation applied consisted of a chip rate of 1 MHz with code length 1024. The results for both interferometers were found to be very similar. Fig.~\ref{fig:point_source} shows the plots for one and two point sources at different angles in sky, for conventional and code-modulated interferometry. Note that CMI imager is able to accurately track the bright point source, as well as resolve two point sources. These simulations, however, assume a perfectly cold background with only one or two bright point source(s) emitting noise-like thermal radiation, and does not included any of the circuit imperfections and implementation errors. These ideal simulation conditions provide a near-perfect match in the simulation results of a CMI imager with conventional interferometer. Background thermal noise, circuit noise, circuit non-linearity, spurious signals, code-skewing, moving scene, etc. are some of the many imperfections that will affect the actual implementation of the system. The affect of circuit noise figure on the CMI imaging is discussed in Chapter \ref{chap-four}.

\clearpage 

\subsection{Two-dimensional CMI}
\begin{figure}
	\centering
	\includegraphics[clip,trim=0 0 0 0,width=1\textwidth]{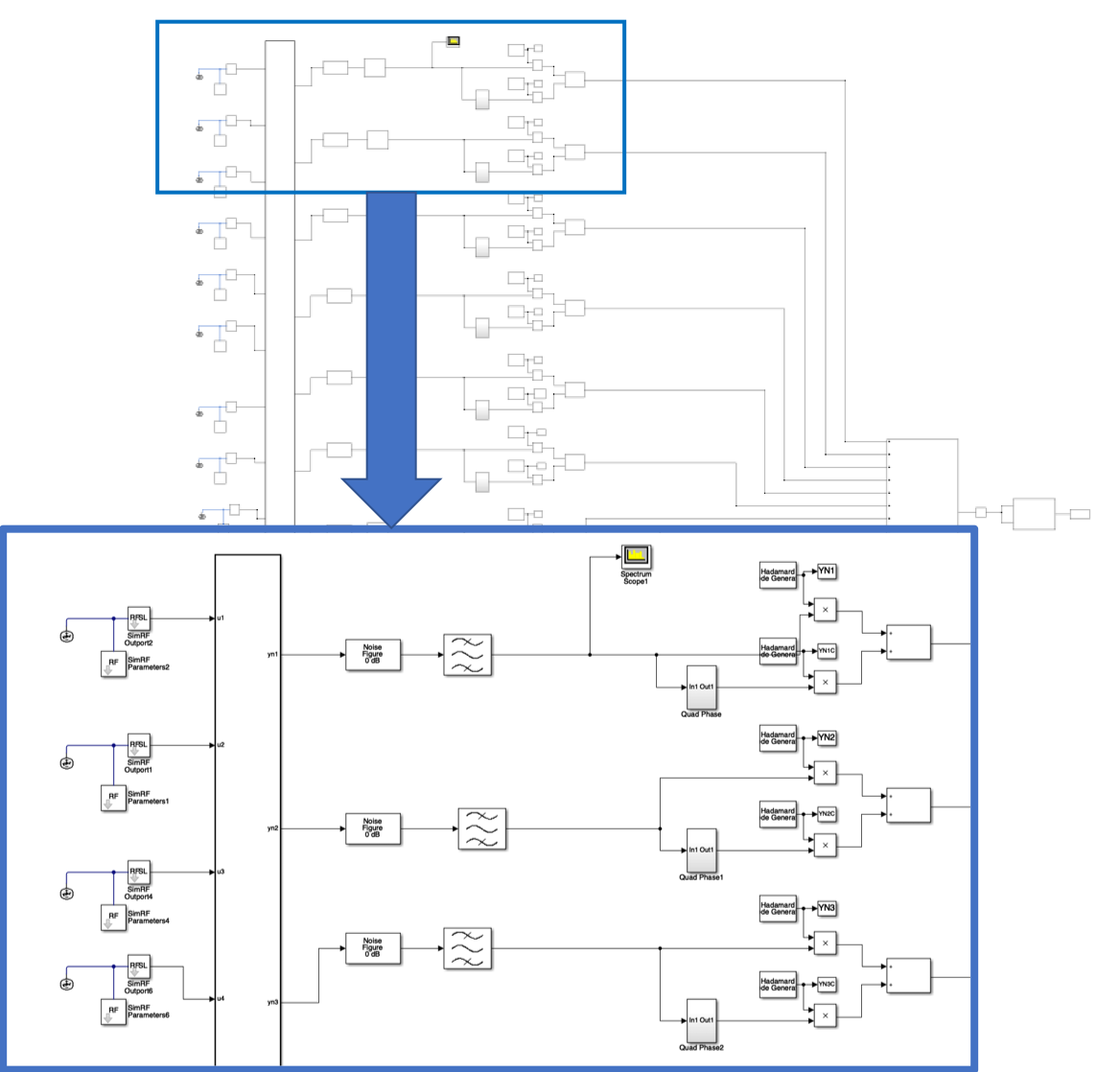}
	\caption{Behavioral model for two-dimensional code-modulated interferometry. }
	\label{fig:Behav_model_full}
\end{figure}

\begin{figure}[h]
   \begin{center}
   \begin{tabular}{c c} %% tabular useful for creating an array of images 
   \rule[-1ex]{0pt}{3.5ex}  {Source} &{Image}\\	\includegraphics[height=4cm]{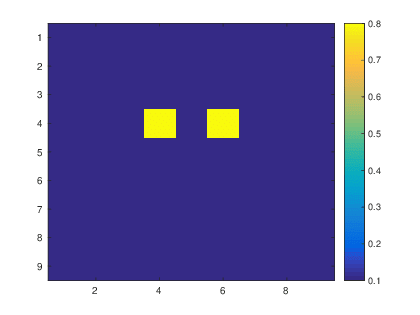}& \includegraphics[height=4cm]{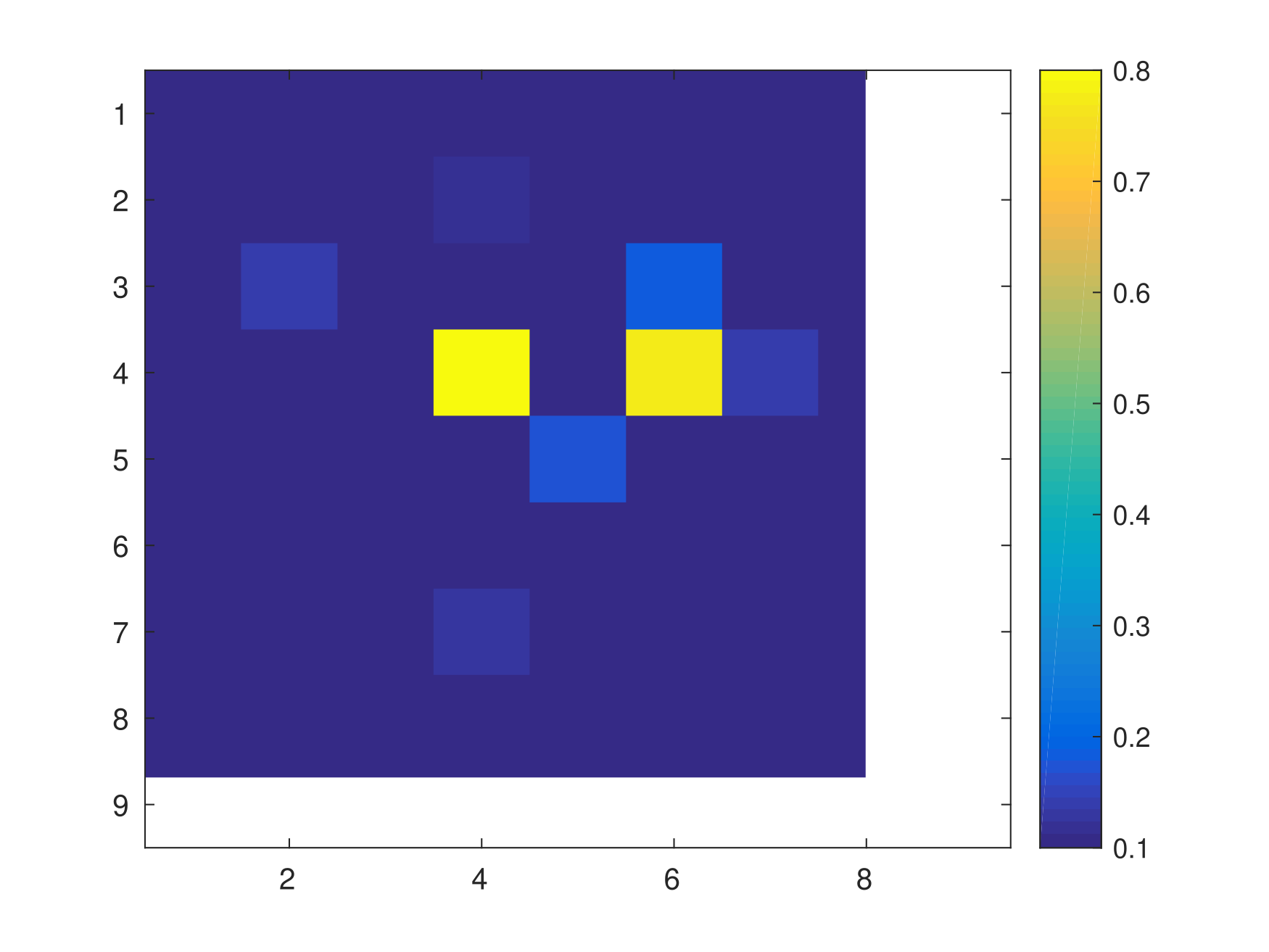}\\ \includegraphics[height=4cm]{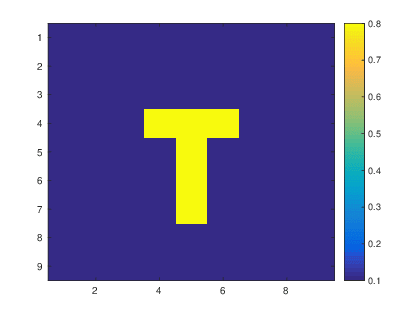}& \includegraphics[height=4cm]{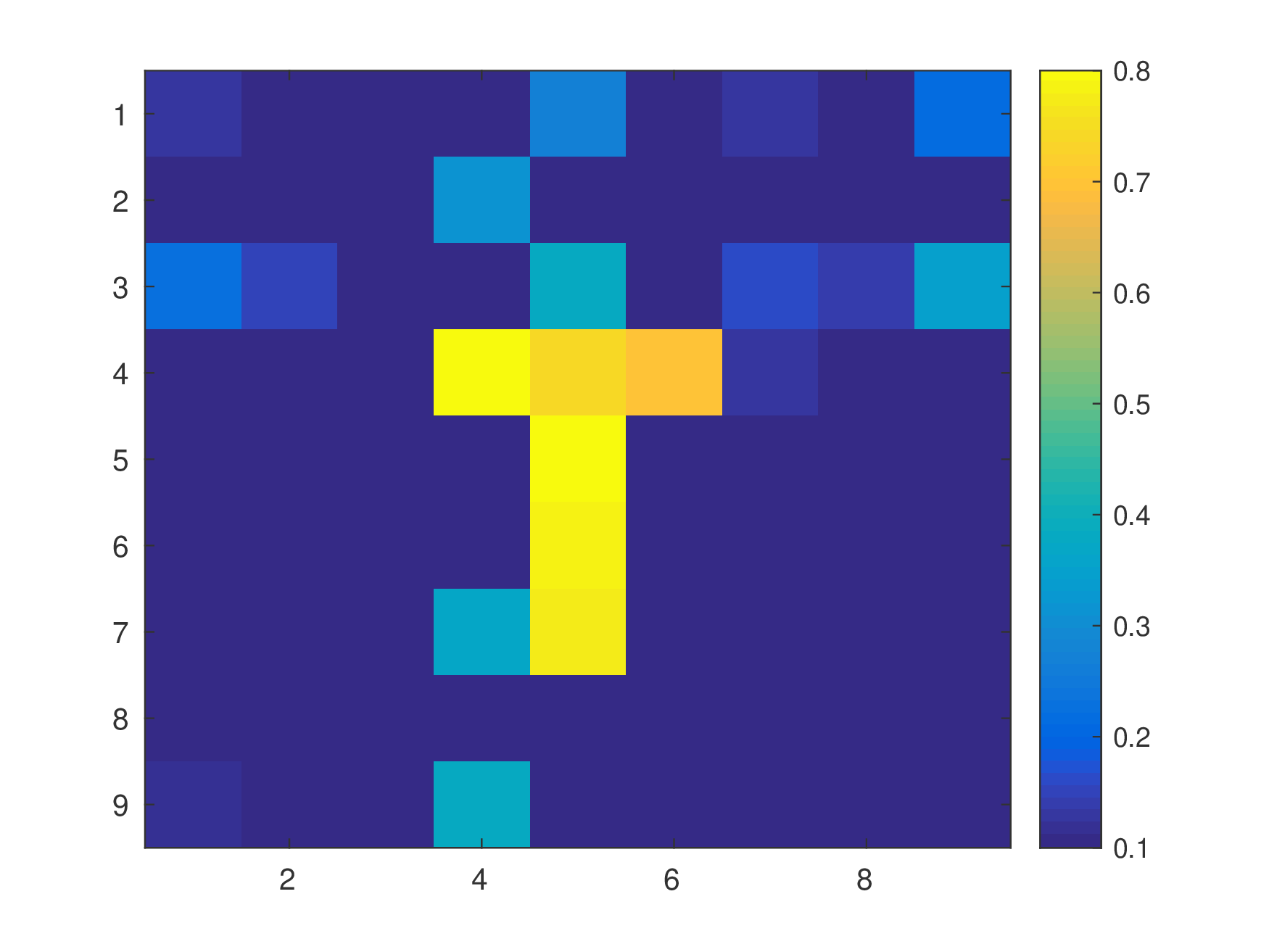}\\  
 \includegraphics[height=4cm]{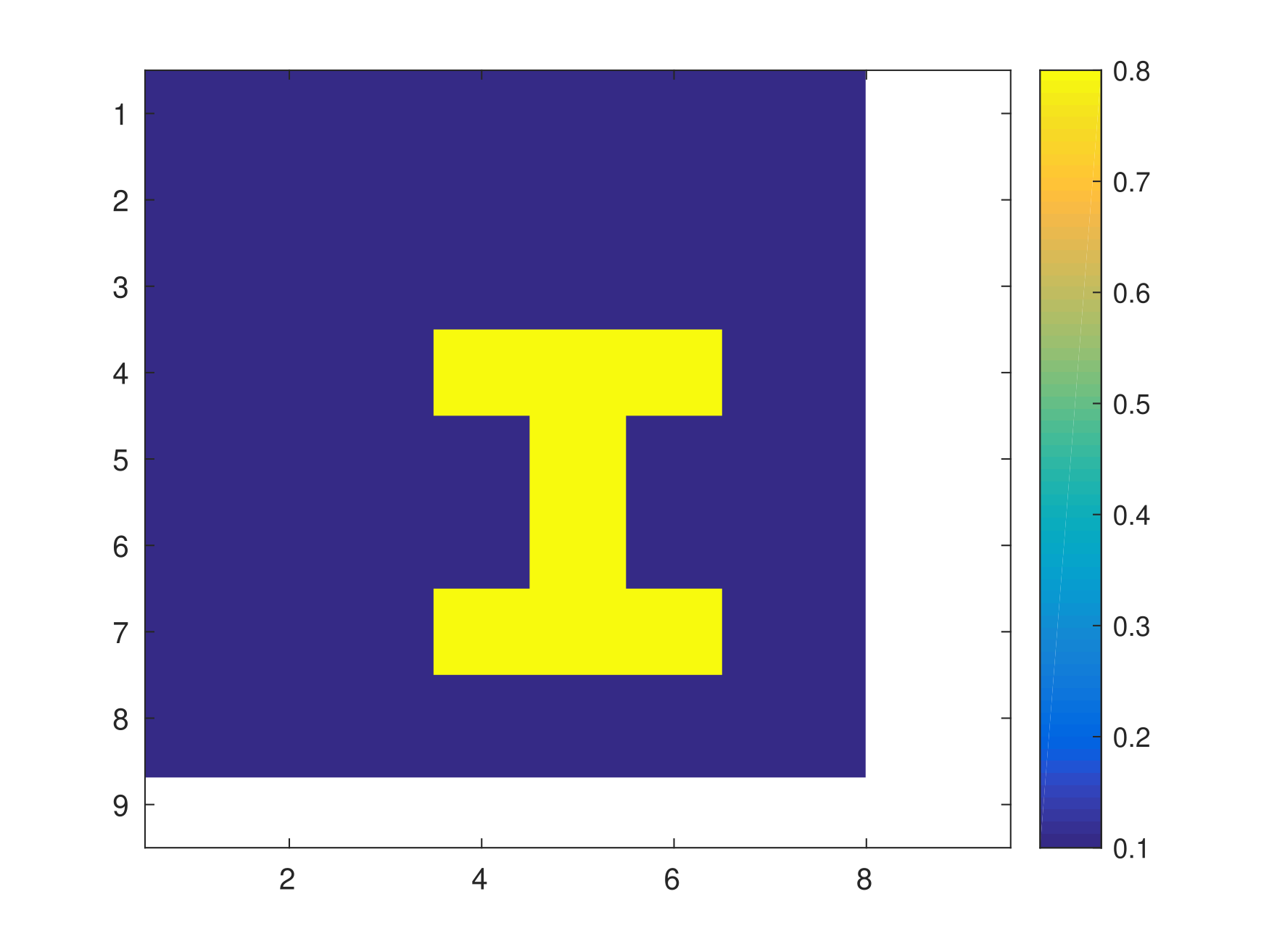}& \includegraphics[height=4cm]{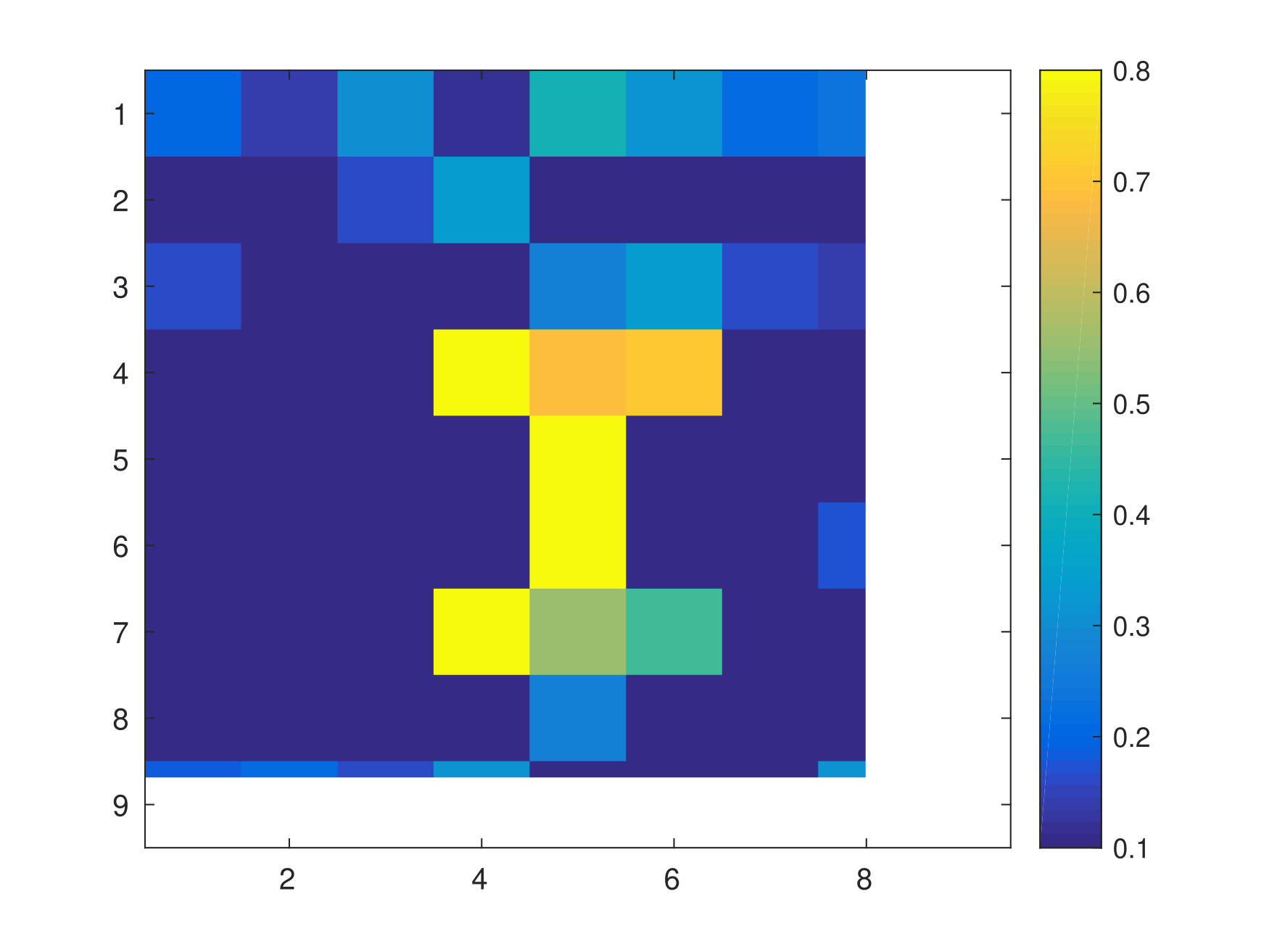}\\ \includegraphics[height=4cm]{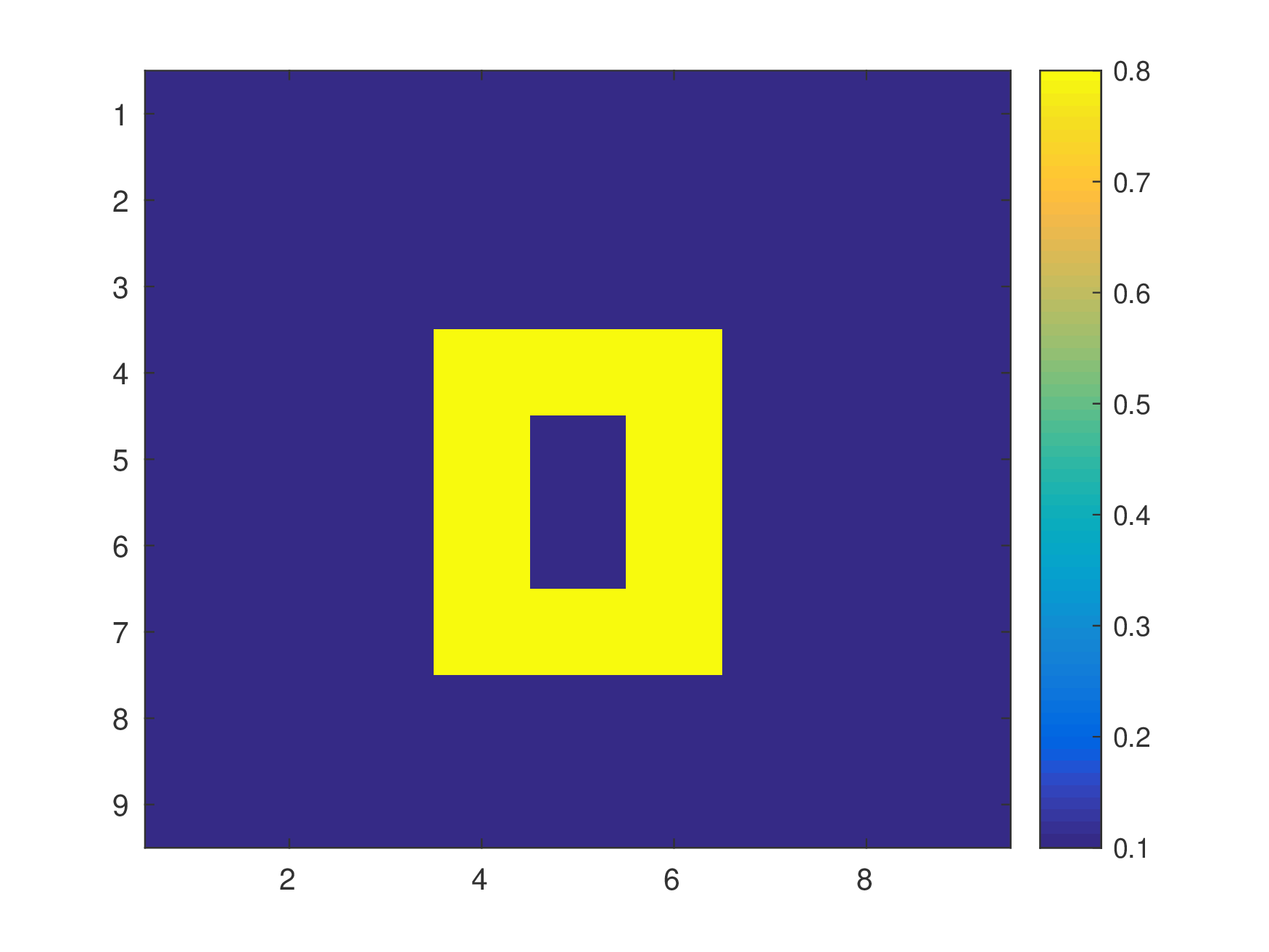}& \includegraphics[height=4cm]{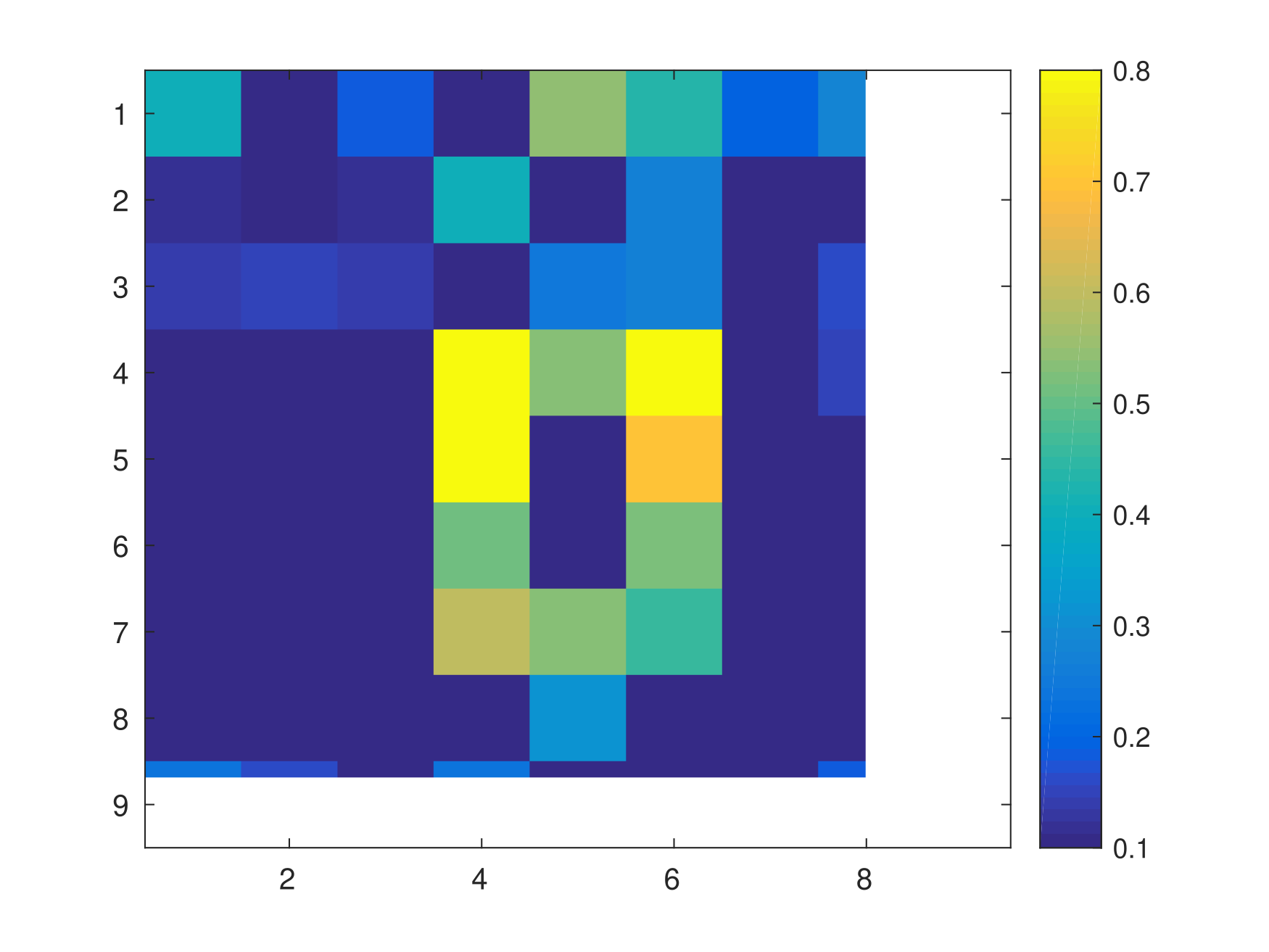}\\ 
   \end{tabular}
   \end{center}
   \caption{Simulation results for imaging 2-D sources with code modulated interferometric imaging system \cite{Spie16}.}
   \label{fig:2D_result} 
   \end{figure}

To test the code-modulated interferometry for extended sources, the behavioral models are expanded to a two-dimensional scene and a two-dimensional imaging array. Fig. \ref{fig:Behav_model_full} shows the behavioral model for two-dimensional CMI. Extended objects are modeled as a combination of multiple independent point sources emitting uncorrelated noise. This makes the object spatially uncorrelated, an important requirement for interferometric imaging systems. The noise signals then pass through a MATLAB code that models the shape of object, its distance from imager, and relative path difference of the signals from different parts of the object to the different receivers in the imaging array\footnote{The phase array imaging toolbox \cite{Patole13} provided by TxACE (M. Torlak, UT Dallas) was helpful in implementing the 2D behavioral models.}. The MATLAB code to simulate 2-D objects for imaging is given in Appendix \ref{Appendix_matlab_2D_object}. The noise sources are wide-band white-noise block and therefore a band pass filter block is used to limit the bandwidth, and code-modulation is applied to both $I$ and $Q$ signal, similar to 1-D CMI. All of the elements are then power combined and fed into a power detector (squaring). Again, the results are saved in MATLAB workspace and data is processed in MATLAB, including the demodulation of the visibilities and the Fourier transform to obtain the image. Twelve elements are placed in a "T" configuration providing 81 pixels in an image. Fig.~\ref{fig:2D_result} shows results for point sources and 2-D "T", "I" and"O" shaped objects and demonstrates the functionality of the system.

%   \begin{figure} 
%    \begin{center}
%    \begin{tabular}{c c} %% tabular useful for creating an array of images 
%    \rule[-1ex]{0pt}{4.5ex}  {Source} &{Image}    \\	\includegraphics[height=4.5cm]{SPIEobjectpoint} & \includegraphics[height=4.5cm]{SPIEimagepoint} \\ 
%  \includegraphics[height=4.5cm]{SPIEsourceI} & \includegraphics[height=4.5cm]{SPIEimageI} \\  \includegraphics[height=4.5cm]{SPIEobjectO} & \includegraphics[height=4.5cm]{SPIEimageO} \\ 
%  \includegraphics[height=4.5cm]{SPIEobjectT} & \includegraphics[height=4.5cm]{SPIEimageT}\\  
%    \end{tabular}
%    \end{center}
%    \caption[2D_result] 
% %>>>> use \label inside caption to get Fig. number with \ref{}
%    { \label{fig:2D_result} 
% Simulation results for imaging 2-D sources with code modulated interferometric imaging system.  }
%    \end{figure}

\clearpage

%\color{red}

\section{Sensitivity Analysis} \label{sec:sensitivity_deriv}
An important metric for the performance of any radiometer is the sensitivity $(\Delta T_{min})$ of the system, defined as the minimum change in temperature of source that can be detected. Sensitivity of a single radiometer is given by \cite{KrausBook}.
 \begin{equation}
 \label{eqn:sensitivity}
\Delta T_{min} = \frac{K_e T_{sys}}{\sqrt[]{\Delta \nu_{HF} \tau_{LF} }}\ \  ,%\verb|K| \,
\end{equation}
where $K_e$ is sensitivity constant, $T_{sys}$ is the system noise ($T_{sys}$ = antenna temperature $T_{A}$
+ receiver noise temperature $T_{RT}$), $\Delta \nu_{HF}$ is the high-frequency bandwidth and $\tau_{LF}$ is the integration time. $K_e$ is a multiplication factor which depends upon different factors such as the interferometer configuration or phase switching, correlator/ power detector, \textit{etc.} and can vary between $0.5$ and $2\sqrt[]{2}$ \cite{KrausBook} in radio astronomy for single and pairs of antennas. In this section we estimate the sensitivity constant ($K_e$) for an N-element code-modulated interferometer. The derivation follows the approach similar to as given in book titled Radio Astronomy by Kraus, Chapter 7, pages 4-22 \cite{KrausBook}. 

Let us assume a code-modulated interferometer with $N$ number of elements. To simplify the analysis, let us assume only real visibilities are being obtained. Each element is connected to an antenna with antenna temperature $T_{A}$ and each receiver adds a receiver noise equivalent to noise temperature $T_{RT}$. We assume that each element has a frequency bandwidth of $\Delta \nu_{HF}$ centered around a mm-wave frequency of $\nu_{HF}$. Corresponding noise powers are $k T_{A} \Delta \nu_{HF}$ and $k T_{RT} \Delta \nu_{HF}$ in each element from background noise and receiver noise, respectively, where $k$ is the Boltzmann constant. Noise power in each element given by $k T_{A1} \Delta \nu_{HF}$, $k T_{RT1} \Delta \nu_{HF}$, $k T_{A2} \Delta \nu_{HF}$, $k T_{RT2} \Delta \nu_{HF}$, ..., $k T_{AN} \Delta \nu_{HF}$, $k T_{RTN} \Delta \nu_{HF}$ can be assumed to be independent, and hence uncorrelated. This implies that total system noise $T_{sys}$ from each antenna is thus uncorrelated. A discrete source of temperature $\Delta T$ produces a signal power of $k \Delta T \Delta \nu_{HF}$ in each receiver. The $N$ elements of the interferometer receive the correlated signal from the discrete source with different phase shifts due to different path lengths. Let us also assume that this phase shift is $cos\phi_1$, $cos\phi_2$, $cos\phi_3$, ...$cos\phi_N$ for antennas one to N. 

Signals from $N$ elements are code-modulated and power combined using a power combiner. It is important to note here that all the signals from $N$ elements are power combined using a circuit element, such as the Wilkinson power combiner \cite{Wilkinson60}. In a Wilkinson power combiner, when uncorrelated noise is provided at each of the two input ports, half of the power from each port is dissipated in the shunt resistor and half of the power reaches the output port. As such, uncorrelated noise temperature $T_{sys1}$ and $T_{sys2}$ provide a total noise temperature of $(T_{sys1}+T_{sys2})/2$ at the output port of power combiner. When correlated signals are presented at the input ports of Wilkinson power combiner, the signals add in power. 

At any instant in time, the spectral power of the power combined signal (before the power detector) for a total power interferometer can be written as (in watts):

\begin{equation}
    W = C_0 k\Delta \nu_{HF}[\sum{T_{A}} + \sum{T_{RT}} + \sum{\Delta T} + 2\Delta T \sum{ cos\phi_n}]
\end{equation}

where $C_0$ is a proportionality constant that includes the gains and losses in the receiver. For a code-modulated interferometer, the above equation can be re-written as:
\begin{equation}
    W = C_0 k\Delta \nu_{HF}[\sum{T_{A}} + \sum{T_{RT}} + \sum{\Delta T} + 2\Delta T \sum{(\pm) cos\phi_n}]
\end{equation}
where the $(\pm1)$ represents the code-modulation of individual elements. 

For an $N$ element CMI, the above equation can be written as:
\begin{equation}
    W = C_0 k\Delta \nu_{HF}[N T_{A} + N T_{RT} + N \Delta T + 2\Delta T (\pm1cos\phi_1 \pm1cos\phi_2 \pm1cos\phi_3 + ...)]
\end{equation}
or
\begin{equation}
    W = C_0 k\Delta \nu_{HF}[N T_{sys} + N \Delta T + 2\Delta T (\pm1cos\phi_1 \pm1cos\phi_2 \pm1cos\phi_3 + ...)]
\end{equation}

The combined signal is then fed in to a power detector. During demodulation we obtain a visibility sample, say $2\Delta T cos\phi_1$, using the code-products, whereas the d.c. terms such as $N T_{sys} + N \Delta T$ and the other visibility samples $2\Delta T (cos\phi_2 + cos\phi_3 + ...)$ would vanish. This is similar to an N-element phase-switch interferometry, the difference being that in CMI we use codes to obtain cross-correlations whereas the codes in phase-switch interferometry are mainly to eliminate system non-idealities.  The desired signal power is then given by: 
\begin{equation}
    W =C[ k\Delta \nu_{HF} 2\Delta T cos\phi_1]^2
    \label{Eqn:power_det_vis}
\end{equation}
 where $C$ is the transfer function of the power detector.
 
Along with the desired signal, there exists a noise voltage in the power detector output, originating due to noise-voltage components in the frequency range from $\nu_{HF} - \Delta \nu_{HF}/2$ to $\nu_{HF} + \Delta \nu_{HF}/2$ beating with each other. This noise voltage has a triangular power spectral density with close to d.c. maximum of: 

\begin{equation}
    W_{noise} =2C[k N T_{sys}]^2 \Delta \nu_{HF}
\end{equation}

assuming $N\Delta T <<< N T_{sys}$ which is true for a small bright object in the scene to be imaged. For extended objects, the total power from the object contributes to the reduction in the sensitivity. 
Also, assuming a rectangular passband for the low pass filter (or averaging) of bandwidth $\Delta \nu_{LF}$, the noise in power detector signal can be written as: 

\begin{equation}
    W_{noise} =2C[k N T_{sys}]^2 \Delta \nu_{HF}\Delta \nu_{LF}
    \label{Eqn:power_det_noise}
\end{equation}

The sensitivity can then be defined as the minimum detectable cross-correlation term when it is equal to the noise in power detector output, $W_{min} = W_{noise}$. Substituting $ \Delta \nu_{LF} = 1/(2\tau_{LF}) $, and equating eqns. \ref{Eqn:power_det_vis} and \ref{Eqn:power_det_noise}, we get: 
\begin{equation}
 \Delta T_{vis} = \frac{N T_{sys}}{\sqrt[]{\Delta \nu_{HF} \tau_{LF} }}
\end{equation}

where $\Delta T_{vis} = (2\Delta T cos\phi_n)_{min}$. This is the sensitivity of a single visibility sample. Hence, for code-modulated interferometry, sensitivity constant $K_e = N$. This degradation of sensitivity is a trade-off to reduce the cost of the imaging system. This reduction in sensitivity can be over come by higher bandwidth and longer integration time. 

According to \cite{Levine88}, unlike in radio astronomy where the source is relatively smaller than the FOV, for the extended sources filling the FOV in imaging, the sensitivity is degraded by the total brightness temperature of the source, \textit{i.e.} $N T_{sys}$ is replaced by ($N T_{sys} + N \Delta T$) in the sensitivity equation. Every point in the source contributes to the noise in each visibility point, and therefore to all the pixels in the image. For a simplified scenario, the sensitivity of the image is given by \cite{Levine88}: 

\begin{equation}
 \Delta T_{image} = \frac{T_{sys} \sqrt{N_p}}{\sqrt[]{\Delta \nu_{HF} \tau_{LF} }}
\end{equation}

where $N_{p}$ is the number of pixels in the image. Therefore the sensitivity of a CMI image obtained with $N$-element phased array and $N_p$ number of pixels is:

\begin{equation}
 \Delta T_{image,CMI} = \frac{T_{sys} N \sqrt{N_p}}{\sqrt[]{\Delta \nu_{HF} \tau_{LF} }}
\end{equation}

Sensitivity is also affected by the redundancy in baselines, which is worst for a zero-redundancy thinned array ($K_e = \sqrt[]{N_{p}}$) and improves as the array becomes more filled. Comparing with the phase-switched interferometry ($K_e$ =2 for two antennas), the $K_e$ for code-modulated interferometry can be up to $N$ due to summation and subtraction of $N$ different noise sources. This might seem like a huge reduction in sensitivity but the number of pixels obtained could be up to ${N(N-1)}$ per scan as compared to one pixel per scan in simple interferometer for the same integration time \cite{Levine88}. 

As an example, an N-element receiver array with 6-dB noise figure ($T_{sys} \approx 1200$ K), RF bandwidths of 6 GHz and 1 GHz, and a frame rate of 30 Hz, $\Delta T_{vis} \approx 0.08 \times N$ and $0.21 \times N$, respectively. The image sensitivity can be further reduced by using redundant baselines. The code modulation interferometry provides information about all possible baselines and thus redundant information can also be demodulated and used to average out receiver noise. For an array with $D_{min} = \lambda$ the FOV is $\theta_{max} = \pm1/2\ $rad (approx $\pm30 ^\circ$). For a 60-GHz system with 6-GHz BW, $D_{max} \approx 50 $ mm which gives a resolution of $5.7^\circ$ for a $\pm 30^\circ$ FOV \cite{Lettington03}. The resolution can be further improved by dividing the total bandwidth into narrower bandwidths so that the correlation could be extended over larger baselines. Reducing the bandwidth would reduce the sensitivity or increase the integration time for constant sensitivity. This is a trade-off between sensitivity, resolution, and frame rate. 

Sensitivity and resolution challenges are inherent to using interferometry for radio astronomy and even more stringent for imaging, remote sensing and similar applications. These have been been discussed in detail in \cite{Levine88} with regards to remote sensing. Ample research exists related to interferometry enhancement, including hardware instrumentation, calibration and imaging enhancement algorithms (such as the CLEAN algorithm \cite{Thompson08}). These can be applied as well to code-modulated interferometry.

%\color{black}

\section{Conclusion}
In conclusion, with the help of preliminary results from behavioral models and the analysis for system metrics we presented that the code-modulated interferometry appears to be a promising and cost-effective alternative to costly imaging systems present in the market today. One important challenge for code-modulated interferometry is scalability to larger systems which needs to be investigated and possible improvements  to be implemented. This is the scope of future work. The Walsh BOCP codes place a hard limit on the number of elements that can be used simultaneously in a CMI imager. A possible solution is to divide the scene into smaller sectors and use multiple chips. Several different kinds of codes can be explored, such as m-sequences, Gold codes, complex Walsh codes, or pseudo-random codes to mitigate this scalability limitation.
\chapter{Demonstration using In-house 60-GHz Phased Array}
\label{chap-four}

\section{Overview}
The behavioral models in the previous chapter have successfully demonstrated the feasibility of the proposed code-modulated interferometric imaging system. The behavioral model has been tested with noise from the circuit but does not include the artifacts such as non-linearity, feed through, mismatches and other non-idealities that can occur in any real system. The next natural step is to build a working prototype of the system to test the system performance and image resolution in presence of such issues. The concepts developed can be applied to both commercially available or custom designed phased arrays. The implementation of imaging system at millimeter-wave frequencies is desirable for higher resolution, but we found it rather difficult to procure commercially available millimeter-wave phased arrays. Therefore for the system to be evaluated, we create a custom chip, a custom package, a custom antenna, and then a custom board.

In this chapter, we discuss the implementation of the code-modulated interferometry using an in-house communication phased array and how to re-purpose it into an imager. We present a simple 13-pixel CMI system operating in the license-free 60-GHz band using a four-element phased-array receiver developed for IEEE 802.11ad (WiGig). The array integrated chip is packaged with compact antenna structures. With minimum redundancy, this four elements phase-array can be used to built a four-element imager that can provide up to six different baselines and up to thirteen pixels with complex visibilities. 

\section{Four-element Code-modulated Interferometry} \label{sec:four_element_CMI}
\begin{figure}
\centering
\includegraphics[width=.6\textwidth]{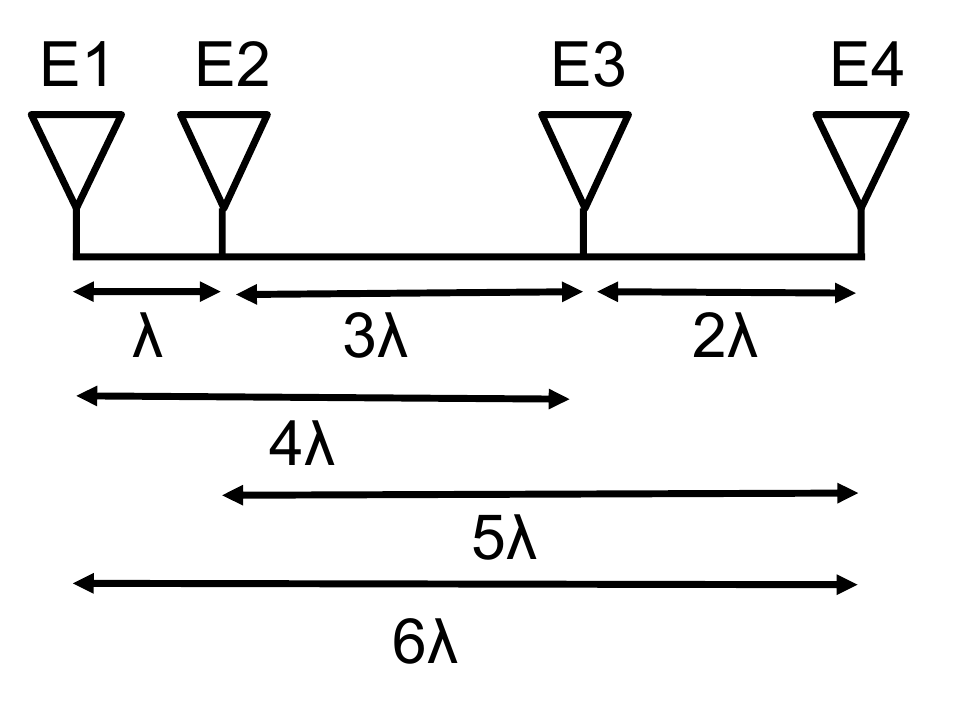}
\caption{Zero redundancy interferometric array using four antennas. The antennas are spaced at $ \lambda $, $ 3\lambda $ and $ 2\lambda $, producing baselines of length of $0 - 6\lambda $.}
\label{fig:4_element}
\end{figure}

This section elaborates the code-modulated interferometry for a four-element phased array. For zero redundancy, one of several preferred placement of four antennas (Elements E1, E2, E3 and E4) is shown in Fig. \ref{fig:4_element}. In this configuration the four-elements provide a complex visibility function with 13 discrete sample points: six discrete points by pair-wise complex cross-correlations of the four elements, their complex conjugates, and one zero-baseline point (\textit{i.e.}, autocorrelation of any element). For example, elements (E1,E2) provide a baselength of  $1\lambda $, elements (E2,E3) provide a baselength of  $3\lambda $, elements (E2,E4) provide a baselength of  $4\lambda $, and so on. The discrete visibility function for these elements can then be represented as:

\begin{equation}
V = [v_{6\lambda}^*,\ v_{5\lambda}^*,\ v_{4\lambda}^*,\ v_{3\lambda}^*,\ v_{2\lambda}^*,\ v_{1\lambda}^*,\ v_{0},\ v_{1\lambda},\ v_{2\lambda},\ v_{3\lambda},\ v_{4\lambda},\ v_{5\lambda},\ v_{6\lambda}]
\end{equation}
where $v_{n\lambda}$ represents the complex visibility for a baselength = $n\lambda$, $n=0-6$ and $v_{n\lambda}^*$ is its complex conjugate. This can then be written in terms of the element number from Fig. \ref{fig:4_element} as:
\begin{equation} \label{four_vis_set}
V = [v_{1,4}^*,\ v_{2,4}^*,\ v_{1,3}^*,\ v_{2,3}^*,\ v_{3,4}^*,\ v_{1,2}^*,\ v_{1,1},\ v_{1,2},\ v_{3,4},\ v_{2,3},\ v_{1,3},\ v_{2,4},\ v_{1,4}]
\end{equation}
where $v_{n,m}$ is the complex visibility for elements ($En, Em$), $n,m= 1-4$. 

The visibility function is obtained in traditional interferometric imaging by cross-correlating the signals received by pairs of antennas using a bank of complex correlators. Since the correlations are typically evaluated in the digital domain, an ideal interferometer array would include complete receive chains in parallel for each antenna. By definition, the visibility is calculated by complex cross-correlation operation, wherein the signals from two antennas are multiplied and averaged for long period of time. This operation can be mathematically represented as $v_{n,m} = \overline{s_{i,n}s_{i,m}\raisebox{2.5mm}{}} + j. \overline{s_{i,n}s_{q,m}\raisebox{2.5mm}{}}  $  and $v_{n,m}^* = \overline{s_{i,n}s_{i,m}\raisebox{2.5mm}{}} - j. \overline{s_{i,n}s_{q,m}\raisebox{2.5mm}{}}  $, where $s_{i,n}$ and $s_{q,n}$ are the in-phase and quadrature-phase components of the signal in the $n^{th}$ element.   

In contrast to this, as we know the phased array consists of multiple phase-shifted elements whose signals are combined and processed through a shared receiver path. To reconfigure a phased array into an interferometer array, CMI applies orthogonal in-phase and quadrature-phase codes to each signal using the already-present phase shifters. Encoded signals are then combined and the aggregate signal is processed coherently in single hardware path. For a four-element phased array the aggregate signal, $s_{sum}$, is represented as
\begin{equation}
\label{eq:summ4}
s_{sum}=\sum_{n=1}^{4} (c_{i,n}s_{i,n}+c_{q,n}s_{q,n}) \ 
\end{equation}
where, $c_{i,n}$ and $c_{q,n}$ represent the codes applied to in-phase and quadrature-phase components $s_{i,n}$ and $s_{q,n}$ of the signal in the $n^{th}$ element, $n=1,2,3,4$. In this example, we assume a vector interpolator phase shifter which allow simultaneous and independent code-modulation of in-phase and quadrature-phase components of each signal. The visibility functions can then be directly detected using a scalar power detector which results in a baseband signal which contains all possible complex cross-correlations of signals (\textit{i.e.}, visibilities), each of which are modulated by code products, as follows:
\begin{equation}
\begin{split}
p = [s_{sum}]^2=\sum_{n=1}^{4}\sum_{m=1}^{4}(c_{i,n}c_{i,m}s_{i,n}s_{i,m} + c_{q,n}c_{q,m}s_{q,n}s_{q,m} +  c_{i,n}c_{q,m}s_{i,n}s_{q,m}) 
\end{split}
\end{equation}

\begin{table}[]
    \caption{Matrix showing the combinations of in-phase and quadrature-phase signals needed to be multiplied to obtain real and imaginary components of visibility samples.}
    \centering
    \begin{tabular}{c}
\includegraphics[clip,trim=0 0 0 0,width=.95\textwidth]{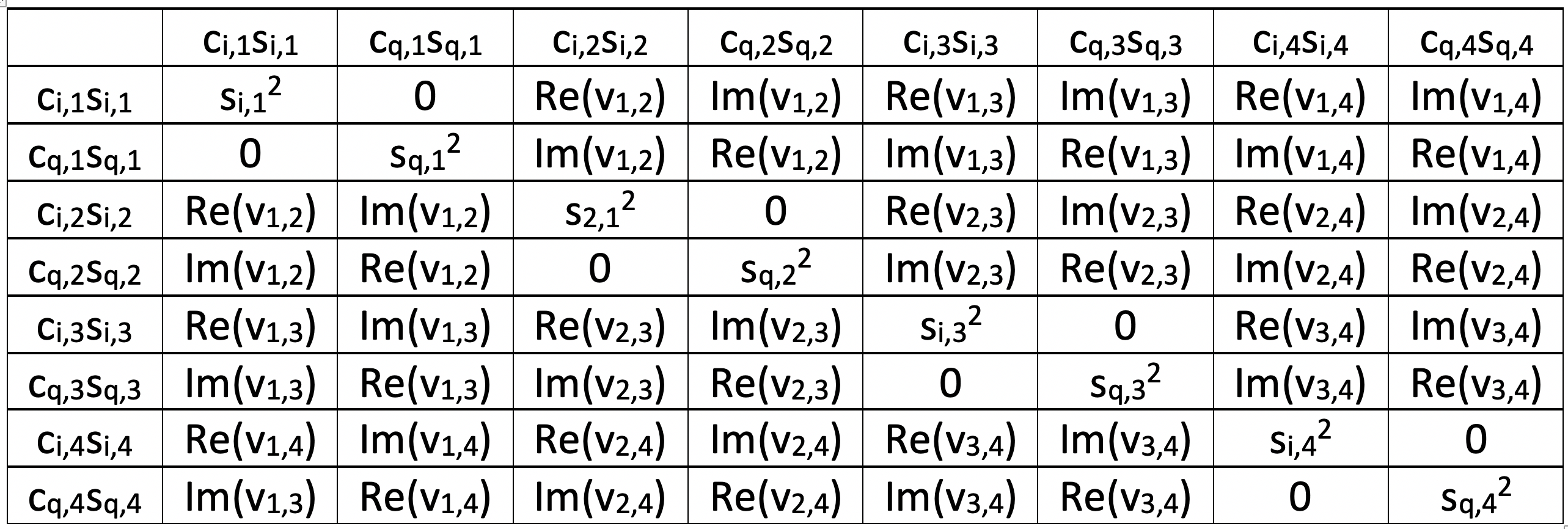}
    \end{tabular}
    \label{tab:vis_table}
\end{table}

The squared output includes a DC term proportional to the total power of all signals and then multiple code-modulated correlation terms. The cross-correlation between $i$ and $q$ signals within each element (\textit{i.e.}, $n$=$m$)
should ideally be zero. For $n$$\neq$$m$, the terms include the real and imaginary cross-correlations (or visibilities) modulated according to \textit{code-products}. All complex visibilities can then be simultaneously demodulated by correlating the squared signal with each code product of interest (\textit{e.g.}, $c_{i,n}c_{i,m}$ for real and  $c_{i,n}c_{q,m}$ for imaginary components of complex visibility $V_{nm}$) and integrating for the time period of code, provided the code-products are orthogonal. For example, demodulation of complex visibility sample $v_{n,m}$ is shown below: 
\begin{equation}
v_{n,m} =\ \overline{(c_{i,n}c_{i,m} + j.c_{i,n}c_{q,m}).p \raisebox{2.5mm}{}} \ =\ \overline{s_{i,n}s_{i,m}\raisebox{2.5mm}{}} + j. \overline{s_{i,n}s_{q,m}\raisebox{2.5mm}{}}   
\end{equation}

Table \ref{tab:vis_table} shows the multiplication pairs to produce the real and imaginary components of the visibilities. As evident from Table \ref{tab:vis_table}, there is redundancy in obtaining the visibility samples. These redundancy can be utilized to improve sensitivity. Note that CMI can be applied to any phased array with at least two-bit programmable phase shifters and does not require any additional hardware. 

The one-dimensional visibility function, $V(u)$ is related to the brightness distribution ($T_{\Omega}$) through a Fourier transform, as follows
\begin{equation}
\label{eqn:vis_1D}
V(u) = \int_{0}^{\pi} T_{\Omega}(\theta) \ e^{j 2 \pi (u \cdot l ) } \  dl ,
\end{equation}
where $l=sin\theta$ and $u$ is baseline vector equal to one-dimensional antenna spacings per unit wavelength. A discrete inverse Fourier transform (DIFT) of the measured visibility samples provide the image $T_{\Omega}$ :

\begin{equation}
        T(i\Delta l) =\sum_{p=0}^{N-1}V(p\Delta u)\cdot e^{j2\pi\frac{ip}{N}}
        \label{eq:fourier_discrete}
    \end{equation} 
    
With a complex visibility function with 13 discrete samples, an image with 13 pixels is obtained.  

%%%%%%%%%%%%%%%%%%%%%%%%%%%%%%%%%%%%%%%%%%%%%%%%%%%%%%%%%%%%%%%%%%%%%%%%%%%%%

\section{60-GHz Hardware Prototype}

\subsection{Circuit Architecture}

A four-element receiver array was fabricated with Dr. Kevin Greene \footnote{North Carolina State University, Raleigh, NC USA 27606} using the TowerJazz 0.13 $\mu m$ SBC18H3 technology, which features NPN transistors with a maximum oscillation frequency ($f_\text{max}$) of 280~GHz and a six metal back-end-of-line (BEOL). This four-element 60-GHz phased-array receiver is then used to demonstrate our approach of code-modulated interferometry.

\subsubsection{Low-noise Amplifier (LNA)}

\begin{figure}
\centering
\includegraphics[width=1\textwidth]{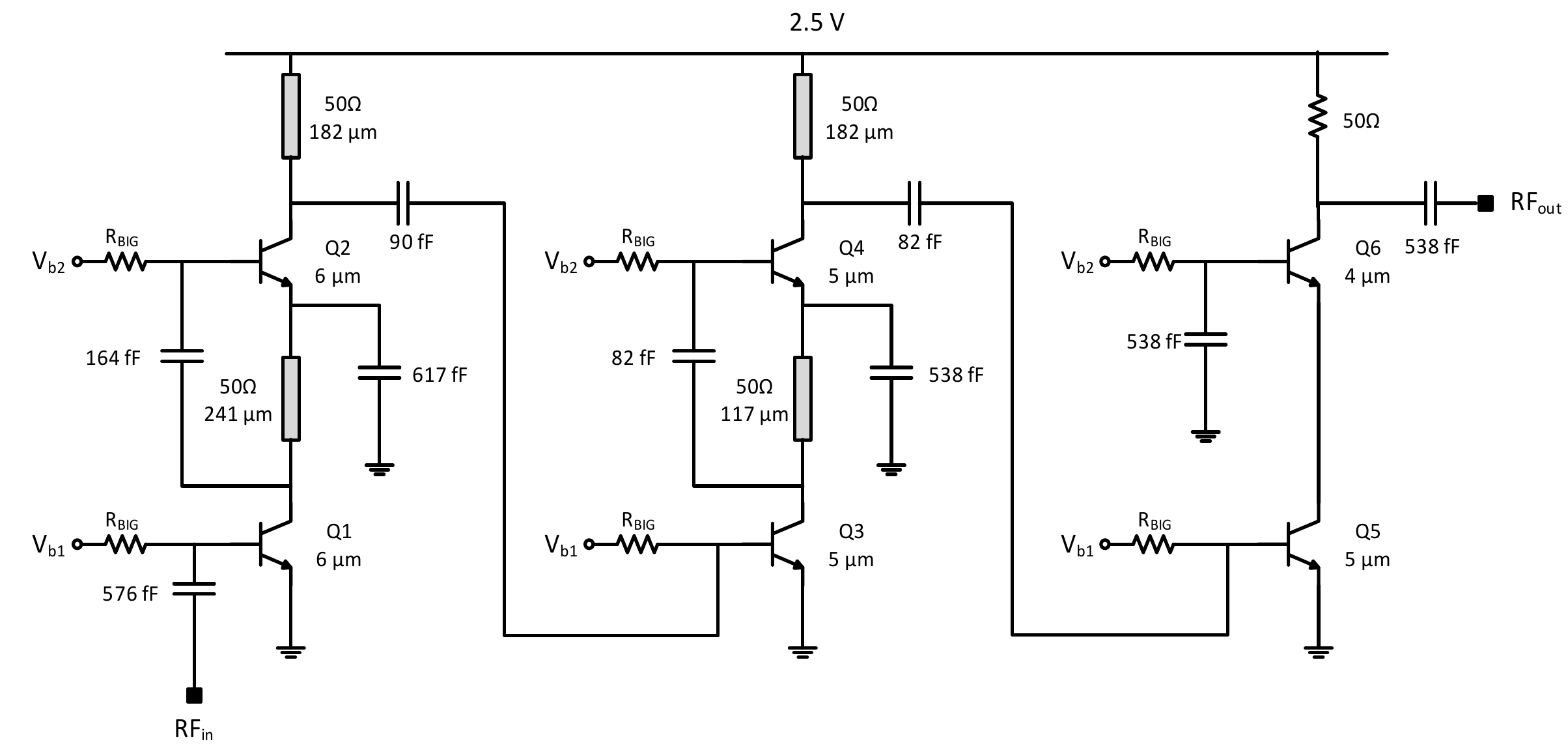}
\caption{Schematic of the 5-stage LNA using stagger-tuned current-sharing common-source amplifiers \cite{Greene2018}.}
\label{fig:LNA_schematic}
\end{figure}

Each element in the phased array includes an LNA as the first stage to limit the noise added by the receiver. A five-stage amplifier shown in \fref{fig:LNA_schematic} is designed and fabricated. The first four stages use a current-sharing topology where two common-emitter amplifiers are stacked in the same DC current path. This allows a 2.5~V supply to be shared between the LNA and phase-shifter, where each common-emitter gain stage sees approximately 1.25~V across the collector and emitter terminals. A transmission line is used for inter-stage matching between each transistor-pair in the current sharing paths and then a large bypass capacitor is placed at the emitter of the top transistor to form a small-signal ground. Finally, a cascode amplifier is used for the fifth stage with a 50~$\Omega$ resistive load to ensure a broadband match to the phase-shifter. Impedance matching throughout the design is stagger tuned to provide a flat gain response from 57 to 66~GHz.

\begin{figure}
   \centering
   \includegraphics[width=.8\textwidth]{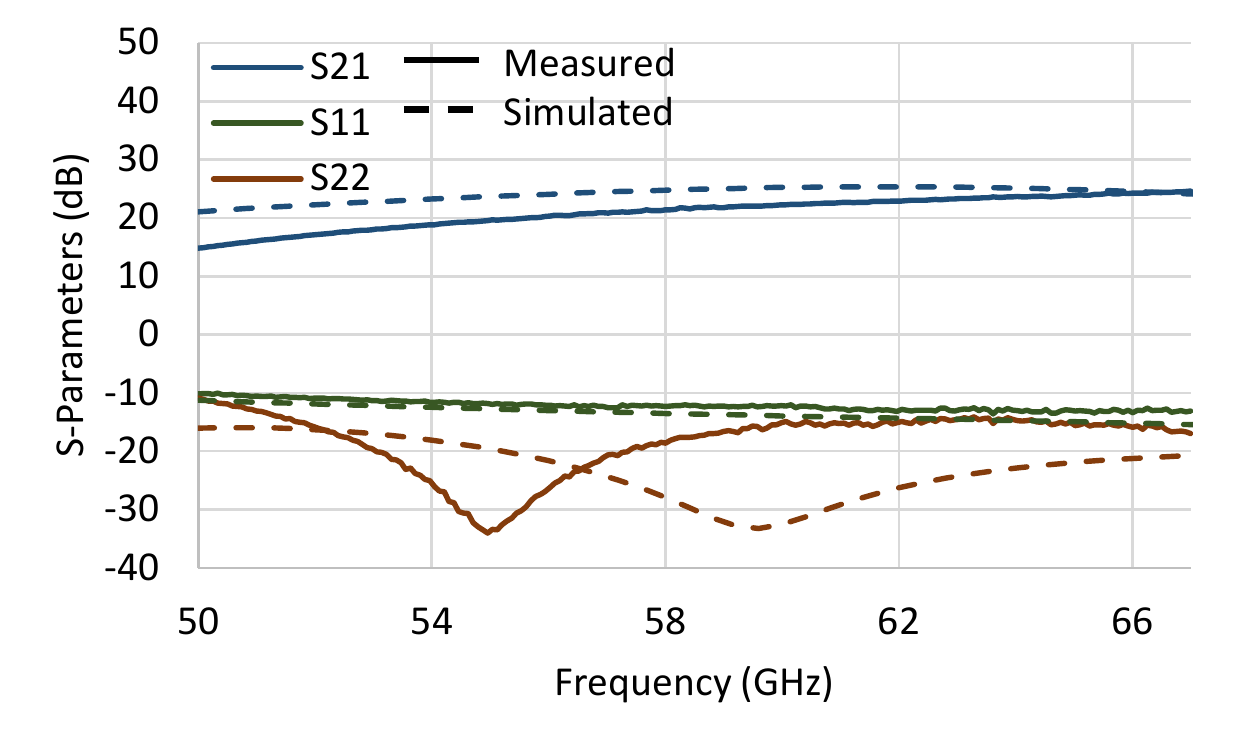}
   \caption{Measured $S$-parameters for the LNA break-out circuit used in the phased array as compared to simulated results. \cite{Greene2018}.}
   \label{fig:LNA_spar}
 \end{figure}

A break-out circuit for the LNA has fabricated and the $S$-parameters have been measured using a VNA with results as shown in \fref{fig:LNA_spar}. The input and output matching shows reasonable agreement compared with simulation, achieving better than 10~dB return loss from 50~GHz to greater than 67~GHz. The peak gain response shifted to approximately 66~GHz, roughly 4~GHz higher than in simulation, owing to this being our first time using the TowerJazz process. The gain at 57~GHz is 21~dB and increases to 24~dB at 66~GHz. Although unintentional, this upward shift in frequency helps to flatten the overall array gain across frequency. Due to equipment limitations the noise figure of the breakout could not be measured, but it is expected to match reasonably well to the simulated value of 5.5~$\pm$~0.1~dB across the band. Power consumption of the LNA is approximately 25~mW \cite{Greene2018, KevinPhD}.

\subsubsection{Phase Shifter}

\begin{figure}
  \centering
  \includegraphics[width=0.5\textwidth]{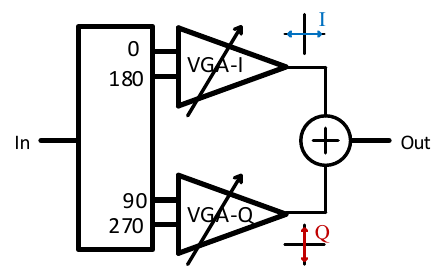}
  \caption{Block diagram of the active vector interpolator topology implemented as each elements' phase shifter \cite{Greene2018}.}
  \label{fig:Vector_Interpolator}
\end{figure} 
\begin{figure}
  \centering
  \includegraphics[width=0.75\textwidth]{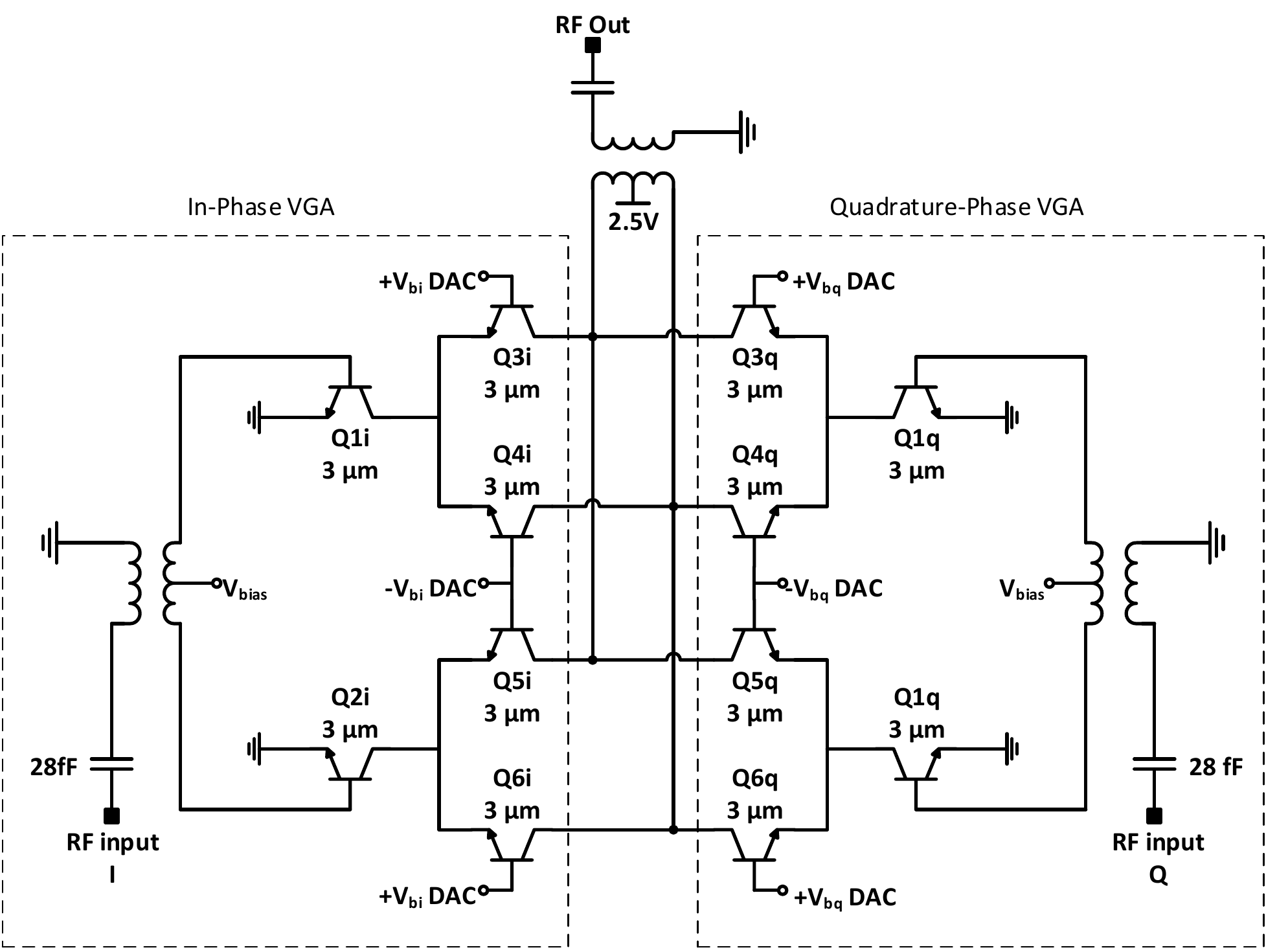}
  \caption{Schematic of the vector interpolator using cross-coupled Gilbert cells to form the variable-gain amplifiers \cite{Greene2018}.}
  \label{fig:PS_schematic}
\end{figure} 

Following the LNA is an active vector interpolator designed by Kevin Greene, whose block diagram and schematic are shown in \fref{fig:Vector_Interpolator} and \fref{fig:PS_schematic}. Simulated NF for the phase-shifter is 15--16 dB and power consumption is approximately 22~mW. This phase shifter provides six-bit phase resolution for normal phased-array operation and Cartesian encoding capability for code-modulation. The input signal to the phase shifter is first split into differential-quadrature signals and then fed into I- and Q-path variable gain amplifiers (VGAs). The gain of each VGA is adjusted to provide a desired phase and amplitude response. This architecture leads to a straight-forward application of  Cartesian encoding as the signs of the $I$- and $Q$-path VGAs need only be commutated according to their respective codes \cite{Greene2018, KevinPhD}.

\subsubsection{Power Detector}
\begin{figure}
\centering
\includegraphics[width=0.8\textwidth]{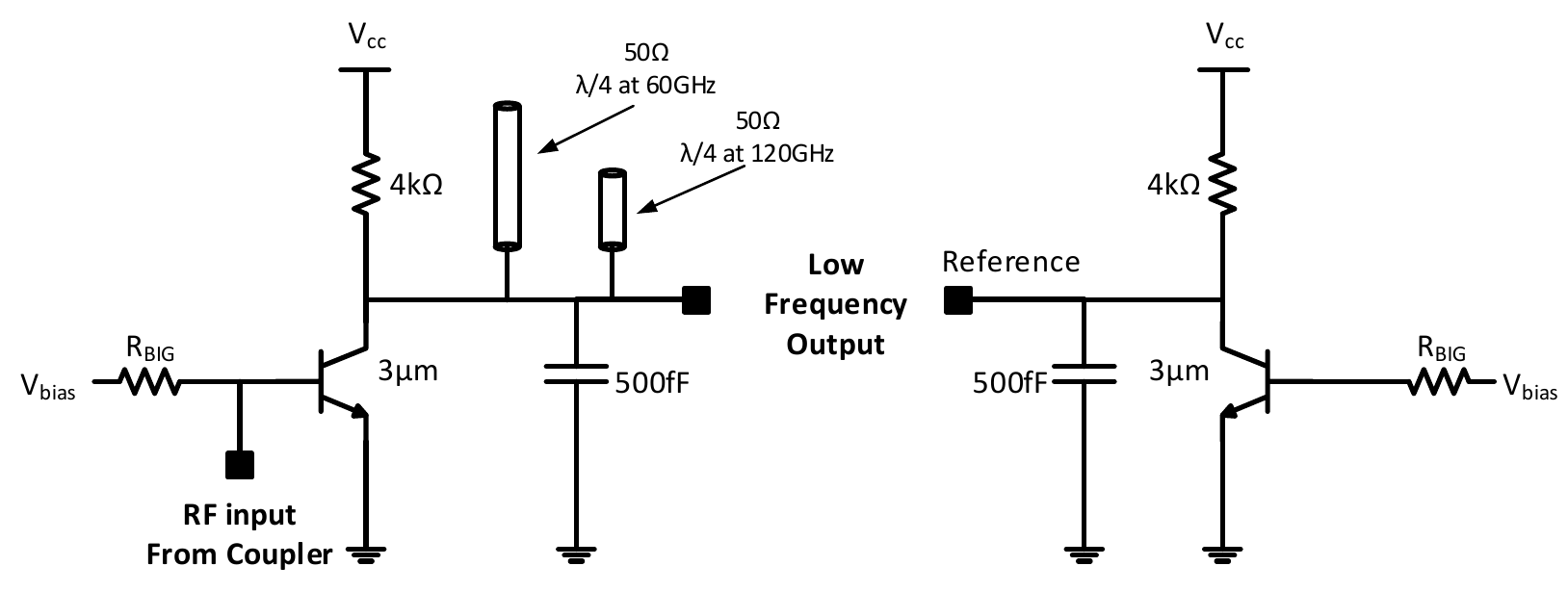}
\caption{Power detector schematic used to square the combined code-modulated signals and generate a low-frequency signal to be captured by an external data converter \cite{Greene2018}.}
\label{fig:PD_schematic}
\end{figure}

One key attribute of code-modulated interferometry is the use of direct detection with a single power detector to provide the squaring of the code-multiplexed signal. This avoids the need to provide coherent quadrature down-conversion with on-chip mixers and local-oscillator (LO) signals, hence reducing the complexity. The RF out signal is coupled into a common-emitter power detector circuit, biased at low current density to perform the squaring function \cite{May2010} as shown in Fig. \ref{fig:PD_schematic}. The power detector must maintain a linear output voltage relative to input power for the highest possible output power of the array to avoid distortion in the down-converted signal. The highest output power occurs when all element are operating in-phase resulting in an $N$ times higher output power of a single element, where $N$ is the number of elements in the array.

\subsubsection{Four-element Array}
\begin{figure}
 \centering
\includegraphics[width=1\textwidth]{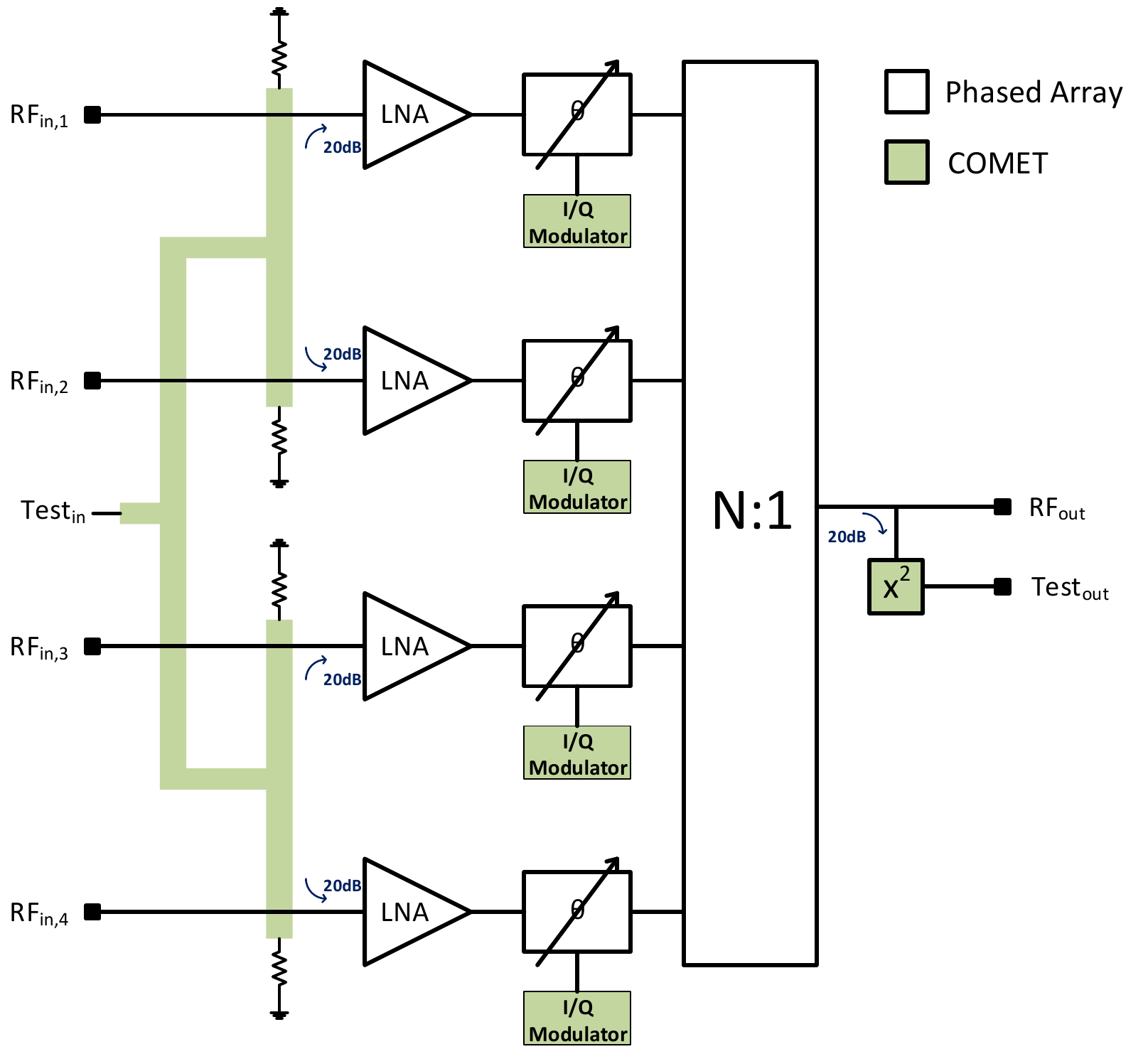}
	\caption{Block diagram of the four-element phased array receiver \cite{Greene2018, KevinPhD}.}
	\label{fig:Phased_Array_sche}
\end{figure}

\begin{figure}
\centering
\includegraphics[width=.8\textwidth, angle=0]{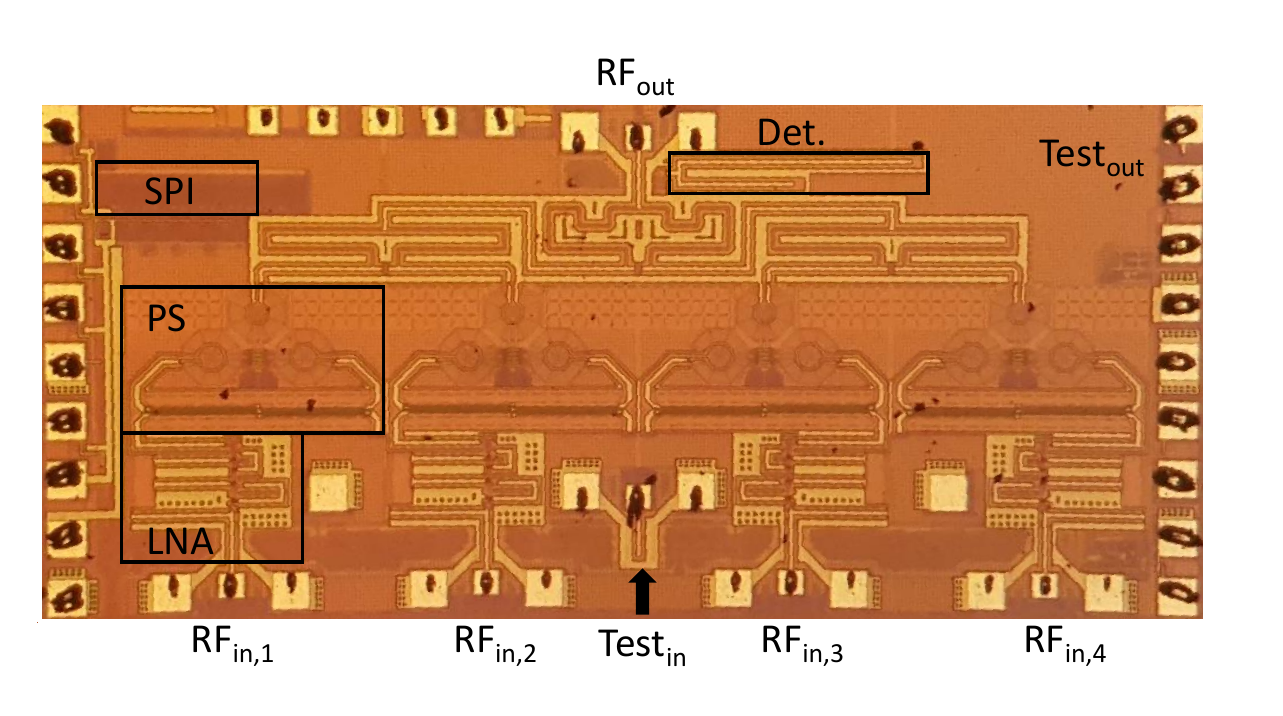}
	\caption{ Die photo. The total area is 3 mm x 1.3 mm \cite{Greene2018, KevinPhD}.}
	\label{fig:Phased_Array_die}
\end{figure}

A four element phased array receiver is built as shown in the block diagram in Fig. \ref{fig:Phased_Array_sche} and die photo in Fig. \ref{fig:Phased_Array_die}.
Each element is comprised of an LNA and an active vector-interpolator phase-shifter whose phase setting is controlled using an on-chip serial programmable interface (SPI) located in the upper left corner of the die. On-chip modulators are also included which can be used to code-modulate the signals by controlling the phase in the phase shifters. Using these modulators, Rademacher codes can be initially applied to the $I$ and $Q$ components of each phase-shifter and their known code products can then be applied to digitally demodulate the cross-correlations at baseband. An interesting property of the Rademacher codes is that the frequency of alternating $+1$ to $-1$ differs by a factor of two (hence also known as divide-by-two codes) making it simple to generate on-chip using a frequency divider. The output of each element is then combined using a cascaded 4:1 Wilkinson power combiner, where the composite output can be measured directly using RF probing. Also, the combined output is weakly coupled to the power detector. The phase shifter provides six-bit phase resolution for normal phased-array operation and Cartesian encoding capability for code-modulation. The phase shifter, power combiner and the power detector were designed by Dr. Kevin B. Greene and more details can be found in \cite{KevinPhD}. Note that this array has two outputs, a 60GHz mm-wave output that can then be used for stacking multiple arrays for larger systems as well as a power detector DC output which can be used for four-element one-dimensional interferometry.

 \begin{figure}
 \centering
\includegraphics[width=0.8\textwidth]{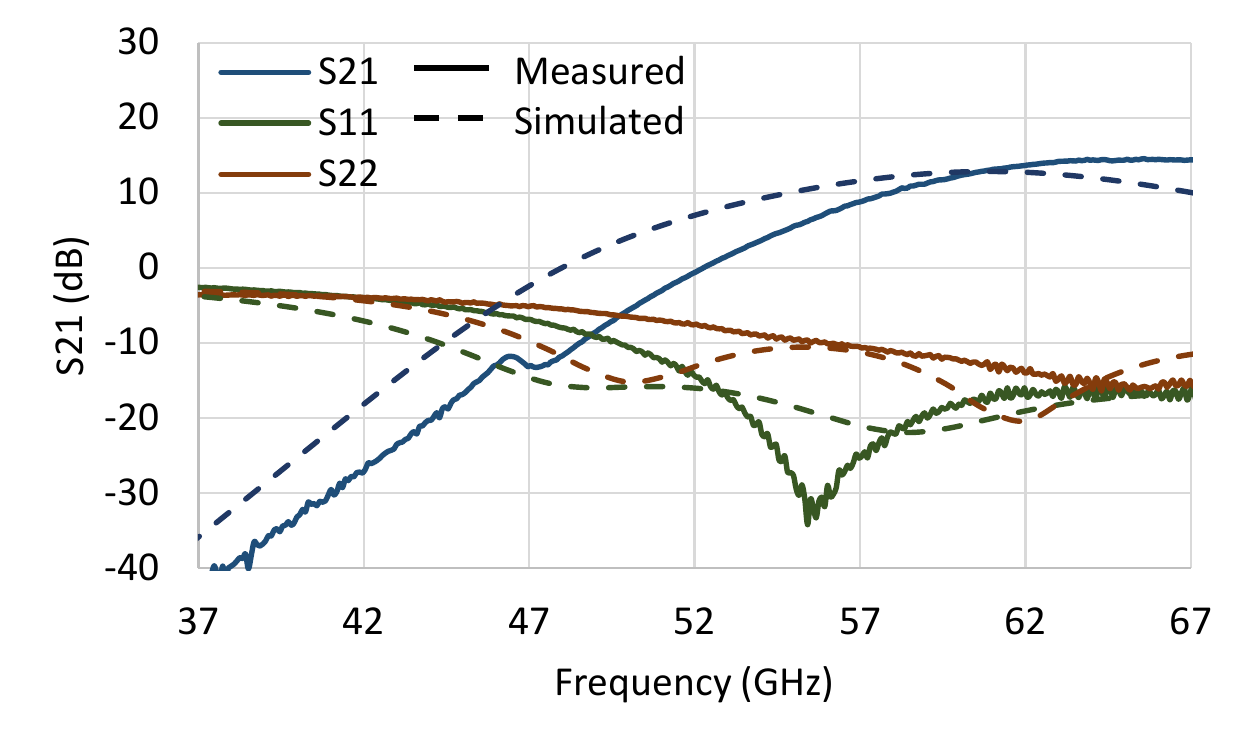}
   \caption{Measured $S$-parameters compared to simulated results for element one of phased array for the 0-degree phase setting \cite{Greene2018, KevinPhD}.}
   \label{fig:Array_spar}
 \end{figure} 

The chip is measured in lab using wafer probing techniques.  The measured results of phased array are shown in \fref{fig:Array_spar}, and then the key performance is summarized in Table \ref{tab:phased_array_perfo}. Each element within the array achieves a gain of approximately 14~dB. Power consumption for each element is 52~mW. The coding can be done by either using the on chip frequency dividers or serial programmable interface (SPI). The chip also features circuitry for built-in-test technique called as Code Modulated Embedded Test, or CoMET \cite{Greene2016}. CoMET uses the code-modulated interferometry and BOCP codes to characterize and calibrate the phased array, and is a secondary contribution of this research \cite{Greene2016, CometIMS19}. This technique has been demonstrated for this particular 60 GHz phased array in \cite{Greene2018}. 

Using CMI, this phased array is repurposed into an interferometer to create a simple imaging system at 60 GHz. Although on-chip modulators (Rademacher codes implemented using frequency dividers) have been included to allow automatic code generation, here we choose to apply the codes directly using the serial interface. 

\begin{table}
	\caption{Measured performance summary for a single channel of the 60GHz phased array}
	\label{tab:phased_array_perfo}
	\begin{center}
		\begin{tabular}{c c}
        
        		\hline
				\multicolumn{1}{|c|}{Technology} & \multicolumn{1}{c|}{ $130nm$ SiGe}  \\	
                 \hline
				\multicolumn{1}{|c|}{3dB bandwidth} & \multicolumn{1}{c|}{$10GHz$ ($60-70GHz$)}  \\	
                \hline

				\multicolumn{1}{|c|}{Gain} & \multicolumn{1}{c|}{$14 dB$}  \\	
                \hline
				\multicolumn{1}{|c|}{RMS amplitude error} & \multicolumn{1}{c|}{$< 0.4dB$}  \\	
				\hline
				\multicolumn{1}{|c|}{Noise Figure} & \multicolumn{1}{c|}{$6 dB$*}  \\	
                \hline
				\multicolumn{1}{|c|}{Power consumption} & \multicolumn{1}{c|}{$52 mW$ per element}  \\	
               \hline
				\multicolumn{1}{|c|}{Number of bits} & \multicolumn{1}{c|}{$6$}  \\	
                \hline
				\multicolumn{1}{|c|}{Phase error} & \multicolumn{1}{c|}{$<6^\circ$}  \\	
                \hline
				\multicolumn{1}{|c|}{oP1dB} & \multicolumn{1}{c|}{$-19.5dBm$}  \\	
                \hline
				\multicolumn{1}{|c|}{Chip area (with pads)} & \multicolumn{1}{c|}{$3.9mm^2$}  \\	
                \hline

		\end{tabular}
	\end{center}
  \hspace{10.5em} \textsuperscript{*}\footnotesize{Simulated}
\end{table}

%%%%%%%%%%%%%%%%%%%%%%%%%%%%%%%%%%%%%%%%%%%%%%%%%%%%%%%%%%%%%%%%%%%%%%%%%%%%%

\subsection{Packaging}
\label{sec:baord}

The 60-GHz phased array IC is wire-bonded onto a multilayer printed-circuit board (PCB) which includes Rogers 5880 on an FR4 substrate, as shown in Fig. \ref{fig:stack}. The board design showing all metal layers is given in Appendix \ref{Appendix_board_layers}. The low dielectric losses of Rogers 5880 substrate is suitable for high-frequency operations. The top and bottom copper metal layers are 1 oz. thick while the middle two layers are 0.5 oz. thick. To reduce the length of RF wirebonds, the diced chip is placed in a 10 mil cavity on board. 

\begin{figure}
\centering
    \includegraphics[width=.6\textwidth]{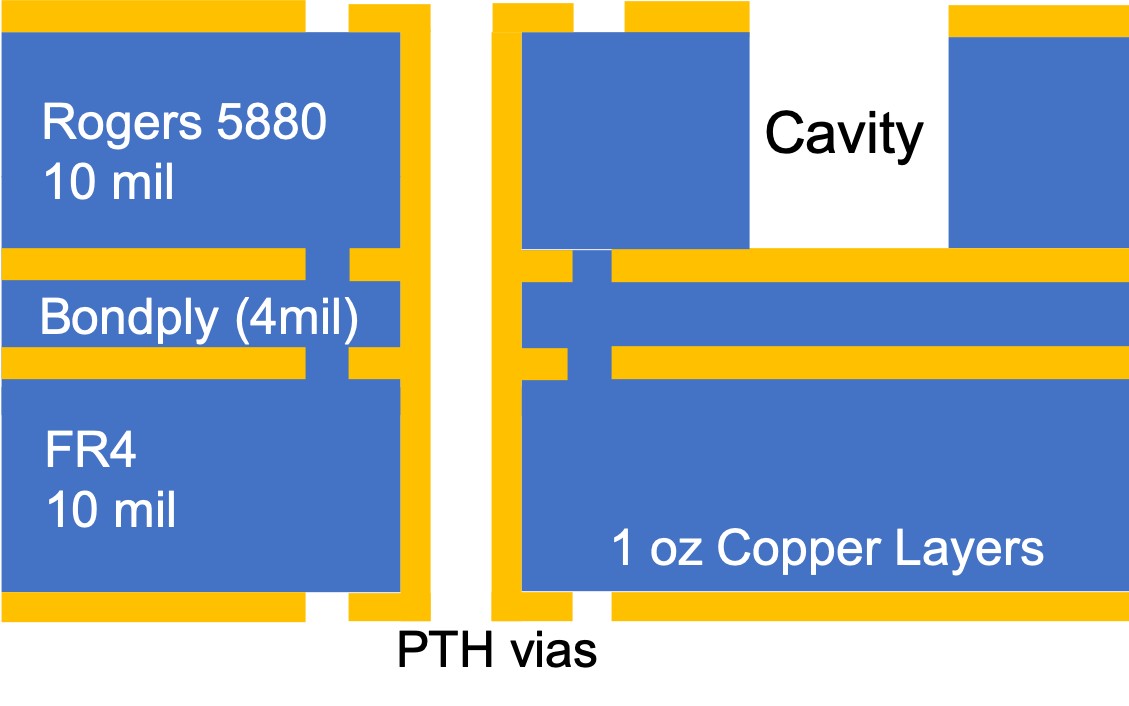} 
\caption{Stack-up of printed-circuit board.}
    \label{fig:stack}
\end{figure}

\begin{figure}
\centering
    \includegraphics[width=.6\textwidth]{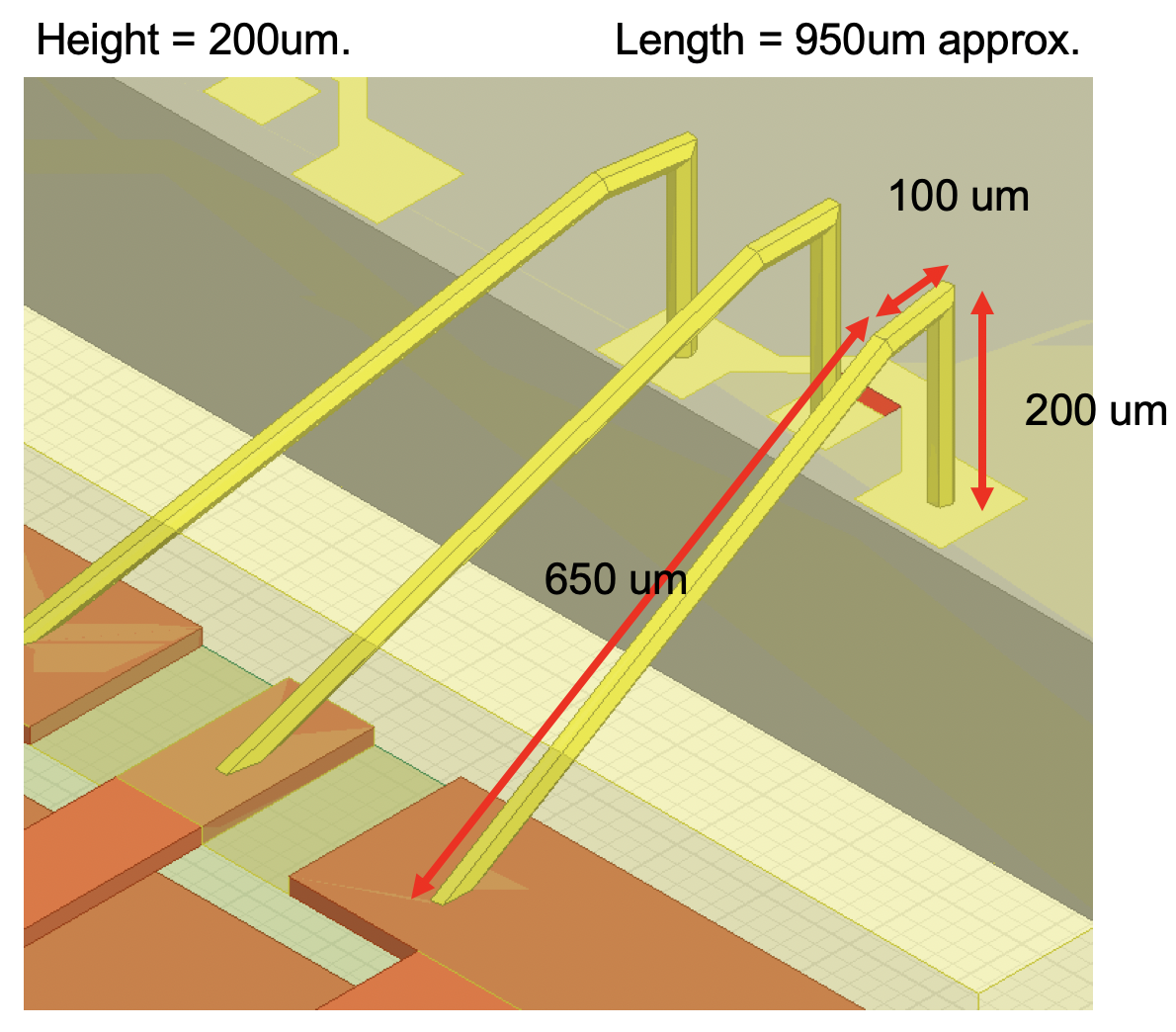} 
\caption{HFSS simulation of G-S-G wirebonds.}
    \label{fig:wirebondHFSS}
\end{figure}

\begin{figure}
 \centering
	\subfloat[]{\includegraphics[clip,trim=0cm 0cm 0cm 0cm, width=0.7\textwidth]{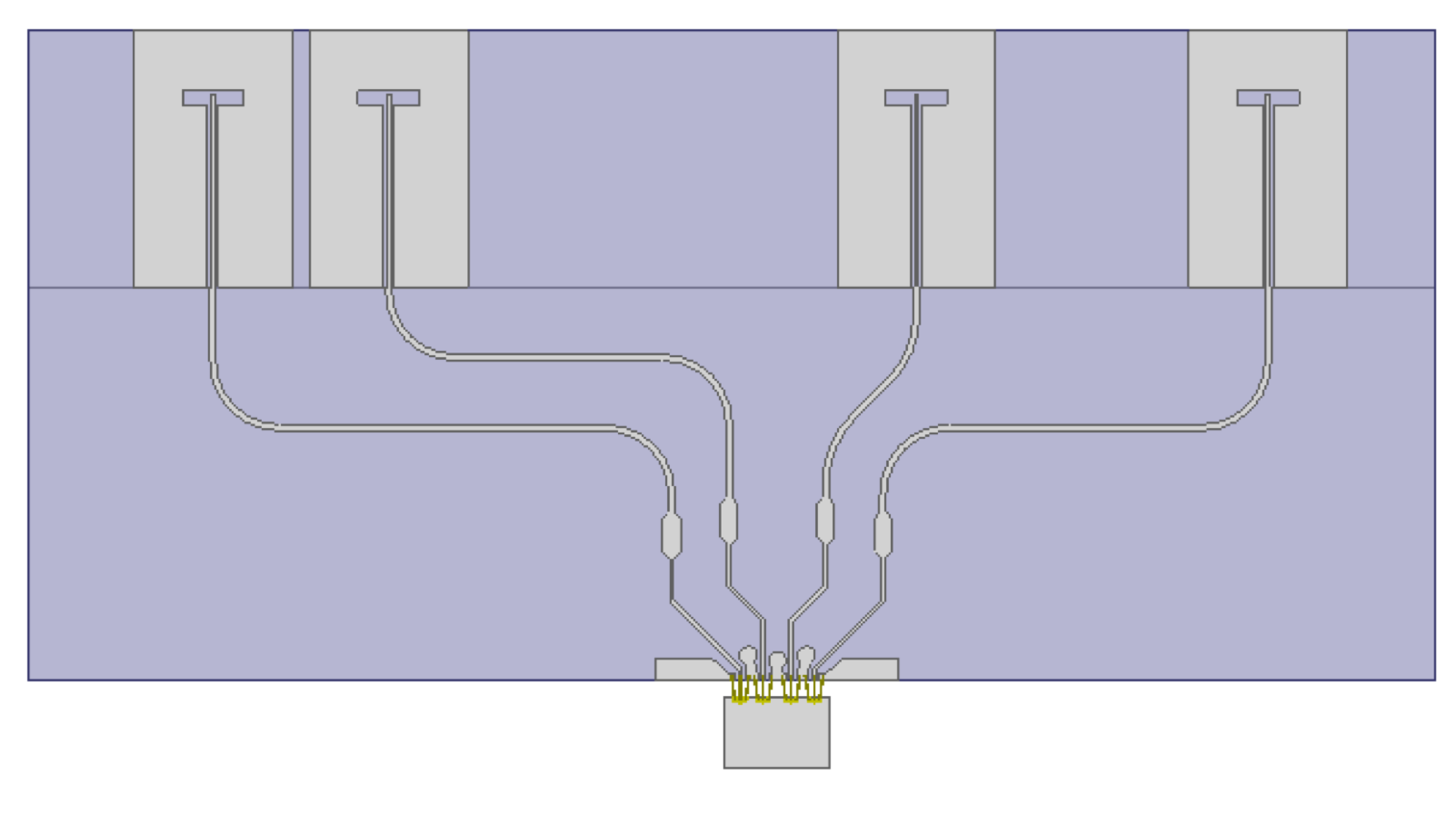}
		\label{fig:feedlines}} \\
	\subfloat[]{\includegraphics[clip,trim=0cm 0cm 0cm 0cm, width=0.6\textwidth]{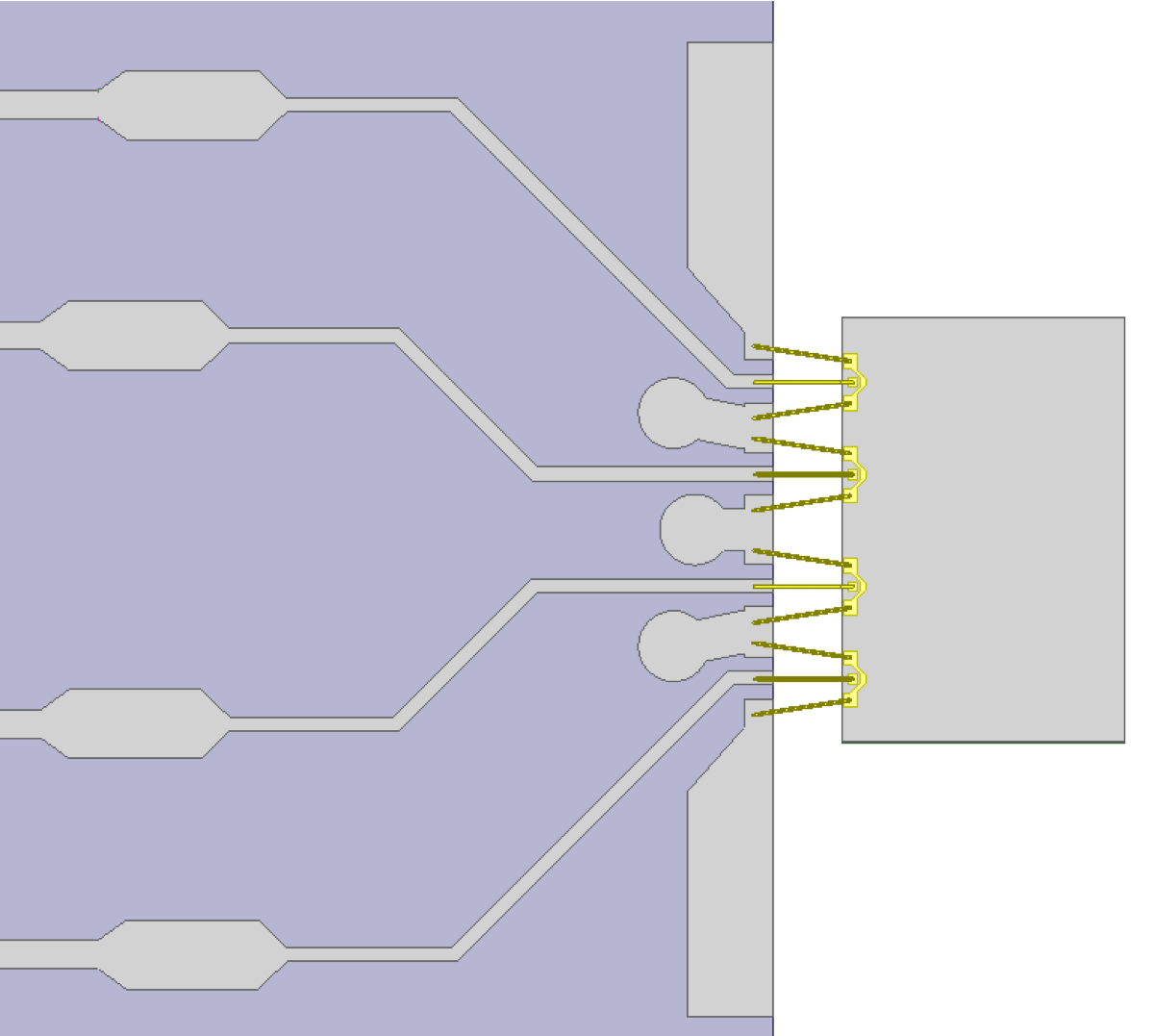}
		\label{fig:wirebonds}} 
	\caption{EM structures for (a) antennas with feedlines; and (b) wire-bonds, compensation, and CPW-to-microstrip transition.}
	\label{fig:boardEM}
\end{figure}

The small pad sizes on the chip prevented us from a low loss flip-chip transition suitable of mm-wave signals, therefore wire-bonding is used for all mm-wave and DC signals. However, to reduce the lengths of the mm-wave wire-bonds, and therefore reduce losses, following steps are taken: 

\begin{itemize}
\item The chip is placed in a cavity on board. The depth of cavity is approximately 10 mil, close to the chip thickness of 11mil to bring the pads on board and pads on chip at the same horizontal level. 
\item The fabricated die originally (4.75 mm X 4.75 mm) consisted of multiple test circuits and break-outs surrounding the phased array. To bring the die edge close to mm-wave pads on chip, the phased array (1.3 mm X 3 mm) is carefully diced out using a low tolerance 45um curve ZH05-5D1500-NI-50 blade and ADT 7100 dicing saw \footnote{Chip dicing performed at Novus Energy Technologies, Raleigh, NC, USA}. 
\end{itemize}

The board was design with the help of Dr. Dong Gun Kam and Haekya Seo \footnote{Department of Electrical and Computer Engineering, Ajou University, Yeongtong-gu, Suwon, 443749 South Korea.}. Ground-signal-ground (G-S-G) wirebonds are used to connect the chip to the PCB. To capture the affects of bond-wire, 1 mil Aluminum G-S-G wire bonds with an encapsulant are EM simulated in Ansys HFSS. The encapsulant (Nusil R-2186) around the wire-bonded chip provides physical protection, as well as its known dielectric constant allows more accurate simulation of wire-bonding. The loop height and length are carefully controlled during wire-bonding process to match the simulated conditions. A compensation network is included on the PCB to resonate wire-bond parasitic inductance. Several techniques are available to resonate the wire bond inductance in literature \cite{Zhang09,Sun11}. A fan-out structure and coplanar waveguide (CPW) to microstrip transitions are implemented on board, as shown in Fig. \ref{fig:boardEM}. Figs. \ref{fig:wirebondHFSS} and \ref{fig:wirebonds} show the simulated wire-bond structure, sizes, and the compensation network designed to resonate out wire-bond parasitic inductance. Due to the low dielectric constant of Rogers' substrate ($ \epsilon_r \approx 2.2$), and the minimum trace width/spacing requirements from board manufacturer, the coplanar wave-guide transmission lines tend to be wider than G-S-G pads on chip. Hence, the mm-wave G-S-G signal from chip transition onto a micro-strip transmission line while two side grounds via down to ground plane. A fan-out design of feed-lines further help in transition.

%%%%%%%%%%%%%%%%%%%%%%%%%%%%%%%%%%%%%%%%%%%%%%%%%%%%%%%%%%%%%%%%%%%%%%%%%%%%%

\subsection{Thinned Antenna-Array Design}
\label{sec:baselenghts_antenna_spacings}
The antennas are spaced at $ \lambda $, $ 3\lambda $ and $ 2\lambda $, as shown in Fig. \ref{fig:feedlines}, to obtain minimum redundancy in baselines ($\Delta u =1$). With minimum redundancy, four elements can provide six different baselines and up to $N$$=$$13$ unique complex visibility samples, corresponding to a 13-pixel image, as described in detail earlier in section \ref{sec:four_element_CMI}. The minimum and maximum spacing in the $x$-direction are $ \lambda $ and $ 6\lambda $ respectively, corresponding to an azimuthal field of view (FOV) of $\pm sin^{-1} (1/2\Delta u) = \pm30^{\circ}$, a resolution (defined as angular distance between peak and first null) of $sin^{-1} (1/N\Delta u) = 4.4^{\circ}$, and a central beam-width of $8.8^{\circ}$.

\begin{figure}
 \centering
	\subfloat[]{\includegraphics[clip,trim=0cm 0cm 0cm 0cm, width=0.7\textwidth]{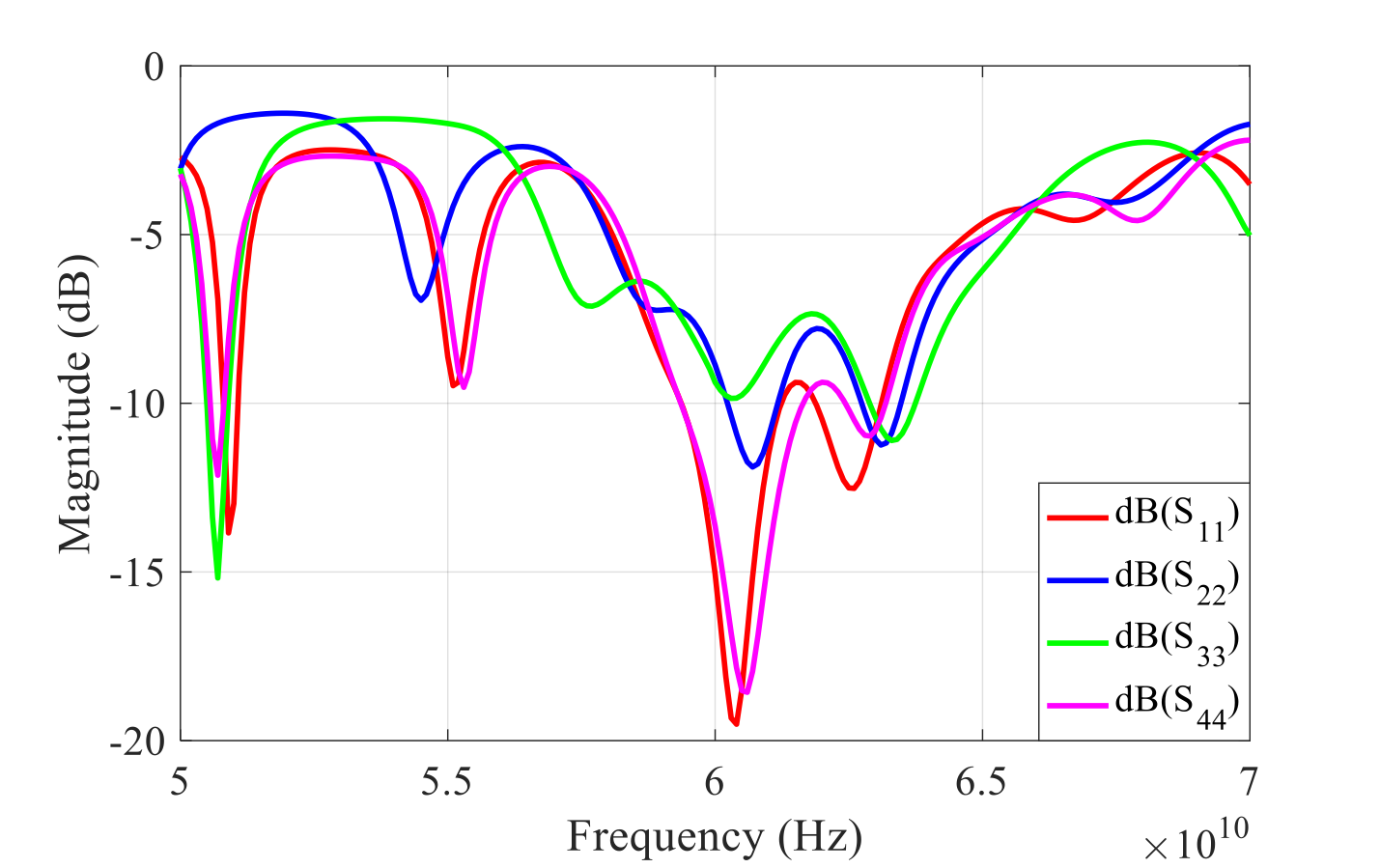}
		\label{fig:antennaS11}} \\
	\subfloat[]{\includegraphics[clip,trim=0cm 0cm 0cm 0cm, width=0.6\textwidth]{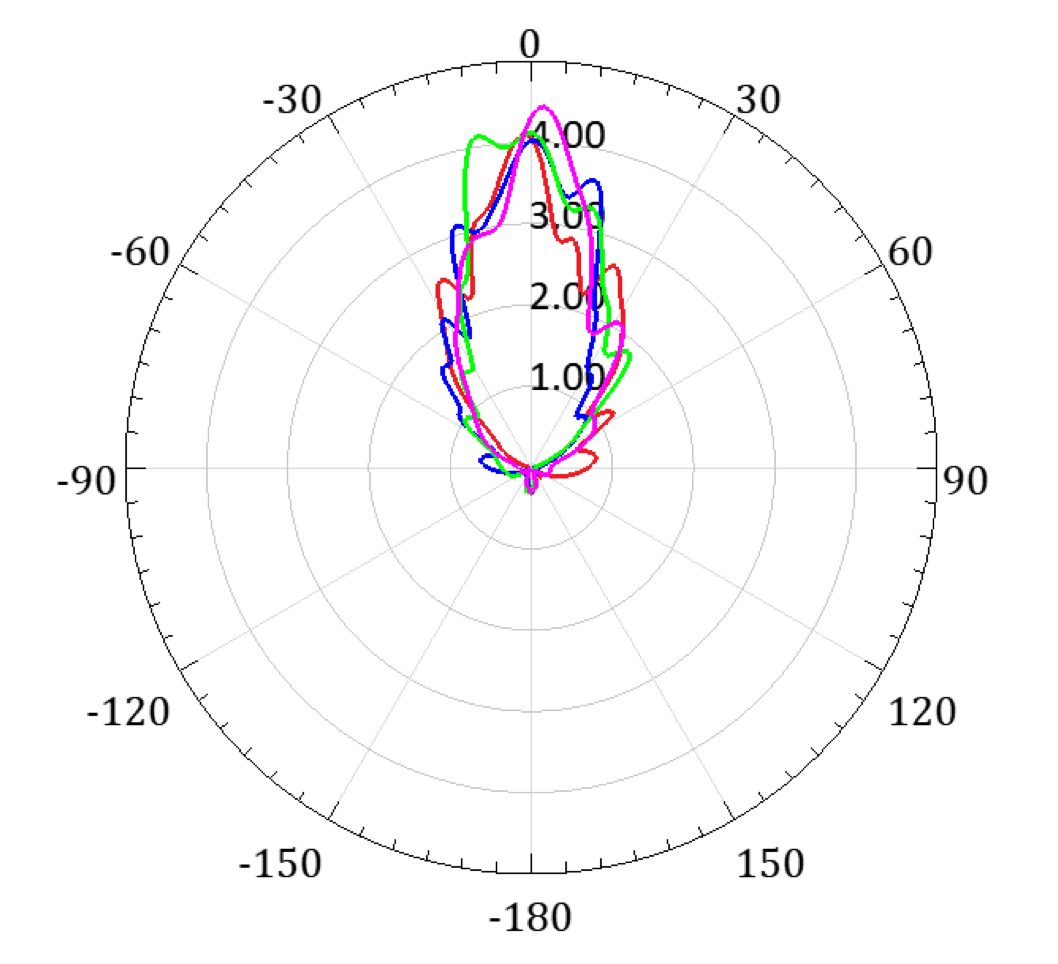}
		\label{fig:antennaGain}} 
	\caption{Simulated (a) input return loss and (b) gain radiation pattern (in dB, \textit{vs} theta at phi =0 deg., 60 GHz) for antennas.}
	\label{fig:4antenna}
\end{figure}

Each antenna is a CPW-fed slot with separate ground planes, designed to take into account the impact of ground-plane size on performance \cite{Haekyo2016}, carefully suppressing surface waves. The antenna performance is especially sensitive to ground plane sizes and feed-line lengths  at these frequencies.
Fig. \ref{fig:4antenna} shows the Ansys HFSS simulated input match and the E-plane antenna-gain patterns. These indicate that each antenna achieves a uniform performance with gain matching to within 1 dB across the pattern. 

\clearpage

%%%%%%%%%%%%%%%%%%%%%%%%%%%%%%%%%%%%%%%%%%%%%%%%%%%%%%%%%%%%%%%%%%%%%%%%%%%%%

\section{Imaging Experiment and Results}

\subsection{Demonstration Setup}
\begin{figure}
\centering
\includegraphics[clip,trim=0cm 0cm 0cm 0cm, width=0.8\textwidth]{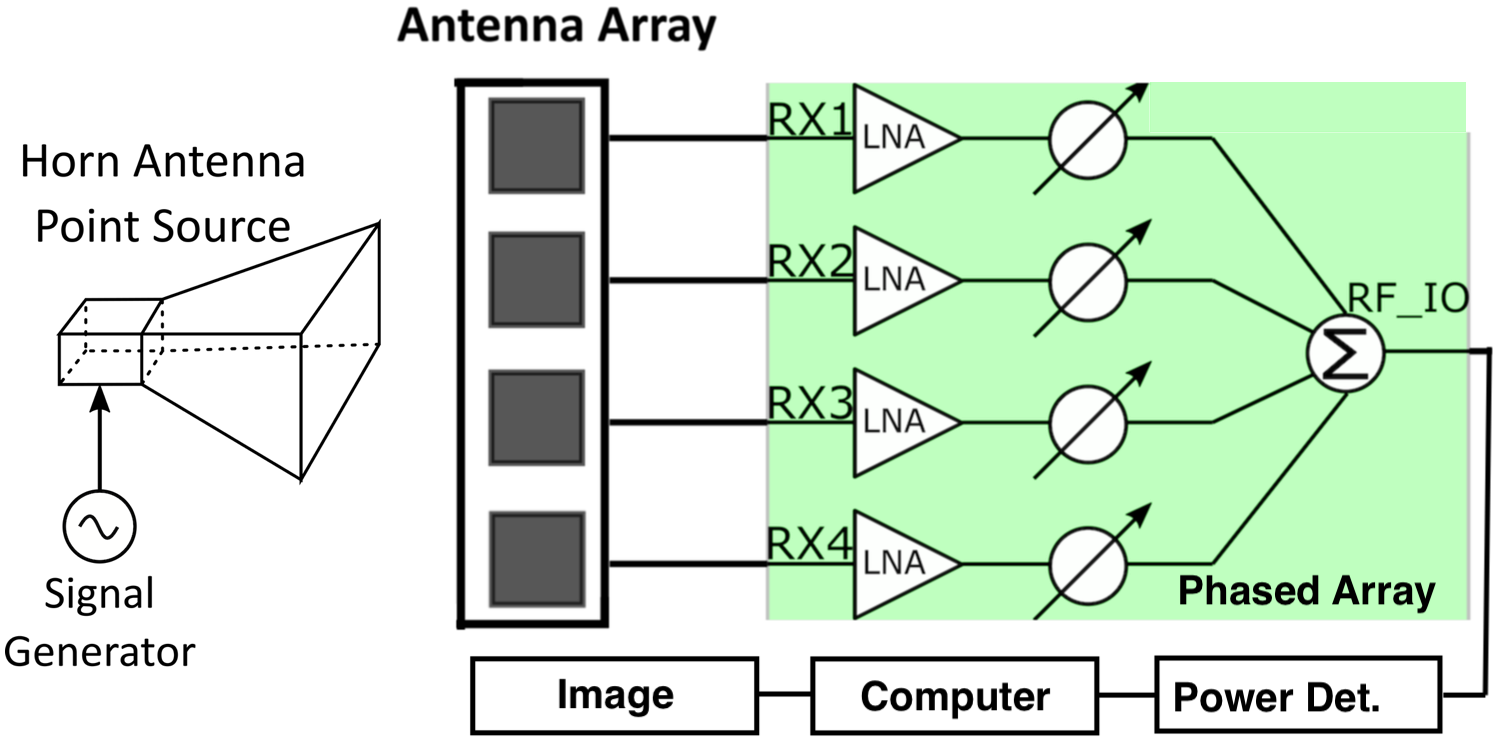}
\caption{Block diagram of imaging setup.}
\label{fig:block}
\end{figure}

\begin{figure}
\centering
\includegraphics[width=.9\textwidth]{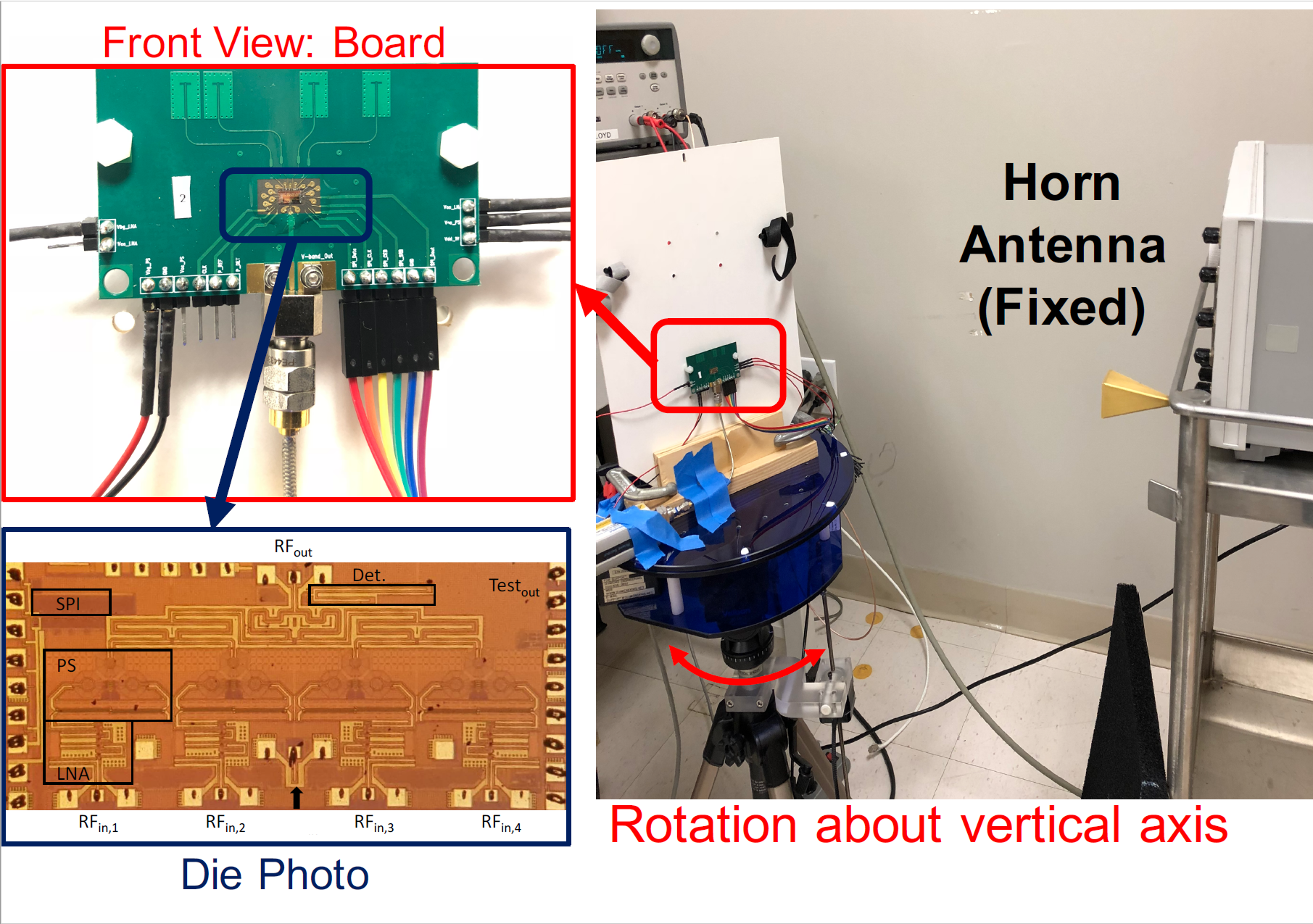}
\caption{Photographs of board, 60-GHz phased array, and overall experimental setup. The board size is 60 x 36 mm\textsuperscript{2}.}
\label{fig:setup}
\end{figure}

Fig. \ref{fig:block} shows the block diagram of the  overall measurement setup. Due to the peaking of array gain around 62 GHz, the experiment is performed at 62 GHz instead of 60 GHz. A radiating horn antenna is used as a point source to obtain point spread function of the setup, detected using the on-board mm-wave antenna array and phased array receiver chip. Fig. \ref{fig:setup} shows the photographs of the imaging setup, as well as the high frequency board with encapsulant-covered chip in cavity, feed-lines, antennas, and power-supply and serial-interface connections. The 60-GHz output on board is connected through a clamped V-band connector (Pasternack Enterprises Inc, PE45402: 1.85mm female connector clamp/solder with captive contact, 0.35 width) with an insertion loss of 0.12 dB. The 60 GHz output is measured using Agilent N1913A power meter and an Agilent N8488A power sensor. The 60-GHz imager is placed on a tripod with the ability to rotate about the vertical axis to simulate a point source at various azimuthal angles within the FOV of the image. The fixed point source is realized using a horn antenna (Pasternack Enterprises Inc, PE9881-24: WR-15 waveguide standard gain horn antenna operating from 50 GHz to 75 GHz with a nominal 24 dBi gain with UG-385/U round cover flange) and a signal generator providing a 15 dBm 60-GHz monotone (Agilent E8257D). The horn antenna is connected to signal generator using a WR-15 waveguide to V(F) coax adapter with an insertion loss of 0.5 dB (by Sage Millimeter Inc. SWC-15VF-R1). 

\subsection{Code-Modulation, Data Acquisition, and Processing} \label{subsec:data_aquisition_4element}
\begin{figure}
 \centering
	\subfloat[]{\includegraphics[clip,trim=0cm 0cm 0cm 0cm, width=0.6\textwidth]{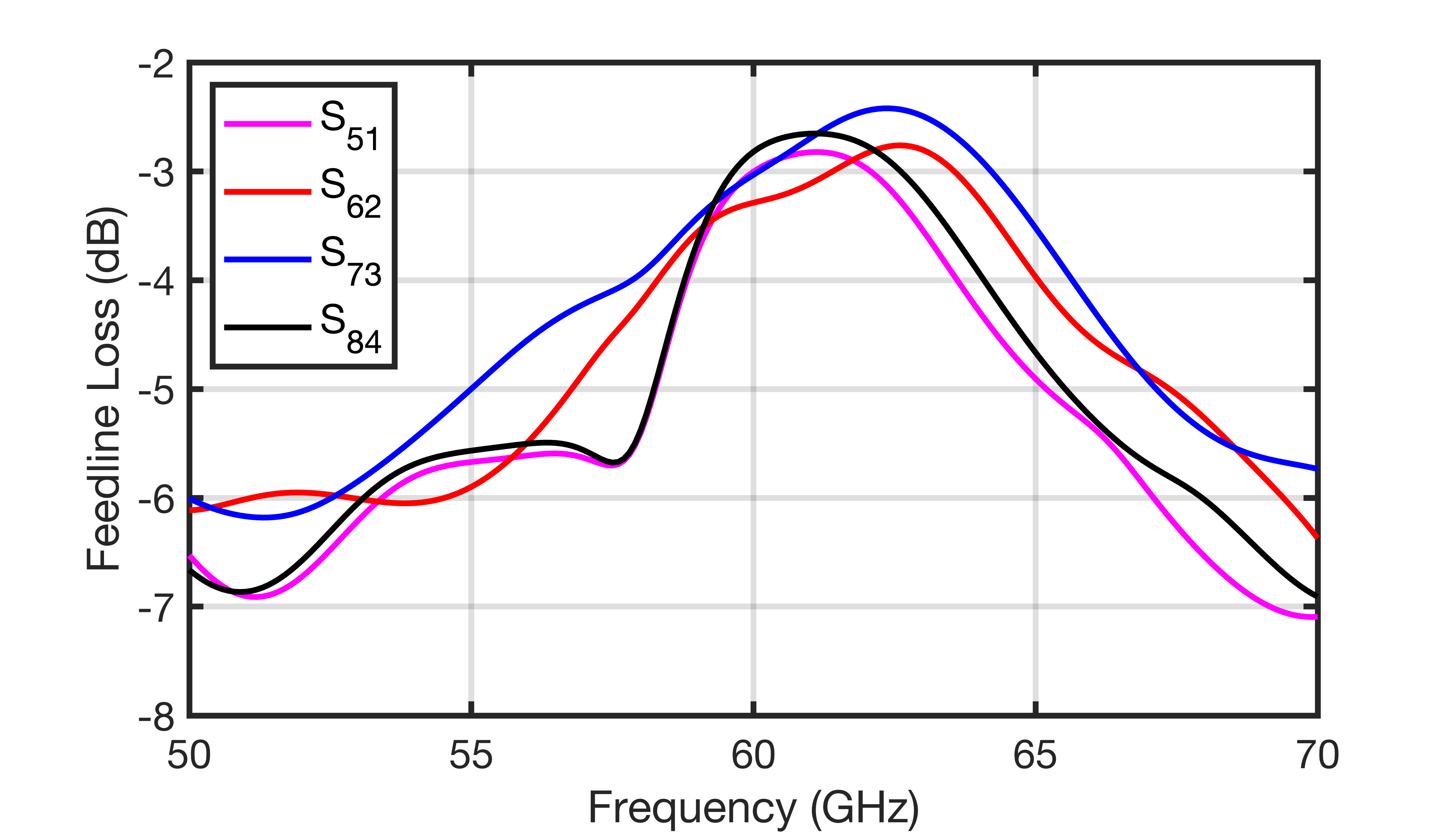}
		\label{fig:Feedline_loss}} \\
	\subfloat[]{\includegraphics[clip,trim=0cm 0cm 0cm 0cm, width=0.6\textwidth]{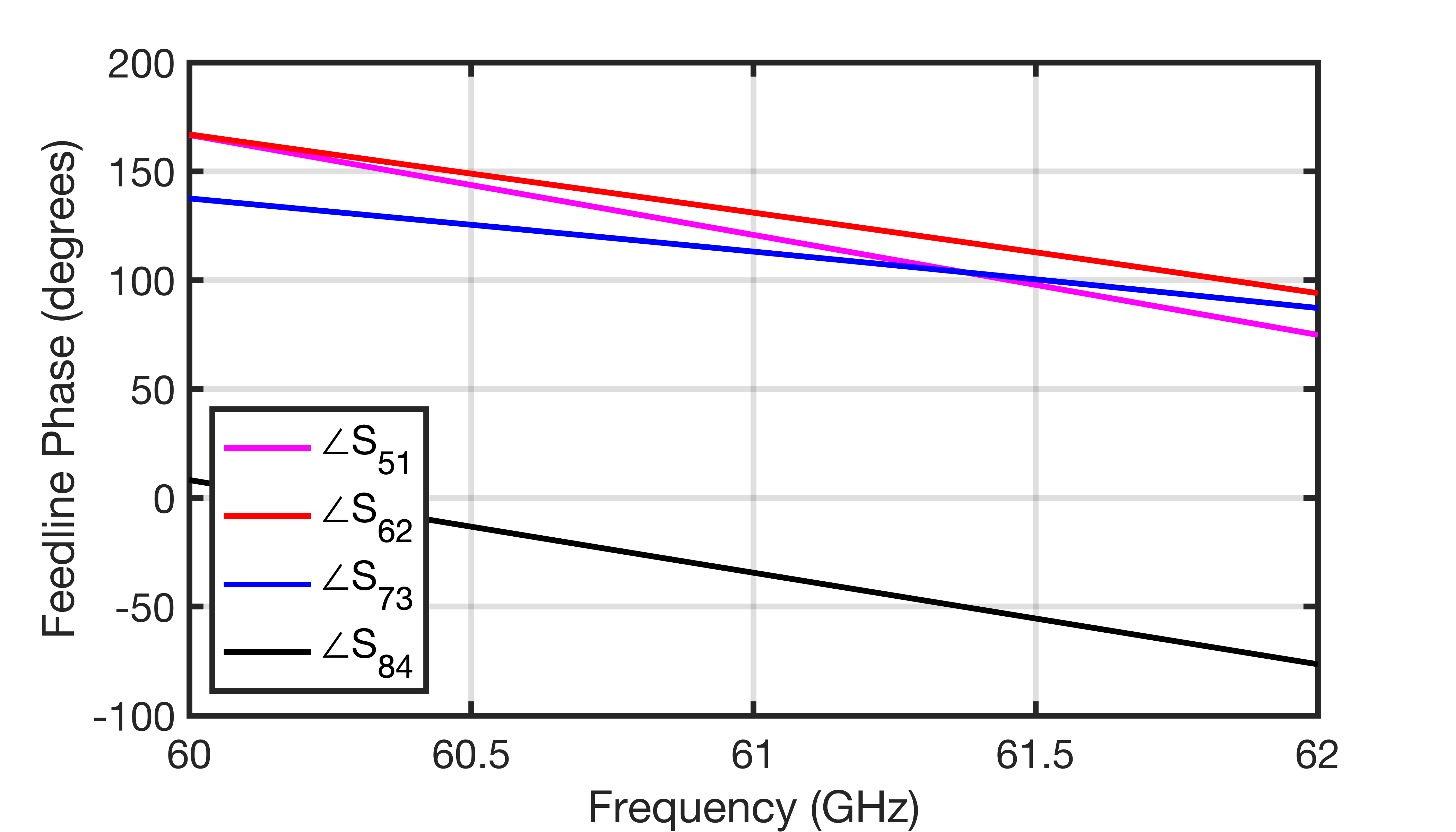}
		\label{fig:Feedline_phase}} 
	\caption{Simulated (a) feedline loss in dB and (b) phase. $S_{51}$ for E1,$S_{62}$ for E2, $S_{73}$ for E3, and $S_{84}$ for element E4. }
	\label{fig:Feedline_cal}
\end{figure}

The phase shifters inside the chip are programmed using a serial peripheral interface (SPI) and they are first used to calibrate the phase and gain differences between channels due to different feed-line lengths. As seen in Fig. \ref{fig:feedlines}, the lengths of feedlines are different (22.6 mm, 19.6 mm, 13.0 mm, and 20.5 mm for elements E1-E4 respectively). The differences in gains and phases of four elements due to different feedlengths need to be calibrated for proper function of the imaging system. The phase differences and the gain differences in the elements are estimated using the HFSS simulation of the feedlines. Fig. \ref{fig:Feedline_cal} show the simulated loss and phase differences of the four feedlines. The losses are similar for all four elements and therefore ignored. The phase differences, however, are significant and therefore calibrated using the phase shifters. Note that such an imaging system can be calibrated using the over-the-air code-modulated interferometric built-in-self test or CoMET, as given in \cite{CometIMS19}. The phase shifters are also used to code-modulate the RF signals with balanced, orthogonal, binary ($\pm1$) 256-bit Walsh-Hadamard codes (easily achievable by a $0^\circ$ and $180^\circ$ phase shift).
In \cite{Greene2018}, we used the chip's integrated power detector to demonstrate chip-level CMI for built-in test. In this imaging work, we switched to using an external 60-GHz power detector to overcome an ESD reliability issue we identified. This allows a single board to be confidently used for repeated experiments. 

As discussed earlier in Eqn. \ref{four_vis_set}, both real and imaginary visibilities are to be measured. Real visibility is the cross-correlation obtained when in-phase signals from two elements are multiplied and integrated. When in-phase signal of one element is cross-correlated with quadrature phase signal of other element, imaginary visibility is obtained. Real and imaginary visibilities are obtained as follows. First, all elements are placed in phase and CMI produces all real components, i.e. the set of cross-correlations obtained are $[Re(V_{1,2}),$ $Re(V_{1,3}),$ $ Re(V_{1,4}),$ $ Re(V_{2,3}),$ $ Re(V_{2,4}),$ $ Re(V_{3,4})]$, see Table \ref{tab:vis_table}. Second, elements three and four are placed in a 90 deg. state and CMI produces a first set of imaginary visibilities along with some redundant real visibilities, i.e.  $[Re(V_{1,2}),$ $ Im(V_{1,3}),$ $ Im(V_{1,4}),$ $ Im(V_{2,3}),$ $ Im(V_{2,4}),$ $ Re(V_{3,4})]$. Then elements two and four are placed in a 90 deg. state and CMI produces a second set of imaginary visibilities, i.e. $[Im(V_{1,2}),$ $ Re(V_{1,3}),$ $ Im(V_{1,4}),$ $ -Im(V_{2,3}),$ $ Re(V_{2,4}),$ $ Im(V_{3,4})]$. These real and imaginary visibility are components of complex visibility sample and its conjugate, for example, $v_{1,2}= Re(V_{1,2}) + Im(V_{1,2})$, and $v^*_{1,2}= Re(V_{1,2}) - Im(V_{1,2})$.  In runs two and three, certain sets redundant of real and imaginary visibilities are also obtained. These redundant readings are averaged to reduce the noise. Although it is possible to obtain both real and imaginary components in a single run using complex codes, we found that the steps above minimized overall run time. 
All hardware are controlled using LabView (including SPI controller to control phase-shifters to calibrate and code-modulate, reading data from power meter as well as DC supplies) and all data processing (including demodulation of visibilities and DIFT) are performed using MATLAB. Windowing function has not been used in this implementation. The signal generator with attached horn antenna is placed on a cart to allow readings at different distances from antenna array.

\subsection{Results}
\begin{figure}
 \centering
	\includegraphics[width=0.7\textwidth]{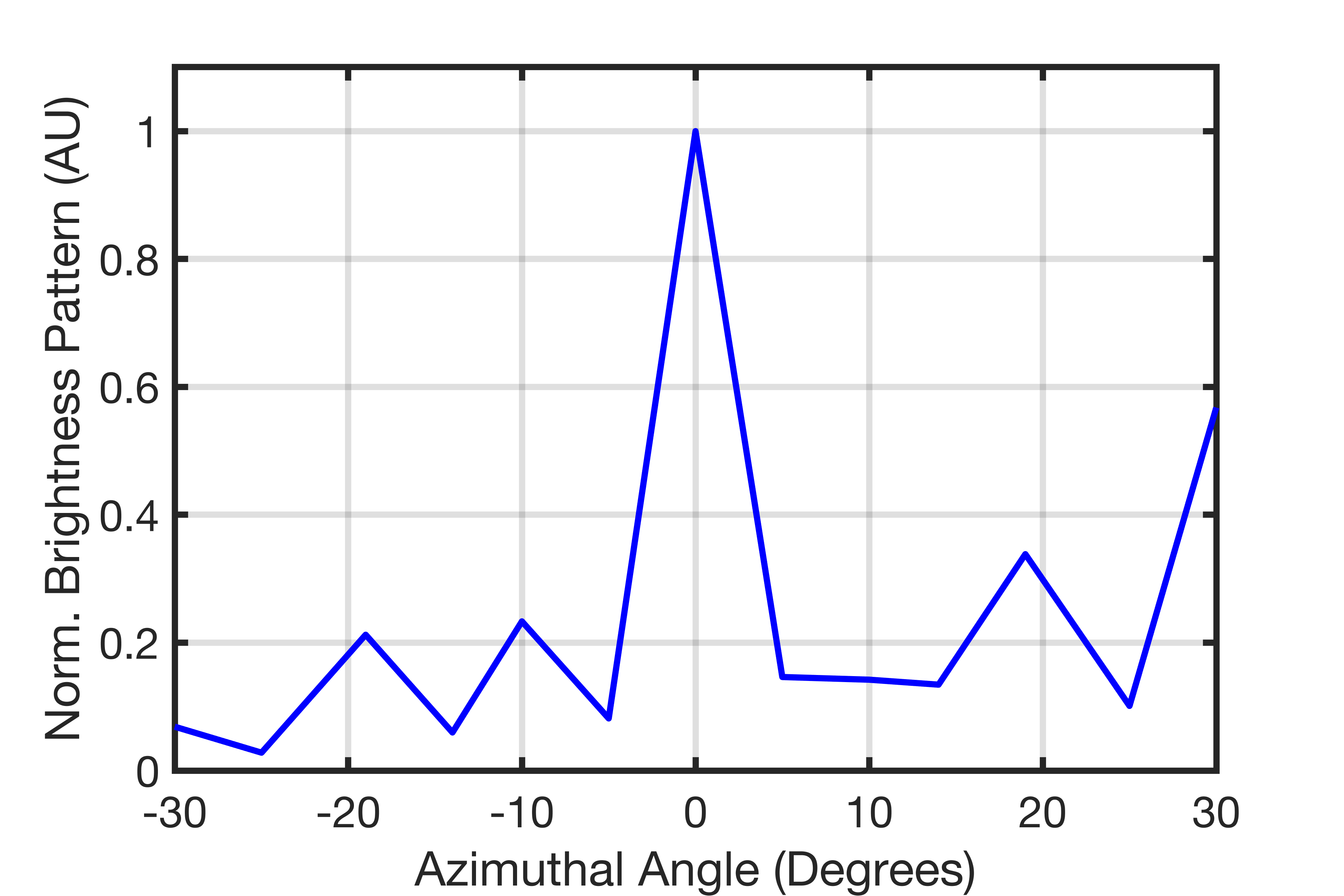}
	\caption{Normalized brightness plots for point source at azimuthal angle of approximately 0$^\circ$.}
	\label{fig:Results_PSF}
\end{figure}

\begin{figure}
%\vspace{-1em}
 \centering
	\subfloat[]{\includegraphics[width=0.3\textwidth,trim={0.4cm 0cm 1.5cm 1cm},clip]{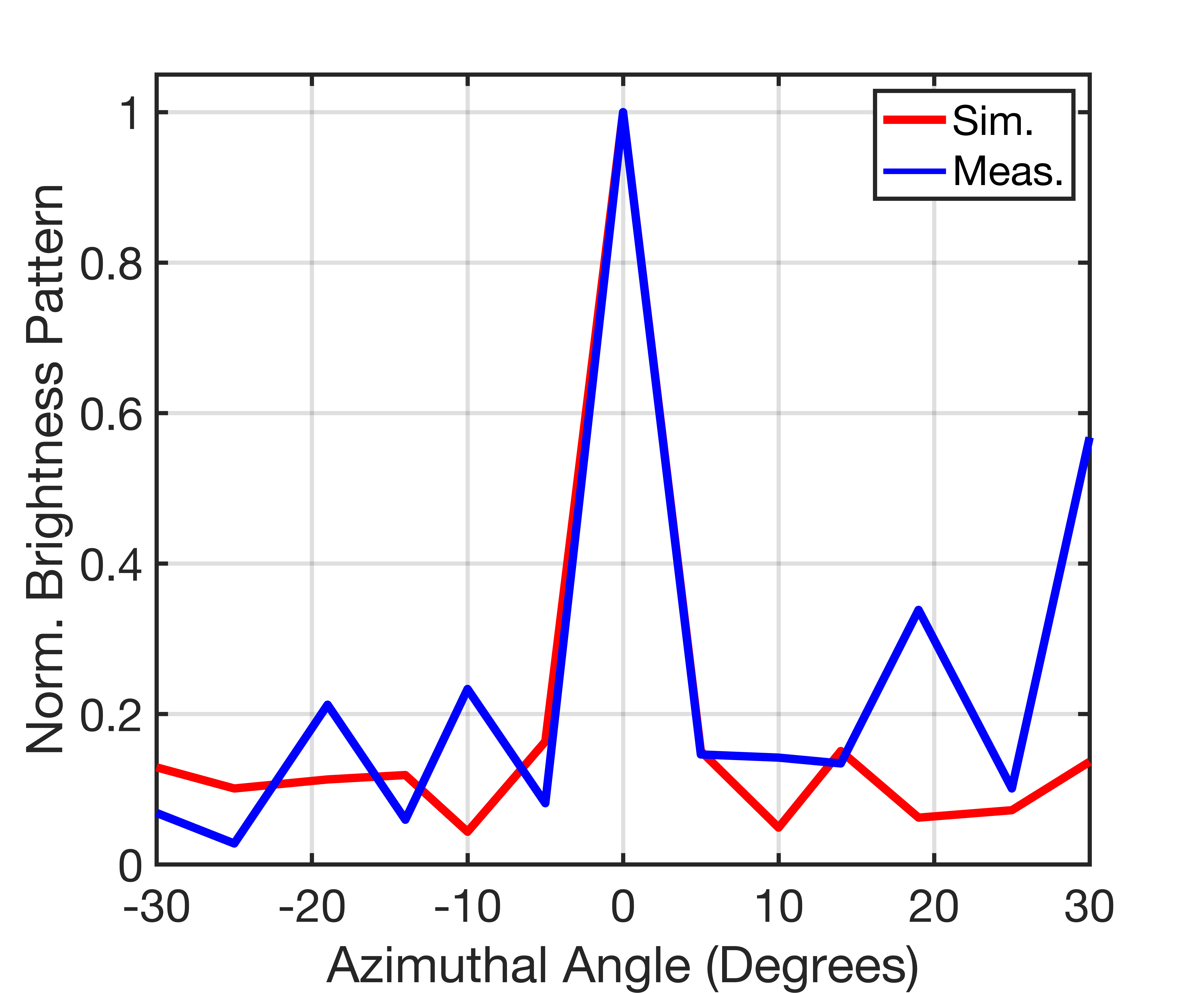}
		\label{fig:R1}}
	\subfloat[]{\includegraphics[width=0.3\textwidth,trim={0.4cm 0cm 1.5cm 1cm},clip]{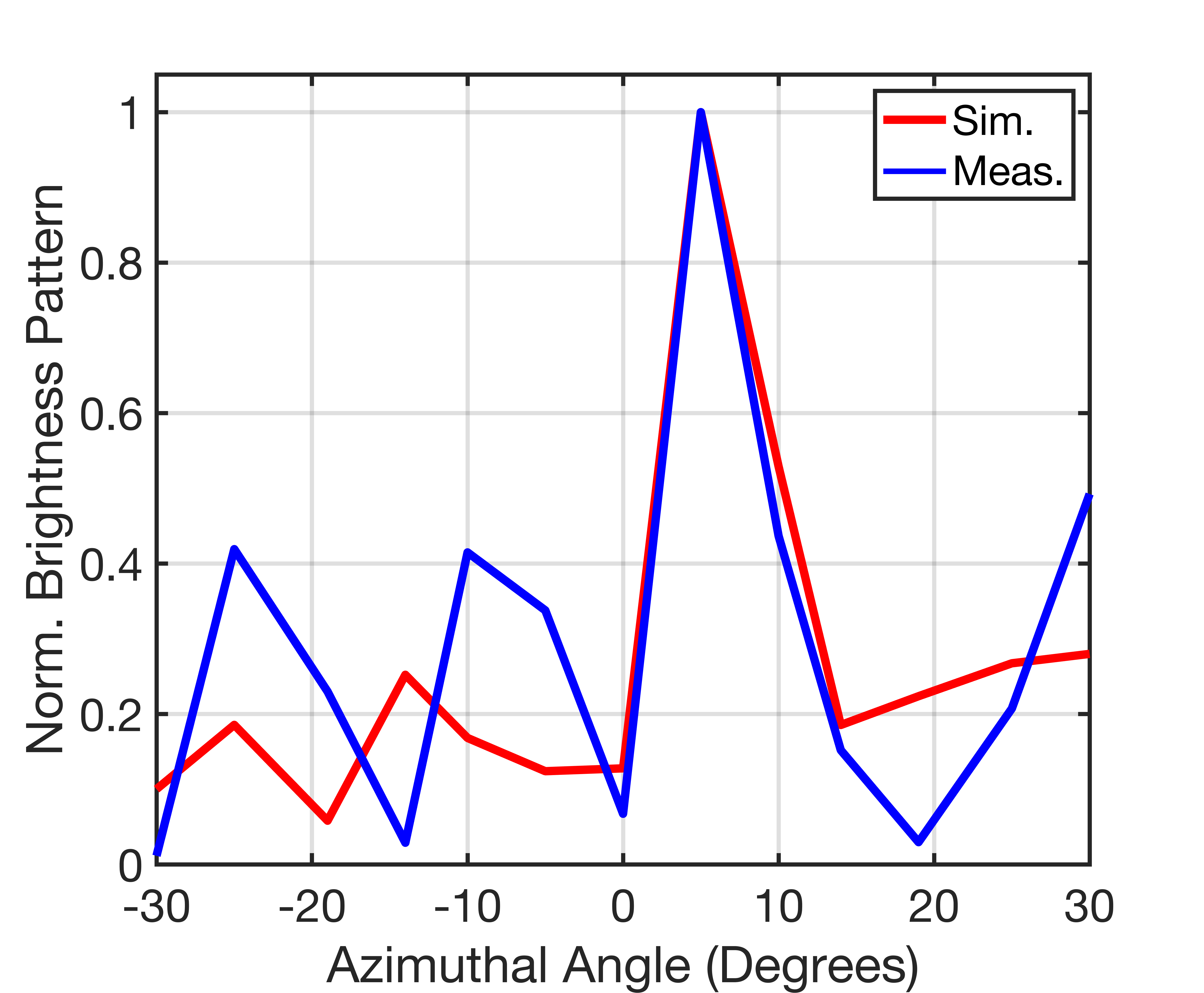}
		\label{fig:R2}} \\ \vspace{-1em}
	\subfloat[]{\includegraphics[width=0.3\textwidth,trim={0.4cm 0cm 1.5cm 1cm},clip]{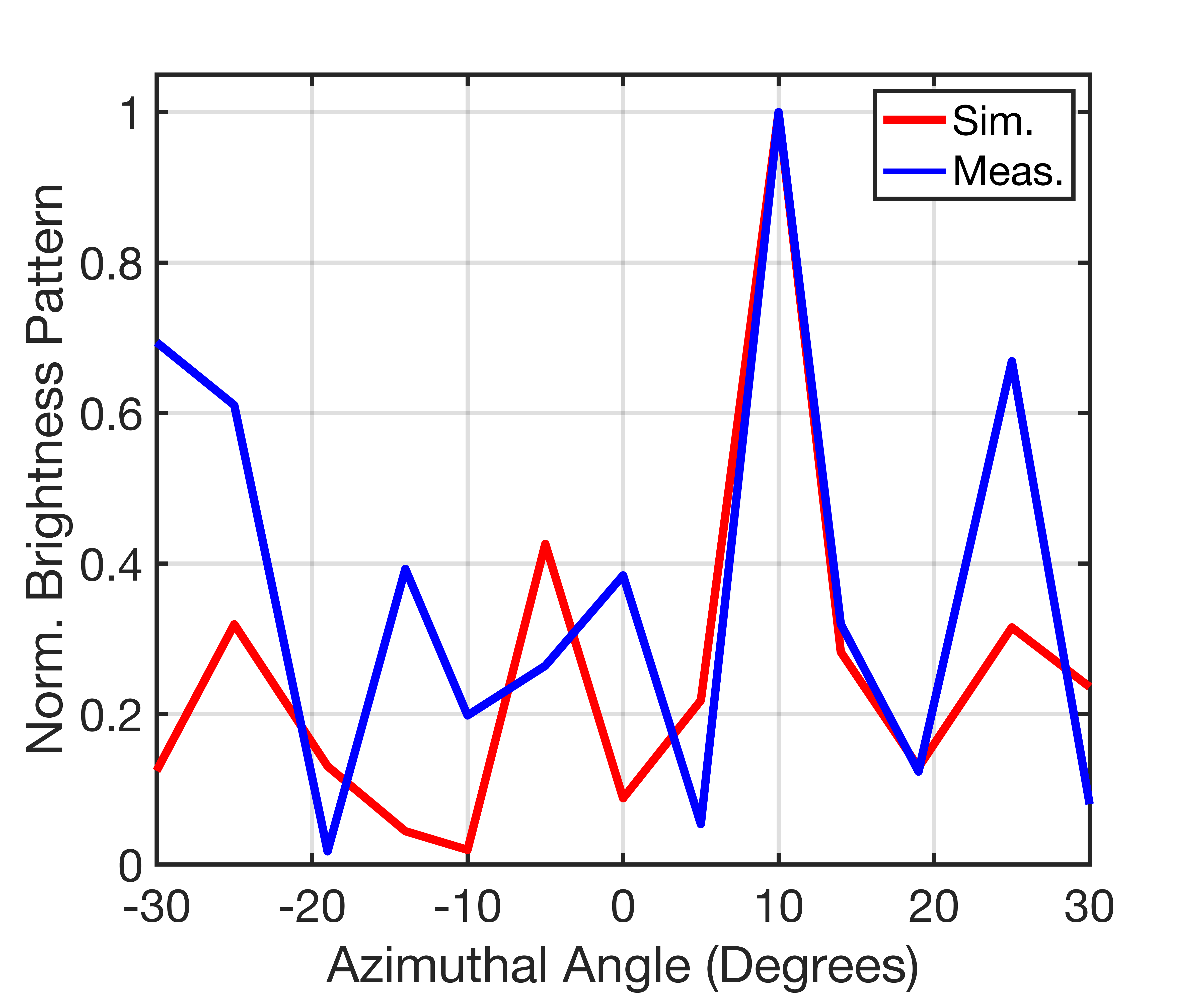}
		\label{fig:R3}}
	\subfloat[]{\includegraphics[width=0.3\textwidth,trim={0.4cm 0cm 1.5cm 1cm},clip]{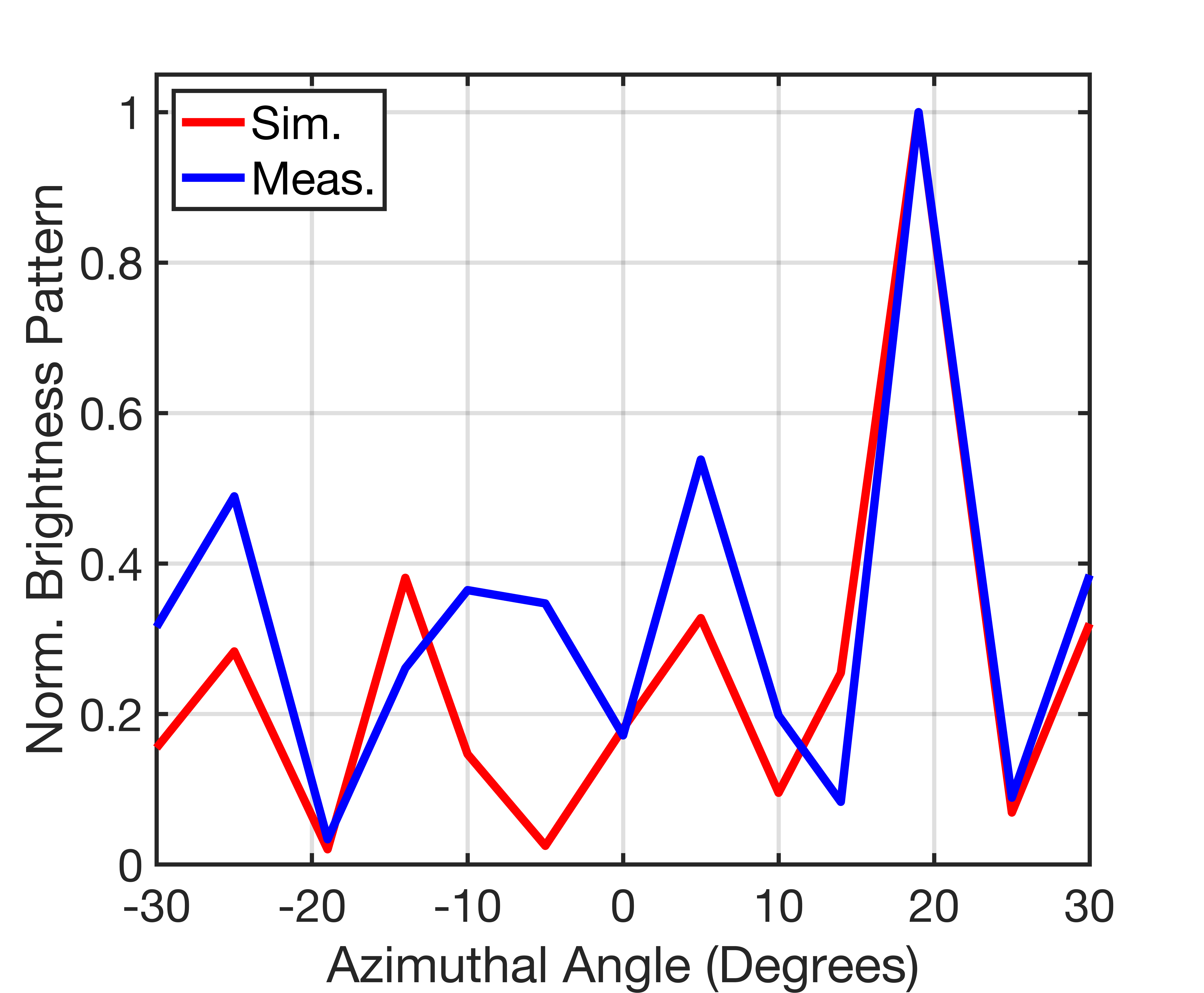}
		\label{fig:R4}} \\ \vspace{-1em}
	\subfloat[]{\includegraphics[width=0.3\textwidth,trim={0.4cm 0cm 1.5cm 1cm},clip]{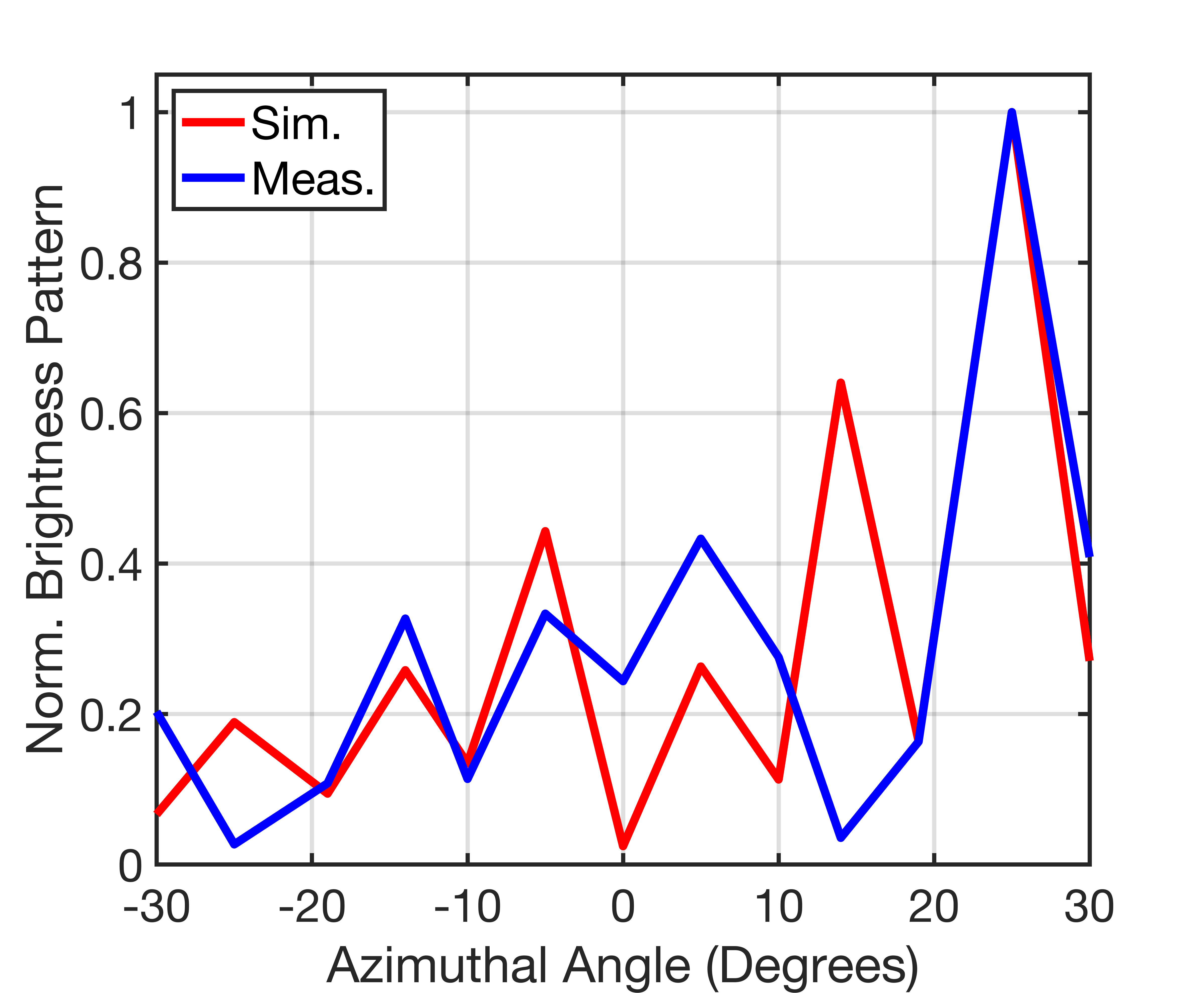}
		\label{fig:R5}}
%	\subfloat[]{\includegraphics[width=0.25\textwidth]{figures/7_30L_Redund_comp_vis_fast_11in_28L2.png}
%		\label{fig:R6}}\\
	\subfloat[]{\includegraphics[width=0.3\textwidth,trim={0.4cm 0cm 1.5cm 1cm},clip]{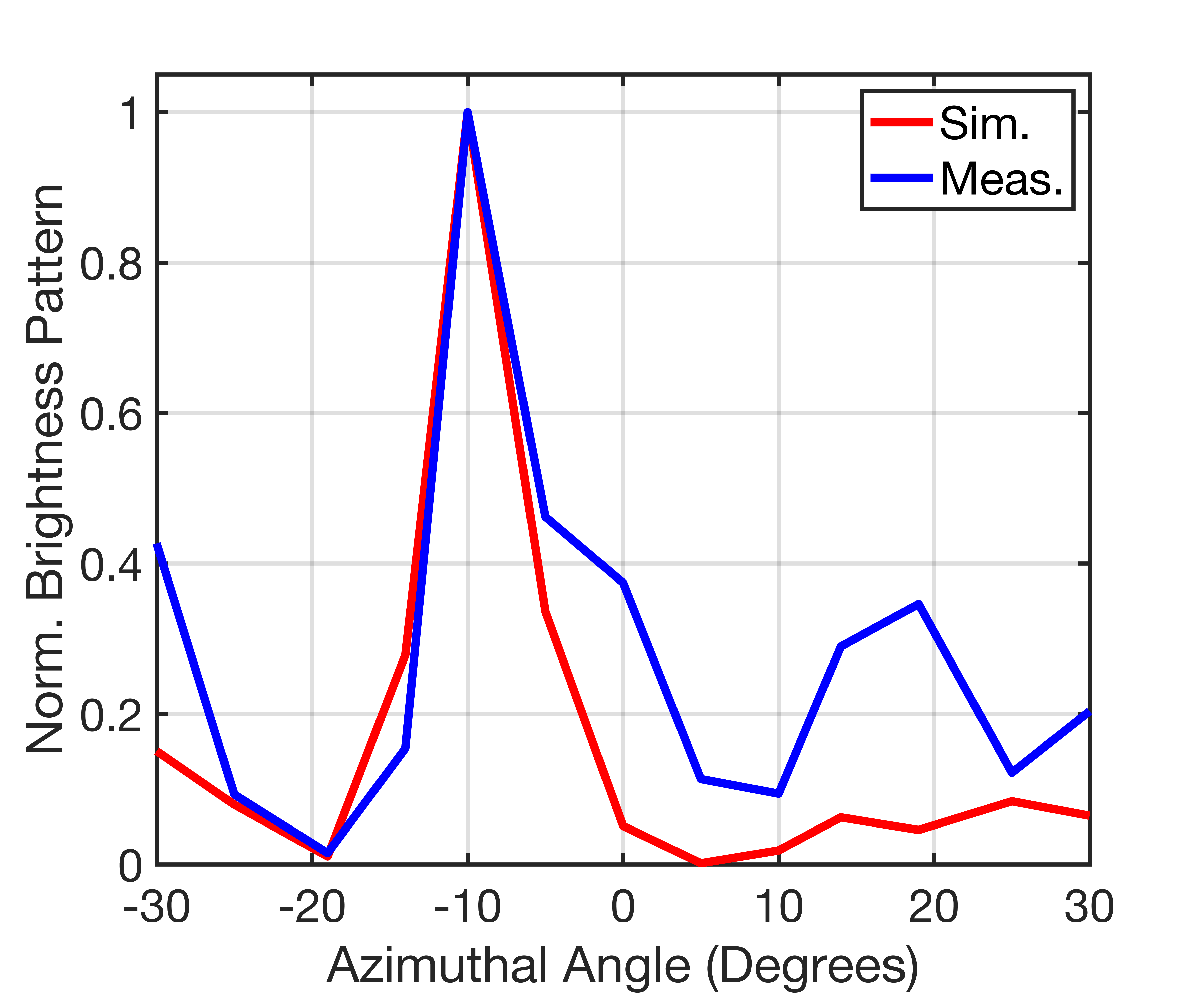}
		\label{fig:R7}} \\ \vspace{-1em}
	\subfloat[]{\includegraphics[width=0.3\textwidth,trim={0.4cm 0cm 1.5cm 1cm},clip]{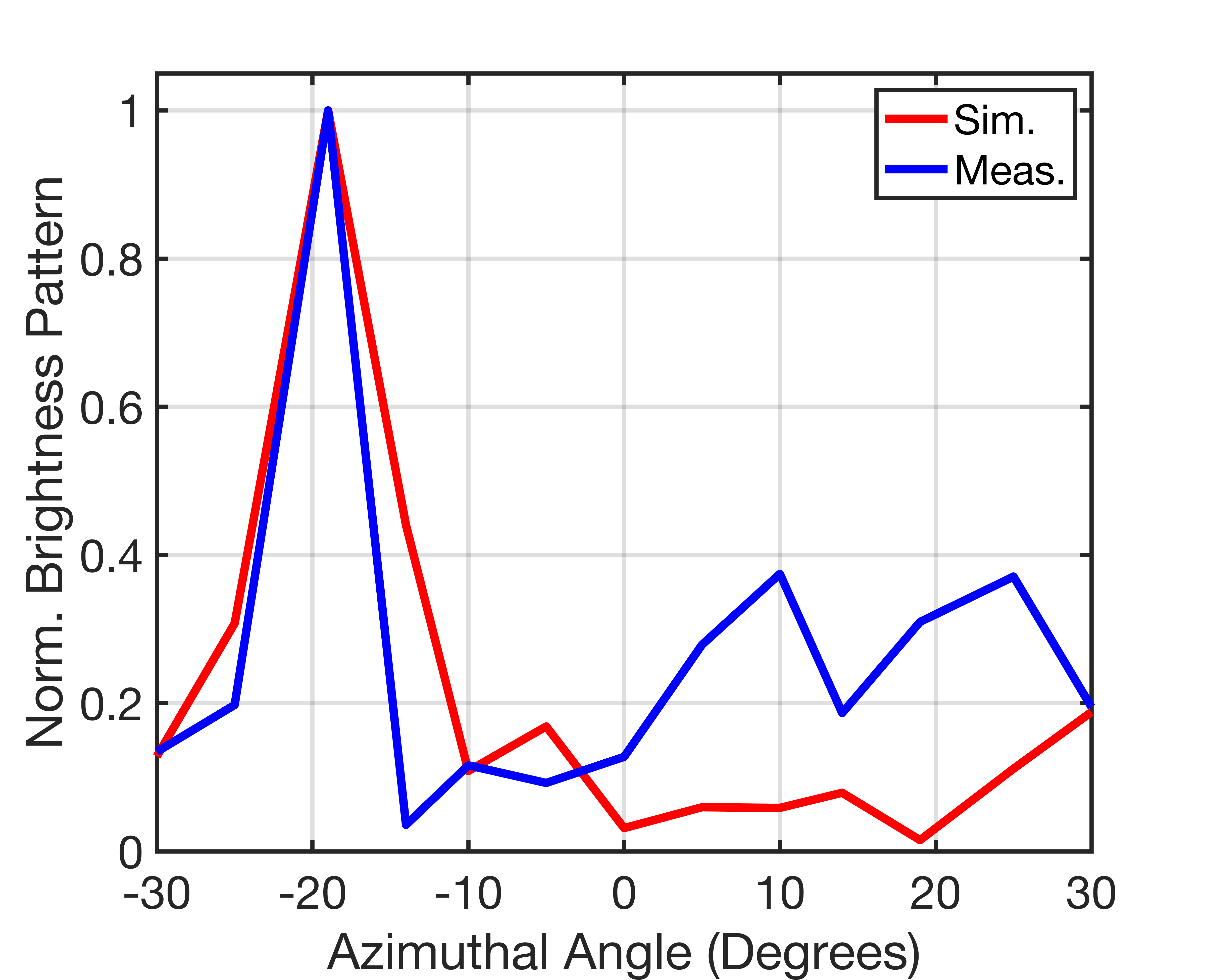}
		\label{fig:R8}}
	\subfloat[]{\includegraphics[width=0.3\textwidth,trim={0.4cm 0cm 1.5cm 1cm},clip]{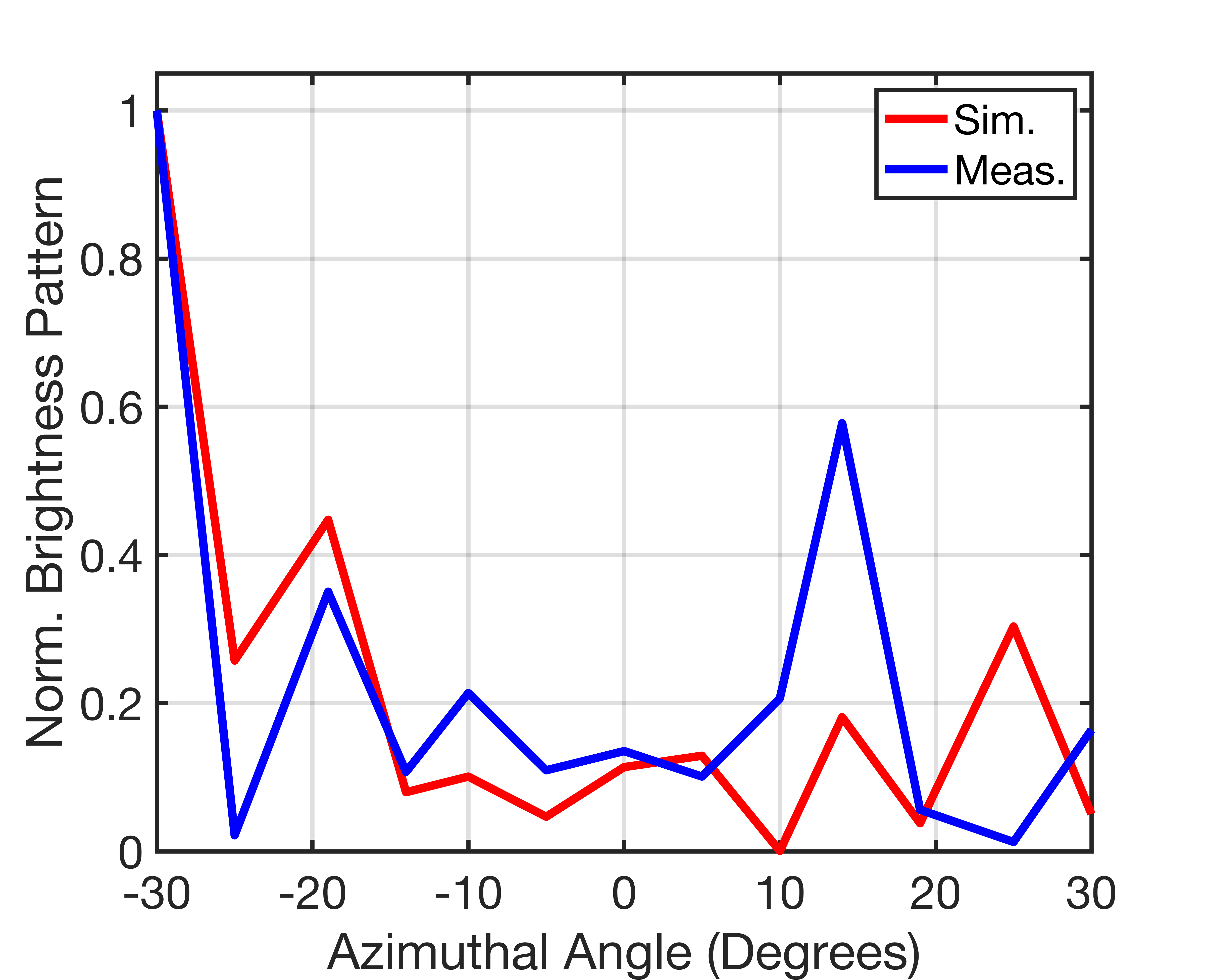}
		\label{fig:R9}}
%	\vspace{1em}
	\caption{Measured and simulated normalized brightness plots for point source at azimuthal angle of approximately (a) 0$^\circ$   (b) 5$^\circ$ (c) 10$^\circ$ (d) 20$^\circ$ (e) 25$^\circ$ % (f) 30$^\circ$ 
	(f) -10$^\circ$ (g) -20$^\circ$ (h) -30$^\circ$ \cite{VikasMWCL}.}
	\label{fig:Results_all}
%	\vspace{-1.5em}
\end{figure}

The 60-GHz CMI imager is demonstrated by detecting a point source at various azimuthal angles. Imaging of a point source is an important experiment as it can be used to verify the functionality of the imaging system, its resolution, FOV, and the quality of its point spread function (PSF). Comparison of ideal and obtained PSF can provide insights about the quality of imaging system. For the demonstration, first, the horn antenna is placed at nadir angle to obtain the point spread function. Fig. \ref{fig:Results_PSF} shows the detected point spread function, or the point source at azimuthal angle of approximately 0$^\circ$. The peak is sharp, easily noticeable and the sidelobe levels are close to that of expected sinc function (except for few points on the right). Then the imager, which is placed on the gimbal is rotated to simulate point source at different azimuthal angles. Fig. \ref{fig:Results_all} shows point-source images for some of the measured angles, at approximately $5^\circ$, $\pm10^\circ$, $\pm20^\circ$, $25^\circ$ and $\pm30^\circ$. The minimum rotation to move the peak to an adjacent bin in the image is approximately $5^\circ$, which is close to the calculated resolution of 4.4$^\circ$ (as shown in section \ref{sec:baselenghts_antenna_spacings}). The observed FOV is approximately $30^\circ$ in each direction, as expected. The detection of active point source in the correct bin in image is a proof of the functionality of the 60-GHz code-modulated interferometric imaging system. 

\begin{comment}
\begin{figure}
 \centering
%	\subfloat[]{\includegraphics[width=0.25\textwidth]{figs4/1_0_Redund_comp_vis_fast_27in_0.png}
%		\label{fig:R1}}
	\subfloat[]{\includegraphics[width=0.35\textwidth]{figs4/2_5L_Redund_comp_vis_fast_14p5in_5L.png}
		\label{fig:R2}} 
	\subfloat[]{\includegraphics[width=0.35\textwidth]{figs4/3_10L_Redund_comp_vis_fast_11in_10L.png}
		\label{fig:R3}} \\ %\vspace{-1em}
	\subfloat[]{\includegraphics[width=0.35\textwidth]{figs4/5_20L_Redund_comp_vis_fast_14p5in_18L.png}
		\label{fig:R4}} 
	\subfloat[]{\includegraphics[width=0.35\textwidth]{figs4/6_25L_Redund_comp_vis_fast_14p5in_24L3.png}
		\label{fig:R5}} \\ %\vspace{-1em}
	\subfloat[]{\includegraphics[width=0.35\textwidth]{figs4/7_30L_Redund_comp_vis_fast_11in_28L2.png}
		\label{fig:R6}}
	\subfloat[]{\includegraphics[width=0.35\textwidth]{figs4/8_10R_Redund_comp_vis_fast_22in_8R.png}
		\label{fig:R7}} \\ %\vspace{-1em}
	\subfloat[]{\includegraphics[width=0.35\textwidth]{figs4/9_20R_Redund_comp_vis_fast_18in_18R.png}
		\label{fig:R8}}
	\subfloat[]{\includegraphics[width=0.35\textwidth]{figs4/10_30R_Redund_comp_vis_fast_11in_30R2.png}
		\label{fig:R9}}
	\caption{Normalized brightness plots for point source at azimuthal angle of approximately (a) 5$^\circ$ (b) 10$^\circ$ (c) 20$^\circ$ (d) 25$^\circ$  (e) 30$^\circ$ 
	(f) -10$^\circ$ (g) -20$^\circ$ (h) -30$^\circ$.}
	\label{fig:Results_all}
\end{figure}
\end{comment}

%trim = left lower right above

\begin{figure}
 \centering
	\subfloat[]{\includegraphics[clip,trim=0cm 0cm 0cm 1.2cm, width=0.45\textwidth]{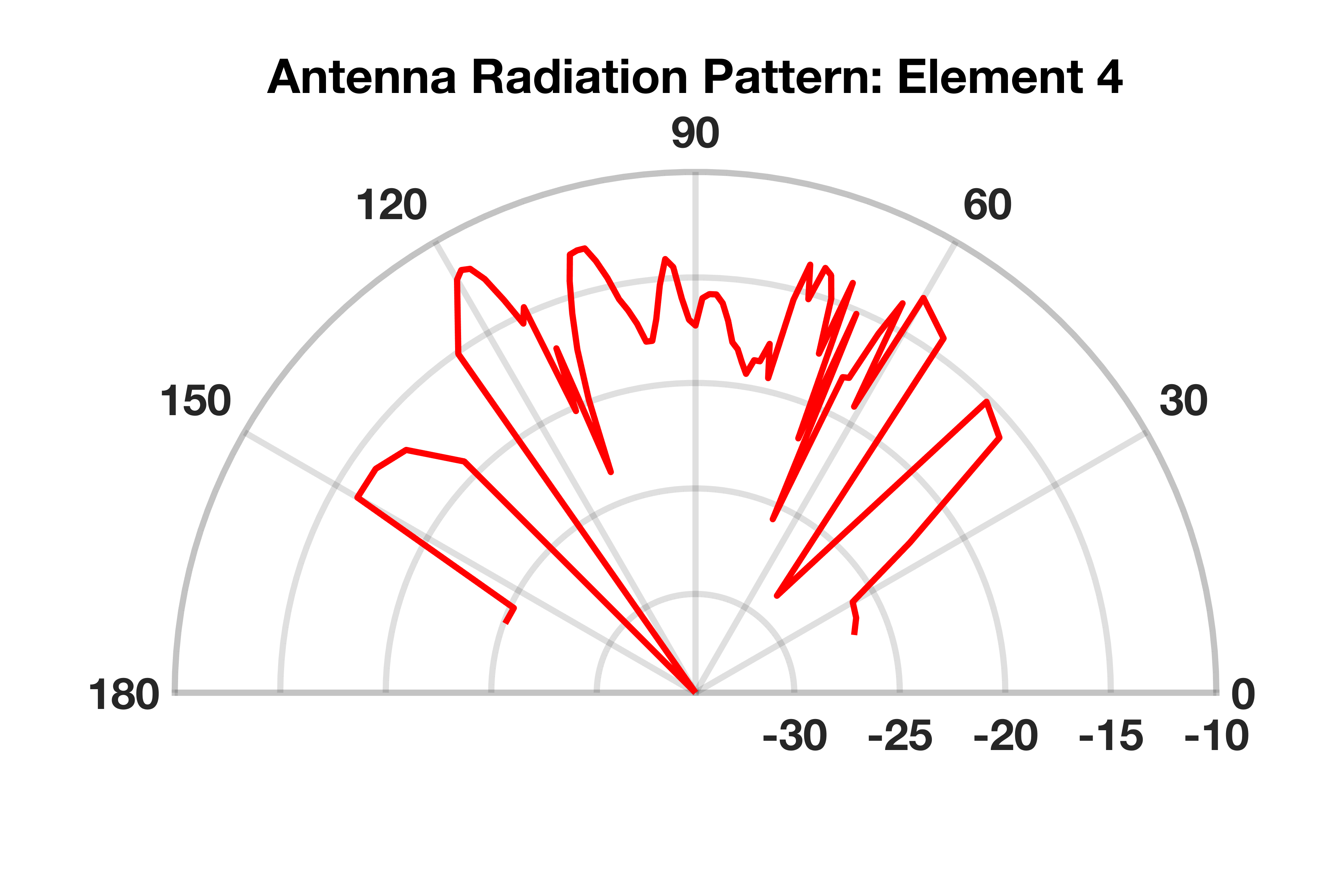}}
	\subfloat[]{\includegraphics[clip,trim=0cm 0cm 0cm 1.2cm, width=0.45\textwidth]{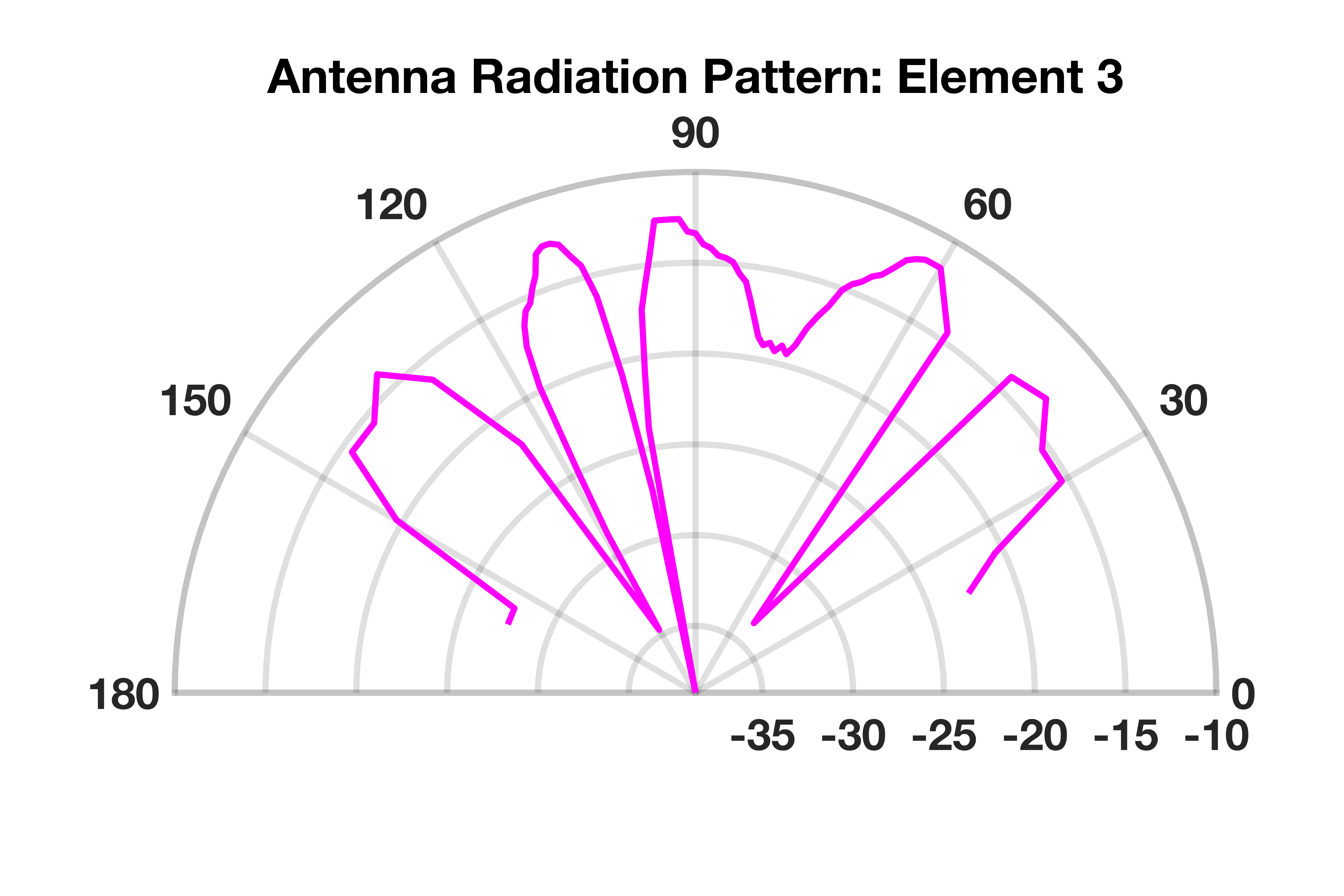}} \\
	\subfloat[]{\includegraphics[clip,trim=0cm 0cm 0cm 1.2cm, width=0.45\textwidth]{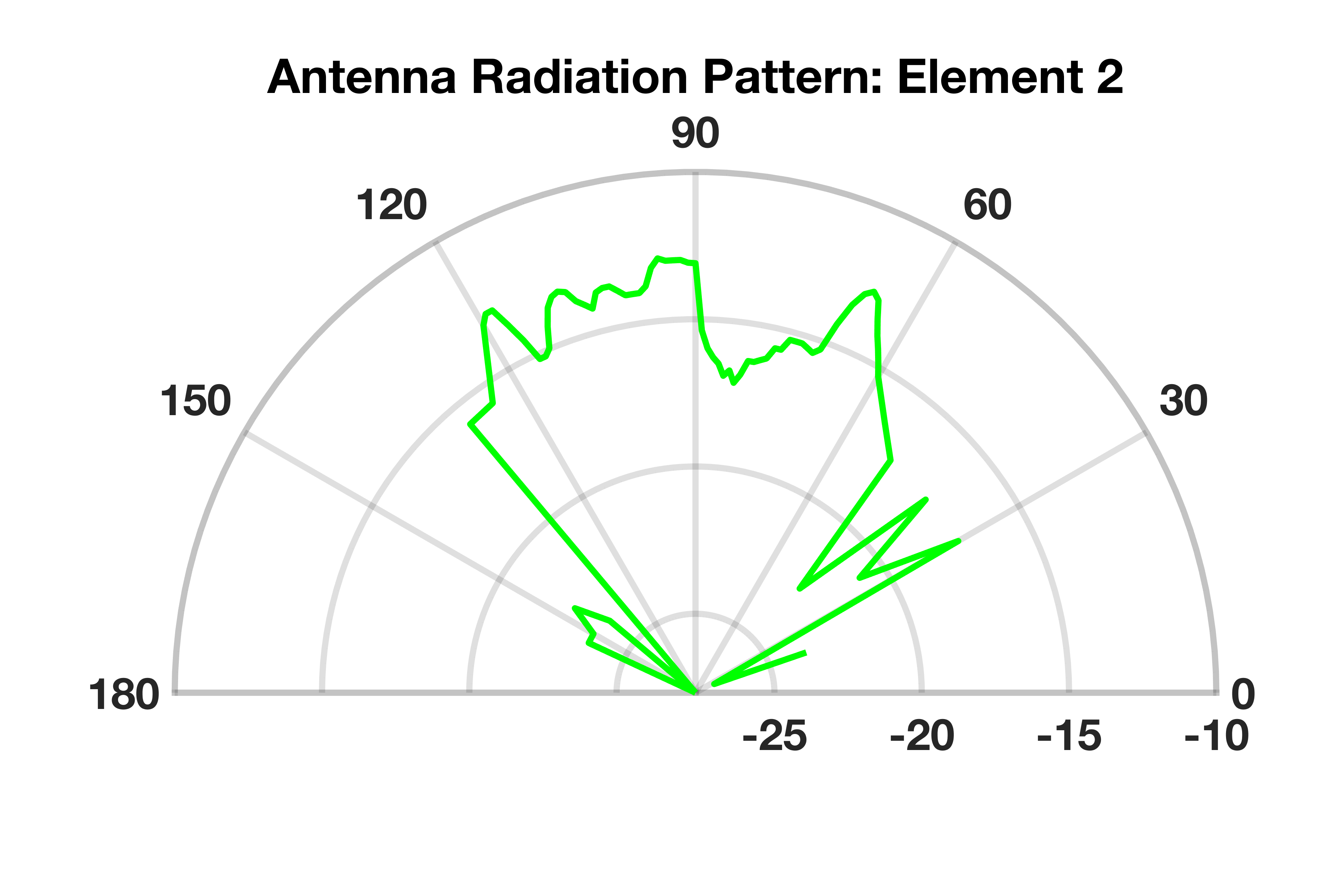}}
	\subfloat[]{\includegraphics[clip,trim=0cm 0cm 0cm 1.2cm, width=0.45\textwidth]{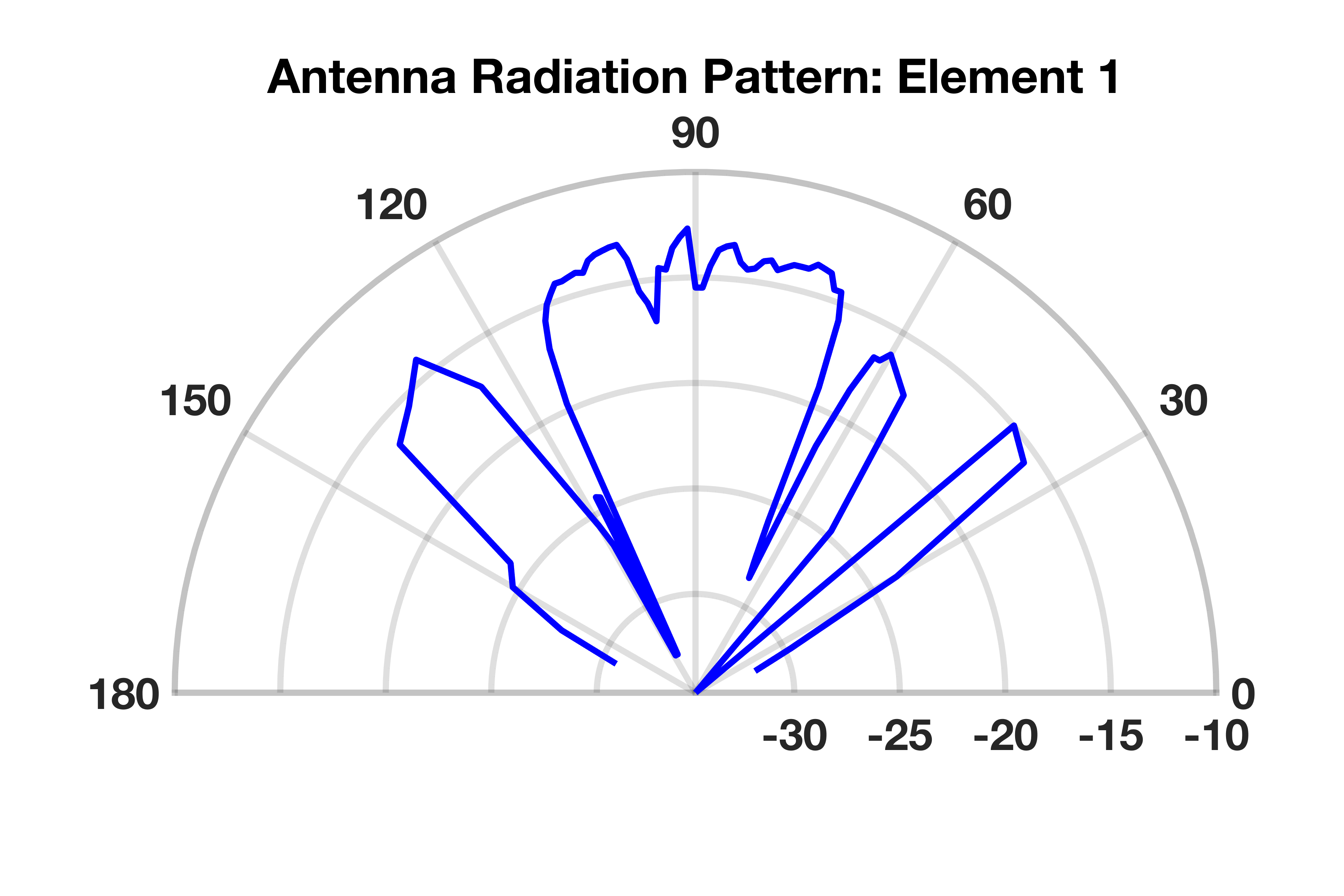}}
	\caption{Gain radiation pattern of elements (a) E1, (b) E2, (c) E3 and (d) E4.}
	\label{fig:Results_BP}
\end{figure}

\begin{figure}
 \centering
	\includegraphics[clip,trim=0cm 2cm 0cm 0cm, width=0.8\textwidth]{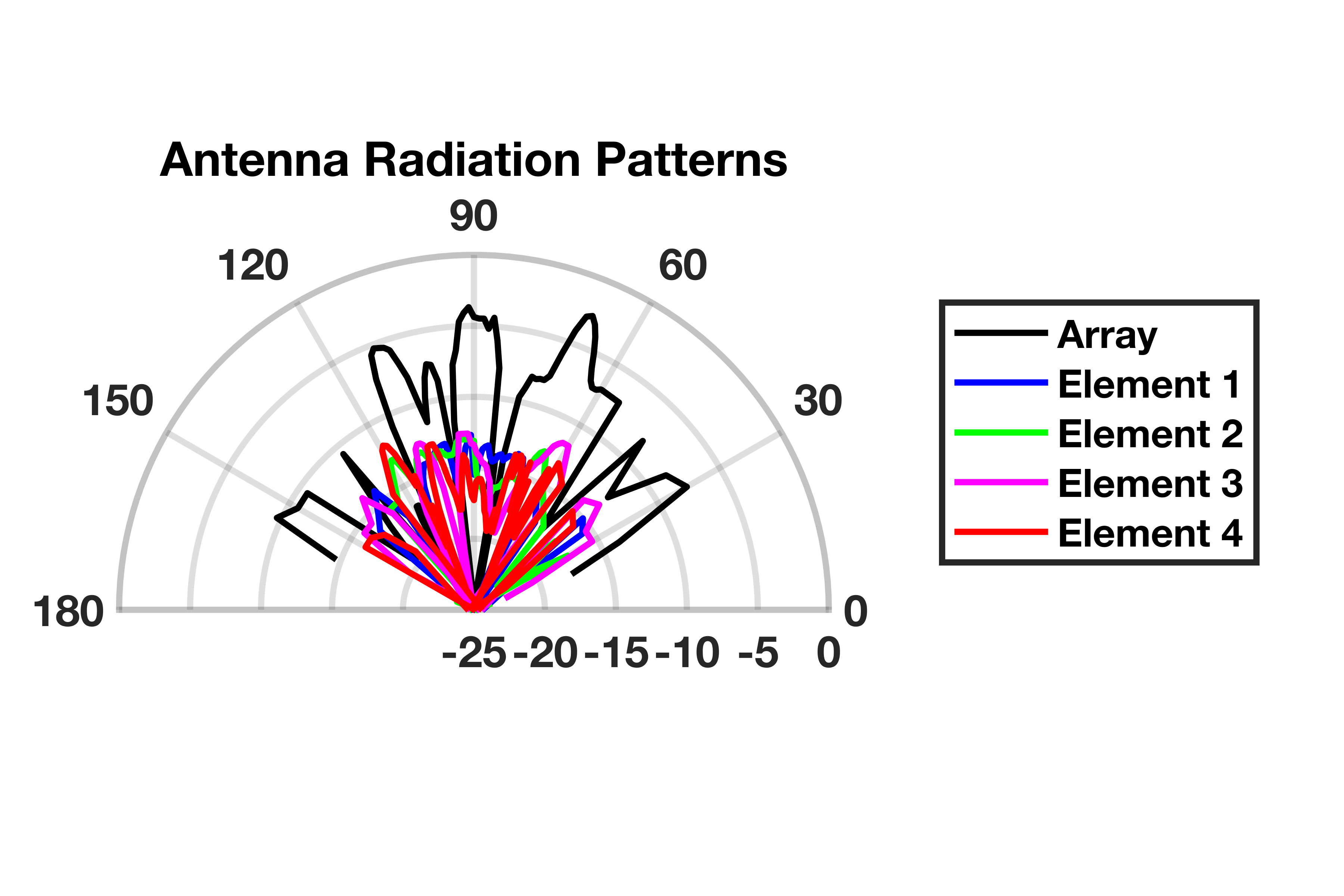}
	\caption{Gain radiation pattern of the four element array.}
	\label{fig:Results_BPA}
\end{figure}

The active points source at few of the measured angles such as $-5^\circ$, $\pm15^\circ$ and $-25^\circ$ is not clearly detected. We attribute this to nulls in the radiation pattern of the antennas and due to power mismatch between antennas and the wire-bonds. A manual measurement of the gain radiation pattern of all four antennas and the full array is carried out by measuring power meter reading for antennas at different angles. Figs. \ref{fig:Results_BP} and \ref{fig:Results_BPA} show the radiation patterns. Nulls  can be seen in the array beam pattern for angles $\pm15^\circ$ explaining the error in detecting point sources at those angles. These nulls are undesirable and should be carefully avoided by better antenna design and matching network design to match antennas to the wire-bonds and the receiver front-end.

\begin{figure}[hbtp]
%\vspace{-1em}
 \centering
	\subfloat[]{\includegraphics[clip,trim=0cm 0cm 1.3cm .8cm, width=0.4\textwidth]{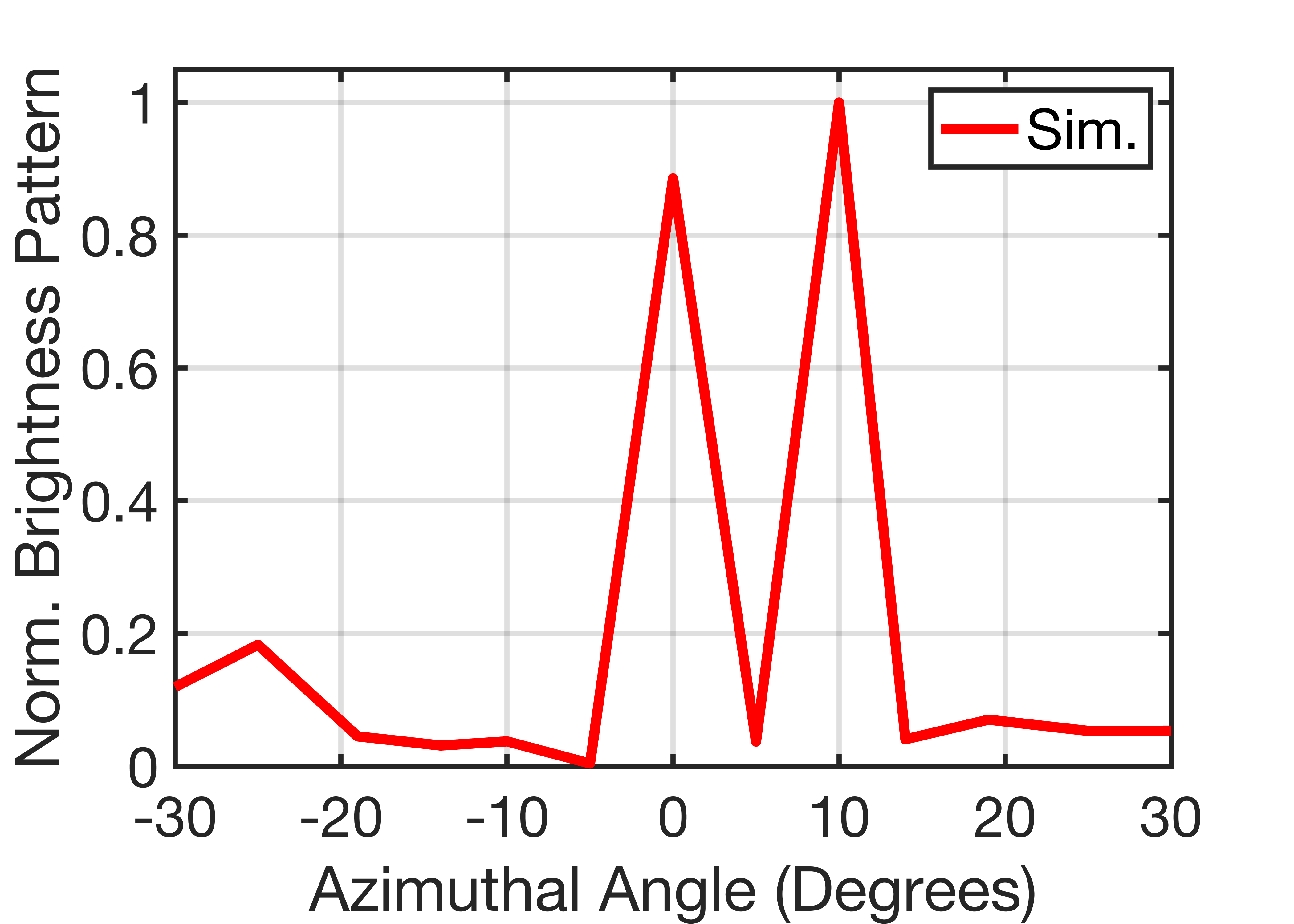}
		\label{fig:2sources}}
	\subfloat[]{\includegraphics[clip,trim=0cm 0cm 1.3cm .8cm, width=0.4\textwidth]{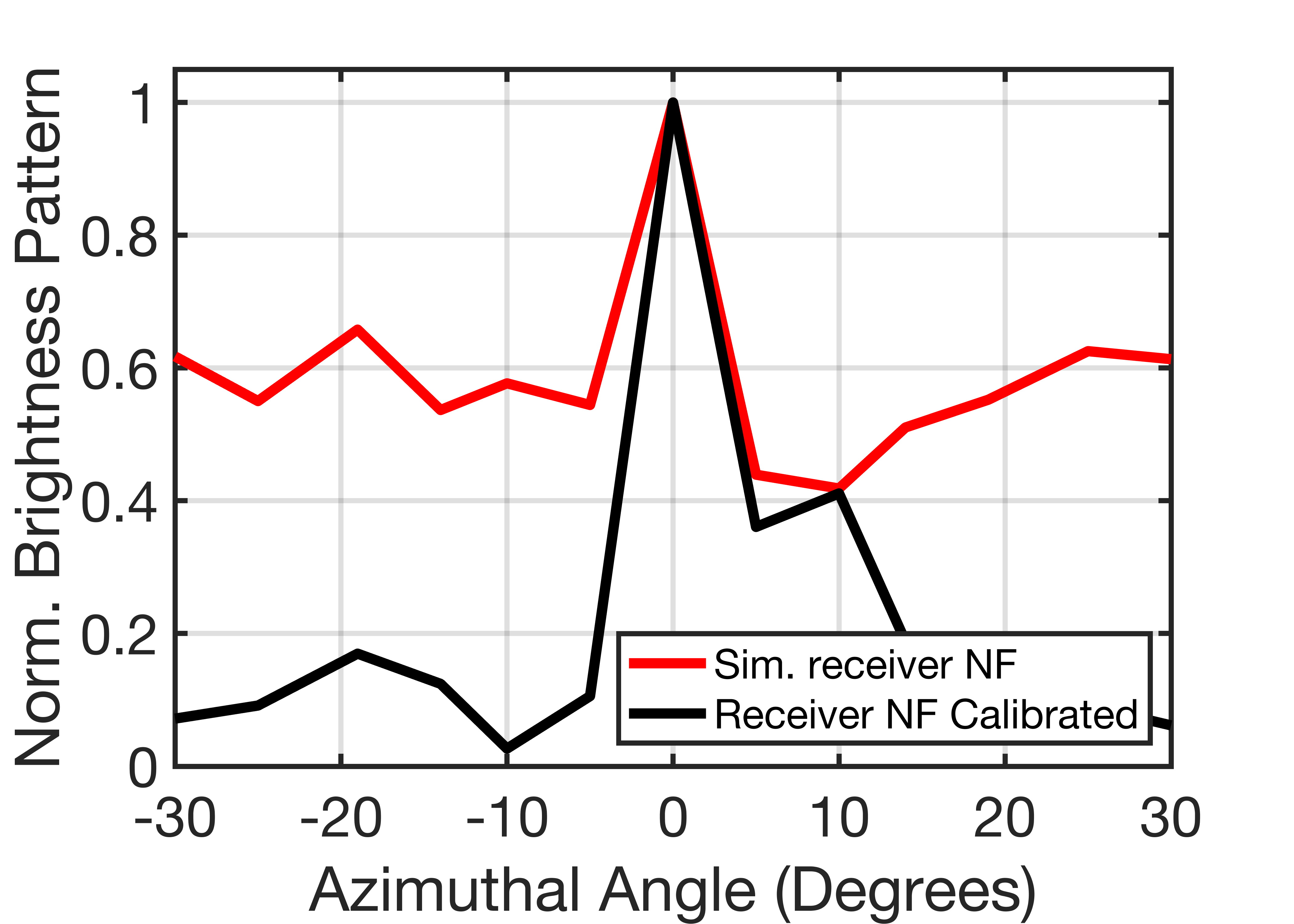}
		\label{fig:noise_cal}} 
	\caption{Brightness simulation with (a) two point sources at  0$^\circ$  and  10$^\circ$; and (b) 22 dB receiver NF before and after noise calibration.}
	\label{fig:simu}
%	\vspace{-1.5em}
\end{figure}

A four-element 60 GHz CMI imager is simulated using a SIMULINK\textsuperscript{\textregistered} behavioral model of the imager and results are compared to measurements in Fig. \ref{fig:Results_all}. The brightness errors are observed to be higher in measurement, which is attributed to different and non-ideal gain patterns of individual antennas, as shown in Fig. \ref{fig:antennaGain} for simulation and Fig. \ref{fig:Results_BP} for measurement, and low resolution of the imager. These artifacts can be removed using known image processing techniques for interferometry, such as deconvolving with the point spread function (PSF) and image cleaning algorithms. Finally, in Fig. \ref{fig:2sources}, it is again shown through simulation that two point sources can be resolved by this 13-pixel CMI imager.
% in spite of low resolution.} 

It is important to note the effect of receiver noise on system performance. In CMI, the receiver noise primarily affects the zero-baselength visibility measurement, causing a DC-like shift in the brightness pattern. This is because the majority of noise generated in each receiver is uncorrelated and does not contribute to cross-correlations of receivers but adds in total power received. This can be calibrated by either measuring the total noise of the system in the absence of a source or setting auto-correlation to zero in data processing. Auto-correlation gives the visibility sample $v_{0}$ which can be set to zero. This improvement is shown in Fig. \ref{fig:noise_cal} for a hypothetical 22 dB noise figure (NF) receiver. The effect of NF in our active imaging experiment is low due to high signal-to-noise power ratio and low enough NF of the array ($\approx$10 dB). Receiver noise, however, will reduce the sensitivity of an interferometer used for passive imaging, as discussed in Sec. \ref{sec:sensitivity_deriv}.

%%%%%%%%%%%%%%%%%%%%%%%%%%%%%%%%%%%%%%%%%%%%%%%%%%%%%%%%%%%%%%%%%%%%%%%%%%%%%

\section{Conclusion}
In this chapter we presented a 60-GHz implementation of a code-modulated interferometric imaging system using a packaged phased array. To use a phased array as an interferometer, incoming signals are code modulated using phase shifters, multiplexed using a power combiner, and processed through a shared receiver chain. Here, a four-element 60-GHz phased array chip is packaged with slot antennas, and a single 60-GHz output is measured using a power detector. This scalar measurement is then demodulated to obtain the interferometric visibilities. The four-element phased array is thinned to obtain a 13-pixel image and the system is demonstrated through the tracking of a point source at different angles, indicating the feasibility of using code-modulated interferometry for low-cost mm-wave imaging. This is the first ever demonstration of the CMI at mm-wave frequency. 

\chapter{Imaging System Prototypes using COTS Phased Arrays}
\label{chap-three}

\section{Overview}
 
The behavioral models as discussed in Chapter \ref{chap-two} and one-dimensional four-element interferometer in Chapter \ref{chap-four} serve as a proof of concept for feasibility of code-modulated interferometry. To further validate and present this technique as a low-cost alternative to existing imaging systems, prototypes with commercially available phased arrays and two-dimensional image capability are built by tiling several phased array chips. Several different hardware options are explored for constructing a two-dimensional imager, such as:
\begin{itemize}
    \item \emph{In-house 60-GHz phased array}: Extending the 60-GHz imager from Chapter \ref{chap-four} into a 2-D imager by stacking multiple chips at board level. For a 16-element imaging system, four 60-GHz phased array would be needed. To apply code-modulated interferometry, the mm-wave output from four phased arrays would need to be power combined at board level before feeding into a power meter. Also, the placing of $16$ antennas and the routing of their respective feed-lines would be challenging. Wire-bonding four different chips at board level and cavities would further complicate the design. Due to these challenges in board design, this option was not pursued.
    \item \emph{In-house 28-GHz phased array}: In-house four-element 28GHz phased array transceivers have been developed under a parallel program at NC State university \cite{DanielPhD}, which can then be tiled to create a larger system. The outputs would need to be power combined at baseband, as the receiver includes a downconversion mixer. Similar to the in-house 60-GHz receivers, the mm-wave wire-bonding of the chips, and their placement/routing on high frequency board was found to be the major challenge and thus this option was not pursued.
    \item \emph{In-house 24-44 GHz wide-band receiver}: One of the major challenge in the above mentioned in-house phased array solutions is that they were not designed with the purpose of packaging and thus flip-chip bonding was not an option. One of the other options explored is to design and manufacture ulta-wide band (UWB) receivers (24 - 44 GHz) \cite{VikasLNA} and optimize the design for flip-chip and packaging. The ultra-wide-band is also advantageous in improving the sensitivity of the imaging system. Appendix.\ref{chap-six} discusses the UWB low-noise amplifier design in detail which was taken as a first step towards realizing this system. The design and validation of the UWB mixer and baseband has been completed, published in \cite{Hari2019}; however, the combination of the LNA and the mixer/receiver was not pursued in the parallel program. Furthermore, this approach would require the creation and packaging/testing of multiple parallel receivers to realize the imager. This was not pursued, and we decided to use commercial-of-the-shelf (COTS) phased arrays instead.
    \item \emph{COTS 60-GHz radios}: Several COTS 60-GHz radios are explored, such as IBM's 60GHz transceivers chip and 60GHz WiHD transmitter and receiver module pair by Lattice Semiconductor (DVDO Air3C pro), 60-GHz phased arrays by imec, etc. We were unable to procure any of these mm-wave phased arrays.
    \item \emph{COTS 10-GHz phased array by ADI}: Since the code-modulated interferometry is independent of the frequency of use, a lower frequency (10-GHz) multi-element imaging system prototype is used for demonstration instead. 8 GHz - 16 GHz (X band and Ku band), four-channel ADAR1000 beamformer by Analog Devices Inc. \cite{Adarchip} is available commercially. This solution has evaluation boards (EVAL-ADAR1000 \cite{AdarEval}), where the array chips are fully packaged and connected to digital interface. Thus, it allowed experiments for imaging and avoided any complications associated with packaging, antenna design, assembly, digital programming. Thus, the 10 GHz approach has been pursued  and been used for several prototypes in this chapter \cite{IRIS2019}. One disadvantage of using 10 GHz hardware is bigger wavelength and thus lower resolution.
\end{itemize}

In this chapter, we present two hardware prototypes for CMI using 10 GHz COTS phased arrays from Analog Devices Inc. (ADI). First, a 33-pixel, eight-element prototype is created using two commercially-available ADAR1000 phased-array receivers from Analog Devices Inc. The chips are connected at board level to a patch antenna array. The serial interface is used to apply codes whereas the on-chip power detector and data converter are used for direct read out of the composite code-multiplexed imaging data. These are then processed off-line in MATLAB to reconstruct the image. The 33-pixel camera is demonstrated in hardware for point-source detection. In second part, the prototype is extended to obtain higher pixel density and to test the scalability of the CMI. Four ADAR1000 boards are connected together to construct a 16-element imaging system, which can produce an image with upto 169 pixels. The system is demonstrated using one and two point sources to obtain point spread function, and demonstrate the resolution of the system, respectively. Here we show that the cost of the imaging solutions can be kept low by re-purposing mass-market produced communication phased arrays and using interferometry to reduce the number of channels required for same number of pixels.

Zhangjie Hong\footnote{North Carolina State University, Raleigh, NC, USA 27606} and Simon Simon Sch\"onherr \footnote{Karlsruhe Institute of  Technology, 76131 Karlsruhe, Germany} contributed to this work on imaging system design and the built-in-selt-test (BIST) implementation of code-modulated interferometry (Code-modulated-embedded-test, or CoMET) in both designing the 10-GHz prototypes and performing the experiments \cite{Simon18,CometIMS19, IRIS2019, CometTMTT2021}.

%%%%%%%%%%%%%%%%%%%%%%%%%%%%%%%%%%%%%%%%%%%%%%%%%%%%%%%%%%%%%%%%%%%%%%%%%%%%%

\section{Code-Modulated Interferometry in COTS Phased Arrays}
\label{sec:CMII}
As discussed in Chapter \ref{chap-two}, a sparse antenna array connected to a phased-array can be used to perform interferometry with a few modifications. To repurpose a COTS analog phased array as an interferometer, incoming signals are code-modulated using phase shifters within each front end. Signals are then combined and processed using a shared hardware path. Visibility data is obtained though either (a) generation of code-demultiplexed antenna responses (without integration) which are then cross correlated or (b) squaring of the combined response using a power detector to create an interference response and then demodulation using code products of interest \cite{Spie16}. Method (b) is used here due to availability of power detectors on board. In particular, the demodulated correlation ($v_{ij}$) between elements $i$ and $j$ is extracted from the squared summation as follows:
\begin{equation}
v_{ij}=\int_0^T c_ic_j \left( \sum_{n=1}^{N} A_n cos(\omega_o t + \psi_n + c_n \frac{\pi}{2}+\Theta_n)\right )^2 dt,
\label{eqn:visibility}
\end{equation}
where this assumes the incoming signals are band-limited around $\omega_o$ with particular amplitude ($A$) and phase ($\psi$). Codes $c_n$ are $\pm 1$ sequences chosen such that all code products are orthogonal. A static phase shift, $\Theta$, can be applied to each element to calibrate the hardware and/or place the element in a ``real" or ``imaginary" state, as will be described later. As described in Chapter \ref{chap-two}, the modulation can be generally applied in a Cartesian manner to both in-phase and quadrature-phase versions of the signal, allowing concurrent demodulation of both real and imaginary visibilities. Here, we simply state that the modulation has a unit increment $\pi/2$ and then an offset, $\Theta$. 
All possible visibilities for the array are demodulated in parallel to create the full visibility function.

\begin{figure}
\centering
\includegraphics[width=1\textwidth]{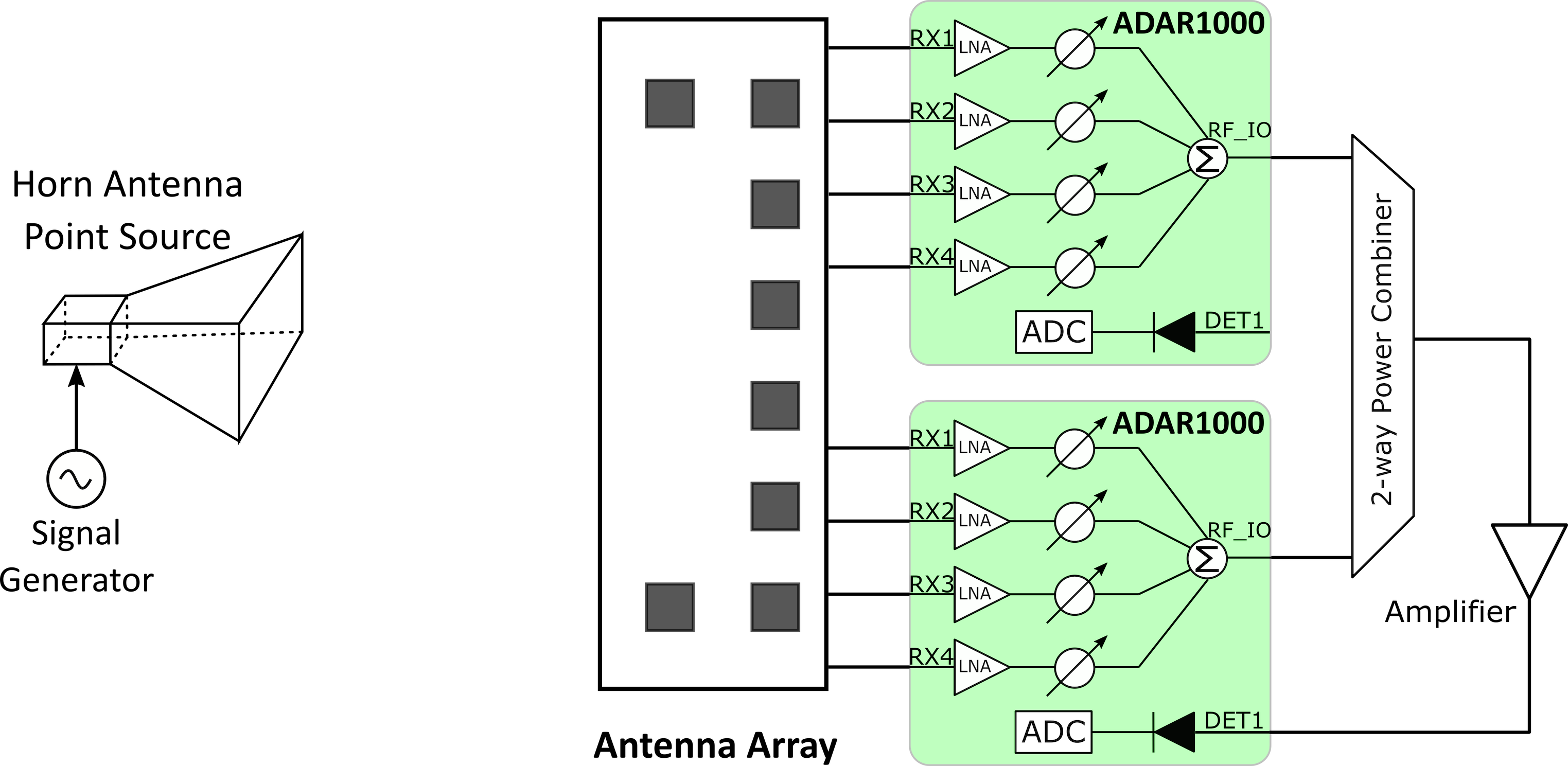}
\caption{Block diagram of interferometric imager using two phased arrays.}
\label{fig:schematic_10G}
\end{figure}

%%%%%%%%%%%%%%%%%%%%%%%%%%%%%%%%%%%%%%%%%%%%%%%%%%%%%%%%%%%%%%%%%%%%%%%%%%%%%

\section{33-Pixel COTS CMI Imaging Prototype}
\label{sec:design}
To test the feasibility of CMI using COTS phased arrays, first a 33-pixel, eight-element 10-GHz eight-element imaging system prototype is built as shown in Fig. \ref{fig:schematic_10G} and \ref{fig:fullsetup} \cite{IRIS2019}. We choose to first create a smaller eight element array for the ease of building the imager.  In the process, we learn to fully control the EVAL-boards, connect with available antenna arrays without modifications, and establish the preferred calibration method of the setup. A working smaller array with COTS hardware would give us the confidence and would serve as a stepping stone to building larger array with multiple EVAL-boards and a custom antenna array.  

%%%%%%%%%%%%%%%%%%%%%%%%%%%%%%%%%%%%%%%%%%%%%%%%%%%%%%%%%%%%%%%%%%%%%%%%%%%%%

\subsection{Eight-Channel Phased Array}
\label{sec:PhasedArray}
\begin{figure}
\centering
\includegraphics[width=.6\textwidth]{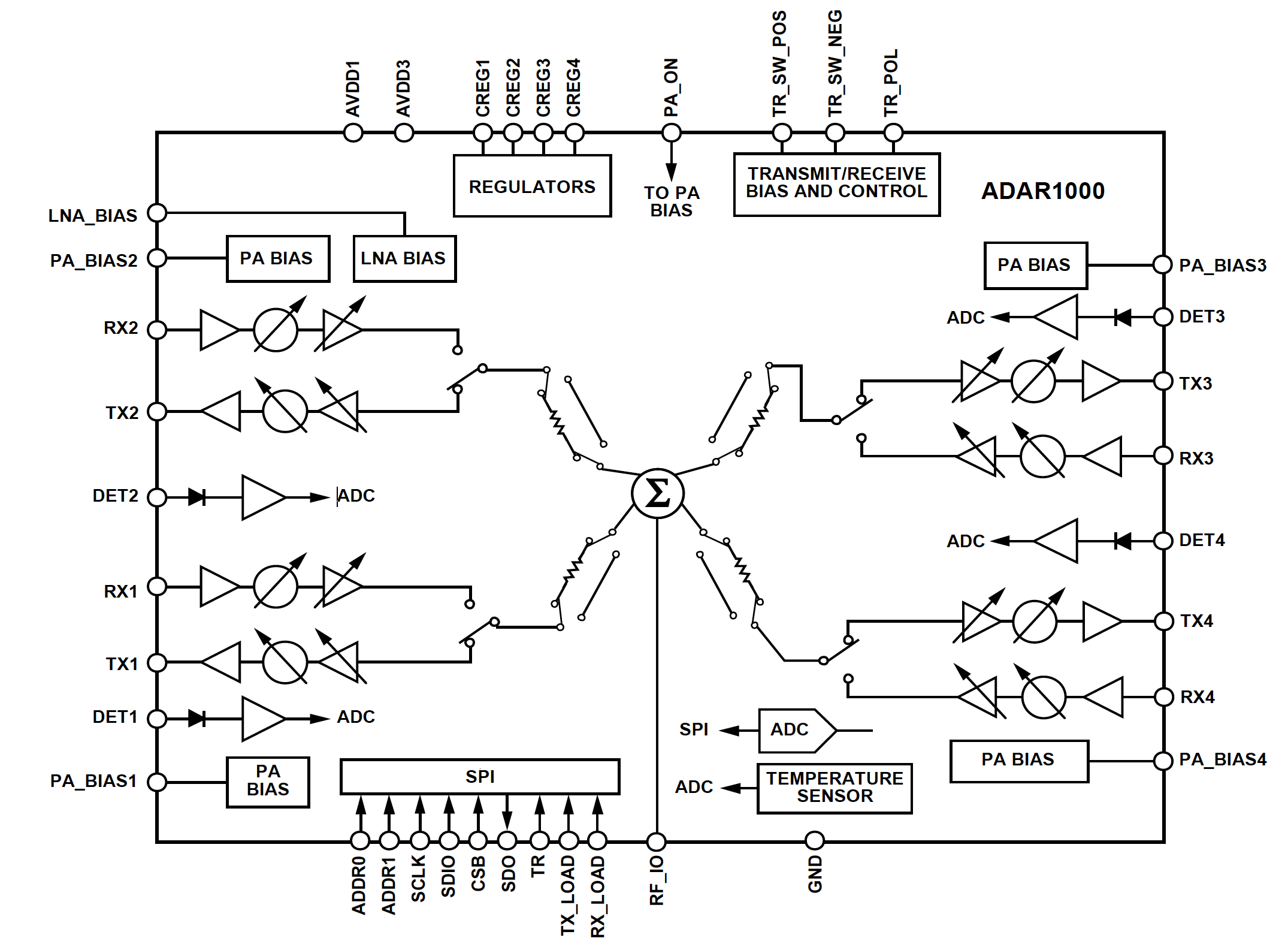}
\caption{Block diagram of ADAR1000 chip showing four TX (TX1 - TX4) and four RX (RX1-RX4) \cite{Adarchip}.}
\label{fig:AdarBlock}
\end{figure}

\begin{figure}
\centering
\includegraphics[width=.6\textwidth]{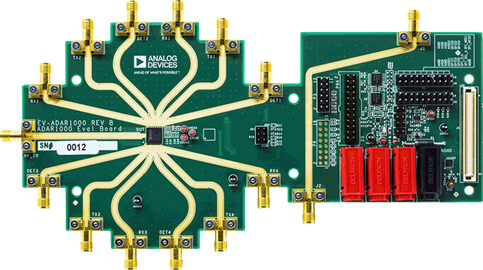}
\caption{Photo of ADAR1000 EVAL board showing four TX (TX1 - TX4) and four RX (RX1-RX4) \cite{AdarEval}.}
\label{fig:AdarEval}
\end{figure}

As shown in Fig. \ref{fig:schematic_10G}, two COTS phased-array ADAR1000 8--16 GHz receiver chips from ADI are used to construct the array \cite{Adarchip}. Each ADAR1000 chip contains four identical TX and RX channels and each TX/RX channel contains a power amplifier (PA)/low noise amplifier (LNA), a variable-gain amplifier (VGA), a switchable attenuator and a phase shifter. The VGA and the switchable attenuator together provide a 31 dB tuning range of channel gain. The phase shifter is implemented with the vector interpolator topology using I and Q VGAs. Each VGA has a five-bit amplitude control and a one-bit polarity control, resulting in 12-bit control and $2.8^\circ$ phase resolution for each phase shifter. Its polarity bit can be used directly to apply the code-modulation. For each chip, an on-chip 4:1 power combiner generates a single summed RF output. Additionally, each ADAR1000 includes four power detectors (PD) along with an ADC to digitize the PD output. Figs. \ref{fig:AdarBlock} and \ref{fig:AdarEval} show the block diagram of the ADAR1000 chip and photograph of the ADAR1000-EVAL board, respectively. In this experiment, two separate ADAR1000 evaluation boards \cite{AdarEval} are used to create an eight-element array-under-test. These evaluation boards contain additional circuitry for power supply management and digital control through serial peripheral interface  (SPI). 

Two four-element arrays are combined using evaluation boards, antennas and cables to create an eight-element array. The two 4:1 combined signals from each RX array are merged using an external 2:1 combiner (ARRA 3-9200-2) and then amplified using an external amplifier (Mini-Circuits\textregistered \ ZVA-213S+). The signal is then fed back into one on-chip PD. All modulations and output power readings are accessed through the SPI.

%%%%%%%%%%%%%%%%%%%%%%%%%%%%%%%%%%%%%%%%%%%%%%%%%%%%%%%%%%%%%%%%%%%%%%%%%%%%%
    \begin{figure}
    \centering
    \includegraphics[width=0.6\textwidth]{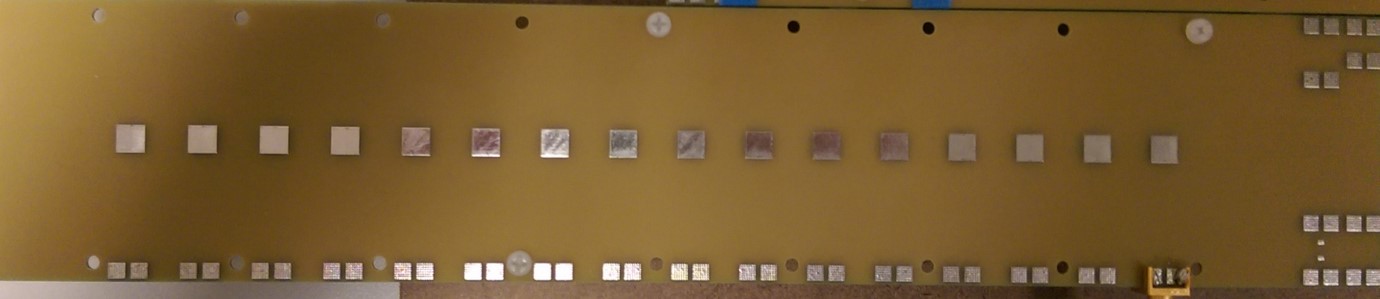}
    \caption{Photo of the available patch antenna board with 16 linearly $\lambda/2$ spaced antenna elements.}
    \label{fig:antenna_board}
    \end{figure}

\subsection{Sparse Antenna Array}
\label{sec:antenna}
\begin{figure}
    \centering
    \includegraphics[width=0.9\textwidth]{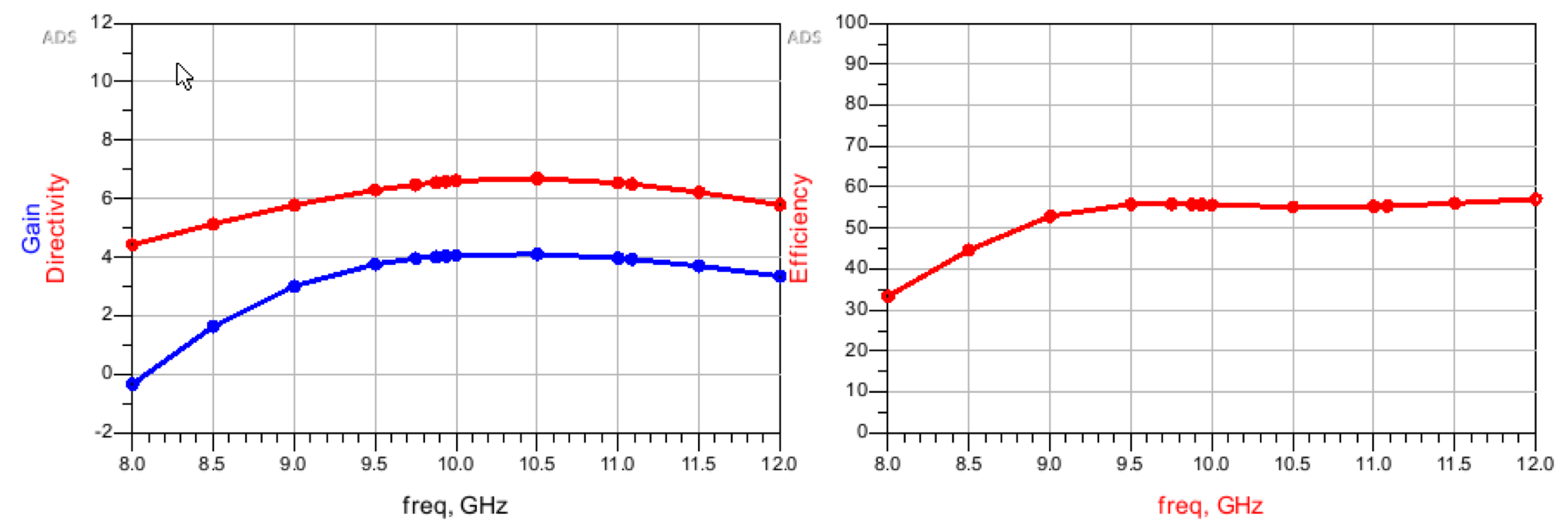}
    \caption{Gain, directivity, and efficiency of a single patch antenna, provided by ADI \cite{ADI}.}
    \label{fig:ADI_ant_gain}
    \end{figure}

\begin{figure}[hbtp]
 \centering
	\subfloat[]{\includegraphics[clip,trim=0cm 0cm 0cm 0cm, width=0.4\textwidth]{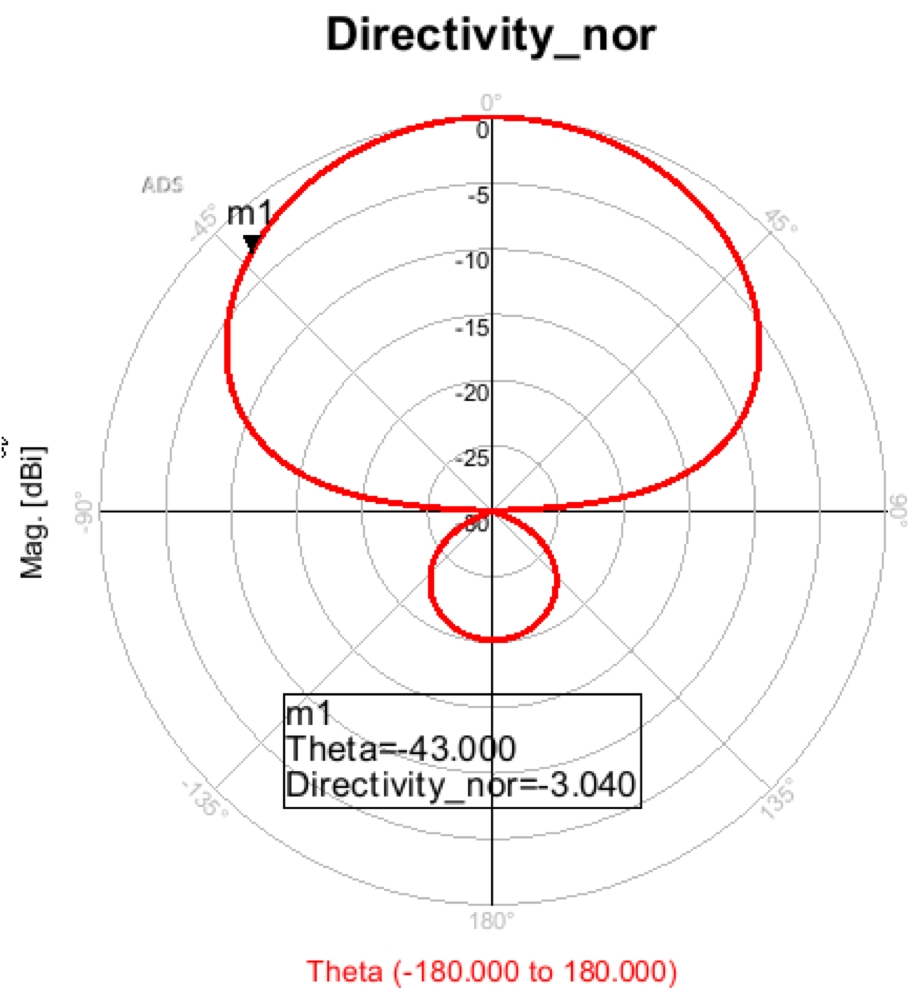}
		\label{fig:ADI_ant_BW}}
	\subfloat[]{\includegraphics[clip,trim=0cm 0cm 0cm 0cm, width=0.5\textwidth]{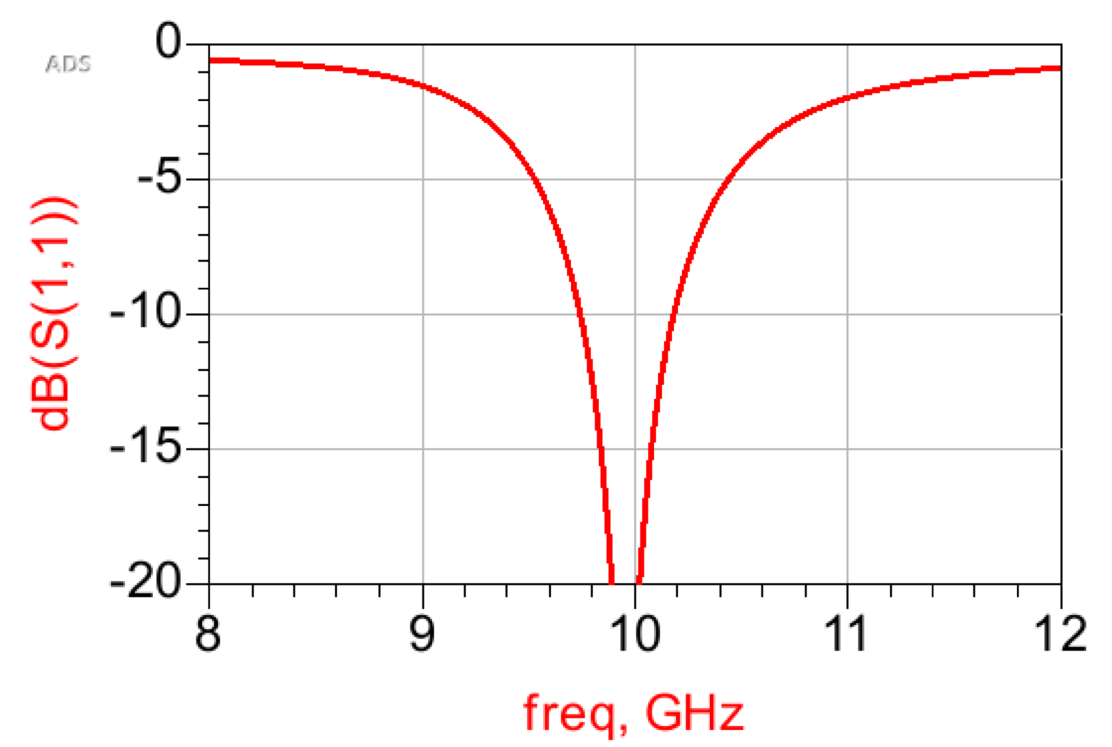}
		\label{fig:ADI_ant_S11}} 
	\caption{(a) Beam-width and (b) input match of a single patch antenna, provided by ADI \cite{ADI}.}
	\label{fig:ADI_ant}
\end{figure}

Patch antennas fabricated on a printed circuit board designed and provided by ADI \cite{ADI} are used for the array, as shown in  Fig. \ref{fig:antenna_board}. The material used for design is FR402 with a dielectric constant of 4.25. Each patch antenna has a gain of 4 dBi and 86$^\circ$ half-power beam-width, and have a narrow input match around 10 GHz, as shown in Figs. \ref{fig:ADI_ant_gain} and \ref{fig:ADI_ant}.

\subsubsection{31-pixels 1-D Imaging System}
\begin{figure}
    \centering
    \includegraphics[width=.9\textwidth]{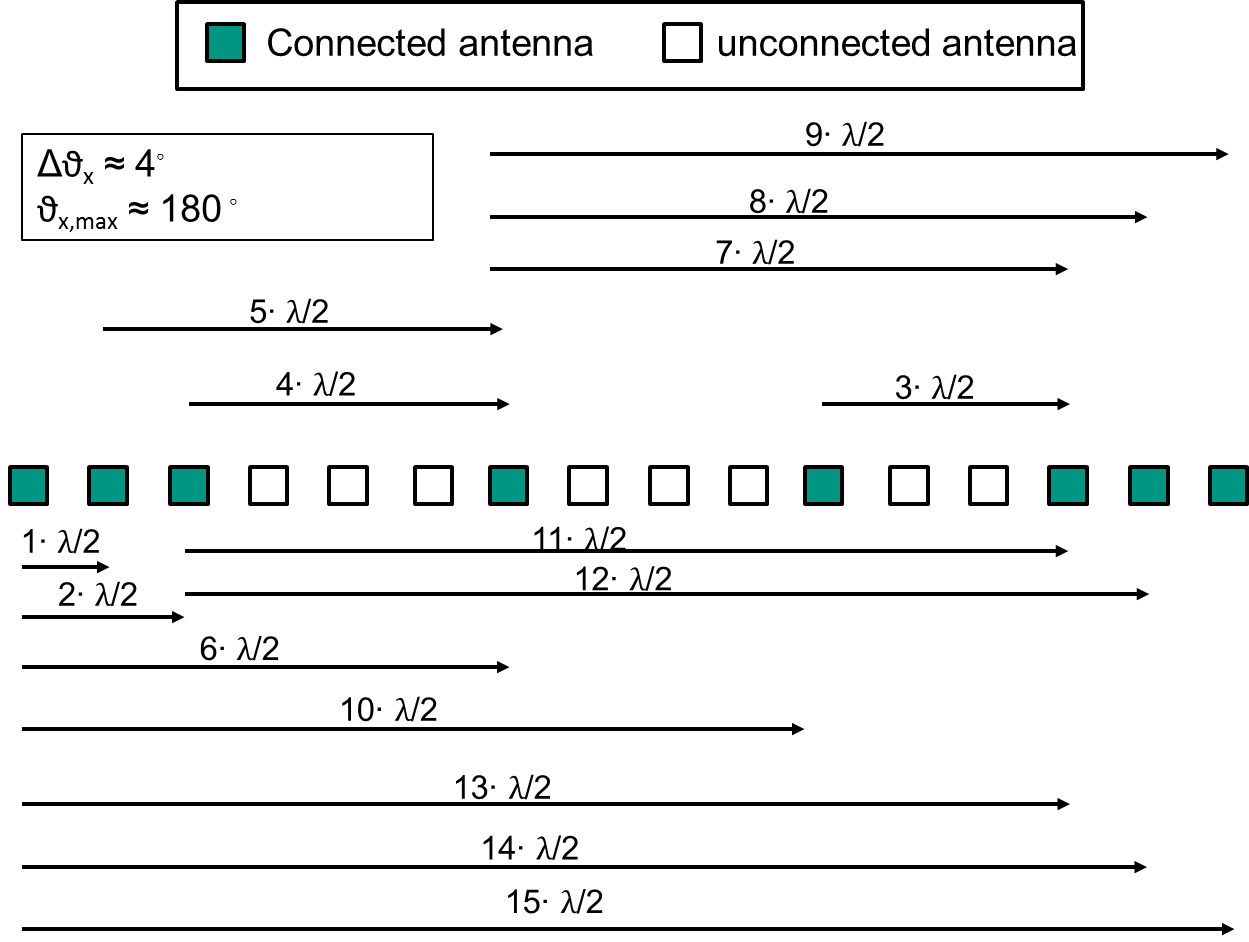}
    \caption{Sparsely sampled antenna configuration for one-dimensional imaging illustrating all available baselines \cite{Simon18}.}
    \label{fig:1_dim_spacing}
    \end{figure}

\begin{figure}
    \centering
    \includegraphics[width=1\textwidth]{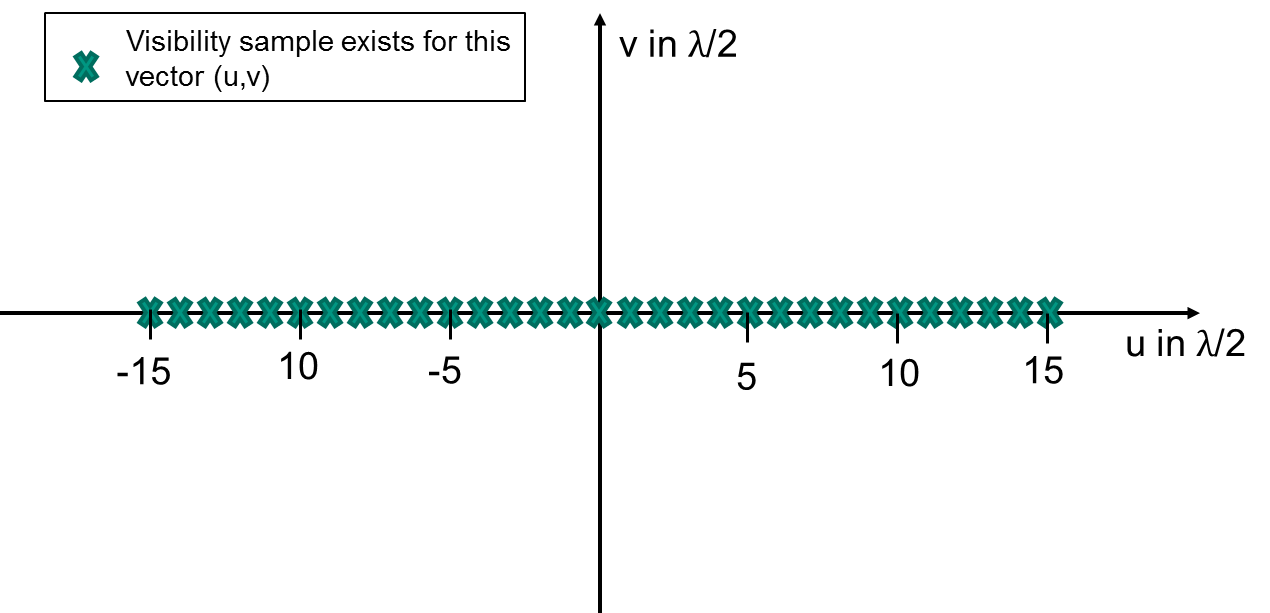}
    \caption{Sampled (u,v)-space for the used antenna configuration in one dimension \cite{Simon18}.}
    \label{fig:1_dim_grid}
    \end{figure}

In continuation of the behavioral models in Chapter \ref{chap-two} and one-dimensional prototype in Chapter \ref{chap-four}, we first demonstrate one-dimensional code-modulated interferometric imaging using the COTS phased arrays. For a sparsely sampled $u-v$ plane, the eight elements in our setup can provide a uniform sampling of $u$ up to $u=\pm15$, a maximum baseline of length $15.\lambda/2$ and up to 31-pixels in image, as shown in Figs. \ref{fig:1_dim_spacing} and \ref{fig:1_dim_grid}. We are limited to a maximum baseline of $15.\lambda/2$ as we connect two antenna boards side-by-side with eight patch antennas each and $\lambda/2$ spacing. The positions of antennas in a minimum redundant interferometric array with eight elements can be found in \cite{Thompson08, Kopil01} and have been adapted for this setup. Fig. \ref{fig:1_dim_grid} shows the baselines that are achieved in this configuration, \textit{i.e.} the distance vectors (per unit half-wavelength) between pairs of antennas for which complex cross-correlations can be obtained in this particular antenna configuration of interferometric array. Note that there are no holes in the $u$-dimension for $u=$ -15 to +15. The complex visibilities for the pairs of antennas shown in Fig. \ref{fig:1_dim_spacing} provide the baselines $u=$ 0 to +15 and their complex conjugate provide the remaining baselines $u=$ -15 to 0. This also tells us the number of pixels achieved in the image after Fourier transform. Since the antennas are placed at a uniform distance of $\lambda/2$, the field-of-view (FOV) is complete $180^\circ$ and primarily limited by the beam-width of antennas. The angular resolution, fixed by the maximum baselength is $\approx 4^{\circ}$ and the center beam width is $\approx 8^{\circ}$ .

\begin{figure}
 \centering
	\subfloat[]{\includegraphics[width=0.7\textwidth]{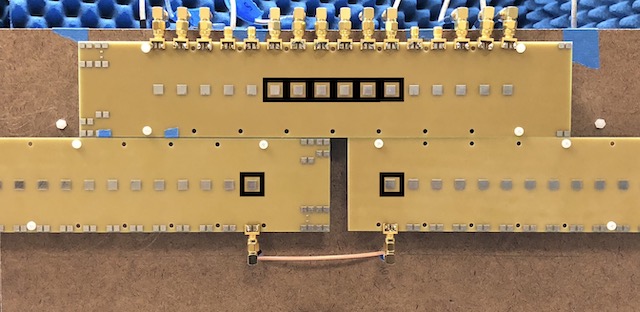}
		\label{fig:antenna}} \\
	\subfloat[]{\includegraphics[width=0.7\textwidth]{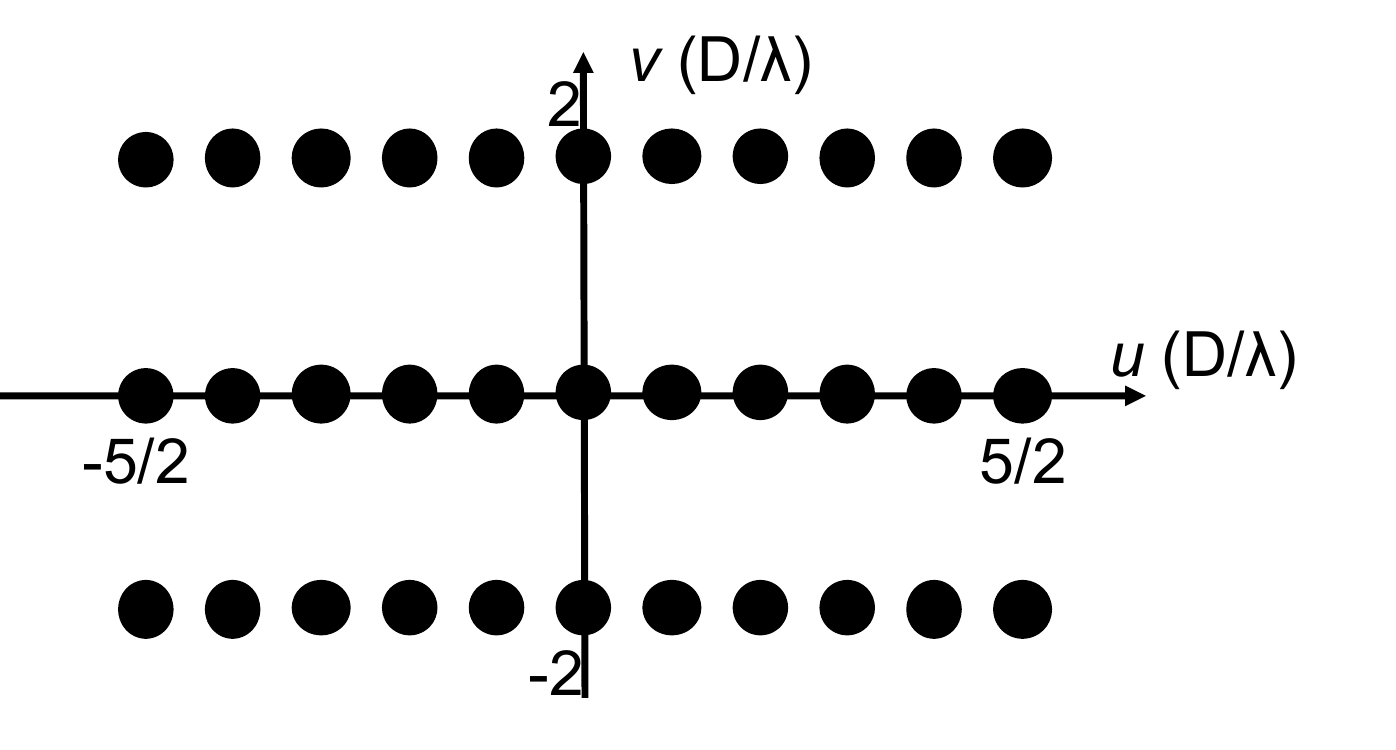}
		\label{fig:UVplane}}
	\caption{(a) Photograph of prototype front view showing antennas. (b) Plot indicating the sampled \textit{u-v} space.}
	\label{fig:ExpSetup}
\end{figure}

\subsubsection{33-pixels 2-D Imaging System}
We then extend the antenna array to sample the $u-v$ space in 2-dimension with the same number of elements by placing three antenna boards (with eight $\lambda/2$ spaced patch antennas). Fig.\ref{fig:antenna} shows the chosen interferometric antenna locations and Fig. \ref{fig:UVplane} shows the sampled \textit{u-v} plane which is obtained. The patch antennas are on three different substrates which might affect the individual antenna performance and therefore would need calibration. Note that two of the antennas have opposite polarity than rest of the six antennas and therefore the cross-correlations with these antennas would have the signs reversed. Eight antennas in this configuration generate 33 samples in the \textit{u-v} plane and 33 pixels in the image. Out of an available 2.$^8C_2=56$ total complex visibilities (or baselines), only 32 baselines are unique and thus $42\%$ (24 out of 56) are for redundant baselines in this configuration, which can be averaged to reduce noise. The minimum and maximum spacing in the $x$-direction are $\lambda/2$ and $15\lambda/2$ respectively, corresponding to an azimuth FOV and resolution of $\pm90^{\circ}$ and 10.5$^{\circ}$, respectively, given by Eqn. \ref{eqn:sampling}. Due to the size limitation of the antenna boards, the minimum spacing in the $y$-direction is $ 2\lambda $, corresponding to an elevation FOV and resolution of $\pm14.5^{\circ}$ and 9.6$^{\circ}$, respectively.

\subsection{Calibration of Array}
\label{sec:calibration}
Calibration of the array is crucial for the performance of imaging systems. Since interferometric imaging works by detecting amplitude and phase differences between incoming signals, any mismatch within the hardware leads to image artifacts since corruption to the signal in any of the antennas will generate noise at all the pixels in image. Although our COTS phased arrays come with pre-defined look-up tables to control the response, errors still exist due to element and chip variations, frequency dependencies, and offsets introduced within packages, boards, cables, and antennas. 

\begin{figure}
\centering
\includegraphics[width=.8\textwidth]{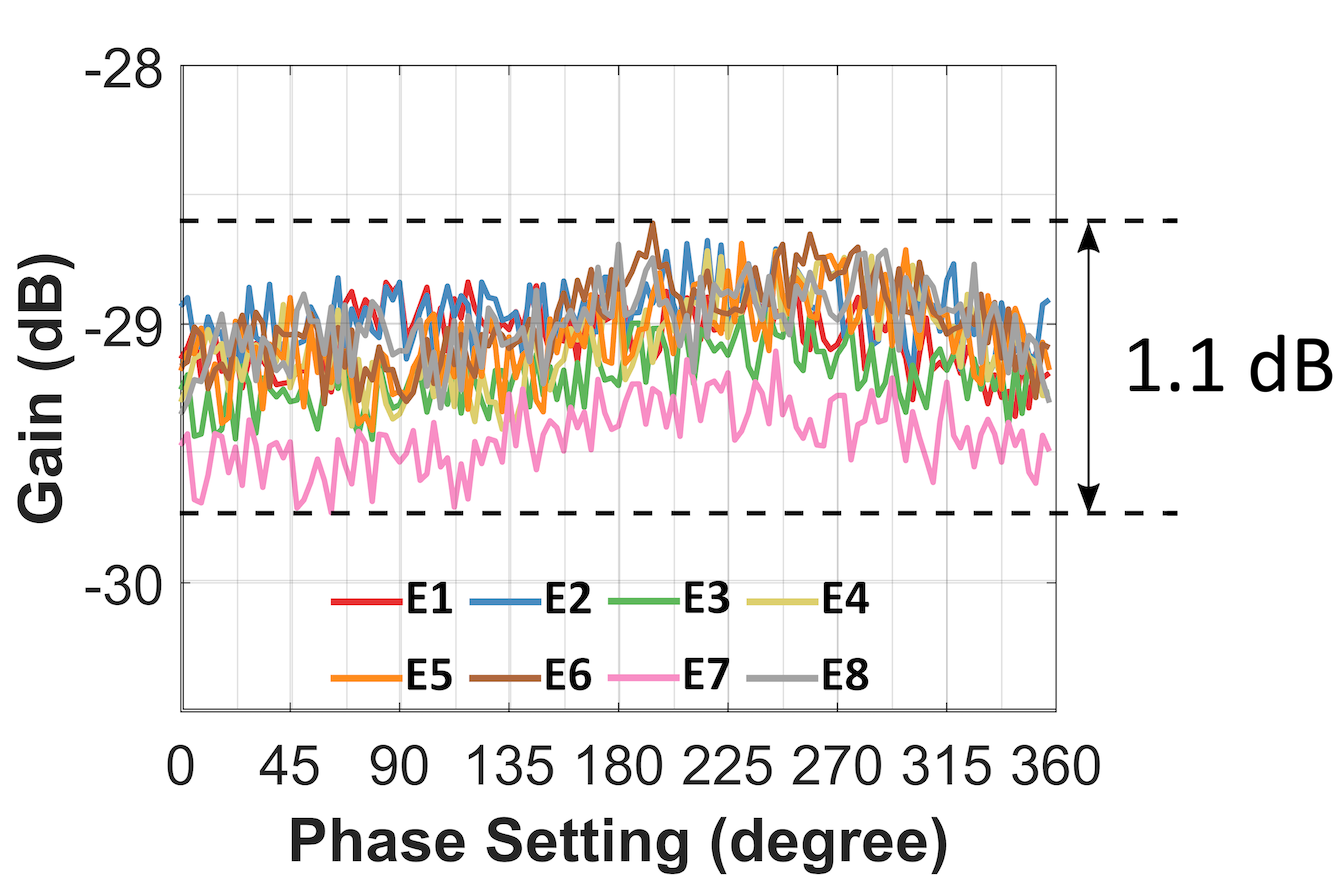}
\caption{Example plot for post-calibration gain of eight RX channels.}
\label{fig:antennagain}
\end{figure}

CMI using a known test signal can be used to calibrate the array, a technique known as Code-Modulated Embedded Test, or CoMET \cite{Greene2016, Greene2018, KevinPhD, CometTMTT2020, CometArxiv2021, CometCICC2018}. In CoMET, a far-field test signal is coupled into the array and the complex cross-correlations between elements are measured and used to extract gain and phase information per element. We use this technique to calibrate and equalize the response of the full array with antennas, with more information provided in \cite{CometIMS19}. 
Fig. \ref{fig:antennagain} shows the resulting ``over-the-air" extracted gain from \cite{CometIMS19} for our eight-element RX array attached to eight adjacent antenna elements after a CMI-based calibration. Here the antennas are oriented in a uniform linear array for demonstration purposes. This extracted gain includes approximately 60 dB of free-space path loss and approximately 25 dB of the TX horn antenna gain. Gain can be equalized with an accuracy of 1.1 dB. 

Code-modulated interferometric imaging and code-modulated interferometric built-in-self test and calibration are operationally similar in a way that both rely on code-modulation at the front-end of the receiver and obtaining cross-correlations of the signals from different antennas. The difference is, however, in the application. In CMI imaging, a calibrated array looks at an unknown scene to obtain an image, whereas in CoMET, an un-calibrated array looks at a known scene (a single bright point source) and feeds the obtained cross-correlations to an equation solver to obtain the array parameters of gain and phase errors between different elements of the array. CMI imaging requires a Fourier transform of the cross-correlations to obtain the image, whereas CoMET uses an equation solver to get the gain and phase errors. 

CMI theoretically may be used to calibrate for phase difference as well, however, here, we use a vector network analyzer for this purpose. In our experiments we notice that phase calibration using VNA and CoMET provide similar results, with VNA being slightly better. This is because the CoMET extraction of phase shifts also includes the phase difference due to path length of source antenna to the antenna array. Decoupling the phase offset due to path length and phase offsets in the array is work under progress in CoMET extraction and calibration. 
The phase calibration is limited to an accuracy of 2.8$^{\circ}$, which is the resolution of phase shifters in the ADAR-1000 chips.  

%%%%%%%%%%%%%%%%%%%%%%%%%%%%%%%%%%%%%%%%%%%%%%%%%%%%%%%%%%%%%%%%%%%%%%%%%%%%%

\subsection{Code-Modulation, Data Acquisition, and Processing}
\label{sec:coding}
As discussed earlier, RF signals are code-modulated using the phase shifters in the array. Binary ($\pm 1$) codes are used, \textit{i.e.} implementation is essentially a mixing of $\pm 1$ (or a series of 0$^{\circ}$ and 180$^{\circ}$ phase shift) to RF signals, hence, the polarity bit control of the VGAs in the vector-interpolating phase shifter is sufficient to apply the modulation. Eight 512-bit long Walsh-Hadamard codes are allocated to the eight channels. Codes are selected to have balanced orthogonal code products (BOCP) for a direct demodulation of cross-correlations from the power detector output. A Walsh-Hadamard code length of 512 bits can provide up to a maximum of 22 unique codes that have BOCP. For an eight element array, we require 16 unique codes to simultaneously code-modulate in both $I$ and $Q$ signal paths for obtaining complex visibilities. A smaller code length 256-bit Walsh-Hadamard code set can provide up to a maximum of 18 unique BOCP codes, but during implementation we found that certain codes from this set were producing high errors at the output. The reason for this is unknown and is a topic of future work. Therefore we decided to go for higher code length of 512-bits that provide more codes to choose from. As mentioned before, a MATLAB code generates the code set of BOCP codes from the 512-bit Walsh-Hadamard code set.

As elaborated in Chapter \ref{chap-two}, in an ideal CMI array both I and Q path in channels are code-modulated simultaneously with separate codes to obtain both real and imaginary visibilities concurrently. In that case, coding is applied in the phase shifter by moving among phase states of $\pm45$ and $\pm135$ based on the code sequence. In our prototype, however, the phase shifters are used to equalize the phases of antenna cables (\textit{i.e.}, using $\Theta$ in \eqref{eqn:visibility}). Therefore, real and imaginary visibilities are obtained sequentially in three runs. The first run provides all real visibilities, whereas the second and third runs are used to obtain imaginary visibilities by placing select elements into a quadrature state, similar to the 60-GHz experiment in Chapter \ref{chap-four}.  

Interestingly, obtaining the data in this sequential manner reduces the length of each required orthogonal codes. Obtaining the data sequentially reduces the number of required orthogonal codes from 16 to 8, hence reducing the BOCP code length by a factor of four (from 256 to 64 in carefully chosen BOCP code set from Walsh-Hadamard code set). Thus, three consecutive runs could effectively results in rather shorter total run time if minimum length codes were to be used. However, we continue to use code length of 512-bit for reducing noise in measurement by averaging.  

%%%%%%%%%%%%%%%%%%%%%%%%%%%%%%%%%%%%%%%%%%%%%%%%%%%%%%%%%%%%%%%%%%%%%%%%%%%%%
\begin{figure}
\centering
\includegraphics[width=.7\textwidth]{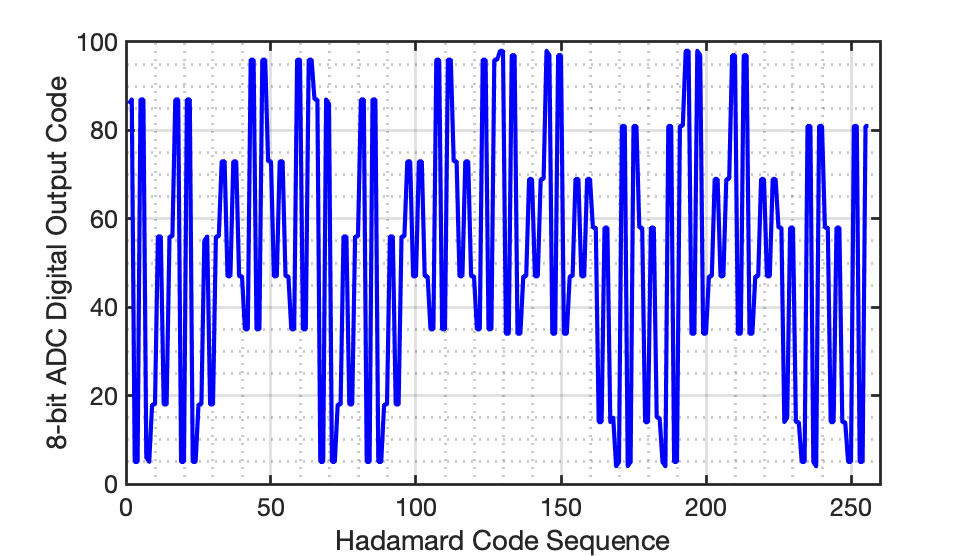}
\caption{Representative ADC output for four-element linear array with code length of 256.}
\label{fig:adcoutput}
\end{figure}

 The aggregate RF signal is squared and digitized using the on-chip PD and ADC. SPI is used to control the phased array, to apply code-modulation, and to read out the scalar power readings from ADC. Fig. \ref{fig:adcoutput} shows a representative data read out of an ADC for a single CMI measurement frame of a four-element array. It is to be noted here that this scalar power measurement frame is sufficient to obtain all possible cross-correlations within the elements. The total run time for a single image is 60 ms corresponding to a frame rate of 16 images per second, given by the time it takes to apply and then read out $3\times 512$ total readings through the SPI and limited by the speed of that interface. In our experiments, however, we control the hardware using a LabView script with embedded Matlab code for demodulation and image processing, resulting in a run time in the order of few seconds due to latency. Total run time is calculate as: \\
  Number of data written/read out = 3 X 512 = 1536\\
  Clock Frequency = 10MHz \\
  Time = 1536 X 16 I/Q settings X 24 Instructions X Clock Time Period = 58.9 ms 
     
A 10 MHz clock frequency with code length of 512-bit was used to for code-modulation with a total run time of approximately 60 ms (three runs), excluding the data processing time. For a continuous image acquisition in video mode, the data processing can be done in parallel with the subsequent image frame acquisition for a faster frame rate. The frame rate is limited by the clock frequency of SPI in this setup. For an integrated solution, the frame rate will be governed by rate of phase switching in phase shifter and/or sensitivity. Increased speed is a subject of future work.

%%%%%%%%%%%%%%%%%%%%%%%%%%%%%%%%%%%%%%%%%%%%%%%%%%%%%%%%%%%%%%%%%%%%%%%%%%%%%

\begin{figure}
\centering
\includegraphics[width=.95\textwidth]{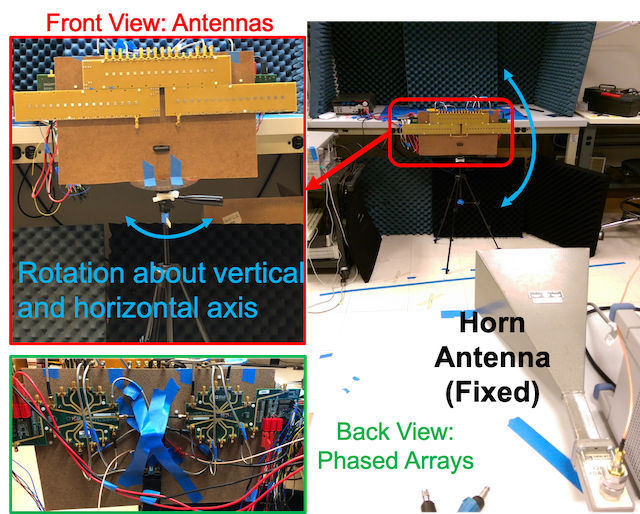}
\caption{Photographs of experimental setup of 10 GHz imager.}
\label{fig:fullsetup}
\end{figure}

\subsection{Imaging Experiment and Results}

We test our imaging system using an active point source placed at various azimuthal and elevation angles, as shown in Fig. \ref{fig:fullsetup}. The fixed point source is realized using a horn antenna connected to a 10 GHz signal generator (HP 83650B). The same setup is used to calibrate the imager, as described in section \ref{sec:calibration}. The imaging array is mounted on a camera tripod with the ability to rotate about both horizontal and vertical axis. 

\clearpage

\subsubsection{Results for 31-pixels 1-D Imaging System}
\begin{figure}
    \centering
    \includegraphics[clip,trim=0cm 0cm 9cm 0cm,width=.95\textwidth]{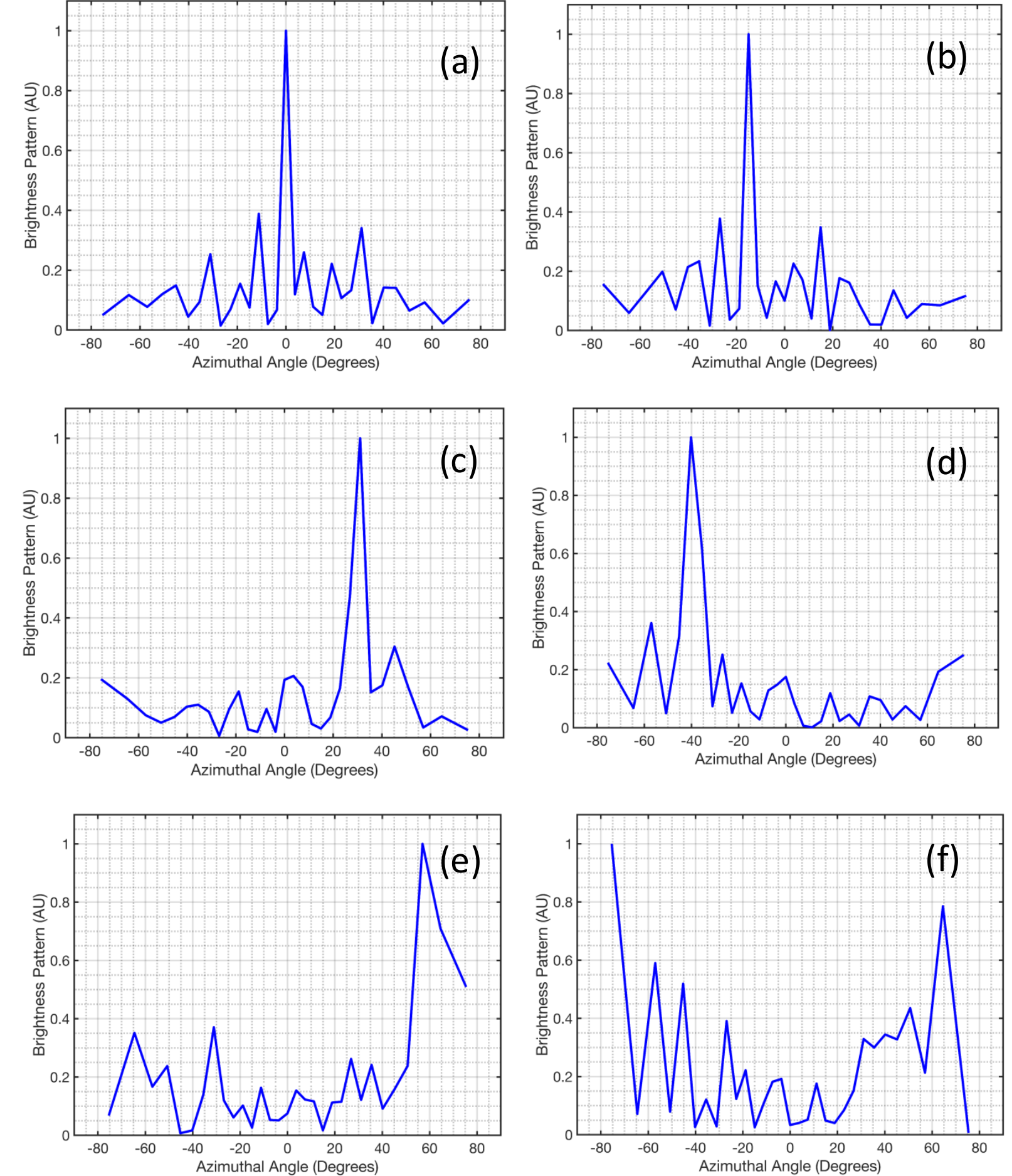}
    \caption{Normalized brightness plots for different point source positions: (a) 0$^\circ$ (b) -15$^\circ$  (c) 30$^\circ$ (d) -40$^\circ$  (e) 60$^\circ$ (f) 70$^\circ$.}
    \label{fig:1_dim_results}
    \end{figure}

Fig. \ref{fig:1_dim_results} shows the results for a point source at different angles correctly imaged and tracked using the one-dimensional eight-element 31-pixel CMI imager. The resolution is approx. 5$^\circ$, verified by moving the point source in steps of 5$^\circ$ and observing it move to next bin in the image.  Since the antennas are placed at a uniform distance of $\lambda/2$, the field-of-view (FOV) is complete $180^\circ$ and primarily limited by the beam-width of antennas, which can be seen in the results for angles 60$^\circ$ and 70$^\circ$. From Fig. \ref{fig:ADI_ant_BW}, it can be seen that the beam-width of patch antennas is around $\pm 45^\circ$ and therefore the imaging results for angle 60$^\circ$ has higher side-lobes and almost indistinguishable peak for 70$^\circ$. The results, however, demonstrate the functionality of the CMI imaging system using commercially available phased arrays at 10 GHz.

\clearpage

\subsubsection{Results for 33-pixels 2-D Imaging System}

    \begin{figure}
    \centering
    \includegraphics[clip,trim=0cm 0cm 0cm 0cm,width=.5\textwidth]{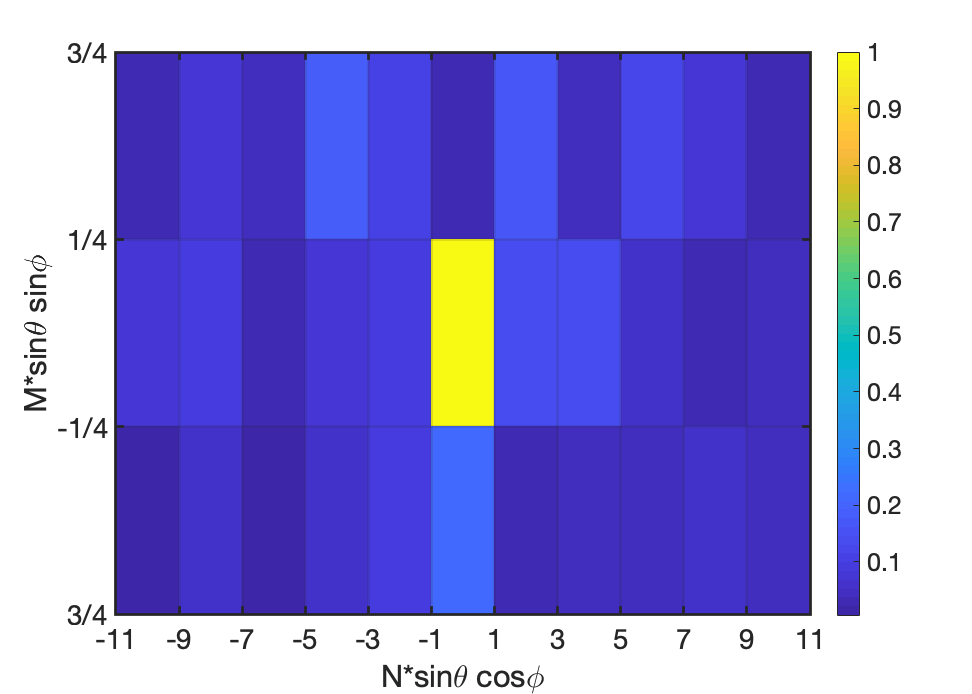}
    \caption{Point spread function: Normalized brightness plots for (elevation, azimuthal) angle 0$^\circ$, 0$^\circ$. $N=11$ and $M=3$.}
    \label{fig:1p5_dim_PSF}
    \end{figure}

\begin{figure}
 \centering
%	\subfloat[]{\includegraphics[clip, trim=0 0 0cm 0cm, width=0.3\textwidth]{figs3/00ref1.png}
%		\label{fig:R1}}
	\subfloat[]{\includegraphics[clip, trim=0 0 0cm 0cm, width=0.35\textwidth]{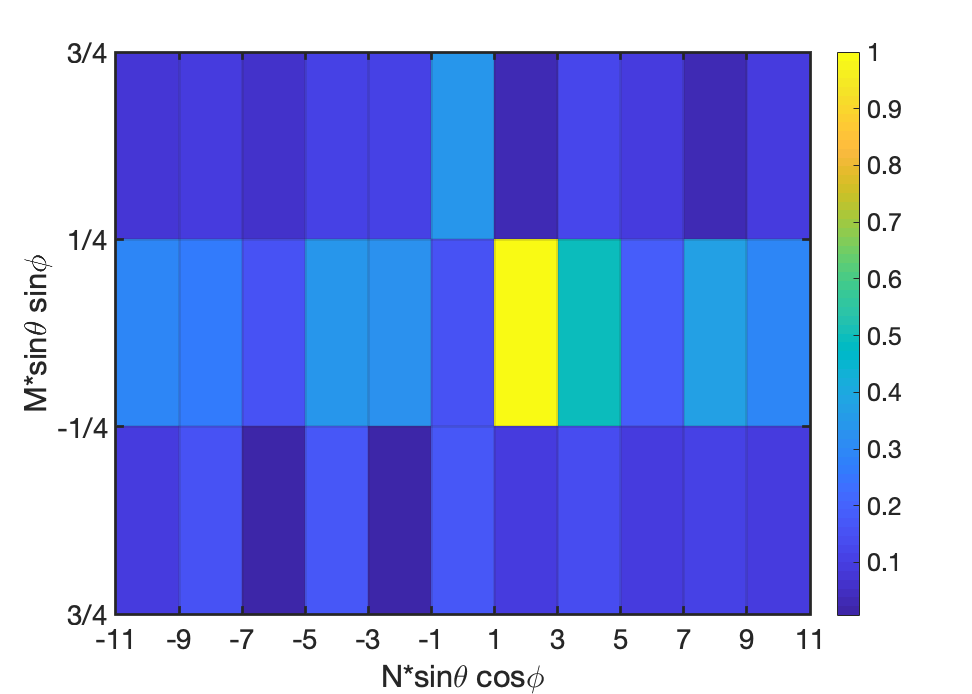}
		\label{fig:R2}} 
	\subfloat[]{\includegraphics[clip, trim=0 0 0 0.5cm, width=0.35\textwidth]{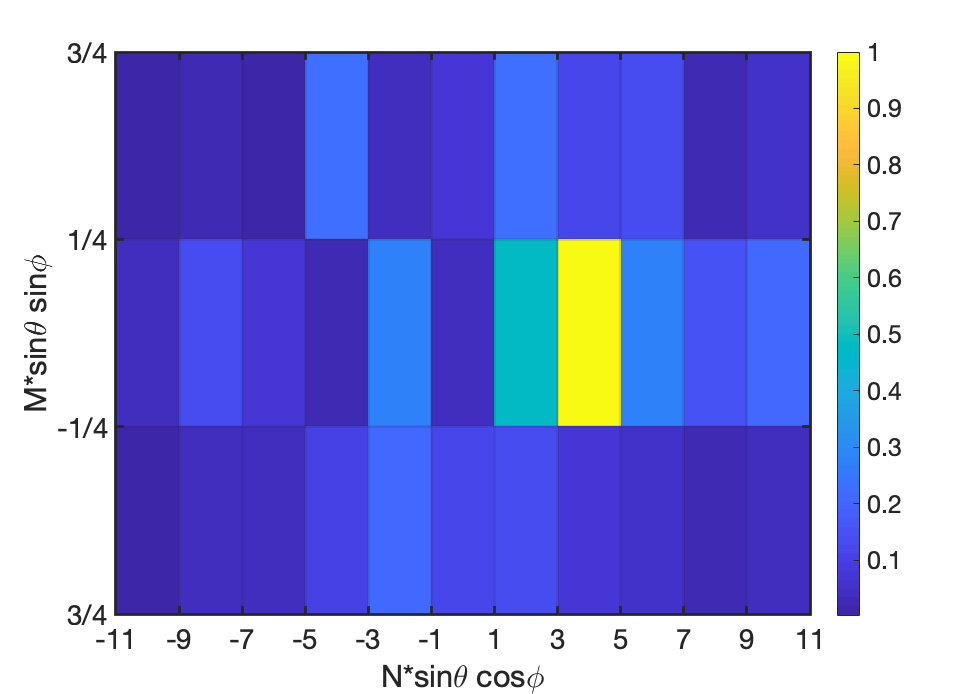}
		\label{fig:R3}} \\
	\subfloat[]{\includegraphics[clip, trim=0 0 0cm 0cm, width=0.35\textwidth]{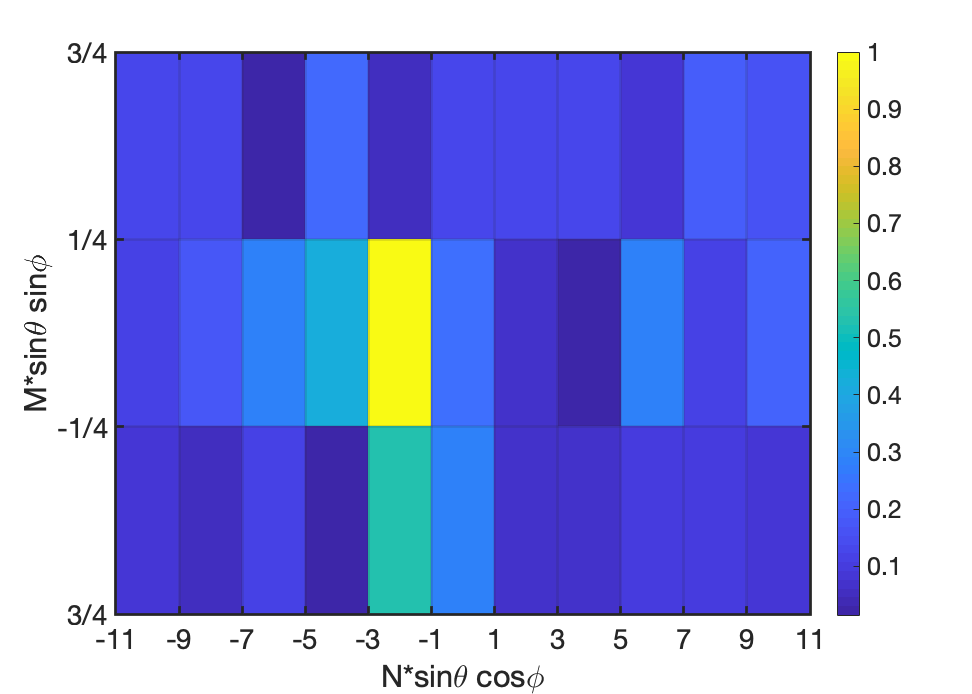}
		\label{fig:R4}} 
	\subfloat[]{\includegraphics[clip, trim=0 0 0cm 0cm, width=0.35\textwidth]{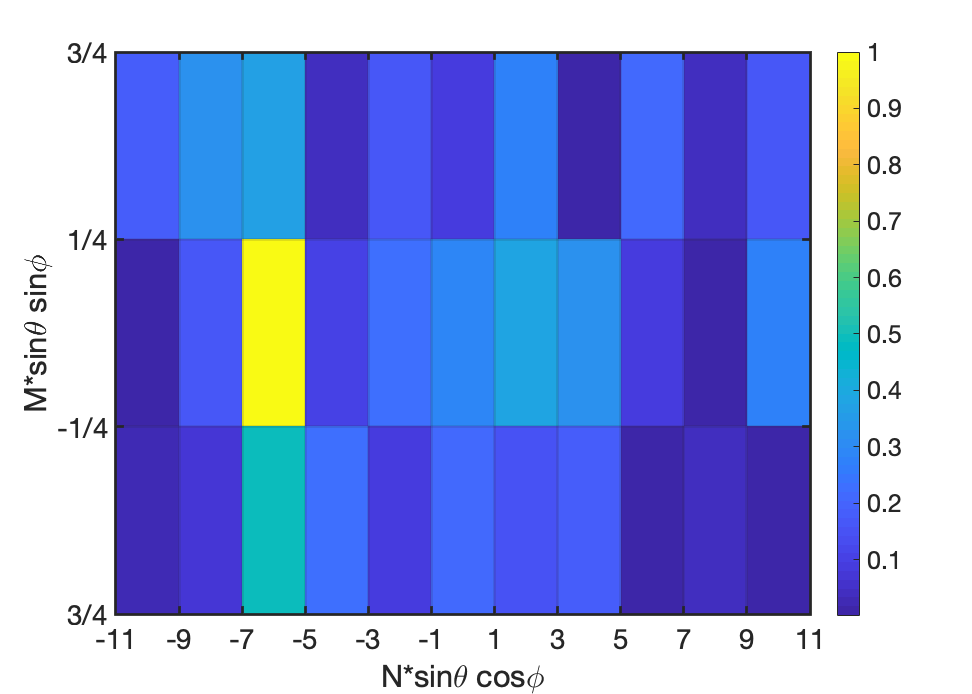}
		\label{fig:R5}} \\
	\subfloat[]{\includegraphics[clip, trim=0 0 0 0cm, width=0.35\textwidth]{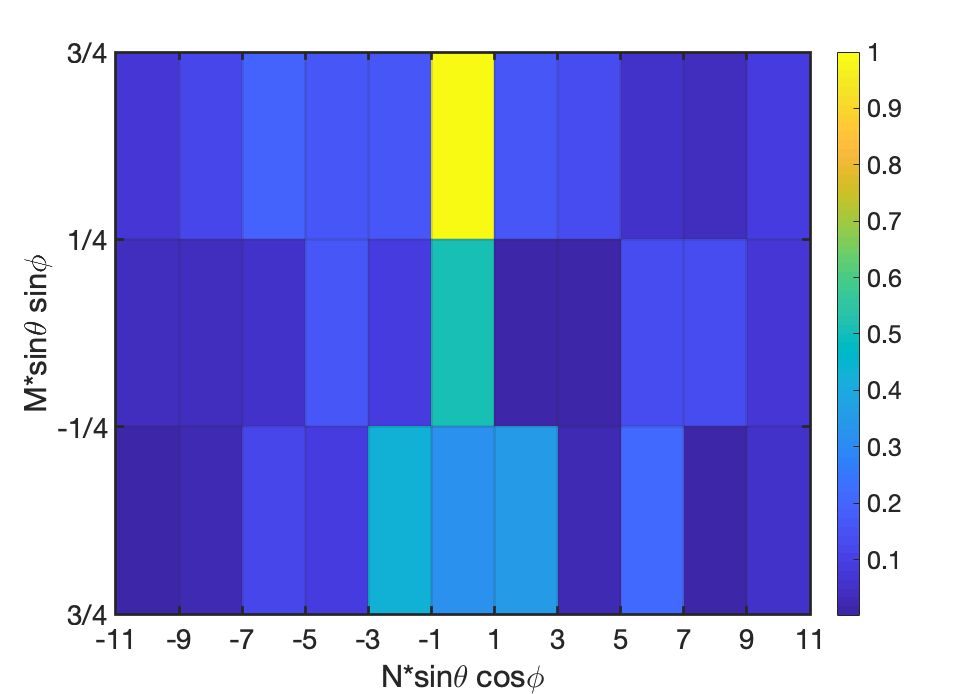}
		\label{fig:R6}}		
	\subfloat[]{\includegraphics[clip, trim=0 0 0cm 0cm, width=0.35\textwidth]{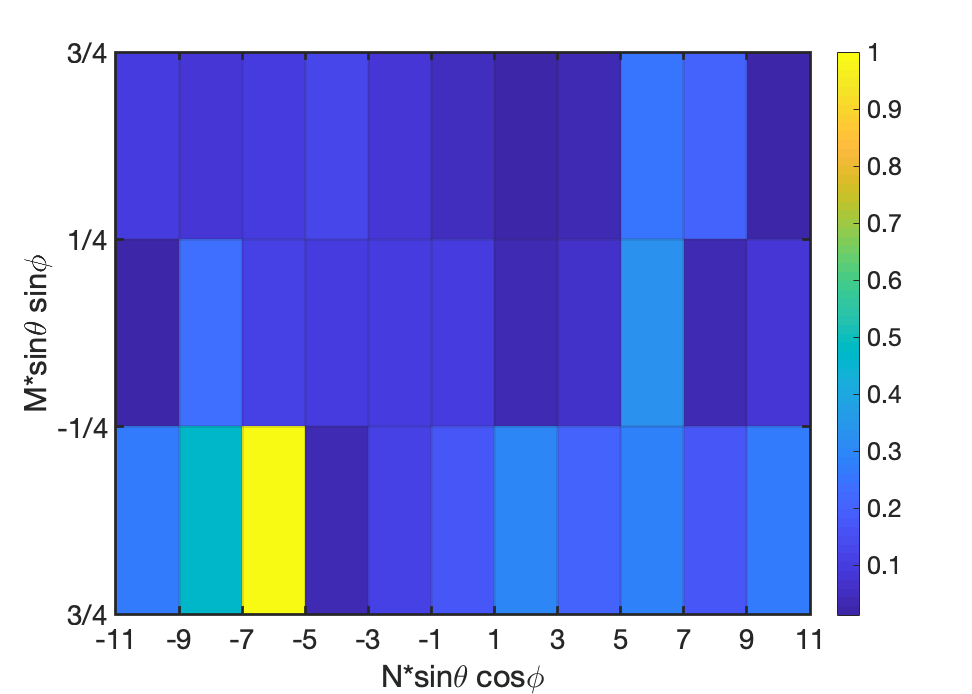}
		\label{fig:R7}} \\
	\subfloat[]{\includegraphics[clip, trim=0 0 0cm 0cm, width=0.35\textwidth]{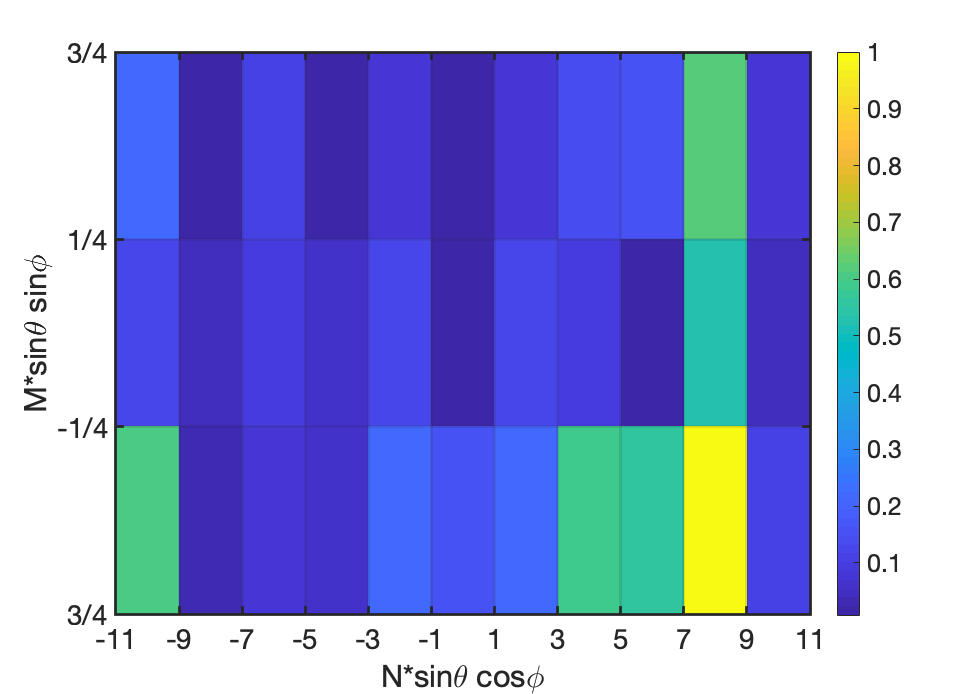}
		\label{fig:R8}} 
	\subfloat[]{\includegraphics[clip, trim=0 0 0 0cm, width=0.35\textwidth]{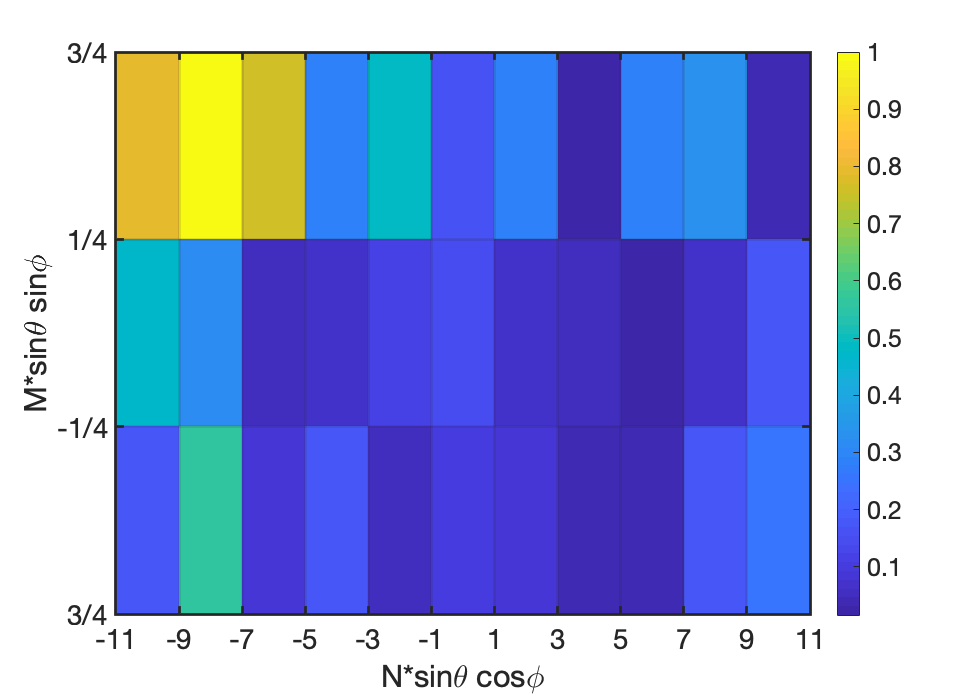}
		\label{fig:R9}}

	\caption{Normalized brightness plots for (elevation, azimuthal) angle (a) 0$^\circ$, 10$^\circ$ (b) 0$^\circ$, 20$^\circ$ (c) 0$^\circ$, -10$^\circ$ (d) 0$^\circ$, -30$^\circ$ (e) 10$^\circ$, 0$^\circ$ (f) -10$^\circ$, -30$^\circ$ (g) -10$^\circ$, 40$^\circ$ (h) 10$^\circ$, -50$^\circ$. $N=11$ and $M=3$.}
	\label{fig:Results}

\end{figure}

Fig. \ref{fig:1p5_dim_PSF} shows the point spread function and Fig. \ref{fig:Results} shows the point source images for various elevation and azimuthal angles for the two-dimensional 33-pixel imaging system, respectively. The PSF is important to visualize the performance of the imaging systems. Higher errors in a PSF correspond to non-idealities and mismatch within the elements and lead to higher imperfections in the image. Fig. \ref{fig:1p5_dim_PSF} shows a clean PSF with low sidelobe. Fig. \ref{fig:1p5_dim_PSF} shows the point source at the center of the image, which moves to the right in Fig. \ref{fig:R2} or to the left in Fig. \ref{fig:R4} with an approximate 10$^\circ$ rotation of the imaging device in either direction along the vertical axis. The point source is detected accurately in the corresponding bins in the image, demonstrating the functionality of the imaging device. This has been used to verify the resolution of the imaging system in $x$-direction. Similarly, an approximate 10$^\circ$ rotation of the imager in either direction along horizontal axis places the point source in corresponding upper or lower bins in the image, verifying the resolution in $y$-direction. For the ease of implementation, we do not use redundant baselines in these results which are available due to three consecutive runs to obtain complex visibilities, as explained in Sec. \ref{subsec:data_aquisition_4element}. We do observe that the point-source detection is erroneous at the corner pixels which we attribute to two phenomenons. Firstly, the limited antenna beam widths of $\pm 45^\circ$ (as discussed in previous section and Fig.  \ref{fig:ADI_ant_BW}) cause the images closer to edges of FOV to have higher errors. Secondly, interferometry assumes a spherical FOV (radially outward) where the position of any point on the spherical FOV is measured in terms of direction cosines $l$ and $m$ (where $l= sin\theta cos \phi$  and $ m= sin\theta sin\phi$, as shown in Fig. \ref{fig:Results}) and this  projection of a spherical FOV onto a planar image causes the corner pixels to have higher errors. The results in Figs. \ref{fig:1p5_dim_PSF} and \ref{fig:Results} are, therefore, plotted in $l$ and $m$ coordinate systems. The unit is arbitrary, where the brightness pattern is normalized to depict the contrast of the bright point source against the background.

%%%%%%%%%%%%%%%%%%%%%%%%%%%%%%%%%%%%%%%%%%%%%%%%%%%%%%%%%%%%%%%%%%%%%%%%%%%%%%%%%%%%%%%%%%
\clearpage

\section{169-Pixel COTS CMI Imaging Prototype}
The imaging prototype thus far show the feasibility of the code-modulated imaging system. Scalability of the system is another important metric. 
For this purpose, the imaging prototype described in the previous section is extended from eight-elemet to 16-elements by using four ADAR1000 EVAL boards. The four 4:1 combined signals from each RX array are merged using an external 4:1 combiner (Mini Circuits ZN4PD-02183-S+) and then amplified using an external amplifier (Mini-Circuits\textregistered \ ZVA-213S+). The amplifier is required to maintain the signal levels withing the dynamic range of the on-chip power detector. The signal is then fed back into one on-chip PD. All modulations and output power readings are accessed through the SPI. Fig. \ref{fig:16_boards} shows the photo of the four boards connected together. 

\subsection{Sparse Antenna Array}
\begin{figure}
    \centering
    \includegraphics[clip,trim=0cm 0cm 0cm 0cm,width=.8\textwidth]{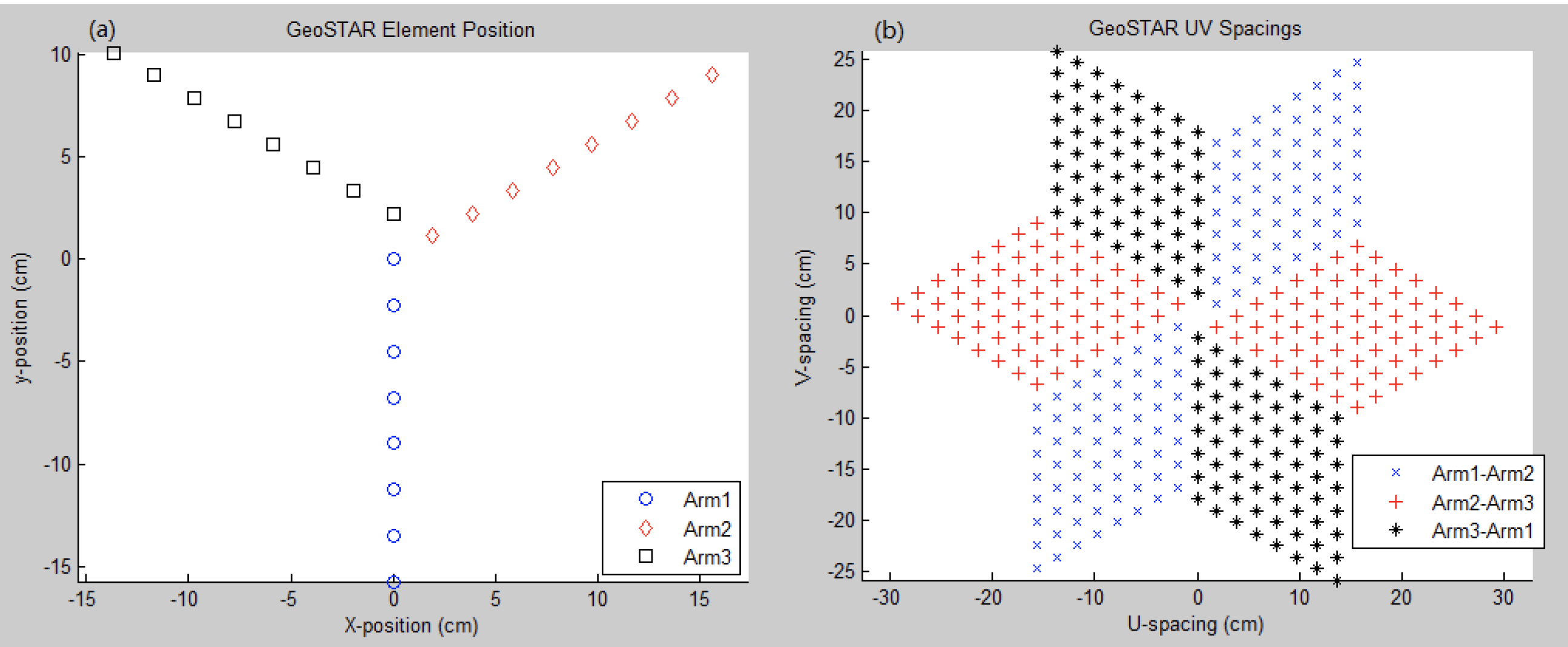}
    \caption{(a) Antenna array layout and (b) the corresponding UV spacing \cite{JasonPhD}.}
    \label{fig:Y_config}
    \end{figure}

Several different configurations for the antenna array are compared, such as ``Y" , ``T" and ``U" configurations. A ``Y" configuration provides maximum number of pixels for a given number of elements but the $u-v$ space is filled (and thus image is constructed) in a star configuration, as shown in Fig. \ref{fig:Y_config}. For $N$ elements in each arm of the ``Y" configuration, \textit{i.e.} ($3N+1$) total elements, the total number of pixels obtained is  $2 (3N^2 + 3N) + 1$. Therefore with 16-elements and an arm of 5-elements, we can obtain up to $181$ pixels. ``T" and ``U" configurations are identical in sampling the $u-v$ plane, and provide a rectangular filled image, but the number of pixels obtained is less due to redundancy. With 16-elements, the number of pixels obtained is $(2N +1)^2 = 121$ in ``T" or ``U" configurations. It is to be noted that the polarization of the antennas has be considered while designing the system. For tiling the same 1-D array thrice to form  a "T", antennas can be designed with circular polarization though with a penalty of 3 dBi lower antenna gain.

\begin{figure}
 \centering
	\subfloat[]{\includegraphics[width=0.45\textwidth]{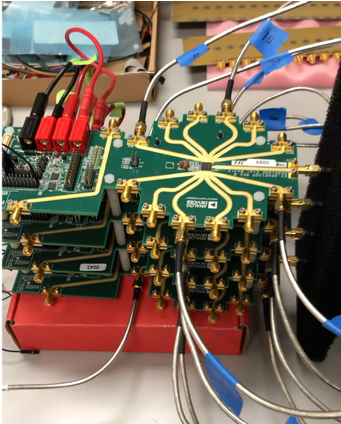}
		\label{fig:16_boards}} 
	\subfloat[]{\includegraphics[width=0.4\textwidth]{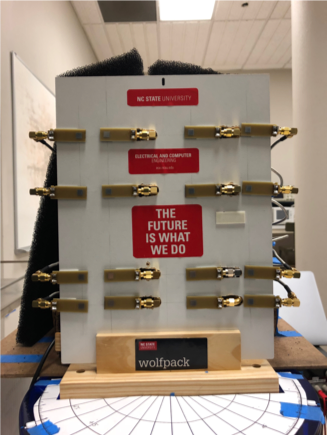}
		\label{fig:16_antenna}}
	\caption{(a) Photograph of four ADAR1000 EVAL boards connected together (b) Sparsely sampled 16-antenna array.}
	\label{fig:16_elements}
\end{figure}

Another option is use the the configuration shown in Fig. \ref{fig:16_antenna} where the antennas are placed at $\lambda$, $3\lambda$, $2\lambda$, spacing in both $x$ and $y$ directions \cite{Kopil01}. This is similar to the configuration of the 60-GHz one-dimensional four-element system in Chapter \ref{chap-four}. This provides an optimum solution of a rectangular sampled $u-v$ plane as well as low redundancy. As explained for one-dimensional 60 GHz interferometer in Chapter \ref{chap-four}, four antennas in this zero redundancy 1-D interferometric array provide 13 unique baselines and thus 1-D image with 13 pixels. The number of pixels obtained in this configuration is thus 169 (\textit{i.e.} $13-by-13$) for two-dimensional interferometric array. The minimum and maximum spacing in both $x$ and $y$-direction are $ \lambda $ and $ 6\lambda $ respectively, corresponding to an azimuthal field of view (FOV) of $\pm sin^{-1} (1/2\Delta u) = \pm30^{\circ}$, a resolution (defined as angular distance between peak and first null) of $sin^{-1} (1/N\Delta u) = 4.4^{\circ}$, and a central beam-width of $8.8^{\circ}$.

The antenna board provided by ADI, and used in the previous section is cut out into separate individual antennas with the help of a saw and placed in the desired configuration. Note that a custom antenna array could be designed with desired configuration, but would lead to longer design time. Also, individual antennas provide us with the flexibility to change the interferometric antenna configuration, if required, for higher resolution or higher number of pixels. This cutting of antenna board into individual antennas, however, could change the performance of individual patch antennas. A difference in gain of antennas can be easily calibrated using CoMET. A difference in antenna beam pattern, will however, affect the quality of the image. This is a trade-off to bring down the cost and reduce the design time of the imager.   

\subsection{Imaging Experiment and Results}
\begin{figure}
    \centering
    \includegraphics[clip,trim=0cm 0cm 0cm 0cm,width=.7\textwidth]{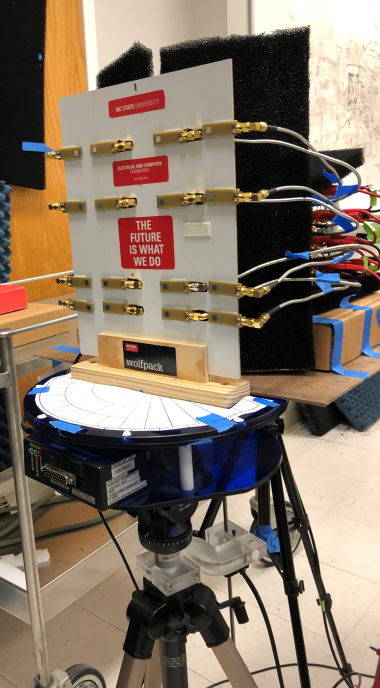}
    \caption{Setup for 16-elements imaging experiment.}
    \label{fig:16_setup}
    \end{figure}

Fig. \ref{fig:16_setup} shows the 16-element imaging setup with an interferometric antenna array by combining four 10 GHz EVAL-boards, and one 4:1 power combiner. The active area of the antenna array is $6\lambda$-by-$6\lambda$, which is 180-by-180 $mm^2$. However, due to the lengths of the feedlines to the patch antennas and the board hoding these antennas, the total size of the imager is approximately 240-by-300 $mm^2$. The sixteen antennas are connected to the receiver inputs of four EVAL boards thorough SMA cables, and the four outputs of these EVAL baords are then connected to the 4:1 power combiner using four SMA cables. These twenty cables in total, are needed to be calibrated for their phase and gain differences. The SMA cable connecting the output of the power combiner to the on-chip power detector, is however, common for all the sixteen channels and therefore does not require calibration. 

The calibration, code-modulation, data acquisition and processing of the 16-element imager is similar to procedure described in the previous section for eight-element imager. The gain calibration is done using CoMET. First, CoMET is used to extract the gains of the elements by two consecutive runs: first, for a set of eight elements (elements E1-E8) and second, for different set eight elements (elements E9-E16). Then eight element CoMET calibration loops are run twice, for E1-E8 and E9-E16, respectively, while equalizing the gain of all the sixteen elements to that of the lowest gain among 16 element. Note that one downside of this gain calibration is that the gain of all elements are now not at their maximum, but equalized to that of the element with highest losses, causing higher noise figure in the system. An alternative is to set gain of each element to it maximum, and calibrate for the inter-element gain difference in data processing of the visibilities. We however, apply gain calibration for ease of implementation. CoMET can be used for phase calibration as well, however, we choose VNA for more accurate phase calibration, as explained in previous section. Code-modulation and data acquisition is done using SPI. Sixteen 512-bit long Walsh-Hadamard codes are allocated to the sixteen channels. Again, codes are selected to have orthogonal code products (OCP) for a direct demodulation of cross-correlations from the PD output. The ADC data is read using SPI and the data is processed in MATLAB, including demodulation of the complex visibilities and Fourier transform. The MATLAB code is given in Appendix \ref{16-ele_code}.

\begin{figure}
 \centering
	\subfloat[]{\includegraphics[width=0.38\textwidth]{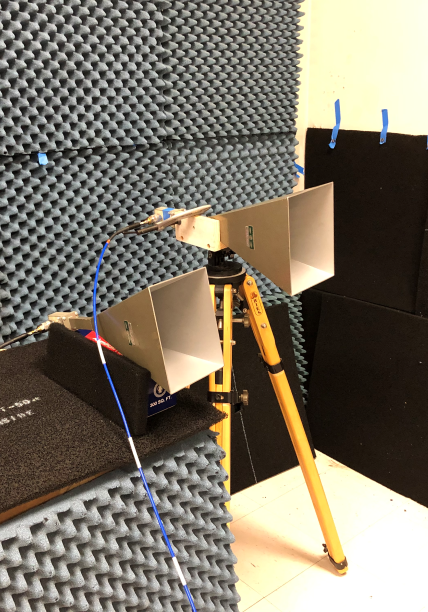}} 
	\subfloat[]{\includegraphics[width=0.38\textwidth]{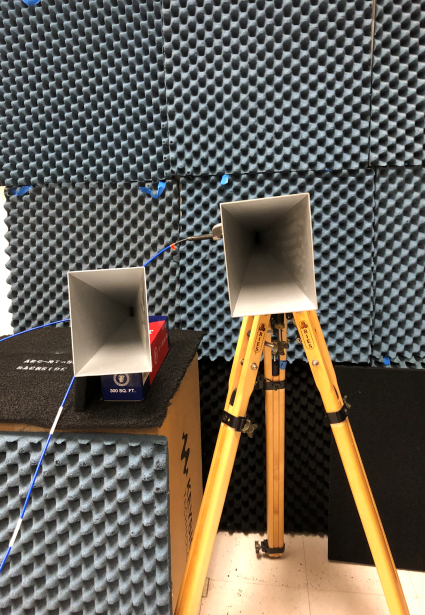}}
	\caption{Photographs of two horn antennas as point sources to demonstrate resolution.}
	\label{fig:two_horns}
\end{figure}

We test this imaging system using one or two active point sources placed at various azimuthal and elevation angles to show the resolution and FOV of the imager, as well as its ability to resolve two point sources, as shown in Fig. \ref{fig:fullsetup}. The fixed point sources are realized using one/two horn antennas connected to a 10 GHz signal generator (HP 83650B) and a two way power splitter, as shown in Fig. \ref{fig:two_horns}. Demonstrating the ability of CMI imager to image, track and resolve to point sources is minimum but satisfactory condition to show that feasibility of the concept. It is to be noted here that spacial incoherence of the scene is an important requirement for interferometry. To force the condition of spatially incoherence on two point sources, we either use two different signal generators with different frequency monotones (within the bandwidth of the antenna/array), or we use long cables to connect the two horn antennas to make them incoherent. In our experiment, however, due to only two point sources, we are able to detect both without major errors in image and the effect of spacial coherence of illumination is relatively low. The horn antennas are placed at a distance of approximately 5 feet from the imager.

\begin{figure}
 \centering
	\subfloat[]{\includegraphics[width=0.45\textwidth]{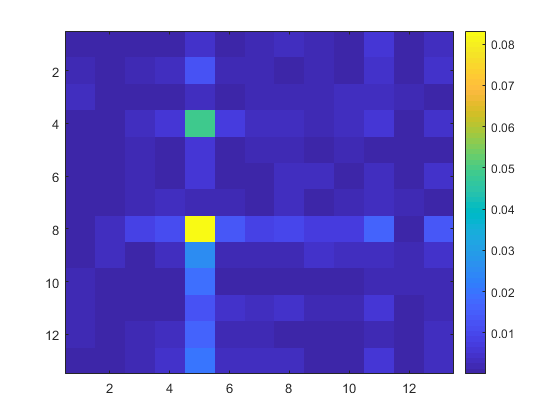}} \\
	\subfloat[]{\includegraphics[width=0.45\textwidth]{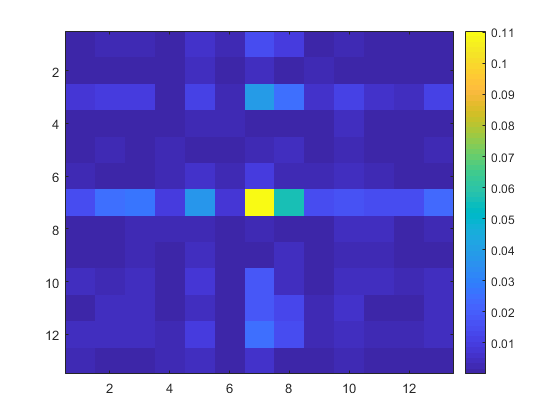}} \\
	\subfloat[]{\includegraphics[width=0.45\textwidth]{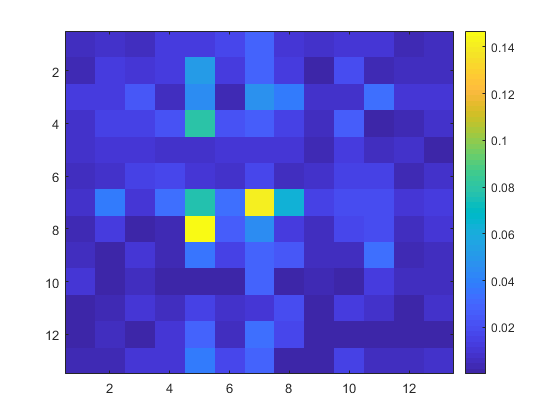}}
	\caption{Images taken for two point sources separately at different locations and together.}
	\label{fig:two_sources}
\end{figure}

\begin{figure}
 \centering
	\subfloat[]{\includegraphics[width=0.45\textwidth]{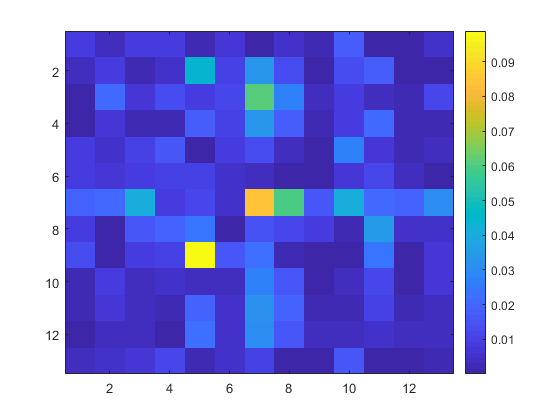}} \\
	\subfloat[]{\includegraphics[width=0.45\textwidth]{figs3/2sourcesC}} \\
	\subfloat[]{\includegraphics[width=0.45\textwidth]{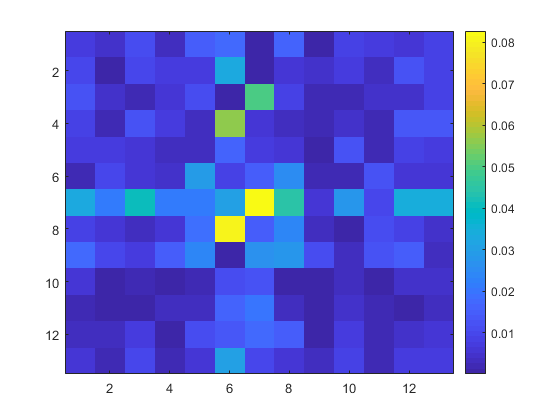}}
	\caption{Images of two point sources moving closer to demonstrate resolution.}
	\label{fig:16_resolution}
\end{figure}

Figs. \ref{fig:two_sources}(a) and \ref{fig:two_sources}(b) show two point sources imaged separately, by turning on horn antennas one at a time, validating the functionality of the imaging system. We observe that the side-lobes in the PSF are higher than the eight-element imaging system. We attribute this to the differences in the patch antenna patters due to cutting of board, and higher number of pixels in the image since the errors in all the visibility points contribute to all the pixels in the image. Fig. \ref{fig:two_sources}(c) then shows the image of two point sources taken together, demonstrating the capability of the system to resolve two point sources. The two horn antennas are then moved subsequently closer to each other to successfully obtain the resolution of the system, as shown in Fig. \ref{fig:16_resolution}. The scale in image is again arbitrary, although normalization has not been performed so as to be able to observe and the compare the brightness levels of individual point sources against two point sources. 

\section{Conclusion}

This chapter presented the prototypes constructed using commercially available phased arrays for the first time. The measurement results for two different imagers with different number of pixels and resolution are presented: an eight-element CMI system using two four-element phased arrays and a 16-element CMI system at 10 GHz using four four-element phased-arrays originally intended for 8--16 GHz communication. The hardware prototype is used to successfully demonstrate CMI, with representative 33-pixels obtained across multiple scenarios, and 169 pixel images used to demonstrate resolution of two point sources for the first time. These results suggest that CMI can be used to create low-cost, compact radio imagers using COTS hardware by avoiding the need for custom integrated circuit design.

%%%%%%%%%%%%%%%%%%%%%%%%%%%%%%%%%%%%%%%%%%%%%%%%%%%%%%%%%%%%%%%%%%%%%%%%%%%%%

%\include{Chapter-4/Chapter-4}
%\include{Chapter-5/Chapter-5}
%\include{Chapter-6/Chapter-6}
\chapter{Conclusion and Future work}
\label{chap-seven}

\section{Summary and Conclusions}
  
Millimeter-wave (mm-wave) imaging provides compelling capabilities for security screening, navigation, and bio-medical applications. The current solutions available in the market are either expensive, bulky due to use of lens, or both. We presented a novel code-modulated interferometric imaging system that repurposes the commercially available communication phased arrays and thus reduces the cost. Results from several SIMULINK and MATLAB behavioral models were presented to test the feasibility of the idea for both one-dimensional and two-dimensional interferometry. 

After several successful behavioral models and simulation results, a 60-GHz implementation of the CMI imaging system using a packaged phased array was presented. Here, a four-element 60-GHz phased array chip was packaged with slot antennas, and a single 60-GHz output was measured using a power detector. This scalar measurement was then demodulated to obtain the interferometric visibilities. The four-element phased array was thinned to obtain a 13-pixel image and the system was demonstrated through the tracking of a point source at different angles, indicating the feasibility of using code-modulated interferometry for low-cost mm-wave imaging. This was the first ever demonstration of the CMI at mm-wave frequency. 

In the following chapter various prototypes constructed using commercially available phased arrays were presented for the first time. We then presented the measurement results for eight element/sixteen element CMI system at 10 GHz using two/four four-element phased-arrays originally intended for 8--16 GHz communication applications. The hardware prototype was used to successfully demonstrate CMI, with representative 33-pixels obtained across multiple scenarios, and 169 pixel images used to demonstrate resolution of two point sources for the first time. These results suggested that CMI can be used to create low-cost, compact radio imagers using COTS hardware, eliminating the need for a custom integrated circuit design.

In Sec. \ref{sec:sensitivity_deriv} the sensitivity analysis of a code-modulated interferometric array is presented. The reduction on the sensitivity of the imager is observed, proportional to the number of antenna elements. To improve the sensitivity, the integration time can be increased which will in turn slow down the frame rate of the camera. It can be concluded that though CMI imaging is a cost-effective alternative to imaging, the trade off is lower sensitivity and/or slower frame rates. Nevertheless, CMI imaging can prove beneficial in constructing low cost imagers for deploying in situations where cost is a major limiting factor, such as schools, bus stops, train stations, etc.  

%Finally, a compact UWB LNA which provides low noise and high gain across 24 to 44 GHz was then  presented and compared to a narrow-band 28 GHz LNA. The UWB LNA achieved a 4.2-5.5 dB NF and 18-20 dB gain across the full range. This UWB LNA could support a potential multi-band 5G operation and avoid the need for a parallel bank of multiplexed LNAs. This wide-band LNA could serve as a first stage to UWB receivers for 5G communications. Such UWB 5G phased arrays could be repurposed into CMI imaging systems for better sensitivity as compared to narrow-band designs.

%In summary, we presented the preliminary results from behavioral models, an analysis for system metrics, and several prototypes that demonstrate that code-modulated interferometry appears to be a promising and cost-effective alternative to costly imaging systems present in the market.

\section{Comparison with Existing Solutions} 
As discussed in Chapter \ref{chap-one}, there are several existing mm-wave imaging solutions in the market. In this section we compare the performance and trade-offs of using a code-modulated interferometric imaging systems against conventional methods. Table \ref{tab:comparison_metric} presents the comparison of important metrics for code-modulated interferometric imaging system against other conventional imaging solutions. Table \ref{tab:comparison_imagers} presents some of the published imaging systems for comparison with the CMI imaging work presented by authors. 

A phased array can be as-is used to image by beam-forming and electronically scanning the scene. With the same number of elements, the resolution of the fully filled phased array scanning system is worse as compared to an interferometer, since interferometer has a larger baselength due to partially filled antenna plane. In a scanning phased array type imager, the resolution is defined by the beam-width of the phased array and thus provides lower the pixel density in the image. The advantage is that the sensitivity in phased array scanning is much higher as the signals in all elements power combine in-phase, increasing the effective signal-to-noise ratio for each pixel. The frame rate would then be a function of scanning speed and desired sensitivity. Each phased array would provide a single pixel per run. 

\begin{table}
	\caption{Comparison metrics for code-modulated interferometric imaging system with conventional imaging systems.}
	\label{tab:comparison_metric}
	\begin{center}
		\begin{tabular}{c c c c c}
        
        		\hline
				\multicolumn{1}{|c||}{Technology}&\multicolumn{1}{c|}{Cost}& \multicolumn{1}{c|}{Sensitivity}&\multicolumn{1}{c|}{Frame Rate}&\multicolumn{1}{c|}{Resolution/Pixels}   \\	
                 \hline
                 \hline
				\multicolumn{1}{|c||}{Phased-Array Scanning}&\multicolumn{1}{c|}{Low}& \multicolumn{1}{c|}{High}&\multicolumn{1}{c|}{Slow}&\multicolumn{1}{c|}{Low}   \\
                \hline
				\multicolumn{1}{|c||}{Raster Scan}&\multicolumn{1}{c|}{High}& \multicolumn{1}{c|}{Good}&\multicolumn{1}{c|}{Slow}&\multicolumn{1}{c|}{Low}   \\	
                 \hline               
				\multicolumn{1}{|c||}{Focal-Plane Array}&\multicolumn{1}{c|}{Highest}& \multicolumn{1}{c|}{Good}&\multicolumn{1}{c|}{Fastest}&\multicolumn{1}{c|}{High}   \\	
                 \hline
				\multicolumn{1}{|c||}{Single-Pixel Camera}&\multicolumn{1}{c|}{Lowest}& \multicolumn{1}{c|}{Good}&\multicolumn{1}{c|}{Slow}&\multicolumn{1}{c|}{Lowest}   \\
			    \hline
				\multicolumn{1}{|c||}{Interferometer}&\multicolumn{1}{c|}{High}& \multicolumn{1}{c|}{Low}&\multicolumn{1}{c|}{Average}&\multicolumn{1}{c|}{Highest}   \\	
                 \hline	
				\multicolumn{1}{|c||}{CMI Imaging}&\multicolumn{1}{c|}{Cheapest}& \multicolumn{1}{c|}{Worse}&\multicolumn{1}{c|}{Slow}&\multicolumn{1}{c|}{Highest}   \\	
                \hline                 
		\end{tabular}
	\end{center}
%  \hspace{10.5em} \textsuperscript{*}\footnotesize{Simulated}
\end{table}

\begin{table}
	\caption{Comparison of CMI imaging systems with published conventional imaging systems.}
	\label{tab:comparison_imagers}
	\begin{center}
		\begin{tabular}{c c c c c c c}
        
        		\hline
				\multicolumn{1}{|c||}{Ref.}&\multicolumn{1}{c|}{Technology}&\multicolumn{1}{c|}{Freq.}&\multicolumn{1}{c|}{Sensitivity/}&\multicolumn{1}{c|}{Frame}&\multicolumn{1}{c|}{Res./}&\multicolumn{1}{c|}{Power}   \\
				\multicolumn{1}{|c||}{}&\multicolumn{1}{c|}{}&\multicolumn{1}{c|}{}&\multicolumn{1}{c|}{NETD}&\multicolumn{1}{c|}{Rate}&\multicolumn{1}{c|}{Pixels (p)}&\multicolumn{1}{c|}{W}   \\
                 \hline 
                 \hline
				\multicolumn{1}{|c||}{\cite{Robert2004}}&\multicolumn{1}{c|}{1 receiver}&\multicolumn{1}{c|}{94GHz}& \multicolumn{1}{c|}{0.6K}&\multicolumn{1}{c|}{3 min.}&\multicolumn{1}{c|}{150X100 p}&\multicolumn{1}{c|}{200}   \\	
				\multicolumn{1}{|c||}{}&\multicolumn{1}{c|}{mech. scan}&\multicolumn{1}{c|}{}& \multicolumn{1}{c|}{}&\multicolumn{1}{c|}{}&\multicolumn{1}{c|}{}&\multicolumn{1}{c|}{}   \\				
                 \hline                 
				\multicolumn{1}{|c||}{\cite{Gold93}}&\multicolumn{1}{c|}{16-element FPA}&\multicolumn{1}{c|}{94GHz}&\multicolumn{1}{c|}{--}&\multicolumn{1}{c|}{33 ms}&\multicolumn{1}{c|}{0.35$^\circ$, 64 p}&\multicolumn{1}{c|}{22}   \\	
                 \hline 
				\multicolumn{1}{|c||}{\cite{JMay10}}&\multicolumn{1}{c|}{Dicke Radio.}&\multicolumn{1}{c|}{94GHz}&\multicolumn{1}{c|}{0.83K}&\multicolumn{1}{c|}{30 ms}&\multicolumn{1}{c|}{--}&\multicolumn{1}{c|}{34.8m}   \\
                 \hline
				\multicolumn{1}{|c||}{\cite{Heyd10}}&\multicolumn{1}{c|}{Dicke Radio.}&\multicolumn{1}{c|}{94GHz}&\multicolumn{1}{c|}{0.3k}&\multicolumn{1 }{c|}{40 ms}&\multicolumn{1}{c|}{--}&\multicolumn{1}{c|}{200m}   \\
                 \hline  
				\multicolumn{1}{|c||}{\cite{MIT07}}&\multicolumn{1}{c|}{1 receiver}&\multicolumn{1}{c|}{77GHz}&\multicolumn{1}{c|}{1.1k}&\multicolumn{1 }{c|}{0.1 s}&\multicolumn{1}{c|}{--}&\multicolumn{1}{c|}{120m}   \\
                 \hline
				\multicolumn{1}{|c||}{\cite{Sheen2009}}&\multicolumn{1}{c|}{3-D, mirror scan}&\multicolumn{1}{c|}{350GHz}&\multicolumn{1}{c|}{--}&\multicolumn{1 }{c|}{10s}&\multicolumn{1}{c|}{1cm lateral}&\multicolumn{1}{c|}{--}   \\
                 \hline
				\multicolumn{1}{|c||}{This work}&\multicolumn{1}{c|}{16-element CMI}&\multicolumn{1}{c|}{10GHz$^*$}&\multicolumn{1}{c|}{1k}&\multicolumn{1 }{c|}{35s$^{**}$}&\multicolumn{1}{c|}{169 p}&\multicolumn{1}{c|}{3.3}   \\
                \hline 
				\multicolumn{1}{|c||}{This work}&\multicolumn{1}{c|}{4-element CMI}&\multicolumn{1}{c|}{60GHz$^*$}&\multicolumn{1}{c|}{1k}&\multicolumn{1 }{c|}{3s$^{**}$}&\multicolumn{1}{c|}{13 p}&\multicolumn{1}{c|}{208m}   \\
                 \hline                   
		\end{tabular}
	\end{center}
  \hspace{10.5em} \footnotesize{$^*$1GHz bandwidth. $^{**}$ Theoretical for passive mm-wave imaging.}
\end{table}

A focal plane array is a faster alternative to CMI imaging due to its "snapshot" method of imaging using a large number of receivers. The number of pixels in the image, the resolution, FOV would depend upon the actual number of receivers and can be higher or comparable to that of CMI. Although the resolution would be worse than a CMI for the same number of elements, the high cost of a focal plane array is the major drawback. Focal plane arrays usually require a lens to focus, making the system even more bulkier and costly.

A single receiver can be used to raster scan the scene to obtain an image, known as single-pixel camera. Such a system is cheap due to smallest required hardware, but is slowest due to minimum pixel density.

Multiple receiver can be used at once to perform a raster scan, which is somewhere between a single-pixel camera and a focal plane array. If N receiver are used at once, N-pixels in the image are obtained in each scan. A raster scan in X and Y direction populates the entire image. Scanning could be either mechanical or electronic and may require a lens.

A traditional correlation type interferometric imaging system can be used for imaging with better sensitivity as compared to CMI imaging system. Though the sensitivity of an N-element CMI imaging system is $N$ times worse, as discussed in Chapter \ref{chap-two}, it provides a cost-effective alternative by repurposing commercial phased arrays. With the same number of antennas, both traditional interferometry and CMI give the same the FOV, resolution and number of pixels. To achieve comparable sensitivity, the frame rate of CMI would be $N^2$ slower than that of a traditional interferometer. 

A code-modulated interferometric imaging system has the worst sensitivity among the list, and therefore has a slow frame rate to obtain quality image, but has the highest resolution and is the most cost-effective approach.

%\color{black}

\section{Key Research Contributions}
There are two key contributions from this research. First key contribution is the theoretical development of the code-modulated interferometry and direct demodulation of complex visibilities by adding a simple squaring circuit. This concept has been successfully implemented in two different research areas by iNCS$^2$ group: design of mm-wave imaging system and built-in-self-test of phased arrays (CoMET). Thus this concept has immense potential and will find applications in several different fields. 

The second key contribution is the design and demonstration (by prototypes and measured results) of a low-cost lens-less alternative to the existing expensive and bulky mm-wave imaging systems; using both in-house and COTS phased arrays. This cost-effective imaging system is suitable for mass deployment in situations where cost is the main limiting factor, such as school, bus/train station, etc.

\section{Future Work}
The scope of the future work is:
\begin{itemize}
 %   \item \emph{Sensitivity Analysis}:  We qualitatively presented  the performance of the code-modulated interferometry. A quantitative analysis with mathematical derivation of sensitivity in presence of code-modulation can be looked into. 
    \item \emph{Imaging Extended Objects}:  A prototype can be constructed with low noise temperature to passively image extended real objects against warm human body such as metal, weapons, etc.
    \item \emph{Active Interferometric Imaging}:  Interferometry is dependent upon spacial in-coherency for imaging. Techniques for active imaging using interferometry can be investigated for better sensitivity of the imagers. One such method is to create active noise-like illumination \cite{Nanzer18}.   
    \item \emph{Scalability}:  One important challenge for code-modulated interferometry is the scalability to larger systems which needs to be investigated and possible improvements are to be suggested. The Walsh BOCP codes place a hard limit on the number of elements per chip that can be used. A possible solution is to divide the scene into smaller sectors and use multiple chips. Several different kinds of codes can be explored, such as m-sequences, Gold codes, Walsh orthogonal codes, or pseudo-random codes to mitigate this scalability limitation.
\end{itemize}
%\restoregeometry

%%---------------------------------------------------------------------------%%
%%  Bibliography 

%%  You can use the bibitem list.
%\bibliographystyle{unsrt}
%\begin{%thebibliography}{99}
%\bibitem{cb02}
%Casella, G. and Berger, R.L. (2002)
%\newblock {\it Statistical Inference, Second Edition.}
%Duxbury Press, Belmont, CA.
%
%\bibitem{t06}
%Tsiatis, A.A. (2006)
%\newblock {\it Semiparametric Theory and Missing Data.}
%Springer, New York.
%
%\end{thebibliography}

%% or use BibTeX
%\bibliography{VC-thesis}{}
%\bibliographystyle{plain}
%\nociterec{*}

%\bibliographystyle{plainnat}%plainnat is necessary to enable the use of citet. Natbib style file.
%\bibliography{VC-thesis}
%\ensureoddstart
\begin{spacing}{1}
 \setlength\bibitemsep{11pt} %22pt = 2*11pt, where fontsize is 11pt
 \phantomsection
 \addcontentsline{toc}{chapter}{{\uppercase{\bibname}}} %\textorpdfstring and \uppercase needed due to hyperref package http://www.latex-community.org/forum/viewtopic.php?f=44&t=16601
 %\vspace{-0.5in}
\titleformat{\chapter}[display]{\bf\filcenter
}{\chaptertitlename\ \thechapter}{11pt}{\bf\filcenter}
\titlespacing*{\chapter}{0pt}{0.0in-9pt}{22pt}

\printbibliography[heading=myheading]
\end{spacing}
%\bibliographystyle{apalike}

%%---------------------------------------------------------------------------%%
% Appendices
%\ensureoddstart
\restoregeometry
\appendix
\newgeometry{margin=1in,lmargin=1.25in,footskip=\chapterfootskip, includehead, includefoot}

\chapter{MATLAB Codes and Design Files}

\lstset{language=Matlab,%
    %basicstyle=\color{red},
    breaklines=true,%
    morekeywords={matlab2tikz},
    keywordstyle=\color{blue},%
    morekeywords=[2]{1}, keywordstyle=[2]{\color{black}},
    identifierstyle=\color{black},%
    stringstyle=\color{mylilas},
    commentstyle=\color{mygreen},%
    showstringspaces=false,%without this there will be a symbol in the places where there is a space
    numbers=left,%
    numberstyle={\tiny \color{black}},% size of the numbers
    numbersep=9pt, % this defines how far the numbers are from the text
    emph=[1]{for,end,break},emphstyle=[1]\color{red}, %some words to emphasise
    %emph=[2]{word1,word2}, emphstyle=[2]{style},    
}

\begin{comment}
\section{SIMULINK Behavioral Model} \label{appendix_simulink}

\begin{figure}
	\centering
	\includegraphics[clip,trim=0 0 0 0,width=.9\textwidth]{figsA/noise_input.png}
	\caption{Behavioral model: Noise sources are input. }
	\label{fig:Behav_model_input}
\end{figure}

\begin{figure}
	\centering
	\includegraphics[clip,trim=0 0 0 0,width=.9\textwidth]{figsA/coding.png}
	\caption{Behavioral model: Code-modulation. }
	\label{fig:Behav_model_CM}
\end{figure}

\begin{figure}
	\centering
	\includegraphics[clip,trim=0 0 0 0,width=.9\textwidth]{figsA/output.png}
	\caption{Behavioral model: Output. }
	\label{fig:Behav_model_output}
\end{figure}
\end{comment}

\section{MATLAB Codes}

\subsection{Code for 3-D Path Length}\label{Appendix_matlab_2D_object}
\begin{lstlisting}
function [R]=antenna_input()
fc=1e9;
c=3e10; % cm/s
lambda=c/fc; %centimeters
% % % -----------Target Initialization--------
% ---------------Image Sampling---------------
object = [0 0 0 0 0 0 0 0 ;
          0 0 0 0 0 0 0 0 ;
          0 1 1 1 1 1 1 0 ;
          0 1 1 1 1 1 1 0 ;
          0 0 0 1 1 0 0 0 ;
          0 0 0 1 1 0 0 0 ;
          0 0 0 1 1 0 0 0 ;
          0 0 0 0 0 0 0 0 ];
      
spct=lambda; % the spacing between the source elements in object
[MT NT]=size(object);
xtb=[-spct(1)*(NT-1)/2:spct(1):spct(1)*(NT-1)/2] ; % target sampling points
ytb=[-spct(1)*(MT-1)/2:spct(1):spct(1)*(MT-1)/2] ; % target sampling points
%------------Square Antenna Array Generation at given Spacing-------------
spc= 1;
spcl=spc*lambda ;                     % lambda*spacing=2mm
%xp=[-spcl*(Na- 1)/2:spcl:spcl*(Na-1)/2]; % antenna locations (square array)
xp = [-6 -5 -4 0 0 0 0 1 2 3];
yp= [-3 -2 -1 0 1 2 3];
zp = 10;
zpl=zp*lambda;
dp = zeros(7,10);
dp(4,:) = xp;
dp(:,7)= 1i* (yp)';
dp = lambda* dp;
zpl=repmat(zp*lambda,size(dp));
[indy indx]=find(object)% indxt reflection points
kc=2*pi/lambda;
ytb;
yt=ytb(indy);
xtb;
xt=xtb(indx);                           % clm-x dimension
dt=complex(yt,xt);						% non zero image points in the complex plane
R=zeros(size(dp));
   for n=1:size(dt,2)  
        Exp1=exp(-1i*kc*(sqrt((abs((dp)-dt(n)).^2)+(zpl.^2)))) ;        
        atmp=object(indy(n),indx(n));
        R=R+single(atmp)*Exp1; % single converts the data type in target which is uint8
end

yn1=R(2,5);
yn2=R(3,5);
yn3=R(4,5);
yn4=R(5,5);
xp1=R(1,6);
xp2=R(1,7);
xp3=R(1,8);
xp4=R(1,9);
xn1=R(1,4);
xn2=R(1,3);
xn3=R(1,2);
xn4=R(1,1);
end
\end{lstlisting}

\subsection{Code for Image Data Processing} \label{16-ele_code}
MATLAB code for 16-element 10 GHz imaging:
\begin{lstlisting}
clear all;
%load("16eleMatrix3a.mat"); % matrix shown I/Q of elements for 4 runs. Loads a, a1,a2,a3.
a =[ 1     1     0     0     1     0     1;
     0     0     0     0     0     0     0;
     1     1     0     0     1     0     1;
     0     0     0     0     0     0     0;
     0     0     0     0     0     0     0;
     1     1     0     0     1     0     1;
     1     1     0     0     1     0     1];
     
a1 =[1     1     0     0     1     0     1;
     0     0     0     0     0     0     0;
     1     1     0     0     1     0     1;
     0     0     0     0     0     0     0;
     0     0     0     0     0     0     0;
     i     i     0     0     i     0     i;
     i     i     0     0     i     0     i]; 

a2 =[1     1     0     0     1     0     1;
     0     0     0     0     0     0     0;
     i     i     0     0     i     0     i;
     0     0     0     0     0     0     0;
     0     0     0     0     0     0     0;
     1     1     0     0     1     0     1;
     i     i     0     0     i     0     i];  
     
a3 =[i     i     0     0     1     0     1;
     0     0     0     0     0     0     0;
     i     1     0     0     i     0     1;
     0     0     0     0     0     0     0;
     0     0     0     0     0     0     0;
     1     1     0     0     i     0     i;
     1     i     0     0     1     0     i];     
     
 % Co is the cross-correlations input from LabView    
  Co1 = Co(1,121); 
  Co2 = Co(122,243);
  Co3 = Co(244,364);
  Co4 = Co(365,485);
  
  b= ones(13);
  
  Run =4; %number of runs
  pixels =nnz(b);    
  a_ud = flipud(a);
  a1_ud= flipud(a1);
  a2_ud= flipud(a2);
  a3_ud= flipud(a3);
  [vis_r, vis_c]= size(b);
  vis_real = zeros(vis_r,vis_c);
  vis_real_redun = zeros(vis_r,vis_c);
  vis_imag = zeros(vis_r,vis_c);
  vis_imag_redun = zeros(vis_r,vis_c);
  redun_real_count = zeros(vis_r,vis_c);
  redun_imag_count = zeros(vis_r,vis_c);
  [Ey,Ex]= find(a_ud); 
  [N, temp]= size(Ex);
  k=1; % k starts with 1 
  
  vis_real((vis_r+1)/2,(vis_c+1)/2)= Co1(1);
  vis_real_redun((vis_r+1)/2,(vis_c+1)/2)= Co1(1)+Co2(1)+Co3(1)+Co4(1);
  redun_real_count((vis_r+1)/2,(vis_c+1)/2)= 16*Run;
  
  
 for i= 1: N
     for j1= i+1:N
         visy= Ey(i)-Ey(j1) + (vis_r+1)/2;
         visx= Ex(i)-Ex(j1) + (vis_c+1)/2;
                   
k=k+1;
vis_real_redun(visy,visx)= vis_real_redun(visy,visx) + Co1(k);
redun_real_count(visy,visx)=redun_real_count(visy,visx)+1;
if vis_real(visy,visx) == 0
vis_real(visy,visx) = Co1(k);
end    

if (imag(a1_ud(Ey(i),Ex(i))) ~=0 && imag(a1_ud(Ey(j1),Ex(j1))) ==0)
vis_imag_redun(visy,visx)= vis_imag_redun(visy,visx) + Co2(k);
redun_imag_count(visy,visx)=redun_imag_count(visy,visx)+1;
if vis_imag(visy,visx) == 0
vis_imag(visy,visx) = Co2(k);
end
                          
elseif (imag(a1_ud(Ey(i),Ex(i))) ==0 && imag(a1_ud(Ey(j1),Ex(j1))) ~=0)
vis_imag_redun(visy,visx)= vis_imag_redun(visy,visx) - Co2(k);
redun_imag_count(visy,visx)=redun_imag_count(visy,visx)+1;
if vis_imag(visy,visx) == 0
vis_imag(visy,visx) = -Co2(k);
 end
 else
redun_real_count(visy,visx)=redun_real_count(visy,visx)+1;
vis_real_redun(visy,visx)= vis_real_redun(visy,visx) + Co2(k);
 end
 
if (imag(a2_ud(Ey(i),Ex(i))) ~=0 && imag(a2_ud(Ey(j1),Ex(j1))) ==0)
vis_imag_redun(visy,visx)= vis_imag_redun(visy,visx) + Co3(k);
redun_imag_count(visy,visx)=redun_imag_count(visy,visx)+1;
 if vis_imag(visy,visx) == 0
 vis_imag(visy,visx) = Co3(k);
 end
                          
 elseif (imag(a2_ud(Ey(i),Ex(i))) ==0 && imag(a2_ud(Ey(j1),Ex(j1))) ~=0)
 vis_imag_redun(visy,visx)= vis_imag_redun(visy,visx) - Co3(k);
redun_imag_count(visy,visx)=redun_imag_count(visy,visx)+1;
 if vis_imag(visy,visx) == 0
 vis_imag(visy,visx) = -Co3(k);
end

 else
redun_real_count(visy,visx)=redun_real_count(visy,visx)+1;
vis_real_redun(visy,visx)= vis_real_redun(visy,visx) + Co3(k);
end  

if (imag(a3_ud(Ey(i),Ex(i))) ~=0 && imag(a3_ud(Ey(j1),Ex(j1))) ==0)
vis_imag_redun(visy,visx)= vis_imag_redun(visy,visx) + Co4(k);
redun_imag_count(visy,visx)=redun_imag_count(visy,visx)+1;
if vis_imag(visy,visx) == 0
 vis_imag(visy,visx) = Co4(k);
 end
                          
elseif (imag(a3_ud(Ey(i),Ex(i))) ==0 && imag(a3_ud(Ey(j1),Ex(j1))) ~=0)
vis_imag_redun(visy,visx)= vis_imag_redun(visy,visx) - Co4(k);
redun_imag_count(visy,visx)=redun_imag_count(visy,visx)+1;
if vis_imag(visy,visx) == 0
vis_imag(visy,visx) = -Co4(k);
 end

else
redun_real_count(visy,visx)=redun_real_count(visy,visx)+1;
vis_real_redun(visy,visx)= vis_real_redun(visy,visx) + Co4(k);
end
 end
 

redun_real_count(~redun_real_count) =1;
redun_imag_count(~redun_imag_count) =1;

vis_real_redun = vis_real_redun./redun_real_count;
vis_imag_redun = vis_imag_redun./redun_imag_count;

vis_complex = complex(vis_real,vis_imag);
vis_complex_redun= complex(vis_real_redun,vis_imag_redun);

vis_complex = vis_complex + rot90(conj(vis_complex),2);
vis_complex((vis_r+1)/2,(vis_c+1)/2) =vis_complex((vis_r+1)/2,(vis_c+1)/2)/2;

vis_complex_redun = vis_complex_redun + rot90(conj(vis_complex_redun),2);
vis_complex_redun((vis_r+1)/2,(vis_c+1)/2) =vis_complex_redun((vis_r+1)/2,(vis_c+1)/2)/2;


fs=fftshift(ifft(ifftshift(vis_complex,2).').',2);
G1=fftshift(ifft(ifftshift(fs,1)),1);
G2=abs(G1);
G3=G2/max(G2,[],'all');

fs=fftshift(ifft(ifftshift(vis_complex_redun,2).').',2);
G1_redun=fftshift(ifft(ifftshift(fs,1)),1);
G2_redun=abs(G1_redun);
G3_redun=G2_redun/max(G2_redun,[],'all');

figure(1)
imagesc(G3)
colorbar

figure(2)
imagesc(G3_redun)
colorbar
\end{lstlisting}

\subsection{Code to Generate BOCP Walsh-Hadamard Codes}\label{Appendix_matlab_BOCP}
\begin{lstlisting}
 function [ X ] = usable_codes()   
%n= input('Length of code? ');
n=16;
UU=[];
X = zeros(n);
start = 2;
H =hadamard(n);
disp('hadamard received')
for i = start :n
   X(1,1)=1; X(start,1) = start;X(1,start)=start; 
   flag =0;
   i;
   for j = start:n
    z= find (X(1,:) == j);
   if(~isempty(z)) 
       for k= 2:n 
          if H(i,:) .* H(j, :) == H(k, :)
             % B= [i j k];
             % A= [A; B];
              w = find (X (2:n, 2:n) ==k);    
              if (isempty(w))
                  X(i,j) =k;
                  X(1,i) = i;
                  X(i,1) = i;
              else 
                  X(i,:)= 0;
                  X(:,i)=0;
                  flag =1; 
                  UU = [UU, i];
              end
          end
       end 
   end
   if(flag) 
       break;
   end
   end
end

disp('codes genrated, deleting empty cells...')
l=n; m=n;e=0;
    for c = 1:l
        ss = size(X);
        if ( (c-e)<= ss(1,1))
            if((X(c-e,1)==0))
            X(c-e,:) = [];
            e=e+1;
            end
        end
    end
    f=0;
    for d = 1:m
        ss= size(X);
        if((d-f)<= ss(1,2))
        if (X(1,d-f)==0)
            X(:,d-f) = [];
            f=f+1;
        end
        end
    end
    
Y = size(X);
size(UU)
end
\end{lstlisting}

\section{Gold Code Circuit Design} \label{Appen:Gold_circuit}
\begin{figure}
	\centering
	\includegraphics[clip,trim=0 0 0 0,width=.7\textwidth]{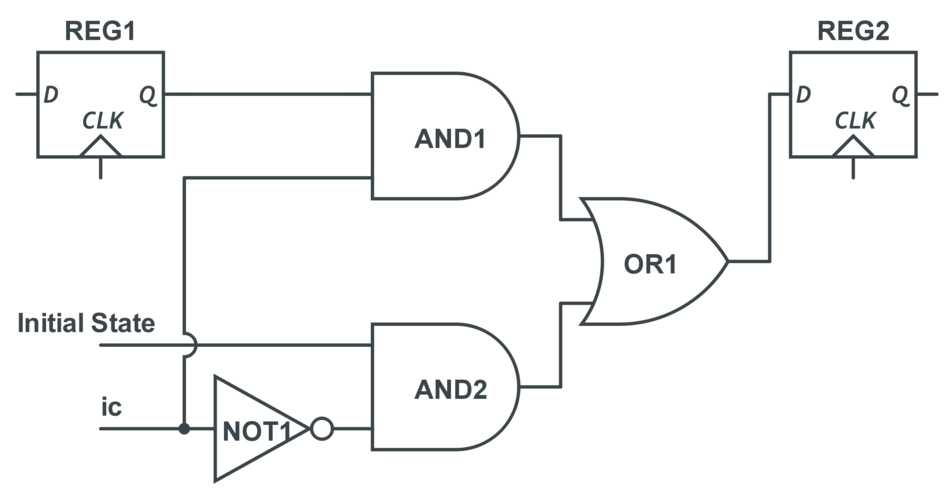}
	\caption{Circuit to initialize the registers.}
	\label{fig:gold_ini_state}
\end{figure}

\begin{figure}
	\centering
	\includegraphics[clip,trim=0 0 0 0, width=.9\textwidth]{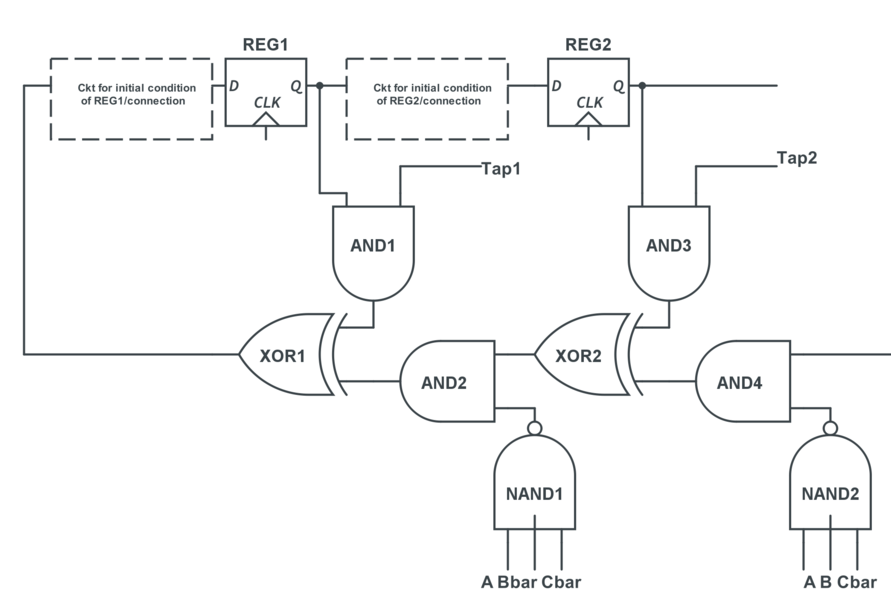}
	\caption{Circuit for choosing the LFSR length and tap position.}
	\label{fig:gold_tap}
\end{figure}

This section describes the circuit design for generating Gold codes as discussed in Sec. \ref{subsec:Gold_codes}. Two eight-stage LFSR are designed to generate m-sequences which are then XORed with certain phase differences to generate the Gold code sequences. Each register input is chosen from either the output of previous register or from external signals used for initialization, as shown in Fig. \ref{fig:gold_ini_state}. The length of LFSR and the position of each tap is programmable as shown in Fig. \ref{fig:gold_tap}.  The signals Tap1 and Tap2 are used to enable or disable a tap. Similarly A, B and C are used to select the length of the LFSR, programmable from one stage to eight stages.  

\begin{figure}
	\centering
	\includegraphics[clip,trim=0 0 0 0,width=1\textwidth]{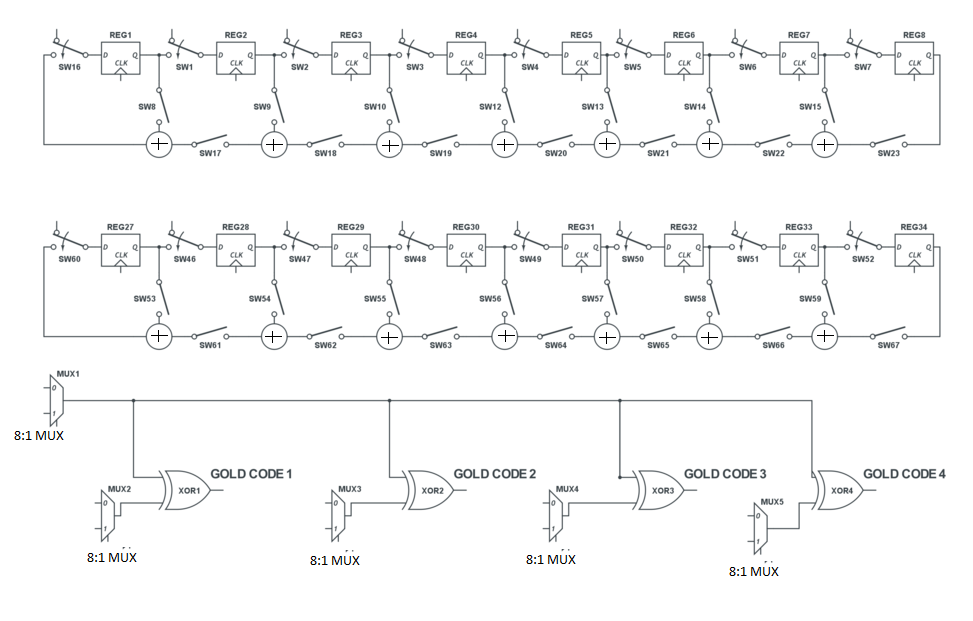}
	\caption{Complete circuit producing four Gold codes. Mux and initial condition of registers can be used to get phase difference in m-sequences.}
	\label{fig:gold_circuit}
\end{figure}

\begin{figure}
	\centering
	\includegraphics[clip,trim=0 0 0 0,width=.8\textwidth]{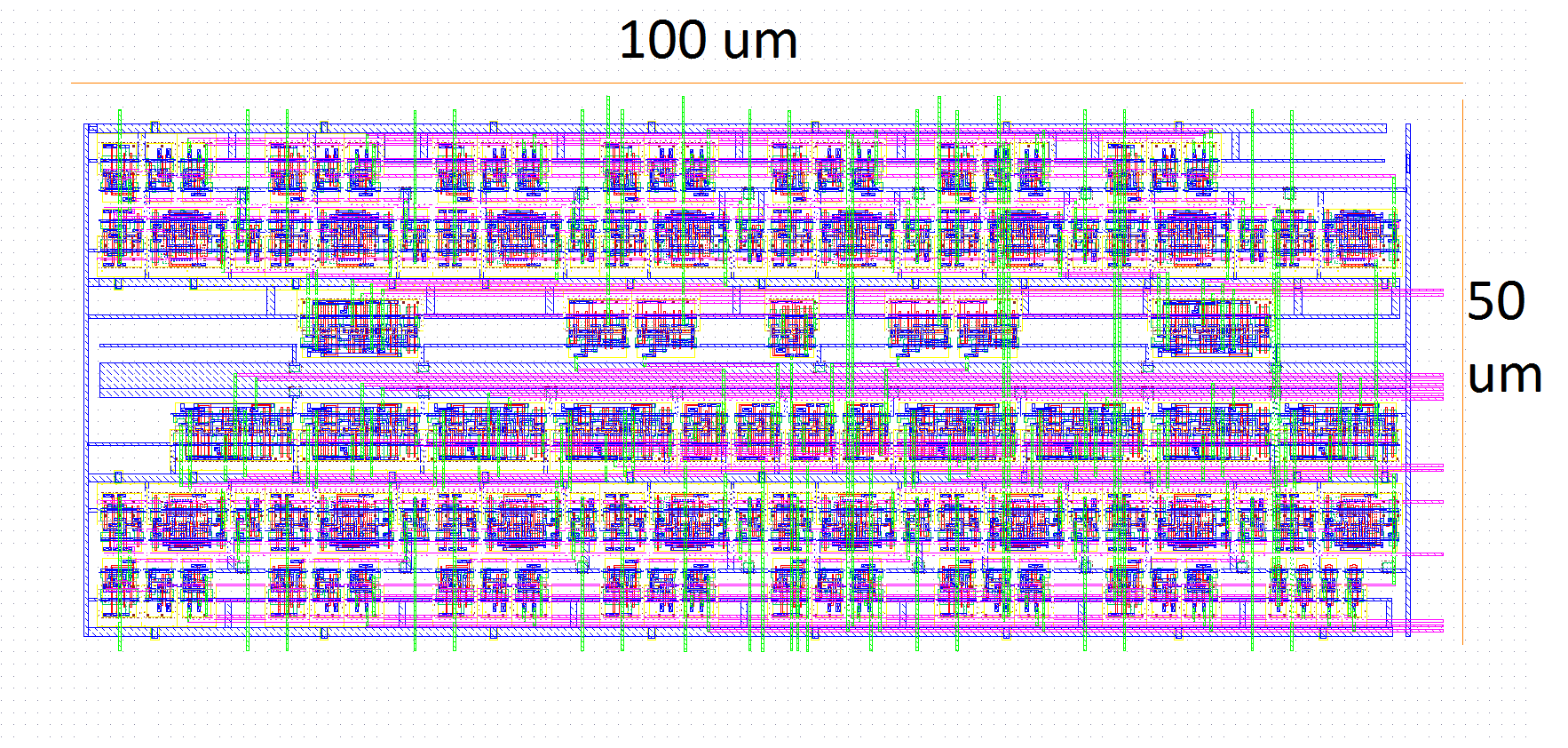}
	\caption{Layout of code generator.}
	\label{fig:gold_layout}
\end{figure}

\begin{figure}
	\centering
	\includegraphics[clip,trim=0 0 0 0,width=1\textwidth]{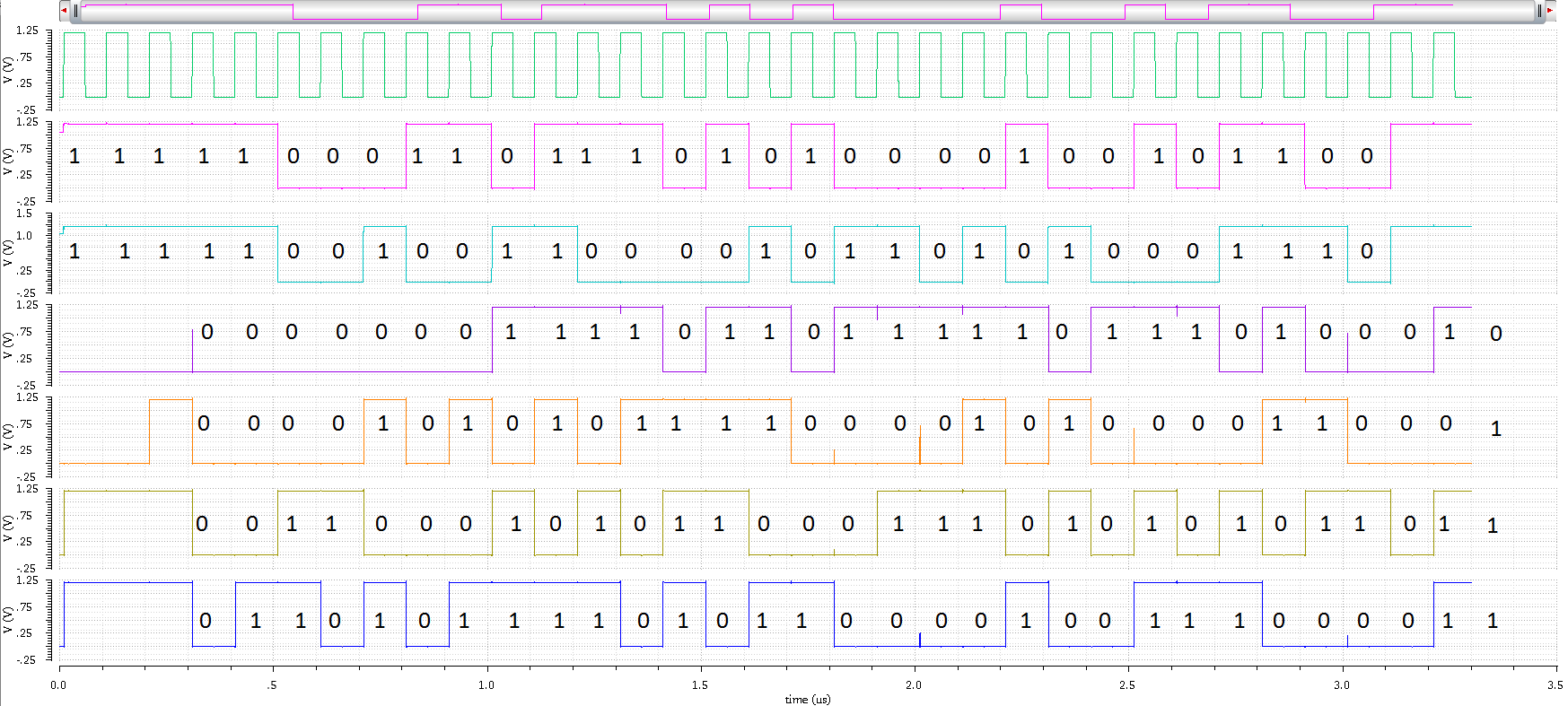}
	\caption{Simulation result showing the Seq1, Seq2 and 0-Shift, 1-Shift, 2-Shift and 3-Shift combinations.}
	\label{fig:gold_sim}
\end{figure}

Fig. \ref{fig:gold_circuit} shows a simplified schematic of the complete code generator. Each m-sequence generator is composed of eight stages with programmable feedback and initialization circuitry to allow the realization of all the sequences summarized in Table \ref{tab:Mseq}. In this schematic, the units depicted as switches are in fact the combinatorial logic presented in Figs. \ref{fig:gold_ini_state} and \ref{fig:gold_tap}. To realize the Gold codes, the two m-sequence registers array outputs are XORed. Programmable phase shifts are introduced by selecting one of the eight possible register outputs using 8:1 multiplexers. Finally, for the particular four-element phased-array implementation planned, all four gold code outputs can be produced locally by the use of different phase delay or multiplexer settings. A layout of the code generator is shown in Fig. \ref{fig:gold_layout}. This circuit was laid out by hand, since the overall size was small enough that digital synthesis was not required. All the control signals are provided using a digital serial interface (SIF).

The circuit is simulated with a LFSR of five stages and initial condition of zero shifts in two m-sequences. The taps were chosen from Table \ref{tab:Mseq} (and Fig. \ref{fig:gold}) and the results were found identical to the list above. Some of the results are shown in Fig. \ref{fig:gold_sim}. The circuit was fabricated as on-chip modulators in a 94 GHz phased array (IBM BiCMOS 8HP technology) \cite{JasonPhD} but due to issues in the fabricated chip, the chip was not tested for code-modulation functionality.

\section{High Frequency Board Design Files} \label{Appendix_board_layers}
\begin{figure}
	\centering
	\includegraphics[clip,trim=0 0 0 0,width=1\textwidth]{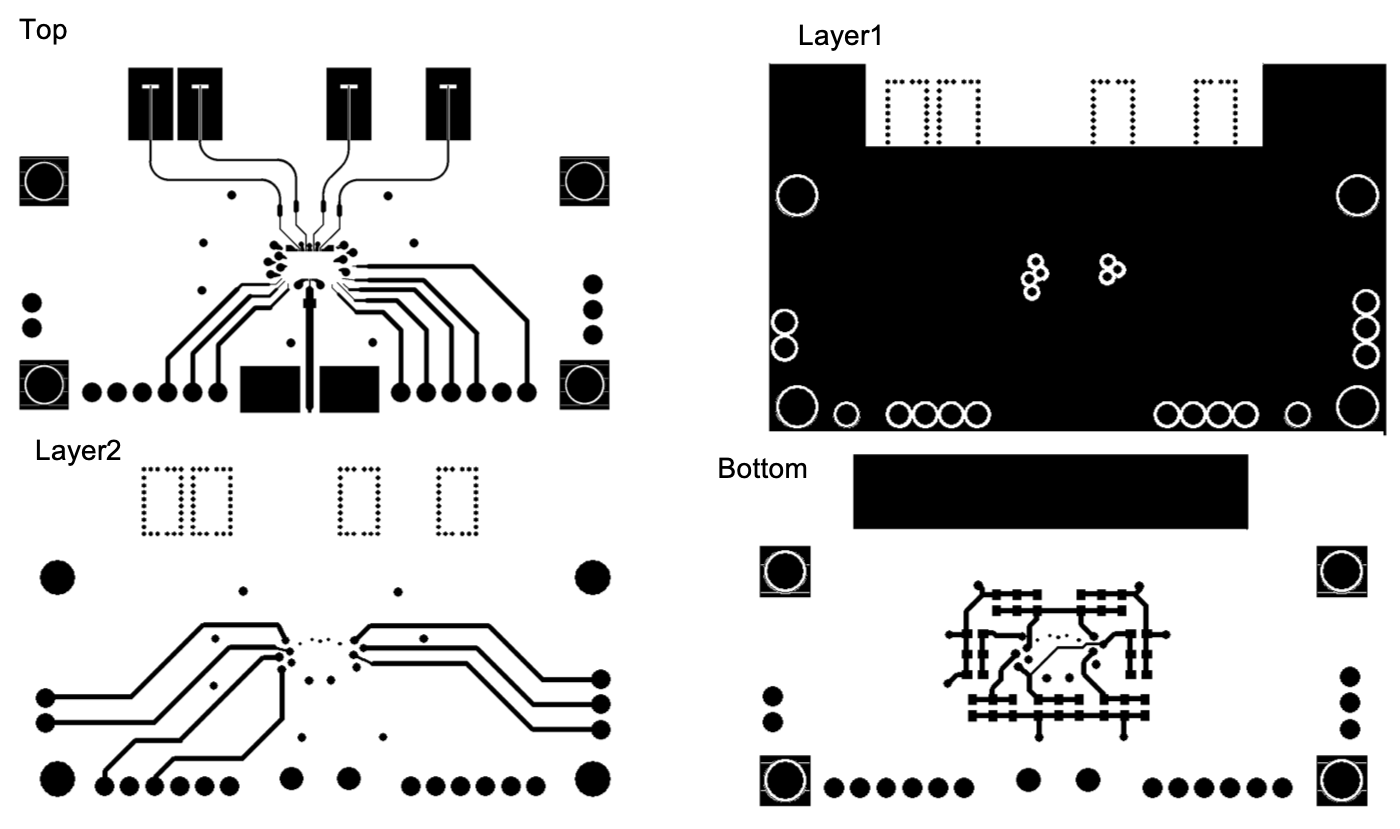}
	\caption{High frequency board design metal layers. }
	\label{fig:Board_layers1}
\end{figure}

\begin{figure}
	\centering
	\includegraphics[clip,trim=0 0 0 0,width=1\textwidth]{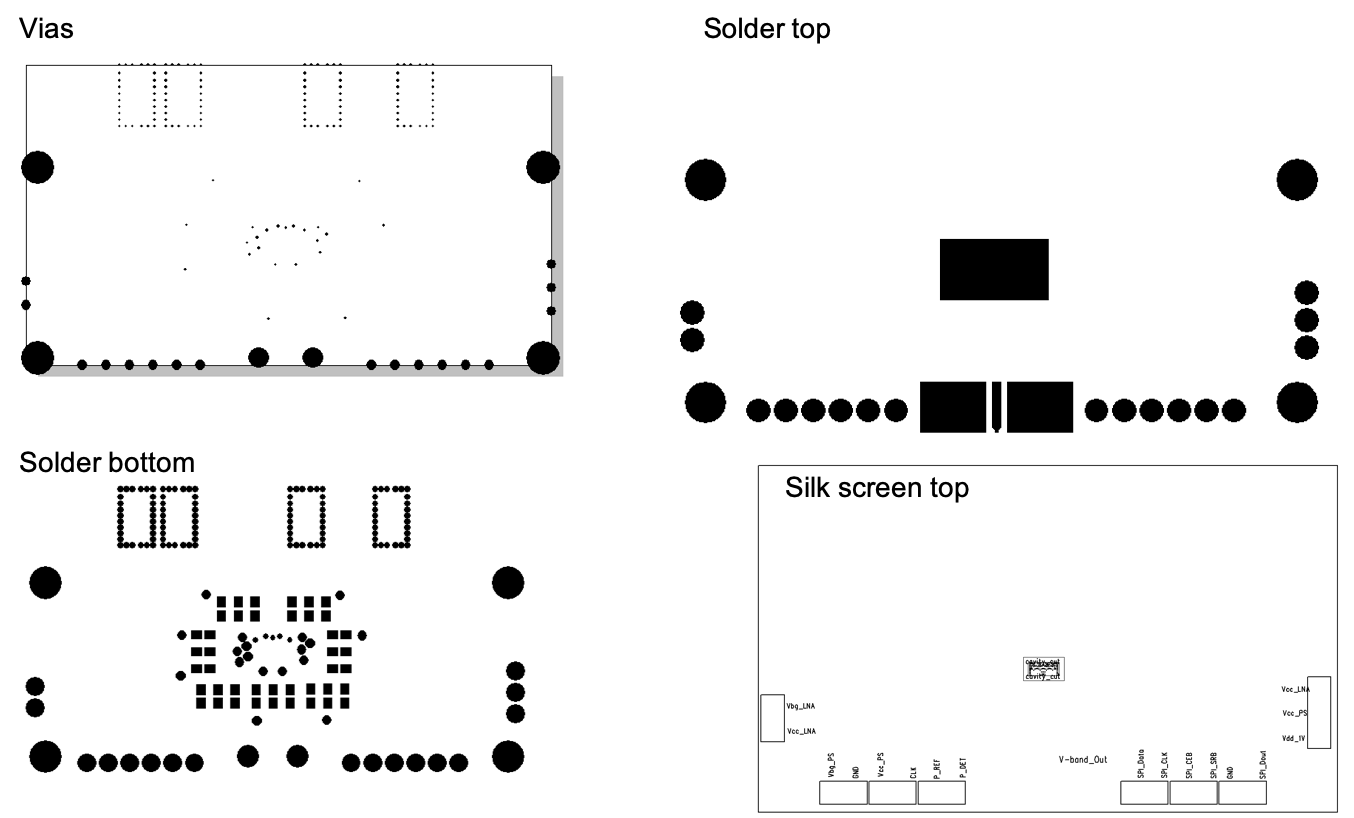}
	\caption{High frequency board design metal layers. }
	\label{fig:Board_layers2}
\end{figure}

Figs. \ref{fig:Board_layers1} and \ref{fig:Board_layers2} show the different metal layers for the high frequency board that was designed for 60-GHz prototype.

\chapter{A 24-44 GHz UWB 5G LNA}
\label{chap-six}

\section{Motivation}
One of the effective methods of improving the sensitivity of an imaging system is to use wider bandwidths. Most communication radios available in market today are generally narrow band because of the lower frequency of operation for fourth generation (4G) mobile communication and satellite communication. Higher frequency bands will be used for fifth generation (5G) communication due to wider useful bandwidths at millimeter wave frequencies. The phased arrays built for 5G communication thus are ideal candidates for repurposing into code-modulated interferometric imagers. Here we present an ultra wide-band LNA design covering all 5G bands.

Sub-6 GHz and millimeter-wave (mm-wave) bands above 24 GHz are the two bands being pursued by academia and industry for the next-generation mobile network ``5G New Radio (NR)" \cite{Rap2013,Qual2017}. The large available contiguous bandwidth along with recent technology development makes the mm-wave spectrum a strong candidate; therefore, regulation agencies around the world have proposed several possible mm-wave frequency bands. These include 28, 37, and 39 GHz by the Federal Communications Commission (FCC) in USA \cite{FCC}; and 24.25-27.5, 31.8-33.4, 37-40.5, 40.5-42.5 GHz, etc., up to 86 GHz by the International Telecommunication Union (ITU) \cite{ITU}. Different service providers and/or different countries could possibly favor different frequency bands for 5G-NR based on spectrum availability. 
Rather than developing custom hardware for each individual frequency band, transceiver systems covering multiples or all of 5G-NR frequency bands could possibly be made to avoid non-recurring development costs and to enable re-usability over carriers and/or different countries. Such systems could be built either using UWB system blocks or switchable narrow-band blocks centered at different frequencies. An UWB system, however, can be of of dual purpose: for 5G communication and imaging systems.

This appendix presents a 24-44 GHz ultra-wideband (UWB) low-noise amplifier (LNA); simultaneously covering all major 5G cellular frequency bands. The LNA has been designed in 45nm CMOS SOI technology, has a maximum gain of 20 dB with more than 65\% 3dB bandwidth (24-47.5 GHz), and a noise figure less than 5.5 dB (typical 4.7 dB) in the band. A narrow-band 28 GHz LNA is presented for comparison and evaluation of merits of a wideband design \cite{VikasLNA}.        

\section{Design}
\begin{figure}
	\centering
	\includegraphics[clip,trim=0 1.5cm 0 0,width=0.9\textwidth]{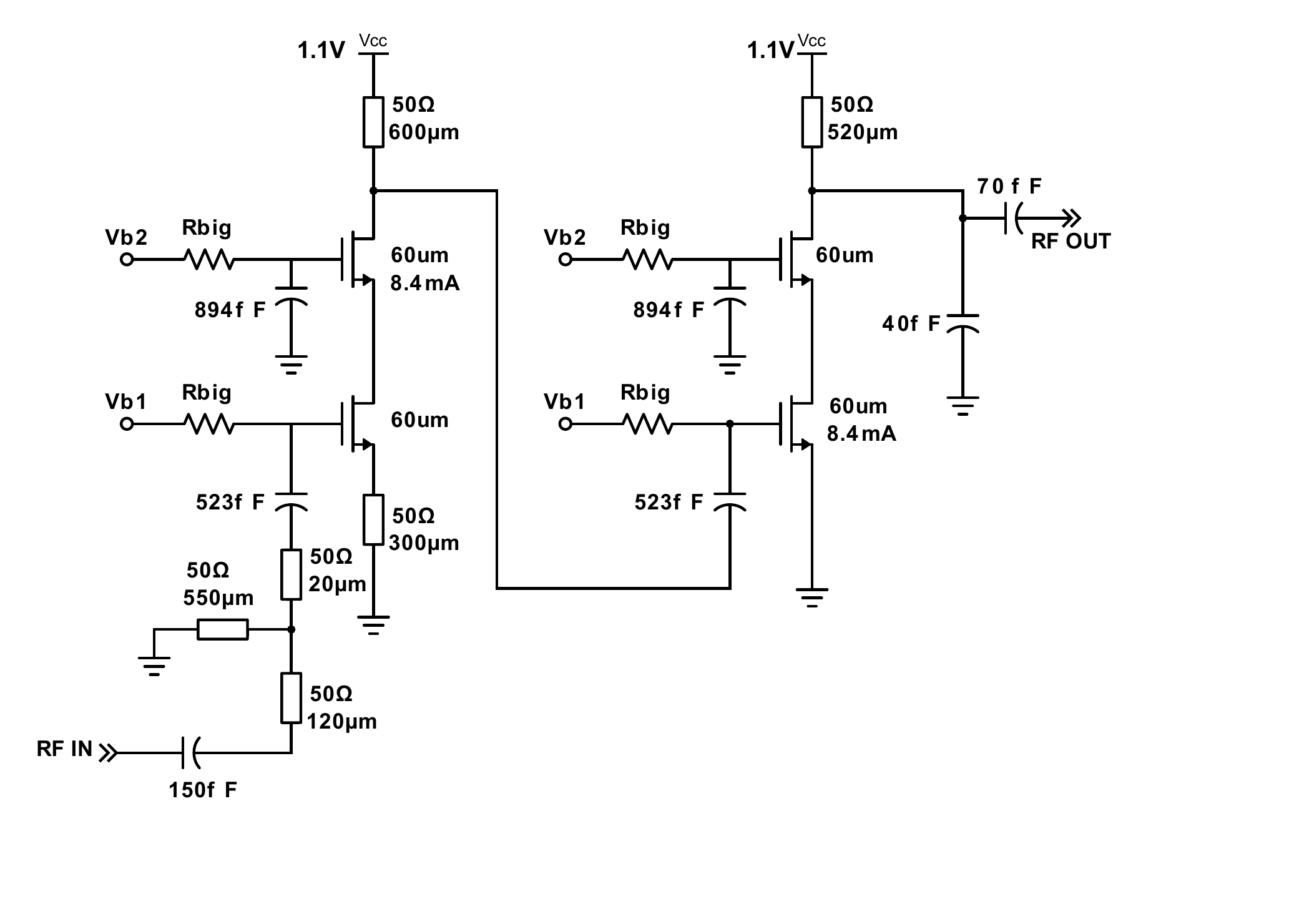}
	\caption{Schematic of the NB LNA (biasing details not shown).}
	\label{fig:schematic_NB}
\end{figure}

\begin{figure}
	\centering
	\includegraphics[clip,trim=0 1.5cm 0 0,width=1\textwidth]{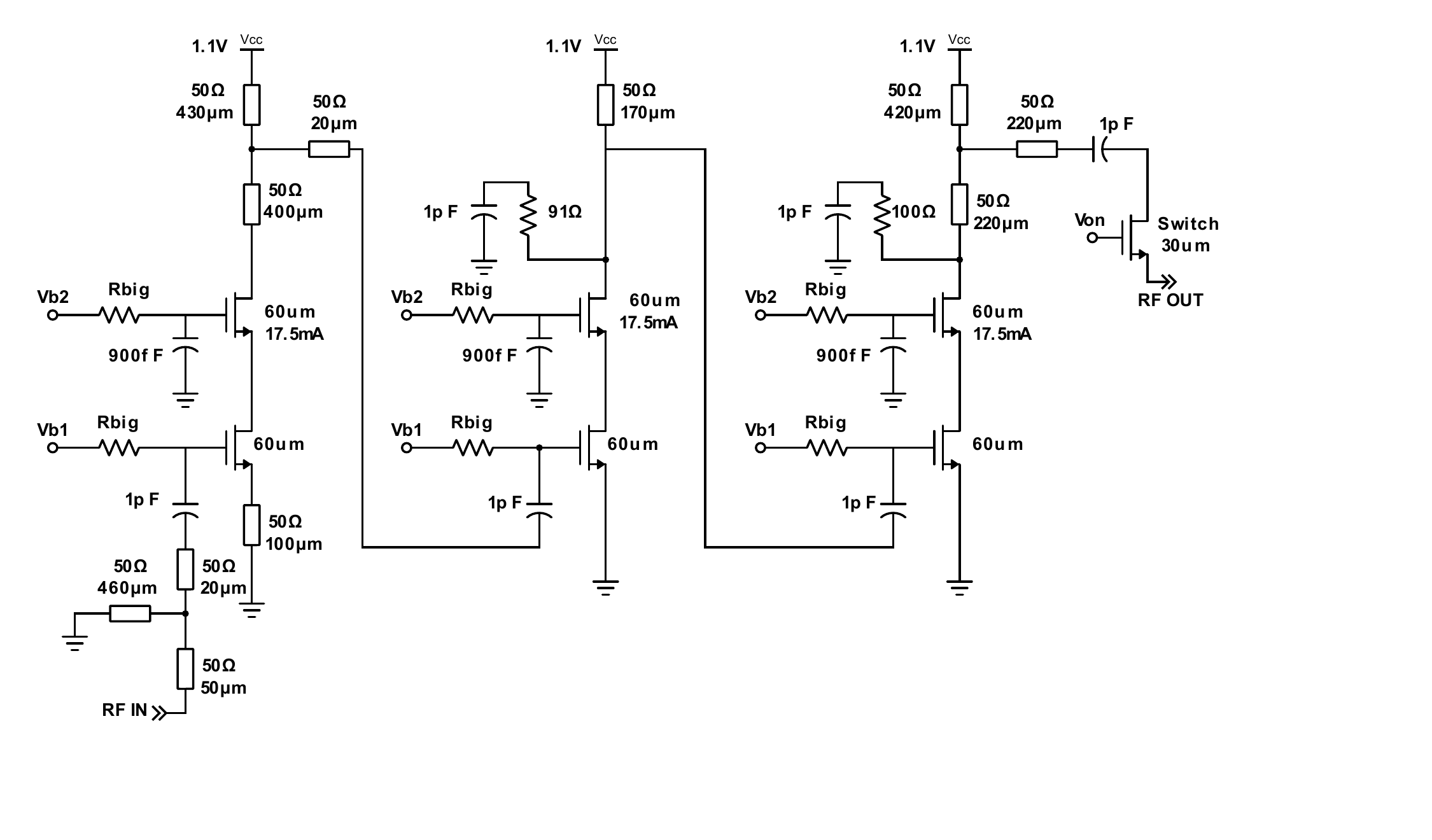}
	\caption{Schematic of the UWB LNA (biasing details not shown).}
	\label{fig:schematic}
\end{figure}

The NB 28 GHz LNA is a two-stage cascode design as shown in Fig. \ref{fig:schematic_NB}. A common-source (CS) with inductive degeneration and input inductor is used for the first stage, designed to achieve a simultaneous noise and input power match, and the second stage further increases gain and provides output matching.
The UWB LNA is a three stage CS cascode design with staggered tuning to achieve a wideband flat gain response. Stage one, two and three have resonances at approximately 24, 40 and 30 GHz respectively. As shown in Fig. \ref{fig:schematic}, ``de-Qing" resistances are placed in parallel with the drain transmission lines (t-lines) of stages two and three to increase bandwidth. Additionally, single-stub shunt tuning is used to further increase bandwidth. Finally, the ``de-Qing" resistance in the third stage is also beneficial for achieving a wideband output match. Although stage three has its resonance in the middle of the band, it only improves the overall absolute gain and does not shape the response due to a very wideband low-Q response of its own.

    \begin{figure}[]
    \centering
   \includegraphics[width=0.5\textwidth]{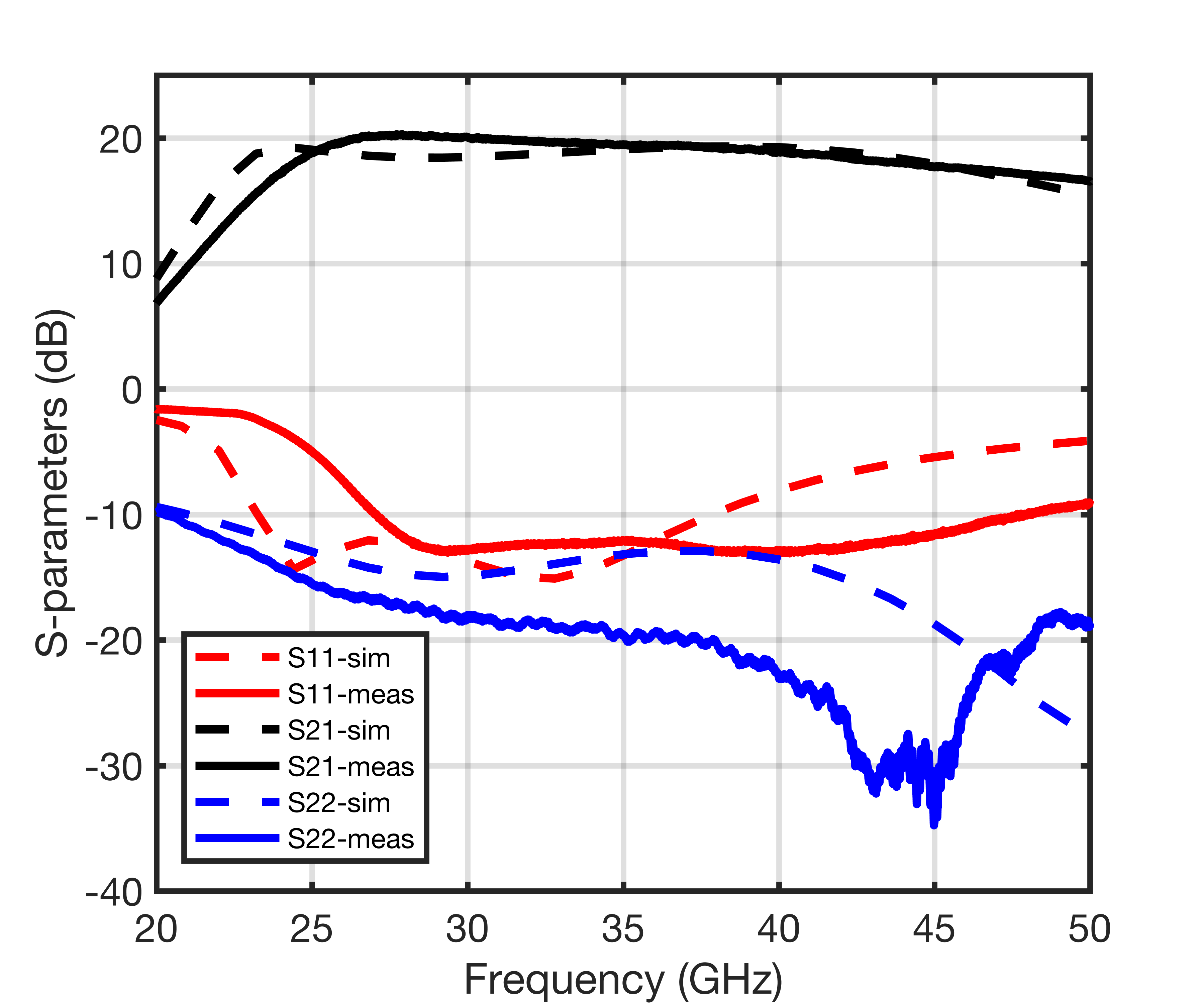} \\ \includegraphics[width=0.5\textwidth]{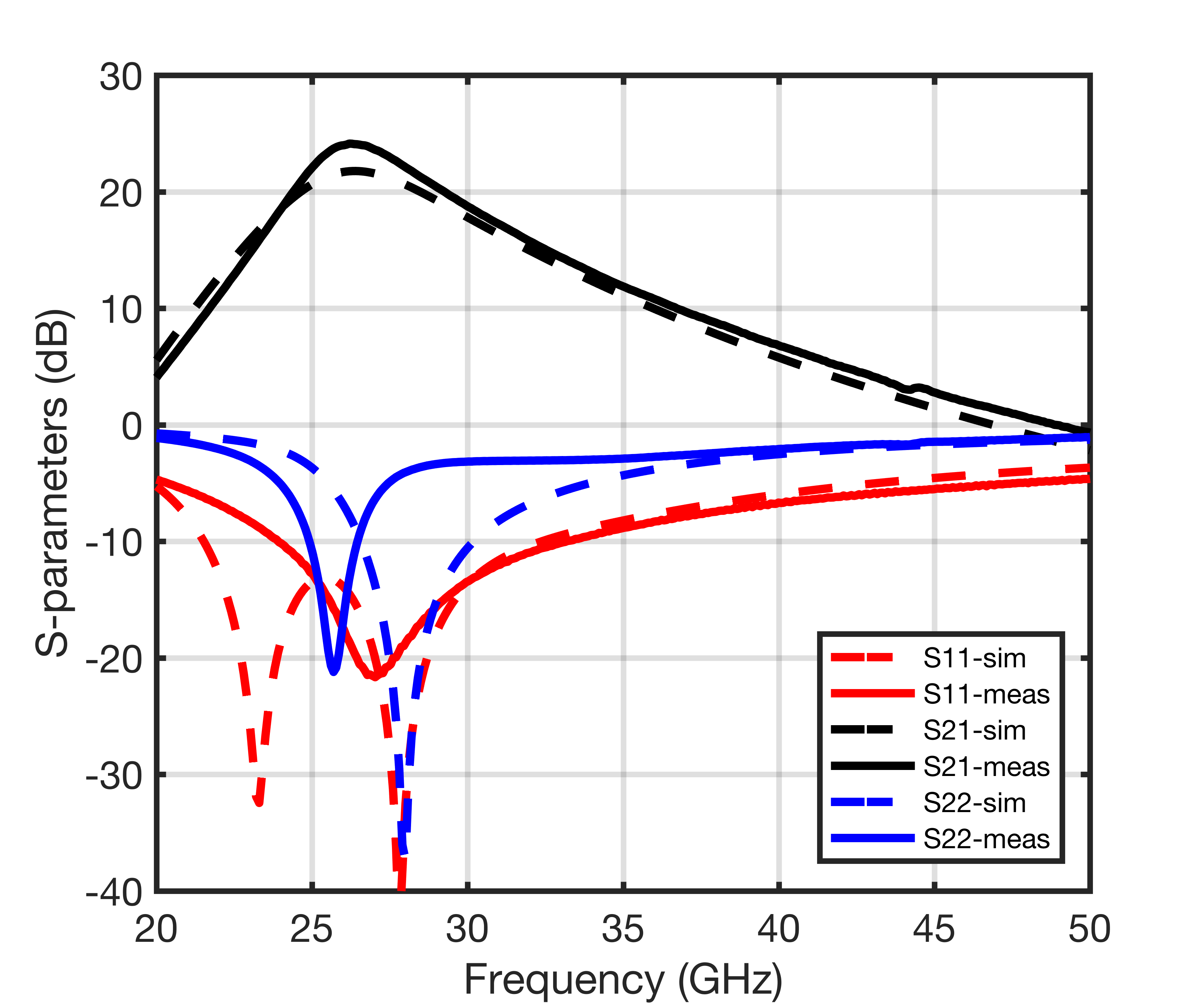} \\ \includegraphics[width=0.5\textwidth]{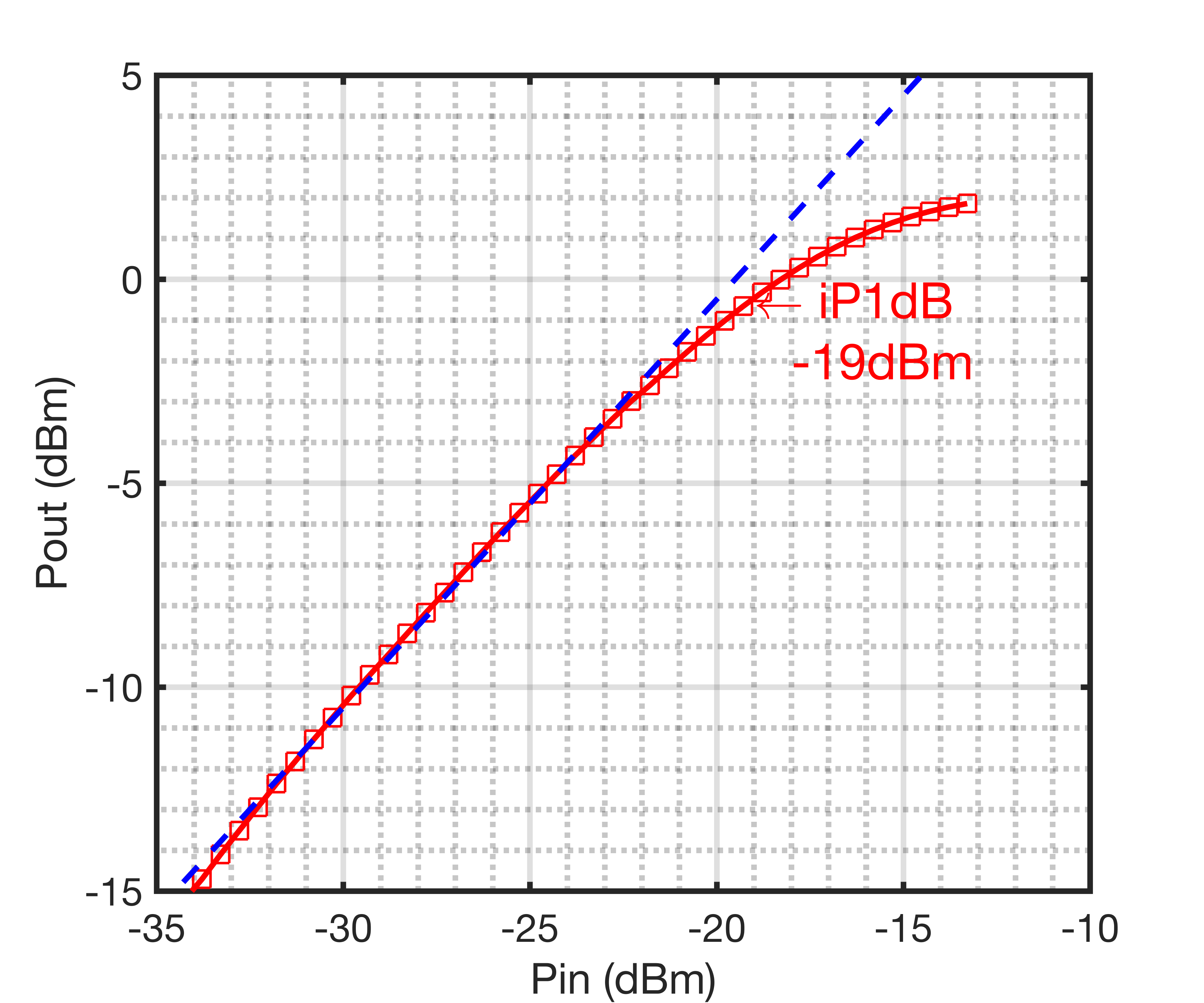}
   \caption[] 
%>>>> use \label inside caption to get Fig. number with \ref{}
   { \label{fig:comparison} 
 Simulated and measured S-parameters of (a) UWB LNA (b) NB 28 GHz LNA. (c) Measured iP$_{1dB}$ of UWB LNA at 28 GHz.}
   \end{figure}

Both LNAs are designed in GlobalFoundries 45nm partially-depleted SOI CMOS technology process, 45RFSOI, featuring transistors with $f_t$/$f_{max}$ $305$/$380$ GHz and 11 metal layers. All transistors in our LNA utilize a floating body, two gate contacts, and the gate-to-gate pitch within each multi-finger device is twice the minimum value. These double-pitch transistors have increased mobility and $f_{max}$. To reduce the parasitic capacitance, the gate and terminal contacts are taken to top metal layers in a stair-step fashion. All of the inductors are realized using grounded coplanar waveguide (GCPW) and custom-modeled using Momentum EM solver.  
%Metal layers LB (topmost), UB \textcolor{red}{(ADD DESCRIPTOR)} and B3 (ADD DESCRIPTOR) are used for signal lines, side shields and ground respectively. 
In both designs, the t-lines have been folded and meandered to make the layout more compact. The side shields of the GCPW are beneficial in decoupling the meandering t-lines. 
%Note that the UWB LNA includes a series output switch which was a custom feature required for the overall multi-band receiver in which the LNA is to operate. This switch has minimal effect on the LNA performance.

\section{Measurement Results}
Both LNAs are characterized through wafer probing. Fig. \ref{fig:comparison}(a) compares the measured and simulated S-parameters results of the UWB LNA across 20-50 GHz. The measured S$_{21}$ is in good agreement with the simulated results across the band.  There is a slight upward shift in the measured peak gain, reducing the overall gain flatness. This has been traced to a frequency shift of the first stage resonance upwards from 24 GHz to 28 GHz. The UWB LNA gain peaks at 20 dB at 28 GHz with a 3dB bandwidth from 24 to 47.5 GHz (65$\%$). There is a similar frequency shift in the input matching. The measured input return loss is above 10 dB from 27-48 GHz. The measured output return loss is greater than 10 dB for the entire frequency band and up to 67 GHz. Finally, the measured S$_{12}$ is less than -45 dB up to 44 GHz (not shown in plot), where high reverse isolation is a natural feature of multi-stage LNAs.

% \begin{figure}
% 	\centering
% 	\includegraphics[width=0.35\textwidth]{images/Spara.png}
% 	\caption{Simulated and measured S-parameters of UWB LNA.}
% 	\label{fig:Spara}
% \end{figure}
% \begin{figure}
% 	\centering
% 	\includegraphics[width=0.35\textwidth]{images/Spara_NB2.png}
% 	\caption{Simulated and measured S-parameters of narrow-band 28GHz LNA.}
% 	\label{fig:Spara_NB}
% \end{figure}

Turning to the NB LNA, Fig. \ref{fig:comparison}(b) compares the measured and simulated S-parameters. The measured peak gain is 24 dB at 28 GHz. The measured input return loss is greater than 10 dB for 24-33 GHz, whereas the output return loss is greater than 10 dB for 25-26.5 GHz which is narrow-band. Note that we observed a downwards 4 GHz shift in S$_{22}$ with respect to simulation which we attribute to our output pad model.

Noise figure (NF) is measured using the Y-factor technique with a spectrum analyzer that works up to 43.5 GHz and a Noise Com Inc. noise source with an ENR of 16-18 in the K-band. A Quinstar amplifier is used between our device-under-test and the spectrum analyzer to reduce the measurement noise floor. The input and output loss of the connectors, cables, bias-Tee at the output of LNA (to ground the source terminal of switch transistor) and the probes are measured using a network analyzer and de-embeded using the Friis equation.

\begin{figure}
	\centering
	\includegraphics[width=0.7\textwidth]{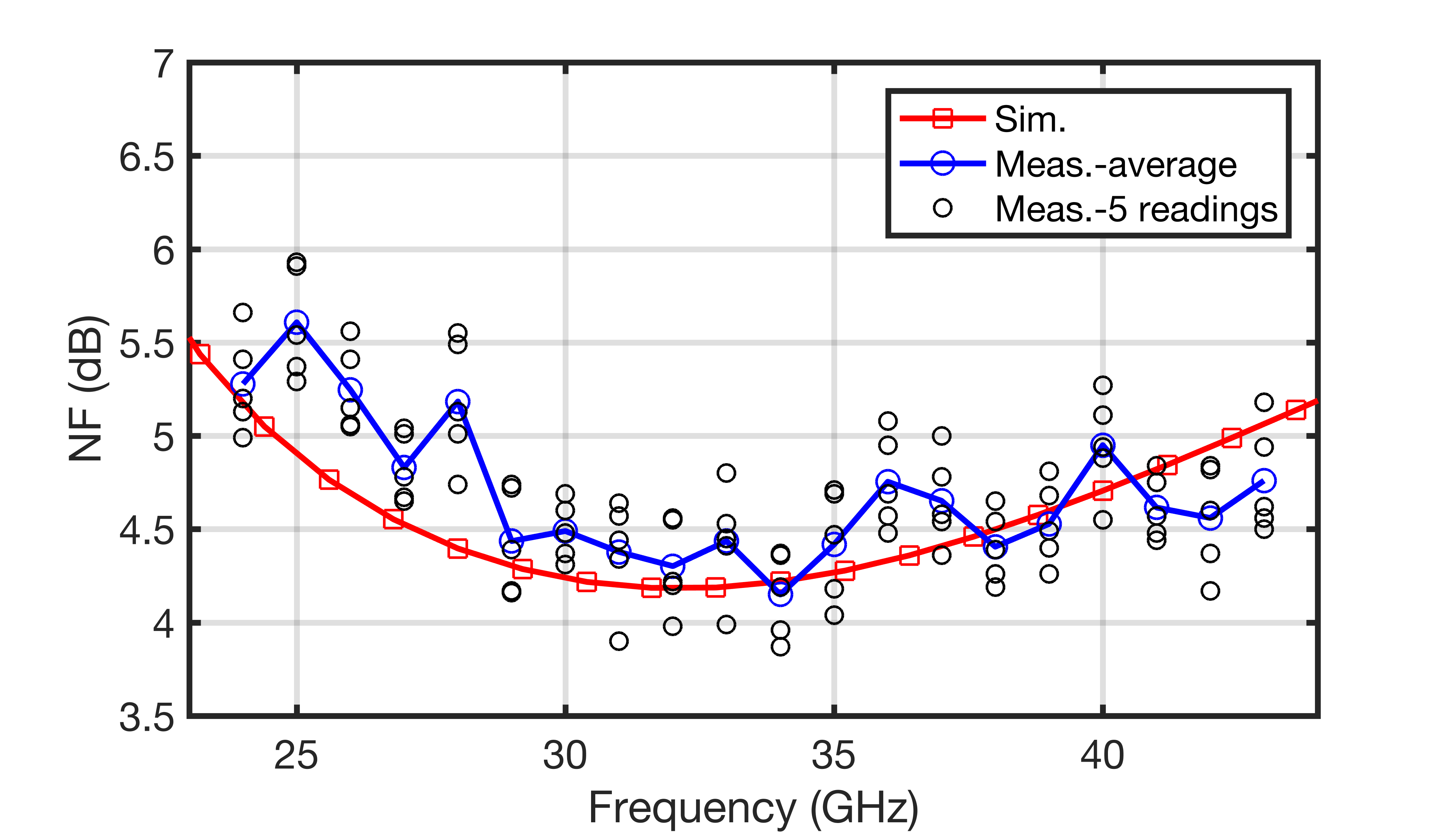}
	\caption{Noise figure simulation and measurement for UWB LNA.}
	\label{fig:noise}
\end{figure}
\begin{figure}
	\centering
	\includegraphics[width=0.7\textwidth]{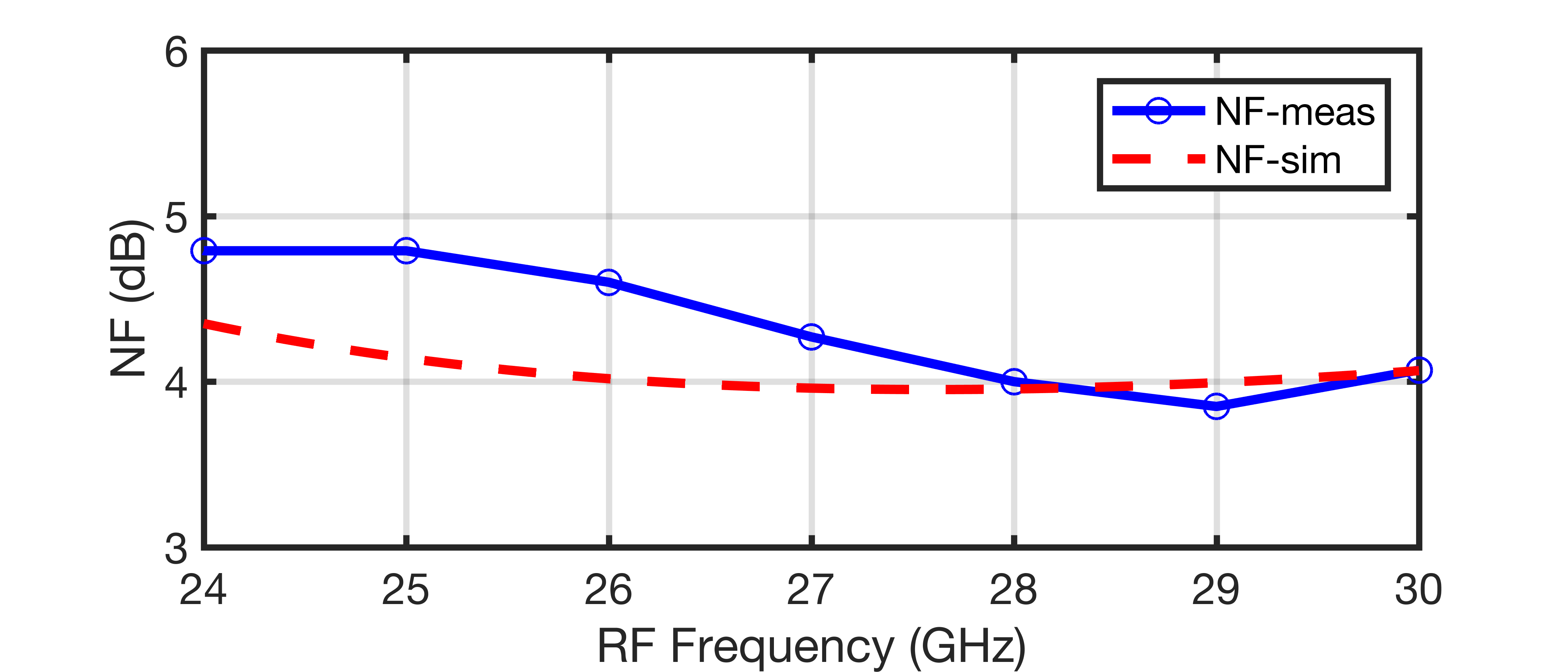}
	\caption{NF simulation and measurement for NB 28 GHz LNA.}
	\label{fig:noise_NB}
\end{figure}

   \begin{figure}
   \begin{center}
   \begin{tabular}{c c} %% tabular useful for creating an array of images 
   \includegraphics[height=5cm]{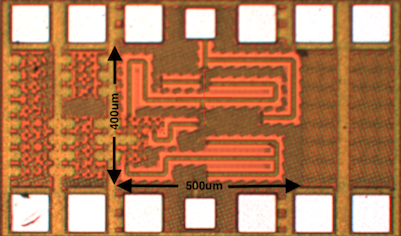} & \includegraphics[height=4cm]{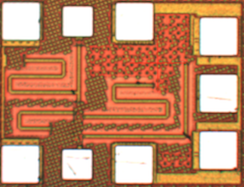}   \\   
   \rule[-1ex]{0pt}{1.5ex}  (a) & (b)  \\	
   \end{tabular}
   \end{center}
   \caption[] 
%>>>> use \label inside caption to get Fig. number with \ref{}
   { \label{fig:die} 
Die photo of (a) UWB LNA (400X500$um^2$ active area) (b) NB LNA (300X500$um^2$ active area) }
   \end{figure}

%\begin{comment}

\begin{table}
\caption{LNA Performance Comparison}
\centerline{
\vbox{\offinterlineskip
\hrule
\halign{&\vrule#&
\strut\quad#\hfil\quad\cr
&\strut &&\multispan3\hfil {\bf This Work}\hfil&&\hfil&&\hfil&\cr
&\omit&&5G LNA&&28 GHz LNA&&\cite{Rebeiz2016}&&\cite{Ellinger2004}&\cr
height2pt&\omit&&\omit&&\omit&&\omit&&\omit\cr
\noalign{\hrule}
height2pt&\omit&&\omit&&\omit&&\omit&&\omit&\cr
&Process&&45nm SOI&&45nm SOI&&45nm SOI&&90nm SOI&\cr
\noalign{\hrule}
&Freq. (GHz)&&24-44&&28&&24-28&&26-42*&\cr
\noalign{\hrule}
&Gain (dB)&&20&&24&&8.5&&11.9&\cr
\noalign{\hrule}
&NF (dB)&&4.2-5.5&&4.0&&3&&3.6-4.2&\cr
\noalign{\hrule}
&NMF** (dB)&&4.2@30GHz&&4.0&&3.3&&3.8&\cr
\noalign{\hrule}
&Power(mW)&&58&&18.5&&12&&40.8&\cr
\noalign{\hrule}
&Area($mm^2$)&&0.2 &&0.15(w/o pads)&&N/A&&0.18&\cr
height0pt&\omit&&\omit&&\omit&&\omit&&\omit&&\omit&\cr}
\hrule}}
\raggedright{*max. $6$ dB input return loss, narrow band S$_{22}$ match. **Noise Measure Figure $=1+\left(F-1\right)/\left(1-Ga^{-1}\right)$.}
\end{table}

%\end{comment}

For the UWB LNA, the results from five different Y-factor measurements and an average of these data are shown in Fig. \ref{fig:noise} along with the simulated NF. Measured and simulated NF agree, and the noise figure remains between 4.2 dB and 5.5 dB across the 24-44 GHz range. This indicates that the UWB can serve as a preamplifier stage for a broadband downconversion mixer stage for any of the 24-44 GHz projected 5G bands and keep the overall receiver NF below approximately 6.5 dB. The NF of 28 GHz NB LNA is also measured using same technique and similar setup. Fig. \ref{fig:noise_NB} reveals a 4 dB NF at 28 GHz. 

The measured input 1-dB compression point (P$_{1dB}$) for the UWB LNA is -17, -19, and -16 dBm at 24, 28, and 39 GHz, respectively. A representative swept-power response at 28 GHz is shown in Fig. \ref{fig:comparison}(c). The NB LNA iP$_{1dB}$ is -23 dBm at 28 GHz. 

Fig. \ref{fig:die} show the die photos of the UWB LNA and the 28 GHz NB LNA, respectively. The die sizes are kept to scale to indicate the relative sizes between the two designs. The UWB LNA chip area is approx. 400X500$um^2$ excluding the pads and by-pass capacitors. The 28 GHz NB LNA occupies 300X500$um^2$ area, excluding the pads. From this, we see that the UWB LNA is approximately 33\% larger than the NB LNA. Finally, from a 1.1V supply the UWB LNA consumes 58 mW power, whereas the 28 GHz NB LNA consumes 18.5 mW.  

Table 1 compares the UWB LNA, and 28 GHz LNA to other published work. Of note is a recent single-stage 24-28 GHz LNA which achieves 3 dB NF \cite{Rebeiz2016}.

\section{Comparison and Discussion}
Although 5G standards are only just emerging, it is useful to consider the performance of various approaches to realize a multi-band 5G system. If a receiver were to be designed to cover multiple bands (e.g., 28, 32-34, and 37-42.5 GHz), a parallel bank of three LNAs could be used, including the 28 GHz version presented here, where a switch would be included at the input and output to allow the sharing of a common antenna and the sharing of a common broadband downconversion stage. Alternatively, a single broadband LNA could be used, such as the UWB one presented here.   

From a performance standpoint, the UWB LNA has roughly 0.5 dB higher NF at 28 GHz compared to the NB LNA, indicating a degradation for the receiver when operating at 28 GHz. The UWB LNA NF does remain low, though, over the 37 and 42 GHz bands, and gain remains high. This should keep the overall receiver NF fairly constant over the full frequency range. For a multiplexed bank of LNAs, the requirement of an input multiplexing switch could certainly degrade NF; however, if the LNA is incorporated into a transmit/receive module, then this switch would already be present in the system; thus, do not include this as a penalty.

From a complexity and power consumption standpoint, the UWB LNA design is more complex and consumes more power. Even though the UWB has one additional gain stage, the increased die size is minimal, owing to an aggressive use of meandering for the t-lines. In comparison, if a multiplexed bank of three narrow-band LNAs were used, the area consumed would be arguably larger. 

\section{Conclusion}
A compact UWB LNA which provides low noise and high gain across 24 to 44 GHz has been presented and compared to a narrow-band 28 GHz LNA. The UWB LNA achieves a 4.2-5.5 dB NF and 18-20 dB gain across the full range. When compared to the narrow-band 28 GHz LNA, we see that the UWB LNA exhibits about 0.5 dB increase in NF and consumes 40 mW additional power. Nevertheless, the UWB LNA could support a potential multi-band 5G operation and avoid the need for a parallel bank of multiplexed LNAs. Also, such a UWB LNA could serve as a useful preamplifier stage for broadband sliding-IF downconversion receivers which are used as the core receiver in Ka-band phased-array systems. This wide-band LNA could serve as a first stage to UWB receivers for 5G commumications. Such UWB 5G phased arrays can be repurposed into CMI imaging systems for better sensitivity as compared to narrow-band designs.

\restoregeometry

%%---------------------------------------------------------------------------%%
%\ensureoddstart
\backmatter

\end{document}